\documentclass[preprint,12pt]{elsarticle}
\biboptions{sort&compress}
\usepackage[utf8]{inputenc}
\usepackage{amsmath, amssymb, graphicx, tikz,tabularx}
\usepackage{esint}
\usepackage[most]{tcolorbox}
\usepackage{comment}
\newcommand{\be}{\begin{equation}}
	\newcommand{\ee}{\end{equation}}
\newcommand{\bk}{\mathbf{k}}
\usepackage{amsmath,bm}
\usepackage{dsfont}
\usepackage{bm}
\usepackage{amsfonts}
\usepackage{appendix}
\usepackage{latexsym}
\usepackage{float}
\usepackage{mathrsfs}
\usepackage{braket}	
\usepackage{color}
\usepackage{xcolor}
\newcommand{\mathbbm}[1]{\mathsf{#1}}
\usepackage{slashed}
\usepackage{mathtools}
\usepackage{hyperref}
\usepackage{comment}
\numberwithin{equation}{subsection}

\newcommand{\CiteMiss}[1]{ \textcolor{red}{(CITATION)}}

\newcommand{\cO}{\mathcal{O}}
\newcommand{\bx}{\mathbf{x}}

\newcommand{\CK}{color-kinematics }
\newcommand{\dd}{\mathrm d}
\newcommand{\ud}{\mathrm{d}}
\newcommand{\mnrs}{{\mu \nu \rho \sigma}}
\newcommand{\eqs}[1]{\begin{align}#1\end{align}}
\DeclareMathOperator{\Sgn}{sgn}

\newcommand{\MT}{\mathbb{MT}}

\newcommand{\C}{\mathbb{C}}
\newcommand{\cp}{\mathbb{CP}}

\newcommand{\h}{\mathbb{H}}
\newcommand{\la}{\left\langle}
\newcommand{\ra}{\right\rangle}
\newcommand{\rd}{\mathrm{d}}
\newcommand{\ri}{\mathrm{i}}
\newcommand{\cint}{\frac{1}{2\pi\ri}\oint_{\gamma}}

\newtcolorbox{mybox}[1][]{
	colback=white,
	colframe=black,
	sharp corners,
	boxrule=1pt,
	width=\textwidth,
	title=#1,
	fonttitle=\bfseries
}

\usepackage{makecell}
\setcellgapes{3pt}\makegapedcells

\usepackage{array,collcell}
\newcommand\AddLabel[1]{%
	\refstepcounter{equation}
	(\theequation)
	\label{#1}
}
\newcolumntype{M}{>{\hfil$\displaystyle}X<{$\hfil}} 
\newcolumntype{L}{>{\collectcell\AddLabel}r<{\endcollectcell}}

\journal{Physics Reports}

\begin{document}
	
	\begin{frontmatter}
		\title{Color-Kinematics Duality and the Double Copy in Curved Spacetimes}
		\author{Mariana Carrillo Gonz\'alez}
		\affiliation{Abdus Salam Centre for Theoretical Physics, Imperial College, London, SW7 2AZ, U.K.}
		\affiliation{School of Physics and Astronomy, University of Southampton, Highfield, Southampton, SO17 1BJ, U.K.}
		
		\begin{abstract}
			The double copy relates gravitational data to gauge-theory data, most sharply through color-kinematics duality for flat-space scattering amplitudes. This review surveys the extent to which this idea can be formulated beyond asymptotically flat backgrounds. After recalling the flat-space ingredients needed for comparison, we describe curved-space constructions for boundary correlators and scattering observables in AdS, dS, and other non-trivial backgrounds. We then review classical and field-level maps, including Kerr--Schild, Weyl, Cotton, twistor, and convolutional double copies. A separate part is devoted to self-dual sectors. Throughout, we emphasize both the common structures that survive from flat space and the obstructions introduced by curvature. 
		\end{abstract}
		
	\end{frontmatter}
	
	\tableofcontents
	
	\pagebreak
	
	\section{Introduction} \label{sec:intro}
	Gravitational physics is highly non-linear and gives rise to many interesting phenomena that, at first sight, seem to lack analogues in gauge theory. For scattering problems on a flat background, it has been known for some time that gauge-theory and gravitational scattering are related by the double copy \cite{Bern:2019prr,Bern:2022wqg,White:2024pve}. The purpose of this review is not to give a new account of this flat-space realization, but to describe the progress and new constructions that have been developed to extend these ideas to curved spacetimes. Although the subject is not as developed as the flat-space case, due to both technical and conceptual challenges, it has matured into an active research area with many examples and promising directions.
	
	\paragraph{Why Double Copy?}
	The double copy is useful for both technical and conceptual reasons. On the scattering-amplitudes side, it has clarified the ultraviolet behavior of perturbative gravity theories and the onset of possible counterterms \cite{Bern:2012uf,Bern:2013uka,Bern:2012cd,Bern:2018jmv,Bern:2017ucb,Carrasco:2021bmu}. It has also become a tool for computing gravitational observables relevant to binary dynamics and gravitational-wave physics in expansions in Newton's constant \cite{Bern:2019crd,Buonanno:2022pgc,Kosower:2022yvp,Chen:2023dcx}. Classical realizations have related gravitational solutions to simpler gauge-theory data, and in favorable cases can be used to construct new exact gravitational solutions from gauge-theory ones \cite{White:2024pve}. Across scattering amplitudes, classical solutions, and field-level constructions, the double copy has also exposed hidden algebraic structures in gravity and their relation to gauge-theory symmetries. This perspective often leads to simpler and more efficient approaches to gravitational computations.
	
	It is therefore desirable to extend these successes to more general situations, including settings that are not simply small perturbations around flat spacetime. This is particularly relevant for cosmology, where graviton perturbations around near de Sitter (dS) spacetimes are technically difficult but central to the study of primordial fluctuations \cite{Maldacena:2011nz,Weinberg:2005vy,Baumann:2020dch}. Similar questions arise for the evolution of our Universe on FLRW backgrounds and for its current near de Sitter phase. Because the full cosmological problem is difficult, simpler theories and more controlled spacetimes provide useful testing grounds for curved-space double copy ideas. Anti-de Sitter (AdS) space is especially useful in this respect, because boundary observables and holographic methods give a sharper handle on questions that are otherwise hard to formulate \cite{Maldacena:1997re,Witten:1998qj,Raju:2012zr}.
	
	\paragraph{This Review's Definition of Double Copy in Curved Spacetimes}
	In this review, double copy or color-kinematics in curved spacetime mostly refers to constructions whose gravitational side is not asymptotically flat. In such cases, an S-matrix may not exist, or may not be as simple as its flat-space counterpart, so one must instead use observables adapted to the background, such as boundary correlators, wavefunction coefficients, or scattering observables defined with respect to special asymptotic regions. For classical solutions, the absence of an S-matrix does not by itself create the same conceptual issue, but it still leads to technical challenges compared with the asymptotically flat case. In some places we use a slightly broader definition and include double copies whose gravitational spacetime still has a region in which asymptotic states define a well-behaved S-matrix. We do this only for cases where the gauge theory lives on a non-trivial background, so that the double copy translates both the background and the perturbations. We note that some related topics not discussed in detail in this review are the CHY double copy \cite{Cachazo:2013iea} and the AdS and cosmological versions of the scattering equations \cite{Eberhardt:2020ewh,Roehrig:2020kck,Gomez:2021qfd,Gomez:2021ujt}.
	
	For boundary observables in AdS and dS, several constructions reproduce double copy structures for three- and four-point correlators in different representations. For classical and field-level double copies, Kerr--Schild, Weyl, Cotton, twistor, and convolutional constructions give controlled maps between families of gauge and gravitational data, including examples on homogeneous spaces and maximally symmetric backgrounds. Self-dual sectors provide another especially coherent arena, where the algebraic structures behind the double copy can often be displayed directly at the level of equations of motion, actions, or twistor data. A few examples also exist where a double copy of both the background and perturbations can be performed at all orders, particularly when the background can be represented by a coherent state and, in some particular classical realizations, when the background is maximally symmetric or self-dual.
	
	There are a few remaining challenges. Curved backgrounds often remove the global translation invariance, simple momentum conservation, and standard S-matrix factorization that make the flat-space double copy so powerful. Boundary terms, contact terms, field redefinitions, and the choice of observable can therefore affect statements that would be automatic for flat-space amplitudes. In AdS and dS, color-kinematics relations can survive in special limits or representations but can fail to give a physical double copy away from them, sometimes producing unphysical singularities or requiring extra contributions not present in the gauge-theory expression alone. Those additional contributions can be fixed by external physical insights and still provide a much simpler expression for gravitational observables than traditional methods produce. These simpler expressions can highlight hidden symmetries, special kinematic limits, and other physical properties. These obstructions do not make the curved-space program less useful; rather, they indicate which pieces of the flat-space story are structural and which depend on the special simplicity of asymptotically flat scattering. Additionally, ongoing progress shows that different representations can avoid these difficulties as we explain below.
	
	One reason the double copy is particularly powerful in flat space is the well-developed understanding of the constraints that fundamental physical principles, such as unitarity, causality, and locality, impose on scattering amplitudes \cite{Dixon:2013uaa,Elvang:2013cua,Brandhuber:2022qbk}. These consistency conditions provide sharp tests of whether a proposed double copy has the expected physical singularities and factorization properties. The corresponding constraints on observables in curved backgrounds are less understood, although substantial progress has been made in recent years, particularly in maximally symmetric spacetimes \cite{Baumann:2022jpr,Jazayeri:2021fvk,Lee:2024sks,Penedones:2016voo,Fitzpatrick:2011hu}. It is therefore desirable to develop these consistency conditions in parallel with curved-space double copy constructions, both as non-trivial checks of existing proposals and as guidance toward more general formulations.
	
	\paragraph{Structure of the Review}
	The review is organized as follows. We first summarize the flat-space ingredients needed later in Sec.~\ref{sec:flatDC}, including color ordering, string-theory origins of KLT relations, BCJ double copy, wavefunction examples, and spinor-helicity variables. We then review curved-space constructions for correlators and scattering observables in Sec.~\ref{sec:curved_observables}, moving through momentum space, Mellin space, embedding space, isometric space, coordinate space, twistor space, Grassmannian methods, and scattering on non-trivial backgrounds. The next section, Sec.~\ref{sec:classical_field_level}, treats classical and field-level double copies, including Kerr--Schild, Weyl, Cotton, twistor, convolutional, and background-plus-perturbation constructions. The final main section, Sec.~\ref{sec:self_dual}, focuses on self-dual double copies, from flat-space heavenly equations to AdS, cosmological, radiative, instanton, coherent-state, boundary-correlator, and higher-spin examples. The appendices collect the conformal-field-theory and spinorial conventions used in the main text.
	
	\section{Double Copy in Flat Space} \label{sec:flatDC}
	
	While the double copy for scattering amplitudes is well-established and useful for loop calculations, here we will focus on the tree-level approaches since these are the first steps that have been taken when attempting a generalization to curved spacetimes.
	
	We first briefly review Yang--Mills color-ordered amplitudes and spinor-helicity variables, then introduce several realizations of the double copy for scattering amplitudes. Useful introductions to scattering amplitudes and double copy techniques include Refs.~\cite{Dixon:2013uaa,Carrasco:2015iwa,Elvang:2013cua,Bern:2019prr,Bern:2022wqg,Brandhuber:2022qbk,Conde:2014bf}.
	
	\paragraph{Conventions and Notation}
	{Throughout this review, unless stated otherwise, we set the relevant couplings to one to avoid cluttering expressions and performing additional color-kinematics replacements; thus, $\kappa^2=8\pi G=1$, $g=1$, and $g_s=1$, where $g$ and $g_s$ denote the Yang–Mills and string couplings, respectively. Equivalently, the amplitude formulas below should be read as coupling-stripped: the usual overall factors such as $g^{n-2}$ and $\kappa^{n-2}$ are suppressed before applying the color-to-kinematics replacement. For flat-space scattering amplitudes, and for vacuum amplitudes, we use the all-outgoing convention, so that $\sum_i p_i=0$. Similarly, for boundary correlators and wavefunction coefficients, the boundary spatial momenta satisfy $\sum_i\mathbf{k}_i=0$, while $\sum_i |\mathbf{k}_i|$ need not vanish. When genuine asymptotic in- and out-states are displayed, we retain their physical labels and state the choice locally. We use the mostly plus signature, which in four dimensions is $(-+++)$. The Mandelstam variables are defined as $s_{ij}=-(p_i+p_j)^2$, and at four points we have $s=-(p_1+p_2)^2$, $t=-(p_1+p_4)^2$, and $u=-(p_1+p_3)^2$.}
	
	\subsection{{Yang--Mills} theory and color-ordered amplitudes}
	
	{Yang--Mills} (YM) theory is a cornerstone of modern quantum field theory; it describes gluons in the adjoint representation of a non-Abelian gauge group and is essential for describing the Standard Model's interactions. In perturbative calculations, {Yang--Mills} scattering amplitudes exhibit remarkable structures, which can be organized efficiently using color ordering. 
	
	The {Yang--Mills} Lagrangian is given by:
	\begin{equation}
		\mathcal{L}_{YM} = -\frac{1}{4} F_{\mu\nu}^a F^{\mu\nu a}, \label{eq:YM_lagrangian}
	\end{equation}
	where the field strength tensor is defined as:
	\begin{equation}
		F_{\mu\nu}^a = \partial_\mu A_\nu^a - \partial_\nu A_\mu^a + \frac{g}{\sqrt{2}} f^{abc} A_\mu^b A_\nu^c.
	\end{equation}
	{Here, $A_\mu=A_\mu^aT^a$ is the gauge field, $g$ is the Yang--Mills coupling, and $f^{abc}$ and $T^a$ are the structure constants and generators of the gauge group, normalized as}
	\be
	\text{Tr} \, T^a T^b = \delta^{ab} \quad \text{and} \quad [T^a, T^b] = i f^{abc} T^c.
	\ee
	Throughout this review, we use amplitude conventions that differ from standard textbook conventions by factors of $\sqrt{2}$ in the structure constants, that is $f^{abc}=\sqrt{2}\tilde{f}^{abc}$. The YM scattering amplitudes naturally decompose into a sum over color-ordered partial amplitudes $A_n$. Starting from the Lagrangian in Eq.~\eqref{eq:YM_lagrangian}, this is done by specializing to a gauge group where one can use ’t Hooft’s double-line notation \cite{tHooft:1973alw}, for example $SU(N)$, to rewrite the structure constants as 
	\be
	i {f}^{abc} = \text{Tr}(T^a T^b T^c) - \text{Tr}(T^b T^a T^c) \ , \label{eq:structure_constants}
	\ee
	and using the completeness relation for the generators 
	\be
	(T^a)_i^{\phantom{i}j} (T^a)_k^{\phantom{k}l} = \delta_i^l \delta_k^j - \frac{1}{N} \delta_i^j \delta_k^l \ . \label{eq:FierzId}
	\ee
	{For a useful graphical depiction in ’t Hooft’s double-line notation~\cite{tHooft:1973alw}, see Fig.~1 of Ref.~\cite{Dixon:2013uaa}.} Thus, the $n$-point amplitude can be expressed as:
	\begin{equation}
		\mathcal{A}_n \left( 1^{a_1}, \dots, n^{a_n} \right) = 
		\sum_{\sigma \in S_{n-1}} 
		\text{Tr} \left[ T^{a_1} T^{a_{\sigma(2)}} \dots T^{a_{\sigma(n)}} \right] 
		{A}_n \left[ 1 \, \sigma(2) \dots \sigma(n) \right]{\ .}
	\end{equation}
	Cyclicity of the trace has been used to fix one leg, leaving an $(n-1)!$ trace decomposition. This set is still overcomplete because further relations, including reflection/photon-decoupling and the KK relations below, reduce the number of independent color-ordered amplitudes~\cite{Bern:2019prr}. These color-ordered amplitudes are gauge invariant \cite{Mangano:1990by}. Utilizing a further relationship known as the Jacobi identity
	\be
	f^{abc} f^{cde} + f^{bce} f^{cda} + f^{cae} f^{cdb} = 0 \ , \label{eq:jacobi}
	\ee
	one can realize that this basis is still over-complete. {This allows us to rewrite the scattering amplitude in the DDM basis, which contains $(n-2)!$ structure-constant elements~\cite{DelDuca:1999rs}.} The existence of this basis tells us that there are additional relations between the color-ordered amplitudes. {These are the Kleiss--Kuijf (KK) relations~\cite{Kleiss:1988ne}, given by}
	\be
	A_n[1, \{\alpha\}, n, \{\beta\}] = (-1)^{|\beta|}
	\sum_{\sigma \in \text{OP}(\{\alpha\},\{\beta^T\})} A_n[1, \sigma, n] \ .
	\ee
	{Here, $\{\beta^T\}$ is the reverse ordering of $\{\beta\}$, $|\beta|$ is the number of elements in that set, and $\operatorname{OP}$ denotes the ordered permutations that preserve the order within each of the sets $\{\alpha\}$ and $\{\beta^T\}$. Everything mentioned so far is a generic characteristic of theories with adjoint particles whose interactions depend only on the structure constants\footnote{If the interaction vertices involve additional symmetric invariant tensors, such as $d^{abc}=\operatorname{Tr}(\{T^a,T^b\}T^c)$, the color basis and its redundancies are different, so the associated partial amplitudes need not obey the standard KK relations and may instead satisfy relations determined by the identities of those tensors.} $f^{abc}$. These observations hold for gauge fields, as well as for pions described by a nonlinear sigma model (NLSM) and for some higher-derivative corrections that arise when these theories are treated as low-energy effective field theories.}
	
	For the specific case of YM and other theories that satisfy the color-kinematics duality and have well-defined double copies, there are additional relations between the color-ordered amplitudes that give an even smaller basis. These are known as the Bern-Carrasco-Johansson (BCJ) relations \cite{Bern:2008qj}. The simplest version, which is referred to as the fundamental {BCJ} relation, reads
	\be
	\sum_{i=2}^{n-1} p_1 \cdot (p_2 + \dots + p_i) A_n[2, \dots, i, 1, i+1, \dots, n] = 0 \ .\label{eq:BCJrelations}
	\ee
	The BCJ relations follow from the color-kinematics duality, which will be reviewed below. These relations further reduce the number of independent color-ordered amplitudes to $(n - 3)!$.
	
	\subsection{String amplitudes, their {low-energy} limit, and the KLT double copy} \label{sec:klt_and_strings}
	The origin of the first double copy relation can be traced back to string amplitudes. Here, we will not give a review of string amplitudes but only refer to useful sources such as \cite{DHoker2018,Schlotterer2021,Berkovits2017,Staessens:2010vi}. Instead, we will directly start from the general form of color-ordered amplitudes in open string theory. Let $(a_1,\ldots,a_n)$ specify the ordering of the insertions along the boundary of the disk. Since the precise measure depends on the choice of bosonic or supersymmetric string, vertex operators, and ghost fixing, the following expression should be understood as a generic ordered disk integral:
	\be
	{A}[a_1, \dots, a_n] = \int \prod_{i=1}^{n} dz_i \frac{|z_{ab} z_{ac} z_{bc}|}{dz_a dz_b dz_c} \prod_{i=1}^{n-1} H(x_{a_{i+1}} - x_{a_i}) \!\!\prod_{1 \leq i < j \leq n} |x_i - x_j|^{2 \alpha' k_i \cdot k_j} F_n. \label{eq:openstring_ampl}
	\ee
	where $H$ is the Heaviside function,  $dz_i=dx_i$ and  $z_{ij}=x_i-x_j$ for the bosonic case and
	$dz_i=dx_i d\theta_i$ and $z_{ij}=x_i-x_j+\theta_i\theta_j$ for
	the supersymmetric case, and $F_n$ is a factor depending on the external helicities and $(x_i,\theta_i)$. Here, $a,b,c\in\{1,\ldots,n\}$ label any three distinct punctures fixed using the residual $SL(2,\mathbb{R})$ invariance.  These amplitudes have been shown to satisfy relationships between them based on monodromy arguments \cite{Bjerrum-Bohr:2009ulz,Stieberger:2009hq}. In fact, in the low-energy limit, these relationships correspond to the BCJ relations described above. The simplest of these relations is observed for the Veneziano amplitude for the scattering of four tachyons of bosonic open string theory with the same mass $m$ \cite{Veneziano:1968yb}. After stripping Chan--Paton factors and forming the colorless crossing-symmetric tachyon amplitude, one may write
	\be
	A_4(s,t,u) = {A}[1,2,3,4] + {A}[1,2,4,3] + {A}[1,3,4,2] \ ,
	\ee
	{where the color-ordered amplitude $A[1,2,3,4]$ is given by}
	\be
	{A}[1,2,3,4] =A_4(s,t) = \int_0^1 dx \, x^{-1-\alpha(s)} (1-x)^{-1-\alpha(t)} \ ,\label{eq:Veneziano}
	\ee
	with the linear Regge trajectories 
	\be
	\alpha(s) = \alpha_0 + \alpha' s, \quad \alpha_0 = -\alpha' m^2 \ .
	\ee
	Here, we have set $x_1=0$, $x_2=x$, $x_3=1$ and $x_4=+\infty$ and the other color orderings can be obtained analogously. Considering the tachyon mass $m^2=-1/\alpha'$ we can further simplify the expressions to
	\begin{align}
		{A}[1,2,3,4] &= \int_0^1 dx \, x^{2\alpha' k_1 \cdot k_2} (1-x)^{2\alpha' k_2 \cdot k_3} \ , \\
		{A}[1,3,2,4] &= \int_1^\infty dx \, x^{2\alpha' k_1 \cdot k_2} (x-1)^{2\alpha' k_2 \cdot k_3} \ .
	\end{align}
	These expressions follow directly from Eq.~\eqref{eq:openstring_ampl}. By deforming the integration contours, taking into account the phases acquired when vertex operators cross each other,
	\be
	(x - y)^{\alpha} = (y - x)^{\alpha} \times 
	\begin{cases}
		e^{+i\pi \alpha} & \text{for clockwise rotation,} \\
		e^{-i\pi \alpha} & \text{for counterclockwise rotation.}
	\end{cases} \ ,
	\ee
	and enforcing the reality condition, one can relate these amplitudes as
	\be
	{A}[1,3,2,4] = \frac{\sin(2\alpha' \pi k_1 \cdot k_2)}{\sin(2\alpha' \pi k_2 \cdot k_4)} {A}[1,2,3,4] \ .
	\ee
	Although we have used tachyons to display the contour deformation in its simplest form, the same monodromy relations hold for open-string amplitudes with other external states. The field-theory limit, $\alpha'\rightarrow 0 $, reduces to
	\be
	{A}[1,3,2,4] = \frac{ k_1 \cdot k_2}{ k_2 \cdot k_4} {A}[1,2,3,4] \ .
	\ee
	This is the four-point BCJ relation in Eq.~\eqref{eq:BCJrelations}. While these monodromy relations were discovered many years after the Kawai-Lewellen-Tye (KLT) relation between open and closed strings \cite{Kawai:1985xq}, the existence of such relations is now understood as a requirement for a theory to have a well-defined double copy.
	
	We will now show an example of the KLT relations by considering again the simplest scenario of the Veneziano amplitude in Eq.~\eqref{eq:Veneziano}. This can be rewritten in terms of the Euler Beta function:
	\be
	A_4(s,t) = B(-\alpha(s), -\alpha(t)) = \frac{\Gamma(-\alpha(s)) \Gamma(-\alpha(t))}{\Gamma(-\alpha(s) - \alpha(t))} \ .
	\ee
	Meanwhile, the Virasoro-Shapiro amplitude \cite{Virasoro:1969me,Shapiro:1970gy} for the scattering of four tachyons of bosonic closed string theory is given as an integral over the sphere
	\be
	\mathcal{M}_4(s,t,u) =
	\int d^2 z \, |z|^{-2-2(1+\alpha' s/4)} |1 - z|^{-2-2(1+\alpha' t/4)} \ ,
	\ee
	Using the $SL(2,\mathbb{C})$ symmetry, we have set $z_1=+\infty$, $z_2=0$, $z_3=z$, and $z_4=1$. Like the Veneziano amplitude, it can be rewritten in terms of Gamma functions as
	\be
	\mathcal{M}_4(s,t,u) =
	\frac{\Gamma(-1-\alpha' s/4) \ \Gamma(-1-\alpha' t/4) \ \Gamma(-1-\alpha' u/4)}{\Gamma(2+\alpha' s/4) \Gamma(2+\alpha' t/4) \ \Gamma(2+\alpha' u/4)} \ .
	\ee
	where we have again considered the tachyonic case. Using properties of the Gamma function, such as the Euler reflection formula, we then find
	\be
	\mathcal{M}_4(1,2,3,4) \propto \sin\!\left(\frac{\pi\alpha' s}{4}\right) A_4^{\text{open}}(1,2,3,4) A_4^{\text{open}}(1,2,4,3) \ .
	\ee
	where the open-string factors are evaluated at half momenta, that is, $s_{ij}(k/2)\allowbreak =s_{ij}(k)/4$. This is the four-point string-amplitude KLT relation. A closely related open--closed string correspondence is provided by the single-valued map. This map acts term by term in the low-energy expansion of the string amplitudes as $\operatorname{sv}\!\left[A_4^{\mathrm{open}}(s,t)/(s+t)\right]=(s+t)\mathcal{M}_4(s,t)$ \cite{Stieberger:2014hba,Schlotterer:2018zce}. In general, the KLT relations give $n$-point closed-string tree amplitudes as sums of products of $n$-point open-string tree amplitudes:
	\be
	\mathcal{M}_n^{\text{closed}} (1, 2, \!\cdots \!, n) = 
	\sum_{\sigma, \rho \ \in S_{n-3}} {A}_n^{\text{open}}[1,\sigma,n-1,n] \, S^\text{string}[\sigma | \rho] \, {A}_n^{\text{open}}[1,\rho,n,n-1] \ ,
	\ee
	where the string theory kernel $S^\text{string}[\sigma | \rho]$ is a function of the Mandelstam variables.

	Taking the $\alpha'\rightarrow0$ field theory limit, the KLT double copy relates color-ordered {Yang--Mills} amplitudes, from possibly different theories, to a gravitational amplitude through the field theory KLT kernel $S[\sigma | \rho]$:
	
	\vspace{0.2cm}
	\begin{mybox}[KLT Double Copy]
		\begin{center}
			\renewcommand{\arraystretch}{1.8}
			\begin{tabularx}{\textwidth}{l X}
				\textbf{Gravity} &   \vspace{-1cm} 
				\begin{equation}
					\hspace{-2.5cm} \mathcal{M}_n^{A \otimes B} (1, 2, \!\cdots \!, n) = 
					\sum_{\sigma, \rho \in S_{n-3}} \mathcal{A}_n^{A}[1,\sigma,n-1,n] \, S[\sigma | \rho] \, \mathcal{A}_n^{B}[1,\rho,n,n-1] \ ,
					\label{eq:KLT_gravity}
				\end{equation} \vspace{-0.5cm}  \\ 
				\textbf{{Yang--Mills}} &   \vspace{-1cm} 
				\begin{equation}
					\mathcal{A}_n^{A}[\sigma]
					\label{eq:KLT_gauge}
				\end{equation} \\ 
				\textbf{BAS} &   \vspace{-1cm} 
				\begin{equation}
					(m_n^{\text{BAS}})_\sigma^\rho =  S[\sigma | \rho] ^{-1}
					\label{eq:KLT_bas}
				\end{equation}  \vspace{-0.5cm} 
			\end{tabularx}
		\end{center}
	\end{mybox}
	\vspace{0.2cm}
	where $\sigma$ and $\rho$ label the permutations of $\{2,3,\dots,n-2\}$. The Lagrangian of the gravitational theory arising as the double copy is given by the universal massless sector of supergravity, sometimes referred to as NS-NS
	gravity, and reads
	\begin{align}
		S_{\text{NS--NS}}
		=\frac{1}{2}
		\int d^{d}x\,\sqrt{|g|}
		\left[
		R
		-\frac{4}{d-2}\,\nabla_\mu\phi\,\nabla^\mu\phi
		-\frac{1}{12}\,e^{-\frac{8}{d-2}\phi}\,
		H_{\mu\nu\rho}H^{\mu\nu\rho}
		\right],
		\label{eq:NSNS_doublecopy}
	\end{align}
	where this is written in Einstein frame, {$\phi$} is the dilaton, $B_{\mu\nu}$ is the Kalb--Ramond two-form, and
	$H_{\mu\nu\rho}=3\,\partial_{[\mu}B_{\nu\rho]}$. The kernel $(m_n^{\text{BAS}})_\sigma^\rho$ refers to a given doubly color-ordered biadjoint scalar amplitude (BAS)
	\begin{equation}(m_n^{\text{BAS}})_\sigma^\rho=m_n^{\text{BAS}}(1, \sigma(2), \dots, \sigma(n-2), n-1, n \mid 1, \rho(2), \dots, \rho(n-2), n, n-1)\ .
	\end{equation}
	The fact that the inverse of the KLT kernel gives rise to BAS amplitudes was first shown in \cite{Cachazo:2013iea}. With this insight, the inverse of the string theory kernel was constructed in \cite{Mizera:2016jhj}, which gives $\alpha'$ corrections to the BAS amplitudes. One should nevertheless note that this $\alpha'$-corrected theory is not a physical string-theory construction since it involves an infinite tower of tachyons.
	
	The biadjoint scalar theory contains scalar fields transforming in the adjoint representations of two, possibly distinct, groups and has cubic interactions. Its Lagrangian is
	\be
	\mathcal{L} = -\frac{1}{2} \partial^\mu \phi^{a \tilde{a}} \partial_\mu \phi^{a \tilde{a}} - \frac{1}{3!} f^{abc} \tilde{f}^{\tilde{a} \tilde{b} \tilde{c}} \phi^{a \tilde{a}} \phi^{b \tilde{b}} \phi^{c \tilde{c}} \ ,
	\ee
	where we have set the coupling strength to one. The {doubly color-ordered} amplitudes for this theory are written as
	
	\begin{align}
		m_n^\text{BAS} \left( 1^{a_1}, \dots, n^{a_n}|1^{\dot{a}_1}, \dots, n^{\dot{a}_n} \right) = 
		\sum_{\sigma, \rho \ \in S_{n-1}} 
		\text{Tr} \left[ T^{a_1} T^{a_{\sigma(2)}} \dots T^{a_{\sigma(n)}} \right] 
		\nonumber \\
		\times \ \text{Tr} \left[ \widetilde T^{\dot{a}_1} \widetilde T^{\dot{a}_{\rho(2)}} \dots \widetilde T^{\dot{a}_{\rho(n)}} \right]   m_n^\text{BAS} \left[ 1 \, \sigma(2) \dots \sigma(n) | 1 \, \rho(2) \dots \rho(n) \right]
	\end{align}
	
	While at first glance this might seem like a very simple field theory, it has many interesting features, including hidden geometric properties of its amplitudes \cite{Arkani-Hamed:2017mur} as well as non-trivial non-perturbative solutions \cite{White:2016jzc,DeSmet:2017rve,Bahjat-Abbas:2018vgo,Armstrong-Williams:2025spu,Armstrong-Williams:2026dmk} relevant for the classical double copy. We will see below that the BAS theory is an important ingredient in all double copy constructions.
	
	Generalizations of the standard KLT double copy have been considered in \cite{Johnson:2020pny} for massive mediators and \cite{Chi:2021mio,Chen:2023dcx,Azevedo:2018dgo} for higher-derivative and stringy corrections. Similar generalizations but in the BCJ context, reviewed next, can be found in \cite{Chiodaroli:2015rdg,Momeni:2020hmc,Momeni:2020vvr}.
	
	\subsection{BCJ double copy} \label{sec:bcj}
	We have already mentioned that color-ordered amplitudes of YM theories satisfy the special relations called BCJ relations. In fact, colored theories that satisfy these relations have been observed to have well-defined double copies. The origin of the BCJ relations can be traced to the color-kinematics (CK) duality of the scattering amplitudes, which we review below.
	
	The $N$-point tree-level amplitude $\mathcal{A}_N$ of a colored theory can be written as
	\begin{equation}\label{eq:An}
		\mathcal{A}_N=c^T D^{-1}n \ , 
	\end{equation}
	Here, $c$ is a vector whose entries are color numerators (structure constants or generators of the group), $D$ is the diagonal matrix whose entries are products of inverse propagators, and $n$ is the column vector of kinematic numerators, which depend on momenta and polarization vectors when the particles have nonzero spin. The number of components of these color and kinematic vectors corresponds to the number of independent cubic graphs contributing to the amplitude. For an $N$-point amplitude, this number is $(2N-5)!!$. Contact terms can be written in this form by multiplying by one in the form $D_i/D_i$ (where $D_i$ is the $i$-th diagonal entry of $D$), so that a contact-term contribution $c_i n_i$ is rewritten as $c_i(n_i D_i)/D_i$. With this notation, the Jacobi identities are
	\be
	\label{eq:Mc}
	Mc=0 \ .
	\ee
	where $M$ is a $(2N-5)!!\times(2N-5)!!$ matrix with entries $\pm1$ and zeros. A theory is said to satisfy the \CK duality if the kinematic numerators satisfy the same relations as the color ones. Here, by the same relations we mean the universal algebraic identities obeyed by the color factors, such as Jacobi identities and the corresponding antisymmetry properties, rather than accidental relations that arise only for special spacetime dimensions, finite group rank, or additional structures such as supersymmetry. In this matrix form, this is written as
	\be
	\label{eq:Mn}
	Mn=0 \ .
	\ee
	Once a representation satisfying \CK duality is found, the double copy is given by exchanging the color factor for the kinematic factor and reads
	\begin{equation}\label{eq:MnDC}
		\mathcal{M}_N=n^T D^{-1} n \ .    
	\end{equation}
	The equivalent, but more familiar, expression of the BCJ double copy written as a sum over graphs, is
	\begin{mybox}[BCJ Double Copy]
		\begin{center}
			\renewcommand{\arraystretch}{1.8}
			\begin{tabularx}{\textwidth}{l X}
				\textbf{Double Copy} &   \vspace{-1cm} 
				\begin{equation}
					\mathcal{M}_N = \sum_{\text{cubic graphs}} \frac{n_i n_i}{\prod_{\alpha_i} p_{\alpha_i}^2}
					\label{eq:BCJ_gravity}
				\end{equation} \\ 
				\textbf{Single Copy} &   \vspace{-1cm} 
				\begin{equation}
					\mathcal{A}_N = \sum_{\text{cubic graphs}} \frac{c_i n_i}{\prod_{\alpha_i} p_{\alpha_i}^2}
					\label{eq:BCJ_gauge}
				\end{equation} \\ 
				\textbf{Zeroth Copy} &   \vspace{-1cm} 
				\begin{equation}
					m_N = \sum_{\text{cubic graphs}} \frac{c_i \tilde{c}_i}{\prod_{\alpha_i} p_{\alpha_i}^2}
					\label{eq:BCJ_bas}
				\end{equation}  \vspace{-0.5cm} 
			\end{tabularx}
		\end{center}
	\end{mybox}
	where we have assumed massless mediators for simplicity, and $c_i$ and $n_i$ denote components of $c$ and $n$, respectively. We have included the zeroth copy, which is found by exchanging the kinematic factor for another color factor in the colored theory. The zeroth copy corresponds to the biadjoint scalar theory introduced above and it can involve higher-derivative corrections when considering double copies for effective field theories. {The kinematic factors in the double copy need not belong to the same single copy;} that is, the double copy can be given as 
	\begin{equation}\label{eq:DC_ntn}
		\mathcal{M}_N=n^T D^{-1} \tilde{n} \ .    
	\end{equation}
	Considering this general form can lead to a large web of theories that are double copy constructible, see the tables in \cite{Bern:2022wqg} for a recent compilation of known examples.
	
	The kinematic factors directly calculated from Feynman diagrams might not satisfy the Jacobi relations, but one can use generalized gauge transformations to change them{~\cite{Bern:2017yxu}}. These transformations leave the amplitude in \eqref{eq:An} unchanged. Examples of such transformations include gauge transformations and field redefinitions. Defining these shifts as
	\be\label{shift}
	n\rightarrow n+\Delta n \ ,
	\ee
	we see that the amplitude in \eqref{eq:An} is unchanged if
	\be
	\label{eq:deltan}
	D^{-1}\Delta n=M^{T}v \ ,
	\ee
	where $v$ is a vector to be determined. In order to satisfy the CK duality, the shifted $n$ must obey the following equation
	\be \label{eq:deltan jac}
	M(n+\Delta n)=0\ ,
	\ee
	which combined with \eqref{eq:deltan} gives
	\be\label{eq:cond v}
	MDM^{T}v=-Mn\ .
	\ee
	The number of non-zero rows of $M$ is the number of independent Jacobi identities, which we denote by $N_j=(2N-5)!!-(N-2)!$. Here $(2N-5)!!$ is the number of cubic color factors, while the Jacobi identities reduce them to a DDM/Kleiss--Kuijf color basis of size $(N-2)!$ \cite{Kleiss:1988ne,DelDuca:1999rs,Bern:2008qj}. Their difference is therefore the number of independent Jacobi rows. Since $M$ is written as a $(2N-5)!!\times(2N-5)!!$ matrix whose only non-zero rows are these $N_j$ Jacobi rows, $MDM^{T}$ is block diagonal with a $N_j \times N_j$ symmetric block that we denote $(MDM^{T})_{N_j \times N_j}$, and all other elements equal to zero.
	$Mn$ will have at most $N_j$ non-zero elements, so we can write it as
	\be \label{eq:U def}
	Mn=((Mn)_{N_j},0, ...,0)\ .
	\ee
	Note that the vector $(Mn)_{N_j}$ measures the violation of the CK algebra. We can see that in order to find the shifts we need to find $v$, for which we need to invert the matrix $(MDM^{T})_{N_j \times N_j}$: 
	\be \label{eq:v}
	v=-((MDM^{T})_{N_j \times N_j}^{-1} (Mn)_{N_j},0,..,0)\ .
	\ee
	The matrix $(MDM^{T})_{N_j \times N_j}$ need not have full rank, as happens for massless Yang--Mills amplitudes. In that case the inverse in Eq.~\eqref{eq:v} should be understood as a restricted inverse on the image of $(MDM^{T})_{N_j \times N_j}$. The equation for $v$ is solvable only if $(Mn)_{N_j}$ lies in this image, or equivalently if it is orthogonal to every null vector of $(MDM^{T})_{N_j \times N_j}$. These orthogonality conditions are the BCJ relations.
	
	Substituting the shifted $n$ back into the double copy formula \eqref{eq:MnDC} we find
	\begin{equation}
		\begin{split}
			\mathcal{M}_N&=(n+\Delta n)^T D^{-1} (n+\Delta n) \\
			&=(n+\Delta n)^T D^{-1} n+(n+\Delta n)^T M^T v\\&=n^T D^{-1} n+\Delta n^T D^{-1} n\ .
		\end{split}
	\end{equation}
	Going from the first to the second line used \eqref{eq:deltan}, and going from the second to the third line we used \eqref{eq:deltan jac}. Now we can replace $\Delta n$ using \eqref{eq:deltan} and \eqref{eq:v} to get
	\begin{equation}
		\begin{split}
			\mathcal{M}_N=n^T D^{-1} n+v^T M n=n^T D^{-1} n-(Mn)_{N_j}^T (MDM^{T})_{N_j \times N_j}^{-1} (Mn)_{N_j} \ .    \label{eq:DCwithnoCK}
		\end{split}
	\end{equation}
	When the block $(MDM^{T})_{N_j \times N_j}$ is not full rank, Eq.~\eqref{eq:DCwithnoCK} is meaningful only with this restricted inverse and only when $(Mn)_{N_j}$ lies in the subspace orthogonal to the null eigenvectors. We can see that the poles of the double copy amplitude, $\mathcal{M}_N$, come from kinematic configurations for which either $D$ or this block become singular. Since $D$ gives rise to the physical poles in the {Yang--Mills} amplitude, unphysical poles could only arise from this block. In fact, this is what happens when attempting to double copy correlators in anti-de Sitter (AdS) or de Sitter (dS) backgrounds, collectively (A)dS correlators, and also in flat space for theories with generic massive mediators \cite{Johnson:2020pny,Momeni:2020hmc,Gonzalez:2021bes}. The double copy relation in Eq.~\eqref{eq:DCwithnoCK} allows us to compute the double copy without explicitly constructing the \CK dual kinematic numerators, which can be a difficult task. This was first realized in \cite{Bern:2017yxu} and applied in \cite{Bern:2018jmv,Bern:2017ucb} to construct double copies at loop level and later in \cite{Momeni:2020hmc,Gonzalez:2021bes} to construct double copies with massive mediators.

	We finish this section with a brief analysis of the two simplest examples. First, we consider the {four-point} amplitude with massless particles in the adjoint representation. We have a single Jacobi identity, $c_s+c_t+c_u=0$ (Eq.~\eqref{eq:jacobi}), so $M$ is a three-by-three matrix with only one non-zero row equal to $(1,1,1)$ and $D=\text{diag}(s,t,u)$. In this case $Mn=n_s+n_t+n_u$ and $(MDM^{T})_{N_j \times N_j}=MDM^T=s+t+u=0$ after using momentum conservation. Thus, all the components of $MDM^{T}$ are zero, and there is no shift that can be performed at {four points} to enforce \CK duality if it is not realized for any given kinematic numerators. {This conclusion changes} when analyzing double copy relations for other objects that are not scattering amplitudes, as we will see below. The {five-point} case is the first non-trivial one for massless particles. We can write the {five-point} amplitude as 
	\be \label{eq:5point}
	\mathcal{A}_5=\sum_{i=1}^{15}\frac{c_{i}n_{i}}{D_i} \ ,
	\ee
	where $D_i$ are the elements of the diagonal propagator matrix $D$ and the 15 color factors given by 
	\begin{eqnarray}
		&& 
		c_{1\phantom{0}} \equiv f^{a_1 a_2 b}f^{b a_3 c}f^{c a_4 a_5}\,, \hskip 0.8cm
		c_{2\phantom{1}} \equiv f^{a_2 a_3 b}f^{b a_4 c}f^{c a_5 a_1}\,, 
		\nonumber \\[6pt] &&
		c_{3\phantom{1}} \equiv f^{a_3 a_4 b}f^{b a_5 c}f^{c a_1 a_2}\,, \hskip 0.8cm
		c_{4\phantom{1}} \equiv f^{a_4 a_5 b}f^{b a_1 c}f^{c a_2 a_3}\,, 
		\nonumber \\[6pt] &&
		c_{5\phantom{1}} \equiv f^{a_5 a_1 b}f^{b a_2 c}f^{c a_3 a_4}\,, \hskip 0.8cm
		c_{6\phantom{1}} \equiv f^{a_1 a_4 b}f^{b a_3 c}f^{c a_2 a_5}\,, 
		\nonumber \\[6pt] &&
		c_{7\phantom{1}} \equiv f^{a_3 a_2 b}f^{b a_5 c}f^{c a_1 a_4}\,, \hskip 0.8cm
		c_{8\phantom{1}} \equiv f^{a_2 a_5 b}f^{b a_1 c}f^{c a_4 a_3}\,, 
		\nonumber \\[6pt] &&
		c_{9\phantom{1}} \equiv f^{a_1 a_3 b}f^{b a_4 c}f^{c a_2 a_5}\,, \hskip 0.8cm
		c_{10} \equiv f^{a_4 a_2 b}f^{b a_5 c}f^{c a_1 a_3}\,, 
		\nonumber \\[6pt] &&
		c_{11} \equiv f^{a_5 a_1 b}f^{b a_3 c}f^{c a_4 a_2}\,, \hskip 0.8cm
		c_{12} \equiv f^{a_1 a_2 b}f^{b a_4 c}f^{c a_3 a_5}\,, 
		\nonumber \\[6pt] &&
		c_{13} \equiv f^{a_3 a_5 b}f^{b a_1 c}f^{c a_2 a_4}\,, \hskip 0.8cm
		c_{14} \equiv f^{a_1 a_4 b}f^{b a_2 c}f^{c a_3 a_5}\,, 
		\nonumber \\[6pt] &&
		c_{15} \equiv f^{a_1 a_3 b}f^{b a_2 c}f^{c a_4 a_5}\,. 
		\label{FivePointColor}
	\end{eqnarray}
	At five points, there are 9 independent Jacobi identities. When written in the form $Mc=0$, where $c=(c_1,{\ldots},c_{15})$, one can find that the matrix $M$ is given by
	\begin{small}
		\begin{eqnarray}
			M=\left(
			\begin{array}{ccccccccccccccc}
				0 & 0 & 1 & 0 & -1 & 0 & 0 & 1 & 0 & 0 & 0 & 0 & 0 & 0 & 0 \\
				-1 & 0 & 1 & 0 & 0 & 0 & 0 & 0 & 0 & 0 & 0 & 1 & 0 & 0 & 0 \\
				-1 & 0 & 0 & 1 & 0 & 0 & 0 & 0 & 0 & 0 & 0 & 0 & 0 & 0 & 1 \\
				0 & -1 & 0 & 1 & 0 & 0 & 1 & 0 & 0 & 0 & 0 & 0 & 0 & 0 & 0 \\
				0 & -1 & 0 & 0 & 1 & 0 & 0 & 0 & 0 & 0 & 1 & 0 & 0 & 0 & 0 \\
				0 & 0 & 0 & 0 & 0 & -1 & 1 & 0 & 0 & 0 & 0 & 0 & 0 & 1 & 0 \\
				0 & 0 & 0 & 0 & 0 & -1 & 0 & 1 & 1 & 0 & 0 & 0 & 0 & 0 & 0 \\
				0 & 0 & 0 & 0 & 0 & 0 & 0 & 0 & -1 & 1 & 0 & 0 & 0 & 0 & 1 \\
				0 & 0 & 0 & 0 & 0 & 0 & 0 & 0 & 0 & 1 & -1 & 0 & 1 & 0 & 0 \\
				0 & 0 & 0 & 0 & 0 & 0 & 0 & 0 & 0 & 0 & 0 & 0 & 0 & 0 & 0 \\
				0 & 0 & 0 & 0 & 0 & 0 & 0 & 0 & 0 & 0 & 0 & 0 & 0 & 0 & 0 \\
				0 & 0 & 0 & 0 & 0 & 0 & 0 & 0 & 0 & 0 & 0 & 0 & 0 & 0 & 0 \\
				0 & 0 & 0 & 0 & 0 & 0 & 0 & 0 & 0 & 0 & 0 & 0 & 0 & 0 & 0 \\
				0 & 0 & 0 & 0 & 0 & 0 & 0 & 0 & 0 & 0 & 0 & 0 & 0 & 0 & 0 \\
				0 & 0 & 0 & 0 & 0 & 0 & 0 & 0 & 0 & 0 & 0 & 0 & 0 & 0 & 0 \\
			\end{array}
			\right) \ ,
		\end{eqnarray}
	\end{small}
	The matrix of propagators, $D$, reads 
	\be\label{eq:D}
	\begin{split}
		D=&\text{diag}\{ D_{12} D_{45}, D_{15} D_{23}, D_{12} D_{34}, D_{23} D_{45}, D_{15} D_{34}, \\
		&D_{14} D_{25}, D_{14} D_{23}, D_{25} D_{34}, D_{13} D_{25}, D_{13} D_{24}, \\
		&D_{15} D_{24}, D_{12} D_{35}, D_{24} D_{35}, D_{14} D_{35}, D_{13} D_{45}\} \ ,
	\end{split}
	\ee
	where $D_{ij}=-(p_i+p_j)^2=s_{ij}$.  In this case, after imposing momentum conservation, which leaves five independent Mandelstam variables, the relevant $9\times9$ matrix is not the zero matrix as it was at four points. To use Eq.~\eqref{eq:DCwithnoCK}, however, one must still use the restricted inverse when necessary and check that the Jacobi-violation vector $(Mn)_{N_j}$ satisfies the image condition described above. When this condition is satisfied, one can construct a double copy without explicitly finding the \CK kinematic numerators by following the prescription in Eq.~\eqref{eq:DCwithnoCK}.

	\subsection{Color-kinematics: Amplitudes vs Wavefunction} 
	
	\label{sec:ckwavefunc}
	In this subsection, we analyze a simple, interesting case that serves as an example of the failure of constructing a double copy in some curved spacetimes. {We consider the four-point BCJ double copy for the nonlinear sigma model (NLSM).} There is a simple, well-known realization of the color-kinematics duality for NLSM scattering amplitudes, which square to the special Galileon \cite{Cachazo:2014xea,Cheung:2016prv,Cheung:2017yef,CarrilloGonzalez:2018ejf,CarrilloGonzalez:2019fzc}. We are going to consider the $U(N)$ NLSM whose Lagrangian (ignoring higher-derivative corrections) reads
	\begin{equation}
		\mathcal{L}=-\frac{F^2}{4}\operatorname{Tr}\left(\partial^\mu \mathcal{U}\partial_\mu \mathcal{U}^\dagger\right)
	\end{equation}
	where $\mathcal{U}\in U(N)$ and $F$ is the pion decay constant and has units of mass. One can utilize standard Feynman rules to compute the scattering amplitudes for the pions. For example, by picking the so-called Cayley parametrization, we have
	\begin{equation}
		\mathcal{U} = 1 + 2 \sum_{n=1}^\infty \left( \frac{i}{\sqrt{2}\,F} \pi^a T^a \right)^n
	\end{equation}
	where $\pi^a$ are the pions. At four points, the relevant leading order contribution to the Lagrangian is 
	\be
	\mathcal{L}^4 = \frac{1}{2F^2} \text{Tr}(T^a T^b T^c T^d) \, \partial_\mu \pi^a \, \partial^\mu \pi^b \, \pi^c \pi^d \ ,
	\ee
	and thus the amplitude of the NLSM can be written as 
	\begin{equation}
		A_4^\text{NLSM}=\frac{1}{F^2}\left(\frac{n_s c_s}{s}+\frac{n_t c_t}{t}+\frac{n_u c_u}{u}\right) \label{eq:4point_bcj}
	\end{equation}
	with 
	\begin{align}
		n_s &= s^2/2,        \quad \quad n_t = -t^2/2,        \quad \quad n_u = -(s^2 - t^2)/2, \\
		c_s &= f^{abe}f^{ecd}, \quad c_t = f^{ade}f^{ebc}, \quad c_u = f^{ace}f^{edb} \, . \label{eq:color_4pt}
	\end{align}
	It is evident that both color and kinematic numerators satisfy the Jacobi identities, and thus we have color-kinematics duality. Performing the usual color-kinematics replacement, one finds
	\begin{equation}
		{A_4^{\mathrm{SGal}}}=-\frac{stu}{4M^6}
	\end{equation}
	where we made the coupling replacement
	\begin{equation}
		\frac{1}{F^2}\rightarrow \frac{1}{M^6}\, .
	\end{equation}
	This is the scattering amplitude of the special Galileon whose quartic Lagrangian is \cite{Hinterbichler:2015pqa}
	\begin{equation}
		\mathcal{L}^{\text{SGal}} = -\frac{1}{2} (\partial \Phi)^2 
		+ \frac{1}{12 M^6} (\partial \Phi)^2 
		\left((\Box \Phi)^2 - (\partial_\mu \partial_\nu \Phi)^2 \right) \ .
	\end{equation}
	
	In this simple scenario, it is also easy to see how the color-kinematics duality gives rise to the BCJ relations. Starting from Eq.~\eqref{eq:4point_bcj}, one can relate this expression to the amplitude expression in terms of the $(n-2)!$ basis of color-ordered amplitudes by using Eq.~\eqref{eq:structure_constants} and Eq.~\eqref{eq:FierzId}:
	\begin{align}
		A_4[1,2,3,4]&=\frac{1}{F^2}\left(\frac{n_{s}}{s}-\frac{n_t}{t}\right)=\frac{s+t}{2F^2}=-\frac{u}{2F^2} \label{eq:AcolorToFactor} \\
		A_4[1,2,4,3]&=\frac{1}{F^2}\left(-\frac{n_{s}}{s}+\frac{n_u}{u}\right)=-\frac{s}{2F^2}+\frac{s-t}{2F^2}=-\frac{t}{2F^2} \ ,
	\end{align}
	{Using $n_t=-(n_s+n_u)$, this can be rewritten as}
	\be \label{eq:DerivBCJ}
	\begin{pmatrix}
		A[1,2,3,4] \\
		A[1,2,4,3]
	\end{pmatrix}
	=
	\frac{1}{F^2}
	\begin{pmatrix}
		\frac{1}{s} + \frac{1}{t} & \frac{1}{t} \\
		-\frac{1}{s} & \frac{1}{u}
	\end{pmatrix}
	\begin{pmatrix}
		n_s \\
		n_u
	\end{pmatrix}
	\ee
	It is easy to note that the determinant of the matrix above vanishes after using momentum conservation in the form of $s+t+u=0$. {Otherwise, one could invert the relation, write the kinematic numerators in terms of an $(n-2)!$-element basis of $n$-point color-ordered amplitudes, and use Jacobi relations to generate the remaining kinematic numerators. This would incorrectly imply a color-kinematics-dual representation for the amplitudes of every theory.} This is just another version of the reasoning observed earlier with the matrix $(MDM^{T})_{N_j \times N_j}$. {The fact that the matrix above does not have full rank implies additional relations among the color-ordered amplitudes, namely the BCJ relations.} In this case, it is easy to see that the BCJ relation is
	\be
	t A[1,2,3,4] = u A[1,2,4,3] =  -\frac{ut}{2F^2} \ .
	\ee
	While we show this explicitly for the NLSM, it holds for any color-ordered amplitudes that satisfy the color-kinematics duality.
	
	We now proceed to look at how this simple realization of the double copy for scalar EFTs works from the point of view of the wavefunction. When working in certain curved spacetimes, there is no clear notion of an S-matrix, or even the existence of one. {Instead, one commonly studies off-shell observables such as the wavefunctional, from which equal-time correlation functions can be computed~\cite{Lehners:2023yrj,Halliwell:1989myn,Kiefer:2008sw,Baumann:2022jpr}.}
	
	Here, we will work with the wavefunctional in the field-basis representation of the ground state, which gives the amplitude for being in a particular spatial
	{configuration $\varphi({\bf x})\equiv \varphi({\bf x},t_*)$ at time $t_*$: $\Psi[\varphi({\bf x}),t_*]\equiv\braket{\varphi|0}$.} As in the Schr\"odinger picture in quantum mechanics, the correlation functions of a field $\varphi$ at a given time $t$ in quantum field theories can be recovered from the wavefunctional as
	\begin{equation}
		\langle \varphi(\vec{x}_1) \cdots \varphi(\vec{x}_n) \rangle =
		\frac{\int \mathcal{D}\varphi \, \varphi(\vec{x}_1) \cdots \varphi(\vec{x}_n) |\Psi[\varphi]|^2}
		{\int \mathcal{D}\varphi \, |\Psi[\varphi]|^2} \ .
	\end{equation}
	{At late times, it is common to write the wavefunctional in momentum space as an expansion in $n$-point wavefunction coefficients $\psi_n(\{\mathbf{k}_i\}_{i=1}^n)$, often called wavefunctions:}
	\begin{equation}
		{
			\begin{aligned}
				\Psi[\varphi,t_f]&=\exp\!\Bigg\{\sum_{n=2}^\infty \frac{1}{n!} \int \frac{\mathrm{d}^3 {\bf{k}}_1 \cdots \mathrm{d}^3 {\bf{k}}_n}{(2\pi)^{3n}} \\
				&\qquad\quad\times \varphi_{{\bf{k}}_1} \cdots \varphi_{{\bf{k}}_n} (2\pi)^3 \delta^{(3)}({\bf{k}}_1 + \cdots + {\bf{k}}_n) \psi_n(\{\mathbf{k}_i\}_{i=1}^n)\Bigg\}\ .
		\end{aligned}}
	\end{equation}
	Solving the corresponding functional Schr\"odinger equation is complicated. {A common approach uses a path integral, written as}
	\begin{equation}
		\Psi[\varphi_f, t_f] = \int \mathcal{D}\varphi_i \braket{\varphi_f|\varphi_i} \braket{\varphi_i|0} 
		= \int \mathcal{D}\varphi_i \int_{\substack{\varphi(t_f) = \varphi_f \\ \varphi(t_i) = \varphi_i}} \mathcal{D}\varphi \, e^{iS[\varphi]} \Psi_i[\varphi_i, t_i],
	\end{equation}
	{The path integral runs over field configurations interpolating from the initial profile $\varphi_i$ at $t_i=-\infty$ to the final profile $\varphi_f$ at $t_f$. We take a Gaussian initial wavefunction at $t_i=-\infty$.} An important fact is that, contrary to amplitudes, wavefunction coefficients are not invariant under field redefinitions and are also dependent on boundary terms. The former can be fixed by demanding that the wavefunction coefficients satisfy soft theorems for massless scalar fields in flat space \cite{Bittermann:2022nfh}. We now proceed to summarize the double copy construction for the wavefunction of the NLSM as proposed in \cite{Bittermann:2022nfh}, which was found to suffer from issues previously observed in curved spacetimes.  One should note that the wavefunction formulas quoted below use the exponential representation of Ref.~\cite{Bittermann:2022nfh}, not the Cayley parametrization used above for the amplitudes. The 4-point color-ordered wavefunction of the NLSM is
	\begin{equation}
		\psi_4^{\text{(nlsm)}}[1,2,3,4] = \frac{1}{2 F^2 E} \left( P_1 \cdot P_2 + P_2 \cdot P_3 + P_3 \cdot P_4 + P_4 \cdot P_1 \right) + \frac{1}{6 F^2} E.
	\end{equation}
	{Here,} $P_i^\mu=(k_i,{\bf{k}}_i)$, $k_i=\sqrt{{\bf{k}}_i^2}$, and $E=\sum_{i=1}^4 k_i$; with spatial momentum conservation, $(\sum_iP_i)^2=-E^2$. Suppressing the overall $1/F^2$ coupling factor in the color-dressed formulas, the wavefunction can be rewritten as
	\begin{align}
		\psi_4^{\text{(nlsm)}} &= 
		- f^{a_1 a_2 b} f^{a_3 a_4 b} \frac{1}{6E} \left( \mathcal{T} - \mathcal{U} + E (q_t - q_u) \right) \nonumber \\
		&\quad - f^{a_1 a_4 b} f^{a_2 a_3 b} \frac{1}{6E} \left( \mathcal{U} - \mathcal{S} + E (q_u - q_s) \right) \nonumber \\
		&\quad - f^{a_3 a_1 b} f^{a_2 a_4 b} \frac{1}{6E} \left( \mathcal{S} - \mathcal{T} + E (q_s - q_t) \right).
	\end{align}
	{The Mandelstam-like variables are}
	\begin{align}
		\mathcal{S} &\equiv -(k_{12} + q_s)(k_{34} + q_s), \\
		\mathcal{T} &\equiv -(k_{14} + q_t)(k_{23} + q_t), \\
		\mathcal{U} &\equiv -(k_{13} + q_u)(k_{24} + q_u).
	\end{align}
	{Here, $k_{ij}=k_i+k_j$, and $q_{ij}=|{\bf k}_i+{\bf k}_j|$, with $q_s=q_{12}=q_{34}$, $q_t=q_{14}=q_{23}$, and $q_u=q_{13}=q_{24}$. The proposal for the double copy is to write the wavefunction as}
	\begin{equation}
		\psi_4^{\text{(nlsm)}} = \frac{1}{E} \left( \frac{c_s \bar{n}_s}{\mathcal{S}} + \frac{c_t \bar{n}_t}{\mathcal{T}} + \frac{c_u \bar{n}_u}{\mathcal{U}} \right),
	\end{equation}
	which does not automatically exhibit the color-kinematics duality. Following previous proposals in curved spacetimes \cite{Armstrong:2020woi}, loop amplitudes \cite{Bern:2017yxu,Bern:2018jmv,Bern:2017ucb}, and amplitudes with massive propagators \cite{Johnson:2020pny,Momeni:2020vvr}, one can perform a generalized gauge shift (which, as reviewed above, leaves the expression invariant) given by
	\begin{align}
		\bar{n}_s &\mapsto {n}_s=\bar{n}_s + \mathcal{S} \chi \, \\
		\bar{n}_t &\mapsto {n}_t=\bar{n}_t + \mathcal{T} \chi \ , \\
		\bar{n}_u &\mapsto {n}_u=\bar{n}_u + \mathcal{U} \chi \ ,
	\end{align}
	which leads to kinematic numerators that satisfy the color-kinematics duality when choosing
	\begin{equation}
		\chi = -\frac{E \mathcal{E}_{\text{nlsm}}}{6 (\mathcal{S} + \mathcal{T} + \mathcal{U})} \ .
	\end{equation}
	Here $\mathcal{E}_{\text{nlsm}}\equiv \mathcal{S}(q_t-q_u)+\mathcal{T}(q_u-q_s)+\mathcal{U}(q_s-q_t)$, which is equivalently defined by $\bar n_s+\bar n_t+\bar n_u=-E\,\mathcal{E}_{\text{nlsm}}/6$. Notice that this is possible only because $\mathcal{S} + \mathcal{T} + \mathcal{U}=- E \left( E + q_s + q_t + q_u \right) \neq 0$, which is not the case in the amplitudes case described above with the usual Mandelstam variables. It is worth highlighting that it reduces to the standard case in the total-energy limit, $E\rightarrow 0$. This is a realization of a known phenomenon where the matrix that relates amplitudes (or wavefunctions in this case) to the kinematic numerators is full rank and therefore invertible. When this happens, the amplitudes or wavefunctions do not satisfy BCJ relations. Losing this important property can obstruct the double copy. This is well-understood when considering amplitudes with massive mediators. In these cases, the lesson is that unless the rank of the KLT matrix \cite{Johnson:2020pny}, or the BCJ one as considered in \cite{Momeni:2020hmc,Gonzalez:2021bes}, is reduced and leads to additional amplitude relations (BCJ relations), then the double copy is not physical, that is, it will exhibit singularities at unphysical kinematics. Being aware of these issues, we can still proceed and construct a double copy as:
	\begin{equation}
		\psi_4^{\text{(dc)}} = \frac{1}{M^6} \frac{1}{E} \left( \frac{n_s^2}{\mathcal{S}} + \frac{n_t^2}{\mathcal{T}} + \frac{n_u^2}{\mathcal{U}} \right).
	\end{equation}
	This will not correspond to the special Galileon wavefunction
	\begin{align}
		{\psi_4^{\text{(SGal)}}} &= \frac{1}{2 M^6 E} \Big( P_1 \cdot P_2 \ P_1 \cdot P_3 \  P_1 \cdot P_4 + \text{perms.} \Big) \nonumber \\
		&\quad + \frac{1}{M^6} \Bigg( \frac{1}{2} k_1^2 k_2 k_3 k_4 + \frac{1}{16} k_1^4 k_2 
		+ \frac{1}{16} k_1^5 + \frac{1}{4} k_1^2 k_2^2 k_3 \nonumber \\
		&\qquad - \frac{1}{16} k_1 q_{12}^4 - \frac{1}{8} k_1^2 k_2 q_{12}^2 \Bigg) + \text{perms.}
	\end{align}
	In fact, it will contain poles at unphysical kinematic values, $E + q_s + q_t + q_u$, as expected.
	
	\subsection{Spinor-helicity variables} \label{sec:spinorhel}
	{For theories with nonzero spin, spinor-helicity variables provide an efficient representation of scattering amplitudes and avoid the need to contract tensor amplitudes with polarization vectors. We begin by explaining why this approach is useful.} Particles are irreducible, unitary representations of the Poincaré group which are labeled by their momentum, $p$, and any additional labels $\sigma$. They transform as
	\begin{equation}
		U(\Lambda) | p, \sigma \rangle 
		= D_{\sigma \sigma'}\!\bigl(W(\Lambda, p; k)\bigr)\,
		| \Lambda p, \sigma' \rangle .
	\end{equation}
	{This relation describes a general Lorentz transformation $\Lambda$, with $U(\Lambda)$ a unitary operator acting on the Hilbert space.} Here, $D_{\sigma \sigma'}(W(\Lambda, p; k))$ is a representation of the little group which is the subgroup of Lorentz transformations that leave the momentum invariant (that is, the stabilizer of the Lorentz group) and $W(\Lambda, p, k)=L^{-1}(\Lambda p; k)\,\Lambda\,L(p; k)$ is a little-group element since the Lorentz transformations act as $p_\mu=L_\mu^{\ \nu}(p; k)\,k_\nu$.  
	{The massive little group is $SO(d-1)$, while the rotational part of the massless little group is $SO(d-2)$\footnote{Here we ignore the $d-2$ translations because finite-dimensional representations transform trivially under them.}.}  
	This can be seen by noting that we can always choose a frame such that the momentum for a massless particle is of the form $p^\mu=(E,E,0,\dots,0)$ and for massive $p^\mu=(m,0,\dots,0)$; the little groups are just the Lorentz transformations that leave these momenta invariant.  
	Thanks to the Poincaré invariance of the S-matrix, we can now infer that it transforms as
	\begin{align}
		\mathcal{M}(\{p_i, \sigma_i\}) 
		&= \delta^D\left({\sum_{i=1}^n p_i^\mu}\right)\,
		M(\{p_i,\sigma_i\}) \ , \nonumber  \\
		M^\Lambda(\{p_i,\sigma_i\}) 
		&= {\prod_{i=1}^n} \bigl(D_{\sigma_i \sigma'_i}(W)\bigr)\,
		M\bigl(\{(\Lambda p)_i,\sigma'_i\}\bigr) \ .
	\end{align}
	
	In the standard quantum-field-theory approach, we compute tensorial objects through Feynman rules and dress them with polarization vectors. These polarization vectors are bifundamentals of the Lorentz and little group, $\epsilon^\mu_\sigma(p)$, and allow us to construct scattering amplitudes with the correct transformation properties. Furthermore, for massless particles, these are not really vectors under the Lorentz group—only their equivalence class under gauge transformations ($\epsilon^\mu\sim\epsilon^{\prime\mu}+\alpha\,p^\mu$) is. When using these representations, Poincaré covariance is not manifest owing to the choice of a reference vector that gives the appropriate little-group transformations of the scattering amplitude. It is possible to avoid these redundancies and the lack of manifest covariance by always working with on-shell objects that have the desired transformations under the little group. These are the spinor-helicity variables.
	
	We will work in four dimensions with a Minkowski signature. {For standard references with more detail on spinor-helicity variables, see \cite{Dixon:2013uaa,Elvang:2013cua,Mangano:1990by,Arkani-Hamed:2017jhn,Dixon:1996wi,Conde:2016izb}; for generalizations to other dimensions, see \cite{Cheung:2009dc,Boels:2012ie,Chiodaroli:2022ssi,Pokraka:2024fao,Caron-Huot:2010nes}.}  
	For massless particles, the little group is $SO(2)$, and for massive particles it is $SO(3)$. It is common to exploit the 2-to-1 homomorphisms and work instead with $U(1)$ and $SU(2)$. A scattering amplitude with $n$ massive particles and $m$ massless particles (all outgoing) transforms as
	\begin{align}
		M&^{\{I_1\!\ldots\!I_{2s_1}\},\cdots,
			\{J_1,\!\cdots\! ,J_{2s_n}\},
			\{h_1\},\cdots,\{h_m\}}  \nonumber\\
		\longrightarrow &\;
		\Bigl((W_1)^{I_1}_{K_1}\!\cdots(W_1)^{I_{2s_1}}_{K_{2s_1}}\Bigr)
		\!\cdots
		\Bigl((W_n)^{J_1}_{L_1}\!\cdots(W_n)^{J_{2s_n}}_{L_{2s_n}}\Bigr)
		(w_1)^{2h_1}\!\cdots\! (w_m)^{2h_m}\!
		\nonumber\\
		&\times
		M^{\{K_1\!\ldots\!K_{2s_1}\},\cdots,
			\{L_1,\!\cdots\! ,L_{2s_n}\},
			\{h_1\},\cdots,\{h_m\}} \, . 
		\label{eq:Smatrix_lg_transf}
	\end{align}
	Here $W_i$ is an $SU(2)$ transformation for the $i$-th massive particle in the spin-${\textstyle\frac12}$ representation and $w_i=e^{i\theta}$ is the little-group phase (helicity $+\tfrac12$) for the $i$-th massless particle. In other words, massive particles of spin $s$ are labeled by symmetric $SU(2)$ tensors of rank $2s$, and massless particles are labeled by their helicity. 
	
	\bigskip
	\noindent\textbf{Massless case.}  
	A representation of the Poincaré algebra $\mathfrak{iso}(1,3)$ can be given in terms of a two-component commuting Weyl spinor $\lambda$ and its complex conjugate $\tilde\lambda$ as\footnote{If we allow complex momenta, $\tilde\lambda$ is independent of $\lambda$.}
	\begin{equation}
		p_{\alpha\dot{\beta}}
		= \lambda_\alpha\,\tilde\lambda_{\dot{\beta}}, \quad
		L_{\alpha\beta} 
		= \lambda_{(\alpha}\frac{\partial}{\partial\lambda^{\beta)}}, \quad
		\tilde L_{\dot{\alpha}\dot{\beta}}
		= \tilde\lambda_{(\dot{\alpha}}
		\frac{\partial}{\partial\tilde\lambda^{\dot{\beta})}} \, ,
	\end{equation}
	where we use the 2-to-1 homomorphism $SL(2,\mathbb{C})\!\to\!SO(1,3)$ to rewrite Lorentz indices in terms of spinor indices.  
	This is done via the $\sigma$-matrices; thus
	\be
	v_{\alpha\dot{\alpha}}
	\;=\;
	v_\mu(\sigma^\mu)_{\alpha\dot{\alpha}}, 
	\qquad
	\sigma^\mu=\{\mathrm{I},\sigma^i\},
	\ee
	with $\sigma^i$ the Pauli matrices and $\mathrm{I}$ the $2\times2$ identity.  
	From this, it follows that
	\begin{equation}
		p^2 = 0\,,
	\end{equation}
	so $\lambda$ and $\tilde\lambda$ furnish a massless representation and are called \textbf{spinor-helicity variables}. These variables allow us to write scattering amplitudes of {Yang--Mills} theory and gravity in a simpler form than using Feynman rules, since they have well-defined Lorentz and little-group transformations:
	\be
	\lambda_\alpha^{(\Lambda p)}
	= w^{-1}\,\Lambda_\alpha^{\ \beta}\,
	\lambda_\beta^{(p)} .
	\ee
	Thus, the little-group transformation in Eq.~\eqref{eq:Smatrix_lg_transf} for a massless particle is generated by
	\begin{equation}
		H \;=\;
		\frac{1}{2}\left(
		\tilde\lambda_{\dot{\alpha}}
		\frac{\partial}{\partial\tilde\lambda_{\dot{\alpha}}}
		-\lambda_\alpha\frac{\partial}{\partial\lambda_\alpha}
		\right)\,,
	\end{equation}
	i.e.\ $\lambda_\alpha$ (\,$\tilde\lambda_{\dot\alpha}$\,) carries helicity $-\tfrac12$ (respectively $+\tfrac12$).
	
	\bigskip
	\noindent\textbf{Massive case{.}}  
	A representation of $\mathfrak{iso}(1,3)$ is given by \cite{Conde:2016izb}
	\begin{equation}
		p_{\alpha\dot{\beta}}
		= \lambda_\alpha^{\ I}\tilde\lambda_{I\dot{\beta}},\quad
		L_{\alpha\beta}
		= \lambda_{(\alpha}^{\ I}
		\frac{\partial}{\partial\lambda_{\beta)I}},\quad
		\tilde L_{\dot{\alpha}\dot{\beta}}
		= \tilde\lambda_{I(\dot{\alpha}}
		\frac{\partial}{\partial\tilde\lambda^{I}_{\dot{\beta})}} .
	\end{equation}
	Here $I=1,2$ is an $SU(2)$ fundamental index. We now have
	\begin{equation}
		p^2 \neq 0\,,
	\end{equation}
	and the traceless part of the little-group generators for massive particles in Eq.~\eqref{eq:Smatrix_lg_transf} are
	\begin{equation}
		J^I_{\ J}=K^I_{\ J}-\frac{1}{2}\delta^I_J K^K_{\ K}\ ,\qquad
		K^I_{\ J}
		= \lambda_\alpha^{\ I}
		\frac{\partial}{\partial\lambda_\alpha^{\ J}}
		-\tilde\lambda_{J\dot{\alpha}}
		\frac{\partial}{\partial\tilde\lambda^{I}_{\dot{\alpha}}}\, .
	\end{equation}
	
	\medskip
	\vspace{0.4em}
	Similar approaches have been used to construct correlation functions in AdS in coordinate space, as reviewed in Sec.~\ref{sec:correll_embedding}. Additionally, spinors are useful for constructing exact solutions in gravitational theories and can expose classical double copy relations (see Sec.~\ref{sec:classical_Weyl}).
	
	\bigskip
	\noindent\textbf{Example: MHV amplitudes.}  
	{Before concluding this brief summary, we give a simple example illustrating the power of these variables.} For the maximally helicity-violating (MHV) configuration, that is, a gluon amplitude with two negative and $n\!-\!2$ positive helicities, the {Yang--Mills} amplitude is given by the {Parke--Taylor} formula \cite{Parke:1986gb}
	\begin{equation}
		A_{n}\!\big(1^{+},\ldots,i^{-},\ldots,j^{-},\ldots,n^{+}\big)
		= i\frac{\langle i\,j\rangle^{4}}{\langle 1\,2\rangle\,\langle 2\,3\rangle\,\cdots\,\langle n-1,\,n\rangle\,\langle n,\,1\rangle} \ . 
	\end{equation}
	{Here, we employ the standard bra--ket notation}
	\begin{equation}
		\lambda_{i\alpha}=|i\rangle_\alpha,\quad
		\lambda_i^{\alpha}=\langle i|^\alpha,\quad
		\tilde\lambda_{i\dot{\alpha}}=|i]_{\dot{\alpha}},\quad
		\tilde\lambda_i^{\dot{\alpha}}=[i|^{\dot{\alpha}}\,,
	\end{equation}
	and the $SL(2,\mathbb{C})$ indices are raised and lowered using
	$\epsilon_{\alpha\beta}$ and $\epsilon_{\dot{\alpha}\dot{\beta}}$ according to $\lambda^\alpha=\epsilon^{\alpha\beta}\lambda_\beta$ and $\lambda_\alpha=\epsilon_{\alpha\beta}\lambda^\beta$, with $\epsilon^{01}=-\epsilon_{01}=1$, and analogously for dotted indices; in particular, $\langle ij\rangle=\epsilon^{\alpha\beta}\lambda_{i\alpha}\lambda_{j\beta}$ and $[ij]=\tilde{\epsilon}^{\dot{\alpha}\dot{\beta}}\tilde{\lambda}_{i\dot{\alpha}}\tilde{\lambda}_{j\dot{\beta}}$.
	
	The simplest three-point example is given by
	\begin{equation}
		A_3\!\bigl[1^-\,2^-\,3^+\bigr]
		\;=\;
		\frac{\langle12\rangle^{3}}
		{\langle23\rangle\,\langle31\rangle} \ ,
	\end{equation}
	where we have considered complex momenta, so that the angle and square brackets are independent, allowing us to obtain a non-vanishing result. The corresponding three-point gravity amplitude reads
	\begin{equation}
		\mathcal{M}_3\!\bigl(1^-\,2^-\,3^+\bigr)
		\;=\;
		\frac{\langle12\rangle^{6}}
		{\langle23\rangle^{2}\langle31\rangle^{2}}\,,
	\end{equation}
	making the double copy structure manifest. {Deriving these expressions at three points requires only little-group covariance and simple kinematics, whereas a Feynman-diagram calculation would be far more tedious. The $n$-point expression can then be derived using BCFW recursion relations~\cite{Britto:2005fq}.}
	
	Since in flat space spinor-helicity variables have proved extremely powerful for uncovering double copy structures, it is natural to ask whether similar ideas might help in curved {spacetimes}. One immediate difficulty is that many existing constructions of spinor-helicity variables on curved backgrounds are not fully covariant, but new approaches might prove fruitful; see \cite{Basile:2024ydc,Maldacena:2011nz,Nagaraj:2018nxq,Nagaraj:2019zmk,Nagaraj:2020sji,Skvortsov:2022wzo,David:2019mos} for various proposals in (A)dS.

	\section{Correlators and Scattering Observables in Curved Space}\label{sec:curved_observables}
	
	In curved spacetimes, the absence of a conventional S-matrix shifts the focus to boundary correlators, wavefunction coefficients, and scattering observables defined on special backgrounds. Double copy structures have been investigated in several representations, each making different aspects of symmetry, factorization, and locality manifest. In this section, we review these constructions in momentum, Mellin, embedding, isometric, coordinate, twistor, and Grassmannian space, before discussing scattering on non-trivial backgrounds.
	
	\subsection{AdS and dS observables}
	We use unit-radius Poincar\'e coordinates throughout this discussion. For de Sitter, with conformal time $\eta<0$, and Euclidean Anti-de Sitter, with radial coordinate $z>0$, the metrics are
	\be \label{eq:PoincarePatch}
	ds^2_{\rm dS}=\frac{-d\eta^2+d\bx^2}{\eta^2}\ ,\qquad
	ds^2_{\rm EAdS}=\frac{dz^2+d\bx^2}{z^2}\ ,
	\ee
	The Wick rotation between these patches is given by $\eta\rightarrow iz$, with the conventional accompanying continuation of the curvature radius and overall sign understood. Dot products are taken in Euclidean signature for Euclidean Anti-de Sitter and in the corresponding Lorentzian signature for de Sitter or Lorentzian Anti-de Sitter.
	
	Before reviewing the momentum-space constructions, it is useful to be clear about which observables are being considered. In flat space the natural on-shell observable is the S-matrix, while in AdS and dS one usually works with boundary quantities. In AdS, the standard object is a boundary CFT correlator. Equivalently, the bulk path integral with boundary condition $\phi(z,\bx)\sim z^{d-\Delta}\phi_0(\bx)$ defines the generating functional~\cite{Maldacena:1997re,Witten:1998qj}
	\be
	Z_{\rm CFT}[\phi_0]
	=Z_{\rm AdS}[\phi\rightarrow\phi_0]
	=\left\langle
	\exp\left(\int d^d\bx\,\phi_0(\bx)\mathcal{O}(\bx)\right)
	\right\rangle ,
	\label{eq:ads_boundary_generating_functional}
	\ee
	and differentiating with respect to $\phi_0$ gives correlators of the dual operator $\mathcal{O}$. In practice the perturbative bulk calculation can be performed using Witten diagrams, and the result is a CFT correlator written in position, momentum, Mellin, or another useful representation.
	
	In dS the analogous object is the late-time wavefunction of the universe. This is the same type of wavefunctional introduced in Sec.~\ref{sec:ckwavefunc} in the flat-space wavefunction discussion, but now the final time is taken near the future boundary of dS and the initial condition is usually the Bunch--Davies vacuum. For a scalar field one writes schematically
	\be
	\Psi_{\rm dS}[\varphi_f,\eta_*]
	=\int_{\rm BD}^{\varphi(\eta_*)=\varphi_f}\mathcal{D}\varphi\,
	e^{iS_{\rm dS}[\varphi]}\ ,
	\label{eq:ds_wavefunction_path_integral}
	\ee
	and expands $\log\Psi_{\rm dS}$ in late-time boundary fields, with coefficients $\psi_n$ as in the earlier wavefunction discussion in Sec.~\ref{sec:ckwavefunc}. These coefficients are often called cosmological wavefunctions or cosmological correlators. They are not equal to expectation values of late-time fields, but they are the basic building blocks from which those expectation values are obtained.
	
	The relation between AdS boundary correlators and dS wavefunction coefficients is an analytic continuation. Roughly, one computes the corresponding Euclidean anti-de Sitter (EAdS) Witten diagram and continues the radial coordinate and external data to dS variables, with possible phases, local counterterms, and normalization factors depending on conventions \cite{Maldacena:2011nz,Sleight:2020obc}. This is why the same momentum-space functions often appear in AdS boundary correlators and dS wavefunction coefficients, and why total-energy singularities have the same flat-space interpretation reviewed below \cite{Raju:2012zr}. In this review we will often use ``(A)dS correlator'' for the common analytic structure, while keeping in mind whether the physical interpretation is an AdS CFT correlator or a dS wavefunction coefficient.
	
	Finally, late-time in-in correlators in dS are obtained by integrating the boundary fields against the probability measure defined by the wavefunction,
	\be
	\left\langle Q[\varphi]\right\rangle_{\rm in-in}
	=
	\frac{\int\mathcal{D}\varphi\,Q[\varphi]\,|\Psi_{\rm dS}|^2}
	{\int\mathcal{D}\varphi\,|\Psi_{\rm dS}|^2}\ .
	\label{eq:inin_from_wavefunction}
	\ee
	For example, at leading order and with the common dS convention for the wavefunction coefficients used in \cite{Chowdhury:2025nnk}, the first few scalar correlators are
	\begin{align}
		\left\langle\varphi(\bk_1)\varphi(\bk_2)\right\rangle_{\rm in-in}
		&=\frac{1}{2\,{\rm Re}\,\psi_2(\bk_1,\bk_2)},\nonumber\\
		\left\langle\varphi(\bk_1)\varphi(\bk_2)\varphi(\bk_3)\right\rangle_{\rm in-in}
		&=\frac{{\rm Re}\,\psi_3(\bk_1,\bk_2,\bk_3)}
		{\prod_{a=1}^{3}{\rm Re}\,\psi_2(\bk_a,-\bk_a)}\ .
		\label{eq:inin_low_point_wavefunction}
	\end{align}
	Perturbatively, the quadratic part of $\log|\Psi_{\rm dS}|^2$ gives the late-time propagator and the higher $\psi_n$ act as vertices in this boundary integral. Thus wavefunction coefficients and in-in correlators carry closely related information, but they are not the same object. This distinction will become important below: some double copy relations are statements about AdS correlators or dS wavefunction coefficients, while the cosmological dressing rules are instead organized directly for dS in-in correlators.

	\subsection{Momentum space} \label{sec:Momentum_space}
	Given that the double copy in flat backgrounds naturally arises for scattering amplitudes in momentum space, one can explore whether a similar feature persists in maximally symmetric backgrounds, or even other less symmetric backgrounds. Momentum space is also particularly well suited to phenomenology, since many cosmological observables are described by correlation functions of matter overdensities in momentum space. More precisely, matter overdensities are not observed directly, but are inferred through tracers such as galaxy number counts and gravitational lensing. In turn, these matter overdensities arise due to the evolution of quantum fluctuations in the early universe. The latter can be computed using field theory methods and are the subject of study here. There have been many explorations regarding momentum space double copies in curved backgrounds which we summarize below before showing an explicit example.
	
	A double copy structure for three-point CFT correlators in $d\geq3$ was found in \cite{Farrow:2018yni} for odd $d$ and \cite{Lipstein:2019mpu} for even $d$. A similar relation for the type-A Euler-anomaly tensor structure in four dimensions was shown in \cite{Bzowski:2017poo} to factorize as the square of the chiral-anomaly structure. Using the complete three-point correlators, this relation was extended in \cite{Coriano:2026zag}: after isolating the anomaly-controlled part of $\langle TTT\rangle$, its flat-space limit is obtained by squaring that of the parity-odd $\langle JJJ\rangle$.In refs.~\cite{Albayrak:2020fyp,Armstrong:2020woi,Jain:2021qcl,Alday:2021odx} it was shown that the residue of the total energy pole of the 4-point graviton (A)dS correlator is given as the square of the residue of the four-point Yang-Mills (A)dS correlator. These relationships, found at the total energy poles, follow from the fact that the residue at this pole is given by the flat space amplitude in one higher dimension \cite{Maldacena:2011nz,Raju:2012zr}. The double copy can be extended beyond this special kinematic limit if one considers a special splitting of the correlators with a momentum-dependent proportionality factor~\cite{Farrow:2018yni,Jain:2021qcl}. At three-points this kernel can be traced to integrations over conformal factors. In inflationary backgrounds, double copy relations for in-in correlators have been studied in \cite{Li:2018wkt,Fazio:2019iit}. KLT-like relations for inflationary four-point correlators were investigated in \cite{Li:2018wkt} where it was shown that even at the energy pole, the double copy fails to reproduce the graviton correlators without adding additional contributions not coming from the gauge theory. Similar results were obtained in \cite{Fazio:2019iit} for three and four-point correlators of gravitons and scalars.
	
	Here, we will briefly review the case of CFT correlators in odd dimensions. For the conserved-current and spinning correlators considered below, odd-dimensional examples are rational functions, whereas scalar seed correlators can contain logarithms; even-dimensional correlators generally have a more involved analytic structure. We will also focus on correlators of conserved currents on the boundary, whose conformal weight is $\Delta_s=d+s-2$, and which are dual to massless spinning fields in the bulk. To make the connection with the bulk scattering amplitudes clear, we proceed to dress the correlators with polarization vectors and interpret them as boundary correlators in the even-dimensional (A)dS bulk. We set the notation and conventions in the following. 
	
	We will work in a four-dimensional (4d) bulk and three-dimensional (3d) boundary where the CFT lives. For the auxiliary four-momentum below, we use Lorentzian signature and set $k_i=|\bk_i|$; Euclidean expressions are obtained by analytic continuation. The 3d momentum of the $i$th particle is $\bk_i$ which allows us to define a 4d null momentum as
	\be
	k_i^\mu=(k_i,\bk_i) \ .
	\ee
	The total {\it energy}, $E$, is defined as
	\be
	E=\sum_i k_i\ ,
	\ee
	and is not vanishing for physical kinematics. While the energy is not conserved, the spatial momentum is 
	\be
	\sum_i\bk_i=0\ .
	\ee
	The polarization vectors that we use to contract the Lorentz indices of the CFT correlators satisfy
	\be
	\boldsymbol{\epsilon}_i \cdot \boldsymbol{\epsilon}_i = 0, \quad \boldsymbol{\epsilon}_i \cdot \mathbf{k}_i = 0 \ ,
	\ee
	and the 4d version that are used to relate to scattering amplitudes read
	\be
	\epsilon_i^\mu=(0,\boldsymbol{\epsilon}_i) \ .
	\ee
	Boundary correlators in (A)dS have a pole at the total energy $E$, which can only be accessed for unphysical momentum choices through analytic continuation. As mentioned earlier, the leading order residue of this pole corresponds to the flat space scattering amplitudes \cite{Maldacena:2011nz,Raju:2012zr}. We proceed to look at several examples of double copy structures. In the following, we will ignore the color factors and the coupling constants for simplicity.
	
	\subsubsection{Three-point correlators}
	For the double copy of three-point correlators in momentum space, we follow the results of \cite{Farrow:2018yni,Jain:2021qcl}. We start by looking at the simpler case of the CFT correlator between two conserved (spin one) currents and a marginal scalar operator is given by \cite{Bzowski:2013sza,Bzowski:2018fql}
	\be
	\langle J J \mathcal{O}\rangle=-\frac{C_1}{2 E^2}\left(E+k_3\right)\left(2 \epsilon_1 \cdot k_2 \epsilon_2 \cdot k_1+E\left(E-2 k_3\right) \epsilon_1 \cdot \epsilon_2\right) \ , \label{eq:JJO}
	\ee
	where $C_1$ is a constant. The flat space limit of the correlator is obtained from the leading contribution in the vanishing total energy limit as
	\be
	\lim_{E \to 0} \frac{E^{2}}{k_3} \langle JJO \rangle \propto \mathcal{A}_{\phi F^2} \ ,
	\ee
	where $\mathcal{A}_{\phi F^2}$ is the on-shell scattering amplitude for a dilaton-gluon-gluon interaction and is given by
	\be
	\mathcal{A}_{\phi F^2} = \boldsymbol{\epsilon}_1 \cdot \mathbf{k}_2 \, \boldsymbol{\epsilon}_2 \cdot \mathbf{k}_1 \ .
	\ee
	
	One can further relate the correlator $\langle J J \mathcal{O}\rangle$ to amplitudes in the bulk without energy conservation and dressed with a conformal time or radial direction integral. In the conformally flat coordinates of Eq.~\eqref{eq:PoincarePatch}, both the field strength $F^{\mu\nu}$ and the Weyl tensor $W^\mu_{\nu\rho\sigma}$ are proportional to their flat space counterparts up to a conformal factor. Given this, we can compute flat space amplitudes without imposing energy conservation, and recover the (A)dS boundary correlators by rescaling with the appropriate conformal factors and integrating over the conformal time/radial direction. For example, for the present case of the dilaton-gluon-gluon amplitude we have that in flat space without energy conservation
	\be
	\mathcal{M}_{\phi F^2} \propto \eta^{\mu\rho}\eta^{\nu\sigma}F_{\mu\nu}^{(1)} F_{\rho\sigma}^{(2)}\propto 2 \, {\epsilon}_1 \cdot {k}_2 \, {\epsilon}_2 \cdot {k}_1 
	+ E(E - 2k_3) \, {\epsilon}_1 \cdot {\epsilon}_2\ , \label{eq:F2phiNoE}
	\ee
	where $\eta^{\mu\rho}$ is the Minkowski metric,
	\be
	F_{\mu\nu}^{(\alpha)} = k_{[\mu}^{(\alpha)} \epsilon_{\nu]}^{(\alpha)}\ ,
	\ee
	and $\alpha$ labels the particle index. To explicitly recover Eq.~\eqref{eq:JJO}, we need to recover the appropriate conformal factors and perform the missing integral. For concreteness, we show this for the case of dS$_4$. The missing conformal factors come from the measure, $\sqrt{-g}$, which get canceled with the factors coming from inverse metrics, and finally a conformal factor from the wave function for the marginal scalar. Thus, we find that the amplitude in Eq.~\eqref{eq:F2phiNoE} has to be dressed with the factor
	\be
	\mathrm{Im} \left( \int_{-\infty}^0 \mathrm{d}\eta \, (1 - i k_3 \eta) e^{i E \eta} \right) \propto \frac{E + k_3}{E^2},
	\ee
	which recovers the leading contribution as $E\rightarrow 0$ of Eq.~\eqref{eq:JJO} 
	\be
	\langle JJO \rangle \propto \frac{E + k_3}{E^2} \mathcal{M}_{\phi F^2} \ ,
	\ee
	
	Now we look at what is expected to correspond to the double copy, that is, the $\braket{TT\cO}$ correlator. This correlator is given by
	\be
	\braket{TT\cO} = C_1 \ k_{1}k_{2}\  \frac{(E + 3k_3)}{E^4} 
	\left( 2 \, \boldsymbol{\epsilon}_1 \cdot \mathbf{k}_2 \, \boldsymbol{\epsilon}_2 \cdot \mathbf{k}_1 
	+ E (E - 2k_3) \, \boldsymbol{\epsilon}_1\cdot \boldsymbol{\epsilon}_2 \right)^2 \  .
	\ee
	In the flat space limit, we can see the double copy relation
	\be
	\lim_{E \to 0} \frac{E^4}{k_1k_2k_3} \langle TTO \rangle \propto \mathcal{A}_{\phi R^2}=\left(\mathcal{A}_{\phi F^2} \right)^2 \propto\left(\lim_{E \to 0} \frac{E^{2}}{k_3} \langle JJO \rangle \right)^2 \ ,
	\ee
	which follows straightforwardly from the flat space double copy. As in the previous case, one can relate the correlator to the flat space scattering amplitude of a dilaton-graviton-graviton without energy conservation which is given by
	\be
	\mathcal{M}_{\phi R^2} \propto \eta^{\mu\alpha}\eta^{\nu\beta}\eta^{\rho\gamma}\eta^{\sigma\xi} W_{\mu\nu\rho\sigma}^{(1)} W_{\alpha\beta\gamma\xi}^{(2)} \ ,
	\ee
	where the linearized Weyl tensor reads
	\be
	W_{\mu\nu\rho\sigma}^{(\alpha)} = k_{[\mu}^{(\alpha)} \epsilon_{\nu]}^{(\alpha)} k_{[\rho}^{(\alpha)} \epsilon_{\sigma]}^{(\alpha)} \ .
	\ee
	The dressing factor (again computed here in de Sitter for concreteness) is obtained by multiplying by one factor from the measure, 4 factors of inverse metrics, two factors from the Weyl tensor ($- i k_\alpha \eta$), and one factor from the marginal scalar wavefunction and integrating over conformal time
	\be
	k_1 k_2 \, \mathrm{Im} \left( \int_{-\infty}^0 \mathrm{d}\eta \, \eta^2 (1 - i k_3 \eta) e^{i E \eta} \right) 
	\propto k_1 k_2 \frac{(E + 3k_3)}{E^4}.
	\ee
	As it can now be seen, in this simple case a double copy relation holds beyond the flat space limit thanks to the double copy holding for flat space correlators without energy conservation
	\be
	\langle TTO \rangle \propto \frac{k_1k_2 (E + 3k_3)}{E^4} \left( \mathcal{M}_{\phi F^2} \right)^2 
	\propto \frac{k_1k_2 (E + 3k_3)}{(E + k_3)^2} \langle JJO \rangle^2 \ .
	\ee
	The structure of this double copy relation is simple and the proportionality factors can be related to conformal factors integrated along the conformal time or radial direction of the bulk. Similar features follow for more complicated cases as we see in the following. Let's now consider the $\langle JJJ \rangle$ and $\langle TTT \rangle$\footnote{Here we have removed a contact term from the $\langle TTT \rangle$ correlator by considering the perturbed metric $g_{\mu\nu}=(e^{\gamma})_{\mu\nu}$ and taking functional derivative now with respect to $\gamma$, which takes $\langle TTT \rangle\rightarrow \langle TTT \rangle - 2 C_{TT} (k_1^3 + k_2^3 + k_3^3) \tilde{\mathcal{A}}_{\text{contact}}$, where $\tilde{\mathcal{A}}_{\text{contact}}= {\epsilon}_1 \cdot {\epsilon}_2 \, {\epsilon}_2 \cdot {\epsilon}_3 \, {\epsilon}_3 \cdot {\epsilon}_1$} correlators \cite{Bzowski:2013sza,Bzowski:2017poo} 
	\begin{align}
		\langle JJJ \rangle &= \frac{C_1}{E^3} \mathcal{M}_{F^3} - \frac{2C_{JJ}}{E} \mathcal{A}_{\text{YM}} \\
		\langle TTT \rangle & = 
		- \frac{960 \, C_1 k_1k_2k_3}{E^6}  \mathcal{M}_{W^3} \nonumber \\
		&+ 2 \frac{C_{TT}}{E^2} \left(k_1k_2k_3+ E(k_1 k_2 + k_2 k_3 + k_3 k_1)- E^3 \right) \mathcal{A}_{\text{EG}}  \ ,
	\end{align}
	where $C_1$, $C_{JJ}$, and $C_{TT}$ are constants, the two latter related to the normalization of the two-point functions. The $C_1$ terms arise from higher derivative bulk terms, either $F^3$ or $W^3$, while the $C_{JJ}$ and $C_{TT}$ ones come from the Yang-Mills and Einstein-Hilbert interactions respectively. The flat space amplitudes without momentum conservation are
	\begin{align}
		\mathcal{M}_{F^3} & =2\mathcal{A}_{F^3} + E \tilde{\mathcal{A}}_{\text{YM}}  \nonumber \\
		&\propto \eta^{\nu\alpha}\eta^{\rho\beta}\eta^{\mu\zeta}F_{\mu\nu}^{(1)} F_{\alpha\rho}^{(2)} F_{\beta\zeta}^{(3)},\\ 
		\mathcal{M}_{W^3} & = \frac{k_1k_2k_3 E}{J^2}\left( \mathcal{A}_{W^3} + \frac{1}{2} E \mathcal{A}_{F^3} \tilde{\mathcal{A}}_{\text{YM}} \right)\nonumber\\
		&\propto \eta^{\rho\alpha}\eta^{\lambda\beta}\eta^{\omega\gamma}\eta^{\sigma\xi}\eta^{\mu\theta}\eta^{\nu\zeta} W_{\mu\nu\rho\lambda}^{(1)} W_{\alpha\beta\omega\sigma}^{(2)} W_{\gamma\xi\theta\zeta}^{(3)}  \ ,
	\end{align}
	with $J^2 = E \, (k_1 + k_2 - k_3)(k_1 - k_2 + k_3)(-k_1 + k_2 + k_3)$, 
	\be
	\tilde{\mathcal{A}}_{\text{YM}} = ({\epsilon}_1 \cdot {\epsilon}_2 \, {\epsilon}_3 \cdot {k}_1) k_3 + \text{cyclic} \ ,
	\ee
	which resembles an amplitude but it is not, and 
	\begin{align}
		\mathcal{A}_{\text{YM}}&=\epsilon_1 \cdot \epsilon_2 \, \epsilon_3 \cdot k_1 + \text{cyclic} \ ,\\
		\mathcal{A}_{F^3} &=\epsilon_1 \cdot k_2 \, \epsilon_2 \cdot k_3 \, \epsilon_3 \cdot k_1 \ , \\
		\mathcal{A}_{\text{EG}} & =(\mathcal{A}_{\text{YM}})^2 \ , \\
		\mathcal{A}_{W^3} & = (\mathcal{A}_{F^3})^2 \ ,
	\end{align}
	which correspond to the 3-point on-shell amplitudes of Yang-Mills, $F^3$, Einstein-Hilbert gravity and Weyl cubed, $W^3$, respectively. 
	
	It is clear again that each contribution from the leading terms and next-to-leading higher-derivative corrections double copy separately up to a factor arising from integrated conformal factors \cite{Farrow:2018yni}.
	\begin{align}
		\langle TTT \rangle |_{C_{TT}=0}&\propto k_1k_2k_3 (\langle JJJ \rangle|_{C_{JJ}=0})^2 \, \\
		\langle TTT \rangle |_{C_1=0}&\propto \left(k_1k_2k_3+ E(k_1 k_2 + k_2 k_3 + k_3 k_1)- E^3 \right)(\langle JJJ \rangle|_{C_1=0})^2 \ .
	\end{align}
	The higher-derivative contributions are sometimes referred to as homogeneous since they do not contribute to the conformal Ward identity, which relates the longitudinal part of the correlator to the two-point function, and are thus identically conserved \cite{Baumann:2020dch,Baumann:2021fxj}. A possible expectation would be that one can double copy both contributions at the same time similar to the flat space case where $\left( \mathcal{A}_{YM} + \alpha' \mathcal{A}_{F^3} \right)^2 
	= \mathcal{A}_{EG} + 2\alpha' \mathcal{A}_{\phi R^2}^{222} + \alpha'^2 \mathcal{A}_{W^3}$. This does not happen in the present $d=3$ case since the $\phi R^2$ contribution to the three-point stress tensor correlator vanishes due to tensor degeneracies, but perhaps the version written above could be rewritten using these degeneracies in a way such that the full double copy holds.

	\subsubsection{Four-point correlators}
	The color-kinematics duality and BCJ double copy for four-point (A)dS correlators was considered in \cite{Armstrong:2020woi,Albayrak:2020fyp} and found to suffer from issues similar to those reviewed in the Subsection \ref{sec:ckwavefunc} for the flat space wavefunction. Here, we give a brief review of these constructions. To start, we need to define the analog of the Mandelstam variables as \cite{Baumann:2020dch}
	\begin{align}
		s &= \left( k_{12} + \underline{k}_{12} \right) \left( k_{34} + \underline{k}_{34} \right), \\
		t &= \left( k_{14} + \underline{k}_{14} \right) \left( k_{23} + \underline{k}_{23} \right), \\
		u &= \left( k_{13} + \underline{k}_{13} \right) \left( k_{24} + \underline{k}_{24} \right),
	\end{align}
	where $\underline{k}_{ij} = |\mathbf{k}_i + \mathbf{k}_j|$ and $k_{ij} = k_i + k_j$. The color-ordered correlators have the same structure as the flat space amplitudes
	\begin{align}
		\langle j_1 j_2 j_3 j_4 \rangle &= \frac{n_s}{s} - \frac{n_t}{t} \\
		\langle j_1 j_3 j_2 j_4 \rangle &= \frac{n_t}{t} - \frac{n_u}{u},
	\end{align}
	but the kinematic numerators do not satisfy the color-kinematics duality $n_s+n_t+n_u\neq 0$. Just as in the flat space wavefunction case, we can perform a generalized gauge transformation to enforce the color-kinematics duality since 
	\be
	s + t + u = \xi, \quad  \xi\equiv E \left( E + \underline{k}_{12} + \underline{k}_{23} + \underline{k}_{13} \right) \ ,
	\ee
	but just as in the wavefunction case, since the matrix $G=MDM^{T}$ from Section~\ref{sec:bcj} has full rank (here $s + t + u  \neq 0$), there are no standard BCJ relations. Instead we have the deformed version
	\be
	u \langle j_1 j_3 j_2 j_4 \rangle - s \langle j_1 j_2 j_3 j_4 \rangle = \xi \, \frac{\tilde{n}_t}{t} \ ,
	\ee
	where $\tilde{n}_t$ is the deformed CK duality satisfying numerator. This tells us that our basis has two instead of just one amplitude. The double copied amplitude will now have unphysical poles at $\xi=0$ as seen from Eq.~\eqref{eq:DCwithnoCK}. Hence, the double copy of Yang-Mills four-point correlators in momentum space does not give the correct gravitational four-point correlator. Nevertheless, it has been shown that the (A)dS graviton four-point function \cite{Bonifacio:2022vwa,Raju:2012zs}, which has a complicated expression, can be largely simplified when writing the wavefunction coefficient as a double copy plus additional terms that can be fixed by physical principles \cite{Armstrong:2023phb}. 
	
	\subsubsection{Differential operators} \label{sec:mom_dif_op}
	Another approach that has been proposed to construct a double copy relation at three-points in momentum space is the use of differential operators \cite{Lee:2022fgr}. The idea consists of writing (A)dS correlation functions of conserved currents by using combinations of weight-shifting operators \cite{Costa:2011dw,Costa:2018mcg,Karateev:2017jgd,Baumann:2019oyu} acting on a scalar correlator as
	\be
	\langle J_s J_s J_s \rangle = \hat{n}_s \langle \Phi \Phi \Phi \rangle \ ,
	\ee
	with $\hat{n}_s$ a combination of weight shifting operators in a specific ordering defined below. The weight-shifting operators in momentum space are given by \cite{Baumann:2019oyu}
	\begin{subequations}
		\begin{align}
			S_{ab} &\equiv \rho_a \rho_b (\mathbf{z}_a \cdot \mathbf{z}_b) 
			+ (\mathbf{z}_b \cdot \mathbf{k}_b) D_{ab} 
			+ (\mathbf{z}_a \cdot \mathbf{k}_a) D_{ba} \nonumber \\
			&\quad + (\mathbf{z}_a \cdot \mathbf{k}_a)(\mathbf{z}_b \cdot \mathbf{k}_b) W_{ab} \ , 
			\\
			D_{ab} &\equiv \rho_a (\mathbf{z}_a \cdot \mathbf{K}_{ab}) 
			- (\mathbf{z}_a \cdot \mathbf{k}_a) W_{ab} \ , 
			\\
			F_{ab} &\equiv (\mathbf{k}_b \cdot \mathbf{K}_{ab} + \Delta_b - d) 
			\mathbf{z}_a \cdot \mathbf{K}_{ab} 
			- (\mathbf{z}_b \cdot \mathbf{K}_{ab})(\mathbf{z}_a \cdot \partial_{\mathbf{z}_b}) 
			\nonumber \\
			&\quad + (\mathbf{z}_a \cdot \mathbf{z}_b) \, \partial_{\mathbf{z}_b} \cdot \mathbf{K}_{ab} 
			- (\mathbf{z}_a \cdot \mathbf{k}_b) W_{ab} \ , 
			\\
			W_{ab} &\equiv \frac12 \mathbf{K}_{ab} \cdot \mathbf{K}_{ab} \ , 
			\qquad \mathbf{K}_{ab} \equiv \partial_{\mathbf{k}_a} - \partial_{\mathbf{k}_b} \ ,
		\end{align}
		\label{eq:weight_shifting_mom}
	\end{subequations}
	where $\mathbf{z}_a$ is an auxiliary null vector encoding the polarizations when taken on-shell, and $\rho_a\equiv\Delta_a+s_a-1$, with $s_a$ the spin of the conformal primary. The action of these operators can be summarized as follows.  Applying the spin operator $S_{ab}$ increases the spin quantum number by one at both locations $a$ and $b$. In contrast, the weight operator $W_{ab}$ decreases the corresponding weights at the same two points by one. 
	The mixed spin–weight operators $D_{ab}$ and $F_{ab}$ each enhance the spin at $a$ by a single unit, but differ in their weight-lowering effect: $D_{ab}$ reduces the weight at $b$, whereas $F_{ab}$ acts on the weight at $a$ itself. To construct the correlators, these operators need to be normalized to take into account the spin and scaling weight combinations for which the correlators become trivial, the normalizations are the following 
	\begin{subequations}
		\begin{align}
			\widehat{W}_{ab} &\equiv - \frac{2 W_{ab}}
			{(\Delta_a + \Delta_b - \Delta_c - 2)(\Delta_a + \Delta_b - \tilde{\Delta}_c - 2)} \ , \\
			\widehat{S}_{ab} &\equiv \frac{1}{\rho_a \rho_b} S_{ab} \ , \\
			\widehat{D}_{ab} &\equiv \frac{1}{\rho_a} D_{ab} \ , \\
			\widehat{F}_{ab} &\equiv \frac{1}{\Delta_a + s_b + s_c - 2} F_{ab} \ .
		\end{align}
		\label{eq:weight_shifting_mom_norm}
	\end{subequations}
	Here $\widetilde{\Delta}_c\equiv d-\Delta_c$ is the shadow weight. Given that in the flat space limit the three-point correlator of a scalar in $d$ dimensions\footnote{In $d=3$ the singularity structure is $\log{E}$.} behaves as
	\be
	\lim_{E \to 0} \langle \Phi \Phi \Phi \rangle 
	= A_{\phi^3} \times (k_1 k_2 k_3)^{\Delta_\Phi - \frac{d+1}{2}} \, E^{\frac{3-d}{2}} \ ,
	\ee
	it was proposed that the appropriate ordering to showcase the double copy structure is given as
	\be
	\widehat{S} \cdots \widehat{S} \widehat{X} \cdots \widehat{X} \widehat{W} \cdots \widehat{W}\ ,  \quad \widehat{X} \in \{\widehat{D}, \widehat{F}\} \ . \label{eq:norm_ord}
	\ee
	The spin operators $S_{ab}$ act last since on-shell they simply turn into a product of polarizations. The weight operators act first since they don't change the degree of the singularity in the total energy. Meanwhile, $F_{ab}$ and $D_{ab}$ turn into products of polarizations and momenta when taken on-shell in the flat space limit while they also act in a non-trivial way to remove logarithmic singularities that can appear in the scalar correlator. The correct combinations of these operators lead to a conserved correlator that satisfies the Ward-Takahashi identity. Having this prescription, the double copy can be made apparent for the three-point functions by writing the correlators as
	\begin{align}
		\langle J_1 J_1 J_1 \rangle
		=& 
		\underbrace{\left( \widehat{S}_{12} \widehat{D}_{31} \widehat{W}_{23} + \text{cyc.} \right)}_{\widehat{n}_1}
		\langle \Phi \Phi \Phi \rangle_{\Delta_\Phi = d} \ , \\
		\langle J_2 J_2 J_2 \rangle 
		=& \underbrace{:\widehat{n}_1^2:}_{\equiv \widehat{n}_2}
		\langle \Phi \Phi \Phi \rangle_{\Delta_\Phi = d+2} \ ,
	\end{align}
	for the leading terms, where operators enclosed in colons are ordered as in Eq.~\eqref{eq:norm_ord} and terms such as $\widehat{D}_{ab} \, \widehat{D}_{cd}$ are ordered such that $a\leq c$, and
	\begin{align}
		\langle J_1 J_1 J_1 \rangle_{F^3} 
		=& \widehat{F}_{12} \, \widehat{F}_{23} \, \widehat{F}_{31} 
		\langle \Phi \Phi \Phi \rangle_{\Delta_\Phi = d} \ ,\\
		\langle J_2 J_2 J_2 \rangle_{W^3} 
		=& \widehat{F}^2_{12} \, \widehat{F}^2_{23} \, \widehat{F}^2_{31}
		\langle \Phi \Phi \Phi \rangle_{\Delta_\Phi = d + 2} \ ,
	\end{align}
	Here $\widehat{F}_{ab}^2$ denotes two ordered applications of $\widehat{F}_{ab}$. For the higher derivative corrections, these expressions for the correlators not only hold in momentum space but also in embedding space, which is described below. While this leads to a simple squaring procedure for the three-point correlators, it is still unclear whether a similar result holds at higher-point and whether it realizes a version of the color-kinematics duality. 
	
	\subsubsection{Cosmological dressing rules}
	The momentum-space correlators reviewed above are wavefunction coefficients or analytically continued AdS boundary correlators. For in-in correlators in dS$_4$, a closely related but technically simpler organization has recently been developed in terms of \emph{cosmological dressing rules} \cite{Chowdhury:2025ohm,Chowdhury:2025nnk}. The starting point is the Schwinger--Keldysh path integral \cite{Schwinger:1960qe,Keldysh:1964ud,Weinberg:2005vy}, which doubles the fields on the forward and backward time contours. After analytic continuation to EAdS$_4$, this doubling can be repackaged into a shadow effective action with two fields, $\phi_+$ and $\phi_-$, whose dimensions, with the (A)dS radius set to one, are
	\be
	\Delta_\pm=\frac{d}{2}\pm\sqrt{\frac{d^2}{4}-m^2}\ .
	\ee
	The external late-time fields are represented by the $\phi_-$ branch, while $\phi_+$ can run internally; the phases and relative signs in the shadow action encode the original in-in contour \cite{Sleight:2019mgd,Sleight:2020obc,Sleight:2021plv,DiPietro:2021sjt}. Thus a physical in-in diagram is obtained by summing the corresponding shadow Witten diagrams. In many examples this sum collapses to an ordinary flat-space Feynman integrand dressed by simple one-dimensional auxiliary propagators \cite{Chowdhury:2023arc,Chowdhury:2025ohm}. An earlier AdS version of this dressing idea appears in BCFW recursion for Witten diagrams, where boundary correlators are recursively assembled from lower-point Witten diagrams with the necessary bulk integrations \cite{Raju:2010by}. This is useful for double copy questions because the numerator algebra can then be kept close to the flat-space one.
	
	The scalar construction can be summarized as follows. For each flat-space diagram one keeps the usual spatial momentum flow and the ordinary flat-space propagators for the internal lines, but one no longer imposes energy conservation at each cosmological vertex. Instead, an auxiliary one-dimensional energy is assigned to the ordinary internal lines. At a vertex, let $k_A$ be the sum of the magnitudes of the external momenta entering that vertex and let $p_A$ be the sum of the auxiliary energies entering through ordinary internal lines. One attaches a one-dimensional auxiliary propagator to the vertex, connects all such auxiliary lines to a common point where the auxiliary energy is conserved, and integrates over the unfixed auxiliary energies. For the conformally coupled $\phi^4$ theory the auxiliary propagator is
	\be
	\Delta_{\rm cc}(k_A,p_A)=\frac{k_A}{p_A^2+k_A^2}\ ,
	\label{eq:cosmo_dressing_scalar_aux}
	\ee
	up to the overall coupling and measure normalizations suppressed here.
	
\begin{figure}[H]
		\centering
		\begin{tikzpicture}[scale=0.82, every node/.style={font=\scriptsize}]
			\coordinate (aL) at (-5.2,0);
			\coordinate (aR) at (-3.0,0);
			\coordinate (bL) at (1.3,0);
			\coordinate (bR) at (3.5,0);
			\coordinate (aux) at (2.4,1.65);
			\draw (-6.4,0.9) node[left] {$\bk_1$} -- (aL);
			\draw (-6.4,-0.9) node[left] {$\bk_2$} -- (aL);
			\draw (aR) -- (-1.8,0.9) node[right] {$\bk_3$};
			\draw (aR) -- (-1.8,-0.9) node[right] {$\bk_4$};
			\draw (aL) to[out=35,in=145] node[above] {$L+K$} (aR);
			\draw (aL) to[out=-35,in=-145] node[below] {$L$} (aR);
			\fill (aL) circle (2pt);
			\fill (aR) circle (2pt);
			\node at (-4.1,-1.55) {flat bubble};
			\draw (0.1,0.9) node[left] {$\bk_1$} -- (bL);
			\draw (0.1,-0.9) node[left] {$\bk_2$} -- (bL);
			\draw (bR) -- (4.7,0.9) node[right] {$\bk_3$};
			\draw (bR) -- (4.7,-0.9) node[right] {$\bk_4$};
			\draw (bL) to[out=35,in=145] node[above] {$L+K$} (bR);
			\draw (bL) to[out=-35,in=-145] node[below] {$L$} (bR);
			\draw[dashed] (bL) -- node[pos=0.55,left] {$\Delta_{12}$} (aux);
			\draw[dashed] (bR) -- node[pos=0.55,right] {$\Delta_{34}$} (aux);
			\fill (bL) circle (2pt);
			\fill (bR) circle (2pt);
			\fill (aux) circle (2pt);
			\node at (2.4,-1.55) {dressed in-in bubble};
		\end{tikzpicture}
		\caption{The $s$-channel one-loop bubble for conformally coupled $\phi^4$ theory and its cosmological dressing. The solid loop is the ordinary flat-space bubble integrand. The dashed lines are auxiliary propagators, with $\Delta_{12}=\Delta_{\rm cc}(k_{12},p)$ and $\Delta_{34}=\Delta_{\rm cc}(k_{34},p)$, and the upper point imposes conservation of the auxiliary energy $p$.}
		\label{fig:cosmo_dressing_bubble}
	\end{figure}
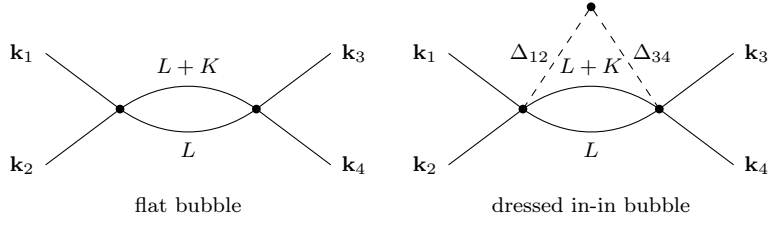

	A useful loop example is the $s$-channel bubble contribution to the four-point in-in correlator in conformally coupled $\phi^4$ theory. With $\phi_a\equiv\phi(\bk_a)$, $\mathbf{y}_{12}=\bk_1+\bk_2$, $y_{12}=|\mathbf{y}_{12}|$, and $K^\mu=(p,\mathbf{y}_{12})$, the dressing rule gives
	\begin{align}
		\left\langle\phi_1\phi_2\phi_3\phi_4\right\rangle_{(s),{\rm bub}}^{\rm cc}
		&\propto
		\int_{-\infty}^{\infty}dp\,
		\Delta_{\rm cc}(k_{12},p)\Delta_{\rm cc}(k_{34},p)
		\int \ud^4L\,\frac{1}{L^2(L+K)^2}\ ,
		\label{eq:cosmo_dressing_cc_bubble}
	\end{align}
	The second integral is the usual flat-space one-loop bubble, while the first line dresses its two vertices by the auxiliary propagators. The loop integral must still be regulated, but the important point for the present discussion is that the non-conservation of cosmological energy is isolated in the auxiliary $p$-integral. Figure~\ref{fig:cosmo_dressing_bubble} shows the corresponding diagrammatic rule. Following \cite{Chowdhury:2025ohm,Chowdhury:2025nnk}, Eq.~\eqref{eq:cosmo_dressing_cc_bubble} is stripped of the standard momentum factor $\eta_*^{\Delta_-}k_a^{2\Delta_--d}$ for each external scalar leg, with the corresponding late-time power also suppressed.  Cubic conformally coupled theories and massless theories have additional auxiliary propagators, often with an extra integral over a positive parameter; these details control infrared behavior and transcendentality, but are not needed for the double copy mechanism.
	
	For gauge fields, the spinning extension of the shadow formalism keeps the longitudinal modes off shell instead of solving their constraint equations, which would give a non-local action. The residual boundary gauge symmetry is fixed by setting the longitudinal boundary field to zero; equivalently, the longitudinal bulk modes obey Dirichlet boundary conditions, while the transverse modes obey the usual shadow boundary conditions \cite{Chowdhury:2025nnk}. This distinction is visible in the dressing rules. For a four-point exchange between external conformally coupled charged scalars, mediated by a photon or gluon, define
	\be
	\pi_{ij}(\mathbf{y}_{12})=\delta_{ij}
	-\frac{(\mathbf{y}_{12})_i(\mathbf{y}_{12})_j}{y_{12}^2}\ ,
	\ee
	and
	\be
	\alpha_i=(\bk_1-\bk_2)_i,\qquad
	\beta_i=(\bk_3-\bk_4)_i \ .
	\ee
	The transverse part of the $s$-channel four-point in-in correlator can then be written as the dressed flat-space exchange diagram
	\be
	\braket{\phi_1\phi_2^\dagger\phi_3\phi_4^\dagger}_{(s)}^{T}
	\propto
	\alpha^i\pi_{ij}(\mathbf{y}_{12})\beta^j
	\int_{-\infty}^{\infty}dp\,
	\Delta_\gamma^T(k_{12},p)\,
	\frac{1}{p^2+y_{12}^2}\,
	\Delta_\gamma^T(k_{34},p)\ ,
	\label{eq:cosmo_dressing_gluon_transverse}
	\ee
	where
	\be
	\Delta_\gamma^T(k,p)=\frac{k}{k^2+p^2}\ .
	\label{eq:cosmo_dressing_gluon_aux}
	\ee
	The full exchange also contains the longitudinal part
	\begin{align}
		\braket{\phi_1\phi_2^\dagger\phi_3\phi_4^\dagger}_{(s)}^{L}
		\propto&
		\alpha^i\frac{(\mathbf{y}_{12})_i(\mathbf{y}_{12})_j}{y_{12}^2}\beta^j
		\int_{-\infty}^{\infty}dp\,
		\Delta_\gamma^L(k_{12},p)\,
		\frac{1}{p^2}\,
		\Delta_\gamma^L(k_{34},p)\ ,
		\label{eq:cosmo_dressing_gluon_longitudinal}
	\end{align}
	where
	\be
	\Delta_\gamma^L(k,p)=\frac{p}{k^2+p^2}\ .
	\ee
	Thus, for this four-point exchange, the object relevant for color-kinematics is the sum of the transverse and longitudinal dressed integrands. The transverse dressing generalizes directly to arbitrary flat-space diagrams. The simple longitudinal rule in Eq.~\eqref{eq:cosmo_dressing_gluon_longitudinal}, however, is special to this exchange topology; at higher points, vertices with more than one internal gauge line produce mixed transverse-longitudinal structures whose dressing has to be worked out separately \cite{Chowdhury:2025nnk}.
	
	Indeed, after stripping color factors one can define a momentum-space kinematic numerator by pulling out the ordinary exchange propagator $(p^2+y_{12}^2)^{-1}$ from the full dressed exchange integrand, with the longitudinal term first rewritten by absorbing the ratio between longitudinal and transverse exchange propagators into its numerator. The $t$- and $u$-channel numerators are obtained by the same permutations as in flat space. They do not obey a pointwise Jacobi identity in the auxiliary energy, but they satisfy the integrated identity
	\be
	\int_{-\infty}^{\infty}dp\,
	\left(n_s(p)+n_t(p)+n_u(p)\right)=0\ .
	\label{eq:cosmo_dressing_integrated_jacobi}
	\ee
	This is the key improvement over the momentum-space wavefunction construction discussed above: the auxiliary-energy integral converts the numerator algebra into the flat-space Jacobi relation with a common total-energy pole, so one does not need to impose color-kinematics duality by a generalized gauge transformation \cite{Chowdhury:2025nnk}. The same logic applies to the four-point Yang--Mills in-in correlator with external currents. In that case the exchange diagram contains both the transverse and longitudinal dressings, while the contact diagram is dressed by the same transverse rule evaluated at zero auxiliary energy; the integrated Jacobi relation is obeyed by the numerator built from all of these pieces. In the scalar-QED example, the corresponding wavefunction coefficients obey the same color-kinematics relation after the auxiliary-energy integral.
	
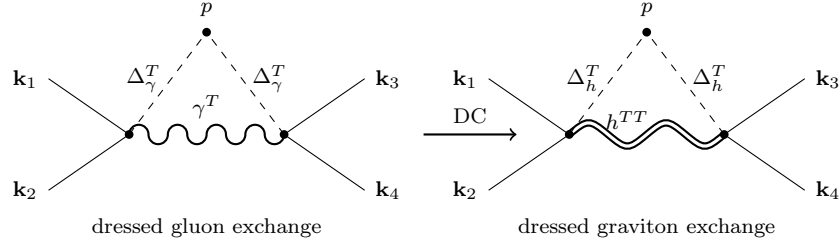
\begin{figure}[H]
		\centering
		\begin{tikzpicture}[scale=0.82, every node/.style={font=\scriptsize}]
			\coordinate (gL) at (-5.5,0);
			\coordinate (gR) at (-3.0,0);
			\coordinate (gAux) at (-4.25,1.65);
			\coordinate (hL) at (1.6,0);
			\coordinate (hR) at (4.1,0);
			\coordinate (hAux) at (2.85,1.65);
			\draw (-6.8,0.9) node[left] {$\bk_1$} -- (gL);
			\draw (-6.8,-0.9) node[left] {$\bk_2$} -- (gL);
			\draw (gR) -- (-1.7,0.9) node[right] {$\bk_3$};
			\draw (gR) -- (-1.7,-0.9) node[right] {$\bk_4$};
			\draw[thick] (gL) arc[start angle=180,end angle=0,radius=0.15625] arc[start angle=180,end angle=360,radius=0.15625] arc[start angle=180,end angle=0,radius=0.15625] arc[start angle=180,end angle=360,radius=0.15625] node[above=2pt] {$\gamma^T$} arc[start angle=180,end angle=0,radius=0.15625] arc[start angle=180,end angle=360,radius=0.15625] arc[start angle=180,end angle=0,radius=0.15625] arc[start angle=180,end angle=360,radius=0.15625];
			\draw[dashed] (gL) -- node[pos=0.55,left] {$\Delta_\gamma^T$} (gAux);
			\draw[dashed] (gR) -- node[pos=0.55,right] {$\Delta_\gamma^T$} (gAux);
			\node[above=2pt] at (gAux) {$p$};
			\fill (gL) circle (2pt);
			\fill (gR) circle (2pt);
			\fill (gAux) circle (2pt);
			\node at (-4.25,-1.5) {dressed gluon exchange};
			\draw[->,thick] (-0.75,0) -- (0.75,0) node[midway,above] {DC};
			\draw (0.3,0.9) node[left] {$\bk_1$} -- (hL);
			\draw (0.3,-0.9) node[left] {$\bk_2$} -- (hL);
			\draw (hR) -- (5.4,0.9) node[right] {$\bk_3$};
			\draw (hR) -- (5.4,-0.9) node[right] {$\bk_4$};
			\draw[thick,double,double distance=1.2pt] (hL) .. controls +(0.3,0.25) and +(-0.3,0.25) .. (2.225,0) .. controls +(0.3,-0.25) and +(-0.3,-0.25) .. node[above=2pt] {$h^{TT}$} (2.85,0) .. controls +(0.3,0.25) and +(-0.3,0.25) .. (3.475,0) .. controls +(0.3,-0.25) and +(-0.3,-0.25) .. (hR);
			\draw[dashed] (hL) -- node[pos=0.55,left] {$\Delta_h^T$} (hAux);
			\draw[dashed] (hR) -- node[pos=0.55,right] {$\Delta_h^T$} (hAux);
			\node[above=2pt] at (hAux) {$p$};
			\fill (hL) circle (2pt);
			\fill (hR) circle (2pt);
			\fill (hAux) circle (2pt);
			\node at (2.85,-1.5) {dressed graviton exchange};
		\end{tikzpicture}
		\caption{Dressed exchange diagrams for the cosmological double copy example. The left diagram is the transverse gluon exchange with auxiliary dressings at the two scalar-gauge vertices. The right diagram is the transverse-traceless graviton exchange obtained by squaring the flat-space numerator tensor and replacing the gluon auxiliary dressings by the graviton ones. The dashed lines carry the auxiliary energy $p$.}
		\label{fig:cosmo_dressing_double_copy}
	\end{figure}

	The corresponding gravitational statement is clearest in the transverse-traceless sector. The graviton exchange dressing is obtained from a derivative dressing,
	\be
	\Delta_h^T(k_i,k_j,p)
	=p\,k_i^2k_j^2\,
	\widehat{\partial}_{p}\widehat{\partial}_{k_i}\widehat{\partial}_{k_j}
	\int_0^\infty ds\,s\,
	\frac{k_{ij}+s}{(k_{ij}+s)^2+p^2}\ ,
	\qquad
	\widehat{\partial}_{x}f\equiv\partial_x\!\left(\frac{f}{x}\right).
	\label{eq:cosmo_dressing_graviton_aux}
	\ee
	Then the transverse-traceless part of the $s$-channel scalar correlator mediated by a graviton has the schematic form
	\be
	\mathcal{M}_{4,(s)}^{TT}
	\propto
	\alpha^i\alpha^j\Pi_{ijnm}(\mathbf{y}_{12})\beta^n\beta^m
	\int_{-\infty}^{\infty}dp\,
	\Delta_h^T(k_1,k_2,p)\,
	\frac{1}{p^2+y_{12}^2}\,
	\Delta_h^T(k_3,k_4,p)\ ,
	\label{eq:cosmo_dressing_graviton_exchange}
	\ee
	where
	\be
	\Pi_{ijnm}(\mathbf{y})=
	\frac12\left(\pi_{in}\pi_{jm}+\pi_{im}\pi_{jn}-\pi_{ij}\pi_{nm}\right)
	\label{eq:cosmo_dressing_tt_projector}
	\ee
	is, with our normalized projector convention, the transverse-traceless projector; each $\pi$ on the right-hand side is evaluated at $\mathbf{y}$. Thus the transverse graviton exchange is obtained from the gluon exchange by squaring the flat-space numerator tensor and replacing the gluon auxiliary dressing by the graviton one,
	\be
	\alpha^i\pi_{ij}\beta^j
	\quad\longrightarrow\quad
	\alpha^i\alpha^j\Pi_{ijnm}\beta^n\beta^m,
	\qquad
	\Delta_\gamma^T\quad\longrightarrow\quad\Delta_h^T\ .
	\label{eq:cosmo_dressing_double_copy_rule}
	\ee
	This diagrammatic replacement is shown in Fig.~\ref{fig:cosmo_dressing_double_copy}.
	The subtraction of the trace term in $\Pi_{ijnm}$ is the analogue of removing the dilaton sector generated by the naive square of a gluon. In flat space this trace contribution cancels after summing channels in the scalar exchange example, but in dS it must be removed explicitly to reproduce the transverse-traceless graviton correlator.
	
	One can also consider external spinning particles. At three points, the double copy reduces to the familiar statement that the tensor part of the stress-tensor correlator is the square of the current correlator tensor structure, with the flat-space cubic amplitude replaced by a dressed cubic diagram. Schematically
	\be
	\braket{JJJ}=\Delta_\gamma^T(E,0)\,A_3^{\rm YM},\qquad
	\braket{TTT}=\Delta_{hhh}(k_1,k_2,k_3)\,\left(A_3^{\rm YM}\right)^2\ ,
	\ee
	where $E=k_1+k_2+k_3$ and $\Delta_{hhh}$ is the three-graviton auxiliary dressing. 
	
	While the cosmological dressing does not yet give a complete all-sector double copy for general in-in correlators, it gives a clean momentum-space mechanism for the transverse sector: start from the flat-space diagram, dress the vertices by auxiliary propagators, integrate over the non-conserved energy, and then apply the usual numerator squaring with the appropriate graviton dressing and trace subtraction.
	
	\subsection{Mellin space}
	The first successful version of a double copy for four-point correlators was found in Mellin space in \cite{Zhou:2021gnu}. We will first introduce the concept of Mellin transforms and Mellin space and then review the double copy construction of \cite{Zhou:2021gnu} and \cite{Drummond:2022dxd} at four-points.
	
	The one–dimensional Mellin transform for $f(x)$ on $(0,\infty)$ and its inverse are defined as 
	\be
	\mathcal{M}\{f\}(z)\equiv \int_{0}^{\infty} x^{z-1} f(x)\,dx,
	\qquad
	f(x)=\frac{1}{2\pi i}\int_{c-i\infty}^{c+i\infty} x^{-z}\,\mathcal{M}\{f\}(z)\,dz,
	\ee
	with $c$ chosen so both integrals converge. Under scalings,
	\be
	\mathcal{M}\{f(\lambda x)\}(z)=\lambda^{-z}\,\mathcal{M}\{f\}(z)\ .
	\ee
	These transformations are used to define the Mellin space for CFT $n$-point functions \cite{Mack:2009mi,Penedones:2010ue}.  For primaries $\Phi_i$ of dimensions $\Delta_i$, the correlator admits the Mellin representation
	\be
	\mathcal{G}({x_i})\equiv \langle \Phi_1(x_1)\cdots \Phi_n(x_n)\rangle
	=
	\int\!\Big[\prod_{i<j}\frac{d\delta_{ij}}{2\pi i}\Big]\,
	\mathcal{M}(\{\delta_{ij}\})\,
	\prod_{i<j}\Gamma(\delta_{ij})\,(x_{ij}^2)^{-\delta_{ij}},
	\ee
	where 
	$ x_{ij}\equiv x_i-x_j$, $\delta_{ij}=\delta_{ji}$ obey
	$\sum_{j\neq i}^n\delta_{ij}=\Delta_i$, and the function $\mathcal{M}$ is the \emph{Mellin amplitude}. There are $n(n-3)/2$ independent integration variables, which is the same number as the independent conformal invariant cross-ratios and independent Mandelstam variables in flat space scatterings at n-points. It is useful to change variables from $\delta_{ij}$ to Mandelstam like variables $s_{ij}$ defined as follows. Consider an auxiliary vector $q_i$ such that
	$\sum_{i=1}^n q_i=0$,  $q_i^2=-\,\Delta_i$ and $\delta_{ij}=q_i\!\cdot q_j$. The Mandelstam–like variables are then defined as
	\be
	s_{ij}\;\equiv\;-\,(q_i+q_j)^2
	\;=\;\Delta_i+\Delta_j-2\,\delta_{ij},
	\ee
	so the inverse map is $\;\delta_{ij}=\tfrac{1}{2}\big(\Delta_i+\Delta_j-s_{ij}\big)$. At four-points we choose
	\be
	s\equiv s_{12},\qquad t\equiv s_{13},\qquad u\equiv s_{14},\qquad
	s+t+u=\sum_{i=1}^4\Delta_i.
	\ee
	which resembles the flat space amplitude relations. Then the four-point amplitude can be written as
	\be
	\mathcal{G}_{\Delta_1\Delta_2\Delta_3\Delta_4}(x_i)= \int_{-i\infty}^{i\infty}  \frac{ds \, dt}{(4 \pi i)^2} \,
	K(x_{ij}^2; s, t, u) \,
	\mathcal{M}(s,t,u) \,
	\Gamma_{\{\Delta_i\}}(s, t, u) \label{eq:Mellin4}
	\ee
	where all the space-time dependence is in 
	\begin{align}
		K(x_{ij}^2; s, t, u) 
		= (x_{12}^2)^{\frac{s - \Delta_1 - \Delta_2}{2}}
		(x_{34}^2)^{\frac{s - \Delta_3 - \Delta_4}{2}}
		(x_{14}^2)^{\frac{t - \Delta_1 - \Delta_4}{2}} \nonumber \\
		\quad \times (x_{23}^2)^{\frac{t - \Delta_2 - \Delta_3}{2}}
		(x_{13}^2)^{\frac{u - \Delta_1 - \Delta_3}{2}}
		(x_{24}^2)^{\frac{u - \Delta_2 - \Delta_4}{2}}  \ ,
	\end{align}
	the gamma factor is 
	\begin{align}
		\Gamma_{\{\Delta_i\}}(s, t, u) 
		= \Gamma\!\left[\frac{\Delta_1 + \Delta_2 - s}{2}\right]
		\Gamma\!\left[\frac{\Delta_3 + \Delta_4 - s}{2}\right]
		\Gamma\!\left[\frac{\Delta_1 + \Delta_4 - t}{2}\right]  \nonumber \\
		\quad \times \Gamma\!\left[\frac{\Delta_2 + \Delta_3 - t}{2}\right]
		\Gamma\!\left[\frac{\Delta_1 + \Delta_3 - u}{2}\right]
		\Gamma\!\left[\frac{\Delta_2 + \Delta_4 - u}{2}\right] \ ,
	\end{align}
	and the Mellin amplitude $\mathcal{M}(s,t,u)$ contains all the dynamical information. In this representation, an exchange of an operator with $(\Delta,J)$ produces $s$-channel poles at
	\be
	s=\Delta-J+2m\ , \qquad m=0,1,2,\ldots \ .
	\ee
	and contact interactions with $2L$ derivatives give rise to a degree-$L$ polynomial in Mellin variables. For spinning external operators, one can build the required tensor structures via weight/spin-shifting operators on top of the same scalar Mellin kernel; in this scalar-kernel representation, the $\Gamma$ factors and Mellin variables are unchanged, while $\mathcal{M}$ carries the spin dependence.
	
	\subsubsection{Double copy of Mellin space amplitudes} 
	A double copy relation was found at 4-points for reduced Mellin amplitudes at tree level in supersymmetric theories \cite{Zhou:2021gnu}. More precisely, the double copy relation is between  $\mathcal{N}=4$ $AdS_5\times S^5$ IIB supergravity, $\mathcal{N}=2$ $AdS_5\times S^3$ super Yang-Mills, and non-supersymmetric $AdS_5\times S^1$ bi-adjoint scalars. The $\mathcal{N}=2$ and $\mathcal{N}=4$ superconformal operators transform under $SU(2)_R$ and $SO(6)_R$, respectively; thus, to write correlators without indices, as in Eq.~\eqref{eq:Mellin4}, we contract with auxiliary $SU(2)$ spinors or $SO(6)$ vectors.  It will also be relevant for the double copy that one can construct an $SO(4)$ vector with two $SU(2)$ spinors, this is just the Euclidean spinor helicity version of what was reviewed in Section \ref{sec:spinorhel}. Super conformal Ward identities imply that the correlators can be written as \cite{Nirschl:2004pa,Eden:2000bk}
	\be
	G^{\mathcal{N}}(x_i)
	= G^{\mathcal{N}}_{0}(x_i) 
	+ {R}^{(\mathcal{N})} H^{\mathcal{N}}(x_i)
	\ee
	where the first term is the protected part of the correlator that is independent of marginal couplings and does not contribute to the Mellin amplitude. In the second term, the ${R}^{(\mathcal{N})} $ factor is crossing-symmetric and fixed by conformal symmetry and $H^{\mathcal{N}}(s,t,u)$ is the reduced correlator that defines the reduced Mellin amplitude via
	\be
	H^{\mathcal{N}}
	= \int_{-i\infty}^{i\infty} \frac{ds \, dt}{(4 \pi i)^2} \,
	K(x_{ij}^2; s, t, \tilde{u}) \,
	\widetilde{\mathcal{M}}^{\mathcal{N}}(s,t,\tilde{u}) \,
	\Gamma_{\{\Delta_i\}}(s, t, \tilde{u}) \ ,
	\ee
	where $\tilde{u}=u-2$ for $\mathcal{N}=2$ and $\tilde{u}=u-4$ for $\mathcal{N}=4$ to compensate for the conformal weight carried by ${R}^{(\mathcal{N})} $.
	
	We now look at the four-point reduced Mellin amplitude of $\mathcal{N}=2$ super gluons. Here super gluons refer to the $1/2$ BPS superprimary states that arise from the Kaluza--Klein reduction of a vector multiplet made of adjoint-valued degrees of freedom localized on an $\mathrm{AdS}_5\times S^3$ space. The reduced Mellin amplitude can be written as \cite{Alday:2021odx,Zhou:2021gnu}
	\be
	\widetilde{\mathcal{M}}^{\mathcal{N}=2}
	= \!\!\! \!\!\!\sum_{\substack{i+j+k = \mathcal{E} - 2 \\ 0 \le i,j,k \le \mathcal{E} - 2}}
	\!\!\!  \!\!\!\mathrm{C}_{i,j,k} \left[
	\frac{n_s^{i,j} c_s}{s - s_M + 2k}
	+ \frac{n_t^{i,j} c_t}{t - t_M + 2j}
	+ \frac{n_u^{i,j} c_u}{\tilde{u} - u_M + 2i}
	\right]  
	\ee
	where $\tilde{u}=u-2$, $\mathcal{E}=(\Delta_1+\Delta_2+\Delta_3-\Delta_4)/2$ if $\Delta_1 + \Delta_4 \geq \Delta_2 + \Delta_3$ and $\mathcal{E}=\Delta_1$ if $\Delta_1 + \Delta_4 < \Delta_2 + \Delta_3$, and $s_M = \min\{\Delta_1 + \Delta_2, \, \Delta_3 + \Delta_4\} - 2 \ , \ t_M = \min\{\Delta_1 + \Delta_4, \, \Delta_2 + \Delta_3\} - 2 \ , \ 
	u_M = \min\{\Delta_1 + \Delta_3, \, \Delta_2 + \Delta_4\} - 2$. The factor $\mathrm{C}_{i,j,k} $ is a universal factor depending on $\Delta_i$ and the dot products $t_a\cdot t_b$ of the arbitrary vectors encoding the R-symmetry structures. The structure within the square brackets resembles the standard split in color and kinematics from flat space. The color numerators are the standard ones from Eq.~\eqref{eq:color_4pt}, and the kinematic numerators are given by
	\begin{align}
		\mathbf{n}_s^{i,j} 
		= \frac{1}{t - t_M + 2j} - \frac{1}{\tilde{u} - u_M + 2i} \ , \\[4pt]
		\mathbf{n}_t^{i,j} 
		= \frac{1}{\tilde{u} - u_M + 2i} - \frac{1}{s - s_M + 2k} \ , \\[4pt]
		\mathbf{n}_u^{i,j} 
		= \frac{1}{s - s_M + 2k} - \frac{1}{t - t_M + 2j} \ ,
	\end{align}
	and satisfy the color-kinematics duality, $\mathbf{n}_s^{i,j} +\mathbf{n}_t^{i,j} +\mathbf{n}_u^{i,j} =0$. It is interesting that these kinematic numerators are non-local, and one might worry that this would lead to unphysical poles when double copying. Nevertheless, this is just an artifact of manifesting all the supersymmetry; similar non-local expressions arise in flat space for the $\mathcal{N}=4$ super Yang-Mills superamplitude\footnote{ The super Yang-Mills amplitude can be written as $\delta^4(Q) \, \delta^4(\tilde{Q}) 
		\left( \frac{c_s n_s}{s} + \frac{c_t n_t}{t} + \frac{c_u n_u}{u} \right)$ with $n_s = \frac{1}{3} \left( \frac{1}{t} - \frac{1}{u} \right) , \
		n_t = \frac{1}{3} \left( \frac{1}{u} - \frac{1}{s} \right) , \
		n_u = \frac{1}{3} \left( \frac{1}{s} - \frac{1}{t} \right)$, where the delta functions impose conservation of the total supermomenta \cite{Elvang:2013cua}.}. The double copy can now be obtained straightforwardly 
	\begin{align}
		\widetilde{\mathcal{M}}^{\mathcal{N}=4}
		= &\!\!\!\!\! \!\!\!\sum_{\substack{i+j+k = \mathcal{E} - 2 \\ 0 \le i,j,k \le \mathcal{E} - 2}}
		\!\!\!  \!\!\!\mathrm{C}_{i,j,k}\!\! \left[
		\frac{(n_s^{i,j})^2}{s - s_M + 2k}
		+ \frac{(n_t^{i,j})^2}{t - t_M + 2j}
		+ \frac{(n_u^{i,j})^2}{\tilde{u} - u_M + 2i} \right]  \nonumber \\
		=& \!\!\!\!\! \!\!\!\sum_{\substack{i+j+k = \mathcal{E} - 2 \\ 0 \le i,j,k \le \mathcal{E} - 2}}
		\!\!\!  \!\!\!\mathrm{C}_{i,j,k}\!\! \left[
		\frac{-9 }
		{(s - s_M + 2k) \, (t - t_M + 2j) \, (\tilde{u} - u_M + 2i)} 
		\right]  
	\end{align}
	where now one has $\tilde{u}=u-4$ and the $SO(4)$ vectors, $t_a$, are replaced by $SO(6)$ vectors. This corresponds to the super graviton amplitudes in $\mathcal{N}=4$ $AdS_5\times S^5$ IIB supergravity\footnote{The super gravitons have to be normalized as $\mathcal{O}_{\Delta_i}\rightarrow\mathcal{O}_{\Delta_i}/\sqrt{\Delta_i}$ to have an exact correspondence without additional proportionality factors.}.

	The biadjoint scalar can be obtained analogously by replacing kinematics by color. One should note that this natural double copy, which in fact holds for all massless ($\Delta_i=2$) and massive $\Delta_i>2$ cases, only arises for the reduced Mellin amplitude. A similar color-kinematics relation was observed at the level of the Mellin amplitude for the massless case and a modified version for the massive one \cite{Alday:2021odx}. Even though the kinematic numerators in that case are simple polynomials, the double copy does not yield the gravitational amplitude. The same happens at five-points \cite{Alday:2022lkk}, where the color-kinematics is displayed at the level of the Mellin amplitude, but no BCJ relations exist and the double copy does not give the gravitational amplitude. It is likely that a generalization of the reduced Mellin amplitudes of \cite{Zhou:2021gnu} to higher-points is needed to showcase the double copy relations. Similarly, insights from \cite{Albayrak:2019yve}, that reduced the computation to scalar factors, could be useful.
	
	Another related approach was taken in \cite{Drummond:2022dxd} where the reduced Mellin amplitude was written in the spirit of the large-$\Delta$ (large-charge) formalism\footnote{Reference~\cite{Drummond:2022dxd} denotes the half-BPS superprimaries by $\mathcal{O}_{p_i}$ and explicitly identifies $p_i$ as their scaling dimension; thus, in the notation of this review, $\Delta_i=p_i$. } of \cite{Aprile:2020luw}.  It explicitly considers both the $SU(2)_L$ and $SU(2)_R$ symmetries of the $\mathcal{N}=2$ superconformal operators and gives a generalized reduced Mellin amplitude as 
	\be
	H^{\mathcal{N}}
	= - \oint ds \, dt \oint d\tilde{s} \, d\tilde{t} \;
	U^{s} V^{t} \tilde{U}^{\tilde{s}} \tilde{V}^{\tilde{t}} \;
	\Gamma \, \widetilde{\mathcal{M}}^{\mathcal{N}}(s,t,u,\tilde{s},\tilde{t},\tilde{u})
	\ee
	where $s + t + u = -\Delta_3 - 1 \ , \  
	\tilde{s} + \tilde{t} + \tilde{u} = \Delta_3 - 2 \ ,$ $\Gamma$ is a universal kernel involving gamma functions and sine function, and the cross-ratios are
	\begin{align}
		\frac{x_{12}^2 x_{34}^2}{x_{13}^2 x_{24}^2} = U = x \bar{x} \ , \quad
		\frac{x_{14}^2 x_{23}^2}{x_{13}^2 x_{24}^2} = V = (1 - x)(1 - \bar{x}) \ ,  \label{eq:CrossRatios}\\
		\frac{y_{12}^2 y_{34}^2}{y_{13}^2 y_{24}^2} = \tilde{U} = y \bar{y} \ , \quad
		\frac{y_{14}^2 y_{23}^2}{y_{13}^2 y_{24}^2} = \tilde{V} = (1 - y)(1 - \bar{y}) \ ,
	\end{align}
	with $y_{ij}^2 = \langle \eta_i \eta_j \rangle \langle \bar{\eta}_i \bar{\eta}_j \rangle$, and  $\eta$, $\bar{\eta}$ the $SU(2)_L$ and $SU(2)_R$ arbitrary spinors that give the index-free correlator. With this generalized Mellin transform the reduced Mellin amplitude is simply given by 
	\be
	\widetilde{\mathcal{M}}^{\mathcal{N}=2}(s,t,u,\tilde{s},\tilde{t},\tilde{u})=\frac{n_s c_s}{\mathbf{s} + 1}
	+ \frac{n_t c_t}{\mathbf{t} + 1}
	+ \frac{n_u c_u}{\mathbf{u} + 1}  \ , \label{eq:gen_reducedMellin_4}
	\ee
	where we have defined the bold Mandelstam variables as
	\be
	\mathbf{s} = s + \tilde{s} \ , \quad
	\mathbf{t} = t + \tilde{t} \ , \quad
	\mathbf{u} = u + \tilde{u} \ , \quad
	\mathbf{s} + \mathbf{t} + \mathbf{u} = -3 \ .
	\ee
	The fact that the amplitude only depends on these new bold Mandelstam variables is a consequence of a hidden $8d$ conformal symmetry of the amplitude. The kinematic numerators are
	\begin{align}
		n_s = \frac{1}{3} \left( \frac{1}{\mathbf{t} + 1} - \frac{1}{\mathbf{u} + 1} \right) \ , \\
		n_t = \frac{1}{3} \left( \frac{1}{\mathbf{u} + 1} - \frac{1}{\mathbf{s} + 1} \right) \ , \\
		n_u = \frac{1}{3} \left( \frac{1}{\mathbf{s} + 1} - \frac{1}{\mathbf{t} + 1} \right) \ ,
	\end{align}
	and satisfy the color-kinematics duality. In fact all the standard flat space amplitude relations follow in this case by considering the massive propagators given in terms of the bold Mandelstam variables in Eq.~\eqref{eq:gen_reducedMellin_4}. For example, the BCJ relation reads
	\be
	(\mathbf{t} + 1) \, \widetilde{\mathcal{M}}[1,2,3,4] 
	= (\mathbf{u} + 1) \, \widetilde{\mathcal{M}}[1,3,4,2] \ ,
	\ee
	where the partial amplitudes are
	\be
	\widetilde{\mathcal{M}}[1,2,3,4] = \frac{n_s}{\mathbf{s} + 1} - \frac{n_t}{\mathbf{t} + 1} \ , \quad  \widetilde{\mathcal{M}}[1,3,4,2] 
	= \frac{n_u}{\mathbf{u} + 1} - \frac{n_s}{\mathbf{s} + 1} \ .
	\ee
	Meanwhile, the double copy is now given by
	\be
	\widetilde{\mathcal{M}}^{\mathcal{N}=4}(s,t,u,\tilde{s},\tilde{t},\tilde{u})=\frac{n_s^2}{\mathbf{s} + 1}
	+ \frac{n_t^2}{\mathbf{t} + 1}
	+ \frac{n_u^2}{\mathbf{u} + 1} =-\frac{1}{(\mathbf{s} + 1)(\mathbf{t} + 1)(\mathbf{u} + 1)}
	\ee
	where in the second equality we used $\mathbf{s} + \mathbf{t} + \mathbf{u} = -3$ to simplify the answer, but to interpret it as a gravitational amplitude one has to instead take the constraint $\mathbf{s} + \mathbf{t} + \mathbf{u} = -4$. In \cite{Drummond:2022dxd}, it was also shown that a KLT version of the double copy holds
	\be
	\widetilde{\mathcal{M}}^{\mathcal{N}=4}
	= (\mathbf{s} + 1) \, \widetilde{\mathcal{M}}[1,2,3,4] \, \widetilde{\mathcal{M}}[1,2,4,3] \ .
	\ee
	This is a very simple, neat example of a double copy relation in AdS that closely resembles the flat space case. It would be compelling to understand if the structure holds at higher points.\\
	
	\subsection{Stringy KLT double copy}  
	A closely related development concerns a \emph{stringy} completion of the Mellin-space story. A KLT-like relation between the building blocks of open- and closed-string amplitudes on AdS was constructed in \cite{Alday:2025bjp}. It is important that this relation has so far been established for the building blocks of the amplitudes, rather than for the complete theory-dependent AdS amplitudes. Here, we review the construction of \cite{Alday:2023mvu,Alday:2022uxp,Alday:2022xwz,Alday:2023jdk}, together with its open-string counterpart \cite{Alday:2024yax,Alday:2024ksp}.
	
	We first consider the tree-level four-graviton amplitude of type IIB superstring theory on AdS$_5\times S^5$, referred to as the AdS Virasoro--Shapiro amplitude. Its holographic dual is the connected four-point correlator of the superconformal primary of the stress-tensor multiplet in $\mathcal{N}=4$ SYM, at leading non-trivial order in the large-$N$ expansion and at strong 't~Hooft coupling. Since a first-principles worldsheet computation in this Ramond--Ramond background is not presently available, the strategy of \cite{Alday:2023mvu,Alday:2022uxp,Alday:2022xwz,Alday:2023jdk} is instead to bootstrap the small-curvature expansion around flat space. The construction combines full crossing symmetry and the correct supergravity and flat-space limits with the pole structure implied by dispersive sum rules, whose derivation relies on Regge boundedness, together with input from integrability. A further ingredient is a worldsheet-like ansatz in terms of single-valued multiple polylogarithms. This is motivated by the fact that the low-energy expansion of the flat-space Virasoro--Shapiro amplitude contains only single-valued multiple zeta values.
	
	For the AdS Veneziano amplitude, the relevant quantities are color-ordered amplitudes rather than a single fully crossing-symmetric amplitude, with crossing relating the different orderings. An analogous construction uses the pole structure, Regge behaviour and flat-space limit, together with a disk-integral ansatz in terms of harmonic polylogarithms \cite{Alday:2024yax,Alday:2024ksp}. Unlike in the closed-string case, the low-energy expansion of the flat-space Veneziano amplitude is not restricted to single-valued multiple zeta values. In both the open- and closed-string cases, the known curvature corrections satisfy several non-trivial checks against the high-energy limit, localization and CFT data.
	
	The AdS Veneziano amplitudes also obey proposed AdS monodromy relations, order by order in the small-curvature expansion \cite{Alday:2025cxr}. These reduce to the usual open-string monodromy relations in the flat-space limit. Whether an appropriate field-theory limit of these relations gives curvature-corrected analogues of the BCJ relations remains a highly-relevant open question.
	
	One can further construct a worldsheet-like representation by taking a Borel transform of the Mellin amplitude \cite{Alday:2023mvu}. This reorganizes its low-energy expansion into universal disk and sphere integrals, whose KLT-like relation will be described below. More concretely, one starts from the low-energy expansion of the Mellin amplitude, defined through the Mellin transform of the reduced correlator as
	\be
	\begin{aligned}
		\mathcal{T}(U,V)=\int_{-i\infty}^{i\infty}\frac{ds\,dt}{(4\pi i)^2}\,U^{\frac{s}{2}+\frac{2}{3}}\,V^{\frac{t}{2}-\frac{4}{3}}\,  \\
		\times \  \Gamma\!\left(\frac{4}{3}-\frac{s}{2}\right)^{2}\,
		\Gamma\!\left(\frac{4}{3}-\frac{t}{2}\right)^{2}\,
		\Gamma\!\left(\frac{4}{3}-\frac{u}{2}\right)^{2}\,
		M(s,t)\ ,
	\end{aligned}
	\ee
	where $U,V$ were defined in Eq.~\eqref{eq:CrossRatios} and the Mandelstam variables were shifted so that they satisfy $s+t+u=0$. The low energy expansion is an expansion in the large t’Hooft coupling
	$\lambda$ defined as
	\be
	\frac{1}{\sqrt{\lambda}}=\frac{\alpha'}{L^{2}} \ ,
	\ee
	where $L$ is the AdS length. One defines the Borel transformed amplitude as
	\begin{equation}
		A(S,T)=2\,\lambda^{3/2}\int_{\kappa-i\infty}^{\kappa+i\infty}\frac{d\alpha}{2\pi i}\,
		e^{\alpha}\alpha^{-6}\,
		M\!\left(\frac{2\sqrt{\lambda}\,S}{\alpha},\frac{2\sqrt{\lambda}\,T}{\alpha}\right),
	\end{equation}
	which reorganizes the low-energy expansion into an object with a direct worldsheet interpretation which is Borel summable. The leading order in the $1/\sqrt{\lambda}$ expansion of $A(S,T)$ gives the Virasoro-Shapiro amplitude for type IIB superstring theory in flat space and higher orders are the AdS corrections \cite{Alday:2023mvu}. One can then make the assumption that each term in the expansion of $A(S,T)$ has a representation as an integral over the Riemann sphere, that is, the
	worldsheet for genus 0 closed string amplitudes \cite{Alday:2023mvu}. Similar ideas can be applied to the open-string amplitude case \cite{Alday:2024ksp,Alday:2024yax} to express the amplitude now as an integral over a disk, that is, the worldsheet for the open string amplitude. In that case, one considers the scattering of open string gluons on
	the worldvolume of D7 branes in type IIB string theory which effectively probe an AdS$_5 \times S^3$ geometry. In the dual theory, the gluon scattering corresponds to a four-point function of flavor multiplets in $\mathcal{N} = 2$ gauge theory, such that one defines the AdS Veneziano amplitude as the leading large $N$ limit of such correlator.

	The key observation~\cite{Alday:2022uxp,Alday:2022xwz,Alday:2023jdk,Alday:2023mvu,Alday:2024yax,Alday:2024ksp,Alday:2025bjp} is that, after this Mellin/Borel transform, both open- and closed-string amplitudes on AdS are built from universal families of integrals depending only on $s$ and $t$. For open strings one considers
	\begin{equation}
		J_{w}(s,t)=\int_{0}^{1}x^{s-1}(1-x)^{t-1}L_{w}(x)\,dx ,
	\end{equation}
	where $L_{w}(x)$ is the multiple polylogarithm  (MPL) associated with a word $w$ in the alphabet $\{0,1\}$.\footnote{Here a word $w$ means any finite ordered string of the letters $0$ and $1$, for example $w=e,0,1,00,01,10,\dots$, with $e$ the empty word. The corresponding functions $L_w(x)$ are iterated integrals built from the kernels $dx/x$ and $dx/(x-1)$.  For example, $L_{01}(x)=\int_0^x \frac{dt}{t}\,L_1(t)
		=\int_0^x \frac{dt}{t}\int_0^t \frac{du}{u-1}$, so the order of the letters keeps track of the order of the integrations.} It is convenient to collect them into the generating function
	\begin{equation}
		L(e_{0},e_{1};x)=L_{e}(x)+L_{0}(x)e_{0}+L_{1}(x)e_{1}+L_{00}(x)e_{0}^{2}+L_{01}(x)e_{0}e_{1}+\cdots ,
	\end{equation}
	so that the generating function for the $J_{w}(s,t)$ integrals is
	\begin{equation}
		\mathcal{J}(s,t;e_{0},e_{1})=\int_{0}^{1}x^{s-1}(1-x)^{t-1}L(e_{0},e_{1};x)\,dx .
	\end{equation}
	
	The generating function $L(e_{0},e_{1};x)$ satisfies the Knizhnik--Zamolodchikov equation, which implies differential relations for the open-string building blocks \cite{Alday:2024ksp,Alday:2024yax}. Together with additional relations under shifts of the Mellin variables, these relations fix $J_{w}(s,t)$ in terms of lower-weight integrals once their $s\rightarrow 0$ behavior is also fixed. At weight zero, $J_{e}(s,t)=\beta(s,t)$ is the beta function, while higher weight integrals are given by its derivatives; for example for weight one, $J_{0}(s,t)=\partial_{s}J_{e}(s,t)$, $
	J_{1}(s,t)=\partial_{t}J_{e}(s,t)$.
	
	The corresponding closed-string building blocks are defined by Riemann sphere integrals with insertions of single-valued multiple polylogarithms (SVMPLs),
	\be
	I_{\omega}(s,t)=\int |z|^{2s-2}\,|1-z|^{2t-2}\,\mathcal{L}_{\omega}(z)\,d^{2}z \ .
	\ee
	The corresponding generating function is given by
	\begin{equation}
		\mathcal{I}(s,t;e_{0},e_{1})=\int d^{2}z\,|z|^{2s-2}|1-z|^{2t-2}\mathcal{L}(e_{0},e_{1};z) ,
	\end{equation}
	where $\mathcal{L}(e_{0},e_{1};z)$ is the SVMPL generating function. It satisfies derivative and shift relations closely analogous to the open-string case, although in the closed-string sector the anti-holomorphic structure involves a deformed variable $e_1'$ and the corresponding identities are slightly more intricate. The deformed variable $e_{1}'=e_{1}-2\zeta(3)\,[e_{0}+e_{1},[e_{1},[e_{0},e_{1}]]]+\cdots$ is obtained by solving $Z^{R}(e_{0},e_{1}')\,e_{1}'\bigl(Z^{R}(e_{0},e_{1}')\bigr)^{-1}=Z(e_{0},e_{1})^{-1}\,e_{1}\,Z(e_{0},e_{1})$, recursively. Here, $Z(e_{0},e_{1})$ is the Drinfeld associator satisfying \cite{brown2006multiplezetavaluesperiods}
	\be
	L(e_{0},e_{1};x)=L(e_{1},e_{0};1-x)\,Z(e_{0},e_{1})\ , \quad Z(e_{0},e_{1})\,Z(e_{1},e_{0})=1\ . 
	\ee
	The zeroth-order term of $I_{e}(s,t)$ is the complex beta function $\beta_{\mathrm{C}}(s,t)$ while the descendants are obtained again by differentiation; for example for weight one we have
	$I_{0}(s,t)=\partial_{s}I_{e}(s,t)$, $I_{1}(s,t)=\partial_{t}I_{e}(s,t)$. Thus the open and closed sectors are governed by parallel hierarchies of functions, built respectively from MPLs and SVMPLs.
	
	\subsubsection{KLT double copy of string amplitude building blocks}
	The main result of \cite{Alday:2025bjp} is that these two towers are related by an AdS version of the KLT formula. 
	At low weights one already finds non-trivial bilinear relations between the closed-string integrals $I_{w}$ and products of the open-string integrals $J_{w}$, with coefficients involving the ordinary flat-space KLT kernel
	\begin{equation}
		\kappa(s,t)=\frac{\sin(\pi s)\sin(\pi t)}{\pi\sin\!\bigl(\pi(s+t)\bigr)} .
	\end{equation}
	For example, the explicit expression for $I_{001}(s,t)$ is a bilinear in the $J_{w}(s,t)$, together with derivatives of $\kappa(s,t)$, which strongly suggests that the entire closed-string tower should be obtainable from the open-string one. 
	This is indeed the case \cite{Alday:2025bjp}: introducing an AdS KLT kernel $K(s,t;e_{0},e_{1})$, one finds the exact generating-function relation
	\begin{equation}
		\mathcal{I}(s,t;e_{0},e_{1})
		=
		\mathcal{J}(s,t;e_{0},e_{1})\,
		K(s,t;e_{0},e_{1})\,
		\mathcal{J}^{R}(s,t;e_{0},e_{1}^{\prime}) ,
	\end{equation}
	where $J^{R}$ denotes the reversed-word generating function appropriate to the single-valued projection. The KLT kernel admits an expansion
	\begin{equation}
		K(s,t;e_{0},e_{1})=\kappa(s,t)+\kappa_{0}(s,t)e_{0}+\kappa_{1}(s,t)e_{1}+\cdots .
	\end{equation}
	Moreover, the kernel is periodic in $s$ and $t$,
	\begin{equation}
		K(s+1,t;e_{0},e_{1})=K(s,t+1;e_{0},e_{1})=K(s,t;e_{0},e_{1}),
	\end{equation}
	and obeys
	\begin{equation}
		\partial_{s}K(s,t;e_{0},e_{1})=\partial_{e_{0}}K(s,t;e_{0},e_{1}),\qquad
		\partial_{t}K(s,t;e_{0},e_{1})=\partial_{e_{1}}K(s,t;e_{0},e_{1}) .
	\end{equation}
	Most strikingly, the inverse kernel can be written to all orders as
	\begin{equation}
		K^{-1}
		=
		1+\frac{e^{2\pi i s}M_{0}}{1-e^{2\pi i s}M_{0}}
		+\frac{e^{2\pi i t}M_{1}}{1-e^{2\pi i t}M_{1}},
	\end{equation}
	where $M_{0}$ and $M_{1}$ are the monodromy matrices of the polylogarithmic system \cite{Brown2004SingleValuedHyperlogarithms}
	\be
	M_{0}=e^{2\pi i e_{0}}\ ,\qquad M_{1}=Z(e_{1},e_{0})\,e^{2\pi i e_{1}}\,Z(e_{0},e_{1})\ .
	\ee
	In terms of the ordinary KLT normalization, this becomes
	\begin{equation}
		\mathcal{K}^{-1}
		=
		\pi\left(\cot\!\bigl(\pi(s+e_{0})\bigr)
		+Z(e_{1},e_{0})\,\cot\!\bigl(\pi(t+e_{1})\bigr)\,Z(e_{0},e_{1})\right),
	\end{equation}
	valid to all orders in the non-commutative variables $e_0, \ e_1$.
	
	Therefore, the four-point KLT double copy extends beyond supergravity correlators and survives in a genuinely stringy form for the building blocks of the amplitudes \cite{Alday:2025bjp}. In other words, the closed-string AdS building blocks are bilinears of the open-string ones, with a curvature-deformed but exact KLT kernel.
	
	Further work has put this relation on a firmer footing. At four points, the AdS KLT relation was derived using a non-commutative generalization of the twisted de Rham framework familiar from flat-space string amplitudes \cite{Kakkad:2025klm}. More recently, the construction was extended to arbitrary multiplicity by dressing the usual disk and sphere integrals with multivariable multiple polylogarithms and their single-valued counterparts. The resulting open-string building blocks satisfy monodromy relations, while the closed-string ones obey a corresponding KLT factorization, extending the four-point construction to general $n$-point kinematics \cite{Nocchi:2026yms}.
	
	\subsection{Mellin-Barnes space}
	So far, we have considered the double copy realizations in Mellin space. Another possibility is to look for a double copy using the Mellin--Barnes, framework for momentum-space AdS and dS correlators introduced and developed in \cite{Sleight:2019mgd,Sleight:2019hfp,Sleight:2021iix,Sleight:2021plv}.  In this framework, the dependence on the boundary momenta is kept explicit while the radial dependence is Mellin transformed. After the external Mellin--Barnes bulk-to-boundary wavefunctions are factored off, the remaining reduced object is the Mellin--Barnes amplitude; this is the object referred to as the Mellin--momentum amplitude in \cite{Mei:2023jkb}. Depending on the application, the radial integral may be retained or evaluated as a delta function imposing a constraint on the Mellin variables. We first summarize this framework and then review the double copy construction of \cite{Mei:2023jkb}.
    
    We work in AdS$_{d+1}$ with unit radius in the Poincaré patch. After Wick rotation this same formalism is directly related to dS wavefunction coefficients \cite{Sleight:2021iix,Sleight:2020obc}.  For a scalar of conformal dimension $\Delta$, the momentum-space bulk-to-boundary wavefunction obeys
	\begin{equation}
		D_k^{\Delta}\,\phi_\Delta(k,z)=0,
		\qquad
		D_k^{\Delta}\equiv z^2 k^2-z^2\partial_z^2+(d-1)z\partial_z+\Delta(\Delta-d).
	\end{equation}
	The Mellin--Barnes representation is obtained by writing
	\begin{equation}
		\phi_\Delta(k,z)=\int_{-i\infty}^{i\infty}\frac{ds}{2\pi i}\,z^{-2s+d/2}\,\phi_\Delta(s,k).
	\end{equation}
	The equation of motion then becomes algebraic in its Mellin-space form
	\begin{equation}
		\left(z^2k^2+\left(\frac d2-\Delta\right)^2-4s^2\right)\phi_\Delta(s,k)=0 \ ,
	\end{equation}
	which plays the role of an on-shell condition in Mellin--Barnes space. For spin one and spin two, after the standard rescalings of the bulk fields, the corresponding equations reduce to scalar equations with effective dimensions $\Delta=d-1$ and $\Delta=d$, respectively. Thus a gluon solution is of the form
	$A_a(k,z)=\epsilon_a\,\phi_{d-1}(k,z)$, and a graviton solution is of the form $h_{ab}(k,z)=\epsilon_{ab}\,\phi_d(k,z)$, with transverse boundary polarizations.
	Multiplication by $z^2$ shifts the Mellin weight by one unit, so the $z^2k^2$ term is understood as the corresponding shift operator in Mellin weight.
	
	An $n$-point boundary correlator admits a Mellin--Barnes representation,
	\begin{equation}
		\begin{aligned}
			\langle O(\mathbf{k}_1)\cdots O(\mathbf{k}_n)\rangle
			=
			\delta^{(d)}\!\left(\sum_{i=1}^n \mathbf{k}_i\right)
			\prod_{i=1}^n\int\frac{ds_i}{2\pi i}\,\phi(s_i)\,A_{n}'(s,\mathbf{k}) \ , \\
			A_{n}'(s,\mathbf{k})=\int \frac{dz}{z^{d+1}}\,\mathcal{A}_n^{\rm MB}(z,\mathbf{k})\,z^{\sum_{i=1}^{n}\left(-2s_{i}+\frac{d}{2}\right)}\ ,
		\end{aligned}
	\end{equation}
	where the reduced object $\mathcal{A}_n^{\rm MB}$ is the reduced Mellin--Barnes amplitude. The radial integral may be performed at each vertex, giving
	\begin{equation}
		\int\frac{dz}{z^{d+1+b}}\,z^{\sum_i(-2s_i+d/2)}
		\propto
		\delta\!\left(d+b+\sum_i\left(2s_i-\frac d2\right)\right),
	\end{equation}
	which is the Mellin analogue of the usual momentum-conserving delta function \cite{Sleight:2021iix,Sleight:2021plv}. The $z^{-b}$ factor is required from scale invariance. An important feature of $\mathcal{A}_n^{\rm MB}$ is that local contact terms in boundary correlators do not contribute to it, so it behaves as an amputated AdS observable, free of field-redefinition ambiguities. This is one of the reasons it is especially natural for on-shell-inspired constructions.
	
	\subsubsection{Double copy and color-kinematics duality}

    Within this Mellin--Barnes framework,\cite{Mei:2023jkb} proposed a double copy construction for Yang--Mills theory and gravity that we review in the following. The three-point color-ordered Yang--Mills amplitude in general dimension takes the flat-space-like form
	\begin{equation}
		A_3^{\rm YM}=z\Big[(\epsilon_1\cdot\epsilon_2)(\epsilon_3\cdot k_1)
		+(\epsilon_2\cdot\epsilon_3)(\epsilon_1\cdot k_2)
		+(\epsilon_3\cdot \epsilon_1)(\epsilon_2\cdot k_3)\Big],
	\end{equation}
	which is precisely the AdS uplift of the usual flat-space cubic numerator \cite{Armstrong:2022mfr,Li:2022tby,Mei:2023jkb}. Squaring this immediately gives the cubic graviton amplitude,
	\begin{equation}
		\mathcal{M}_3
		=z^2\Big[(\epsilon_1\cdot\epsilon_2)(\epsilon_3\cdot k_1)
		+(\epsilon_2\cdot\epsilon_3)(\epsilon_1\cdot k_2)
		+(\epsilon_3\cdot\epsilon_1)(\epsilon_2\cdot k_3)\Big]^2,
	\end{equation}
	so the three-point double copy relation is simply
	\begin{equation}
		\mathcal{M}_3=(A_3^{\rm YM})^2.
	\end{equation}
	As in flat space, the product of two vector polarizations decomposes into a traceless symmetric part, a trace, and an antisymmetric part. The traceless symmetric part gives the graviton, while the trace corresponds to the dilaton sector of the double copy. Thus, the double copy at three-points gives the graviton--graviton--dilaton interaction
	\be
	\mathcal{M}_{3}(1_{h},2_{h},3_{\phi})=(A_{3})^{2}\big|_{\varepsilon_{3}^{\mu}\varepsilon_{3}^{\nu}\to\Pi^{\mu\nu}}
	=z^{2}(\varepsilon_{1}\cdot\varepsilon_{2})^{2}k_{1}^{\mu}k_{1}^{\nu}\Pi_{\mu\nu}\ .
	\ee
	where we exchanged the generic polarization product for the trace projector $\Pi_{\mu\nu}=\eta_{\mu\nu}-\frac{k_{\mu}k_{\nu}}{k^{2}}$. As expected, this vanishes in the flat space limit which can be taken by recovering the explicit factors of AdS length $L$ and taking simultaneously $L \to\infty$, $s_{i}\to\infty$ and replacing $s_{i}\to (z k_{i})/2$ \cite{Mei:2023jkb}.
	
	Thus, to extract pure Einstein gravity at four points one must therefore subtract the dilaton exchange contribution, exactly as in standard flat-space double copy constructions\footnote{In flat space, state projection or an appropriate choice of double copy theories is already needed to isolate pure gravity; the issue becomes especially important at loop level and in the presence of massive matter, whereas in curved space it can arise already at tree level.}\cite{Bern:2019prr,Carrasco:2021bmu}. The explicit four-point color-ordered Yang--Mills amplitude in this representation can be written as
	\begin{equation}
		\begin{aligned}
			A_4^{\rm YM}[1,2,3,4]
			=&\frac{z^2(\epsilon_1\cdot\epsilon_2)(\epsilon_3\cdot\epsilon_4)\,\Pi_{1,1}+z^2W_s}{D_{k_s}^{d-1}}
			+(\epsilon_1\cdot\epsilon_2)(\epsilon_3\cdot\epsilon_4)\,\Pi_{1,0}
			+V_c^{(s)} \\
			&-\big[(12)\leftrightarrow(23)\big],
		\end{aligned}
	\end{equation}
	where
	\begin{align}
		4W_s={}&(\epsilon_1\cdot\epsilon_2)
		\Big[(k_1\cdot\epsilon_3)(k_2\cdot\!\epsilon_4)-(k_2\cdot\epsilon_3)(k_1\cdot\epsilon_4)\Big]
		\nonumber \\
		&+(\epsilon_3\cdot\epsilon_4)
		\Big[(k_3\cdot\epsilon_1)(k_4\cdot\epsilon_2)-(k_4\cdot\epsilon_1)(k_3\cdot\epsilon_2)\Big]
		\nonumber\\
		&+\Big[(k_2\cdot\epsilon_1)\epsilon_2-(k_1\cdot\epsilon_2)\epsilon_1\Big]\cdot
		\Big[(k_4\cdot\epsilon_3)\epsilon_4-(k_3\cdot\epsilon_4)\epsilon_3\Big],
	\end{align}
	and the ``$s$-channel contact term'' is
	\begin{equation}
		4V_c^{(s)}=(\epsilon_1\cdot\epsilon_3)(\epsilon_2\cdot\epsilon_4)
		-(\epsilon_1\cdot\epsilon_4)(\epsilon_2\cdot\epsilon_3).
	\end{equation}
	The subtraction $[(12)\leftrightarrow(23)]$ generates the remaining color ordering in the usual way, see Eq.~\eqref{eq:AcolorToFactor}. The scalar functions $\Pi_{1,1}$ and $\Pi_{1,0}$ are the spin-one polarization sums in Mellin--Barnes space; explicitly,
	\begin{equation}
		\Pi_{1,1}=\frac14(k_1-k_2)\cdot(k_3-k_4)
		+\frac{(k_1^2-k_2^2)(k_3^2-k_4^2)}{4k_s^2},
		\quad
		\Pi_{1,0}=-\frac{(s_1-s_2)(s_3-s_4)}{z^2k_s^2}.
	\end{equation}
	These expressions are the Mellin--Barnes counterparts of the usual exchange and contact terms in the AdS momentum-space gluon amplitude \cite{Armstrong:2022mfr,Mei:2023jkb}.
	
	We can now look for a color-kinematic dual representation to obtain a double copy. As we see now, this requires an unusual replacement along the way, which makes this double copy version not as satisfactory as the flat space well-established case. From the explicit amplitude one can extract a kinematic numerator in the $s$ channel,
	\begin{equation}
		n_s=(\epsilon_1\cdot\epsilon_2)(\epsilon_3\cdot\epsilon_4)
		\Big(z^2\Pi_{1,1}+\Pi_{1,0}D_{k_s}^{d}\Big)
		+z^2W_s+V_c^{(s)}D_{k_s}^{d},
	\end{equation}
	with $n_t$ and $n_u$ obtained by the usual permutations. Note that in the kinematic numerators, one has to replace the operator $D_{k_s}^{d-1}$ of the gluons by $D_{k_s}^{d}$, corresponding to the gravitons. With these definitions, the numerators satisfy the flat-space-like kinematic Jacobi identity
	\begin{equation}
		n_s+n_t+n_u=0,
	\end{equation}
	which is the AdS version of color--kinematics duality in this formalism. Once the numerators are arranged in this form, the naive four-point double copy is immediate,
	\begin{equation}
		\mathcal{M}_4^{\rm naive}
		=\frac{n_s^2}{D_{k_s}^{d}}+\frac{n_t^2}{D_{k_t}^{d}}+\frac{n_u^2}{D_{k_u}^{d}} \ .
	\end{equation}
	However, as mentioned above, this expression is expected to contain the dilaton exchange inherited from the tensor product of two Yang--Mills states. To isolate the pure Einstein-Hilbert contribution one should subtract the trace sector as
	\begin{equation}
		\mathcal{M}_4^{\rm E.H.}
		=\frac{n_s^2}{D_{k_s}^{d}}+\frac{n_t^2}{D_{k_t}^{d}}+\frac{n_u^2}{D_{k_u}^{d}}-\widetilde{\mathcal{M}}_{\rm dilaton} \ .
	\end{equation}
	In \cite{Mei:2023jkb} the subtraction term was chosen to be
	\begin{equation}
		\widetilde{\mathcal{M}}_{\rm dilaton}
		=\big[(\epsilon_1\cdot\epsilon_2)(\epsilon_3\cdot\epsilon_4)\big]^2
		\left(\frac{z^4\Pi_{2,2}^{\rm Tr}}{D_{k_s}^{d}}-\Pi_{2,0}+\Pi_{1,0}^2D_{k_s}^{d}\right)+P(2,3,4),
	\end{equation}
	where $P(2,3,4)$ denotes the $t$- and $u$-channel permutations, and $\Pi_{2,2}^{\rm Tr}$ is the trace part of the spin-two polarization sum. The first term has the structure of a scalar exchange, while the last two terms are part of the trace-sector subtraction chosen in \cite{Mei:2023jkb} to account for the conformal structure of the graviton propagator. This formula has only been verified to give the correct graviton correlator for $d=3$ by converting back to momentum space \cite{Mei:2023jkb}.
	
	A useful way to summarize this double copy proposal is to view it as an uplift map from flat-space amplitudes to AdS amplitudes: one replaces ordinary Lorentz-invariant building blocks by conformally invariant ones, and ordinary propagator denominators by the Mellin--Barnes exchange operators $D_k^{\Delta}$ \cite{Li:2022tby, Mei:2023jkb}.  
	
	\subsection{Embedding space} \label{sec:correll_embedding}
	
	When working with AdS correlation functions in coordinate space, it is often useful to use the embedding-space formalism, in which both bulk AdS isometries and boundary conformal transformations are realized linearly. Embedding-space ideas were developed long ago in the study of conformal symmetry and were later reformulated in a modern way for CFT correlators, in particular for spinning operators~\cite{Dirac:1936fq,Mack:1969rr,Boulware:1970ty,Ferrara:1972cq,Weinberg:2010fx,Simmons-Duffin:2012juh,Costa:2011mg,Costa:2014kfa}. The basic point is that the Lorentz algebra of the ambient space is isomorphic to the conformal algebra of the boundary theory, so the action of the conformal group $SO(d+1,1)$ or $SO(d,2)$ becomes linear when one works on the projective null cone. Here, we will work with an AdS space of unit length $L=1$, embedded in an ambient space $\mathbb{R}^{d,2}$ with mostly plus signature.
	
	In this formalism one embeds $\mathrm{AdS}_{d+1}$ into $\mathbb{R}^{d,2}$ as the hyperboloid
	\begin{equation}
		X^2=-1 , \label{eq:emb_bulk}
	\end{equation}
	and identifies boundary points with projective null vectors $P^A$, satisfying
	\begin{equation}
		P^2=0,
		\qquad
		P^A \sim \lambda P^A . \label{eq:emb_bdy}
	\end{equation}
	Physical boundary operators are therefore homogeneous functions on the null cone, where only the conformal class of $P$ is physical. In the Poincar\'e patch one may parameterize bulk and boundary points as
	\begin{equation}
		X^A=\frac1z\left(x^a,\frac{1-x^2-z^2}{2},\frac{1+x^2+z^2}{2}\right),
		\qquad
		P^A=\left(x^a,\frac{1-x^2}{2},\frac{1+x^2}{2}\right). \label{eq:embedding_coord}
	\end{equation}
	Lorentz transformations in $\mathbb{R}^{d,2}$ then act simultaneously as AdS isometries in the bulk and conformal transformations on the boundary.
	
	For spinning fields it is convenient to package tensors into polynomials using auxiliary null polarization vectors. A bulk spin-$s$ field is represented as
	\begin{equation}
		H(X,W)=W^{A_1}\cdots W^{A_s} H_{A_1\cdots A_s}(X),
		\qquad
		W^2=0,
		\qquad
		W\cdot X=0 \ , 
	\end{equation}
	while a boundary spin-$s$ operator is represented as
	\begin{equation}
		B(P,U)=U^{A_1}\cdots U^{A_s} B_{A_1\cdots A_s}(P),
		\qquad
		U^2=0,
		\qquad
		U\cdot P=0  \ ,
	\end{equation}
	we will sometimes refer to these as polarized versions. For boundary spinning operators, we thus have a gauge redundancy $U \sim U+\alpha P $
	which ensures that only the physical transverse traceless data are retained. For a physical boundary polarization $\epsilon^a$, the corresponding embedding-space polarization is
	\begin{equation}
		U^A = \left(\epsilon^a,-\epsilon\cdot x,\epsilon\cdot x\right).
	\end{equation}
	
	The bulk symmetric traceless tensor can be recovered from $H(X,W)$ using the operator $K_A$
	\begin{equation}
		\begin{aligned}
			K_A =
			\left(\frac{d-1}{2}+W\cdot \frac{\partial}{\partial W}\right)
			\left(\frac{\partial}{\partial W^A}+X_A\left(X\cdot \frac{\partial}{\partial W}\right)\right) \\
			-\frac12 W_A
			\left(
			\frac{\partial^2}{\partial W\cdot \partial W}
			+
			\left(X\cdot \frac{\partial}{\partial W}\right)^2
			\right) \ ,
		\end{aligned}
	\end{equation}
	so that
	\begin{equation}
		H_{A_1\cdots A_s}(X)
		=
		\frac{1}{s!\left(\frac{d-1}{2}\right)_s}
		K_{A_1}\cdots K_{A_s} H(X,W) ,
	\end{equation}
	where $(a)_n$ is the Pochhammer symbol. Similarly, the boundary tensor is recovered from $B(P,U)$ using the operator $D_A$,
	\begin{equation}\label{eq:TodorovOp}
		D_A=
		\left(\frac d2-1+U\cdot \frac{\partial}{\partial U}\right)\frac{\partial}{\partial U^A}
		-\frac12 U_A \frac{\partial^2}{\partial U\cdot \partial U},
	\end{equation}
	through
	\begin{equation}
		B_{A_1\cdots A_s}(P)
		=
		\frac{1}{s!\left(\frac d2-1\right)_s}
		D_{A_1}\cdots D_{A_s} B(P,U).
	\end{equation}
	
	In the following, we will be interested in correlation functions of primary operators \footnote{Primary operators are local operators that are not obtained by acting with translations on another operator. Equivalently, at the origin they are annihilated by the generators of special conformal transformations \cite{Simmons-Duffin:2016gjk}.}, which are irreducible representations of the conformal group labeled by the conformal dimension $\Delta$ and spin $s$. A primary field encoded in an embedding-space polynomial obeys the homogeneity condition
	\begin{equation} \label{eq:scaling_correlator}
		O_i(r_i P^I_i, q_i U^I_i)= r_i^{-\Delta_i} q_i^{s_i} O_i (P_i^I, U_i^I) \ .
	\end{equation}
	A central advantage of the embedding-space formalism is that the generators of bulk isometries and boundary conformal transformations are both just Lorentz generators in $\mathbb{R}^{d,2}$. For bulk fields one has 
	\begin{equation}
		D_X^{AB}
		=
		\frac{1}{\sqrt{2}}
		\left(
		X^A \frac{\partial}{\partial X_B}
		-
		X^B \frac{\partial}{\partial X_A}
		+
		W^A \frac{\partial}{\partial W_B}
		-
		W^B \frac{\partial}{\partial W_A}
		\right),
	\end{equation}
	while for a boundary operator at $P_i$ we have
	\begin{equation}
		D_i^{AB}
		=
		\frac{1}{\sqrt{2}}
		\left(
		P_i^A \frac{\partial}{\partial P_{i,B}}
		-
		P_i^B \frac{\partial}{\partial P_{i,A}}
		+
		U_i^A \frac{\partial}{\partial U_{i,B}}
		-
		U_i^B \frac{\partial}{\partial U_{i,A}}
		\right). \label{eq:bdy_D}
	\end{equation}
	An $n$-point boundary correlator $A(P_i,U_i)$ obeys the conformal Ward identity
	\begin{equation} \label{eq:CWI}
		\sum_{i=1}^n D_i^{AB}\,A(P_i,U_i)=0 ,
	\end{equation}
	which plays a role analogous to momentum conservation in flat space. Meanwhile, the quadratic Casimir built from a single insertion,
	\begin{equation} \label{eq:Casimir}
		-D_i\cdot D_i \cong \Delta_i(\Delta_i-d)+s_i(s_i+d-2)\equiv M_{\Delta_i,s_i}^2 ,
	\end{equation}
	is the embedding-space analogue of the on-shell condition for a boundary state of conformal dimension $\Delta_i$ and spin $s_i$.
	
	Up to three points, any parity-even conformally invariant correlator can be expressed in terms of a small set of basic embedding-space building blocks \cite{Costa:2011mg},
	\begin{equation}\label{invariantseven}
		\begin{aligned}
			P_{ij} &\,= -2 P_i \cdot P_j \,,\\
			H_{ij} &\,= -2\Big[(U_i \cdot U_j)(P_i \cdot P_j) - (U_i \cdot P_j)(U_j \cdot P_i)\Big] \,,\\
			V_i &\,= V_{i,jk}
			= \frac{(U_i \cdot P_j)(P_k \cdot P_i) - (U_i \cdot P_k)(P_j \cdot P_i)}{P_j \cdot P_k} \,.
		\end{aligned}
	\end{equation}
	A general three-point correlator is then obtained by writing the most general scalar combination of these invariants with the correct homogeneity properties, as required by Eq.~\eqref{eq:scaling_correlator}. One must also impose the unitarity bound \cite{Mack:1975je,Ferrara:1974pt,Metsaev:1995re,Minwalla:1997ka,Simmons-Duffin:2016gjk}; in generic dimension $d$, this gives $\Delta \geq (d-2)/2$ for scalars, and $\Delta \geq s+d-2$ for symmetric traceless operators of spin $s\geq 1$. Spinning operators that saturate the bound correspond to a conserved current. For scalars, $\Delta=(d-2)/2$ describes a free scalar, while $\Delta=(d-1)/2$ corresponds to the alternate quantization of a conformally coupled scalar in AdS.
	
	For spinning operators, conservation imposes further restrictions beyond fixing the allowed dimension: it also constrains the admissible tensor structures. In the embedding-space formalism, the divergence of the operator inserted at $P_i$ is implemented by acting on the correlator with $(\partial_{P_i}\!\cdot D_{U_i})$, where $D_{U_i}$ is the operator defined in Eq.~\eqref{eq:TodorovOp} associated with the auxiliary vector $U_i$. Thus, when the $i$-th operator is conserved, \emph{i.e.} when its dimension saturates the unitarity bound $\Delta_i=d-2+s_i$, one must impose
	\begin{equation} \label{eq:embedding_conserv}
		(\partial_{P_i}\!\cdot D_{U_i})\,
		\big\langle \mathcal O_1(P_1,U_1)\cdots \mathcal O_i(P_i,U_i)\cdots \mathcal O_n(P_n,U_n)\big\rangle =0 ,
		\qquad \Delta_i=d-2+s_i ,
	\end{equation}
	up to terms proportional to $U_i^2$ and $U_i\!\cdot P_i$, which are set to zero in the index-free formalism.
	
	\bigskip
	
	\subsubsection{Differential representation} \label{sec:differential_representation}
	After this introduction to embedding space, we now summarize the construction of \cite{Diwakar:2021juk,Herderschee:2022ntr} which shows that, for the tree-level examples considered there, AdS boundary correlators admit a differential representation in which all nonlocality is isolated into inverse quadratic-Casimir operators, one for each internal line, while the remaining factor is local and acts on a single contact $D$-function. This gives an AdS analogue of the flat-space cubic-diagram expansion, with Mandelstam variables replaced by suitably ordered conformal generators.
	
	The starting point is to realize that, given Eq.~\eqref{eq:CWI} and Eq.~\eqref{eq:Casimir}, it is natural to define Mandelstam-like operators
	\begin{equation} \label{eq:embedding_mandels}
		\mathcal{D}_{ij}\equiv (D_i+D_j)_{AB}(D_i+D_j)^{AB},
		\qquad
		\mathcal{D}_I \equiv \left(\sum_{i\in I} D_i\right)_{AB}
		\left(\sum_{i\in I} D_i\right)^{AB},
	\end{equation}
	which satisfy the same linear relations as flat-space Mandelstam invariants on the support of the conformal Ward identities. At four points one has
	\begin{equation}
		\mathcal{D}_{12}\cong \mathcal{D}_{34},
		\qquad
		\mathcal{D}_{13}\cong \mathcal{D}_{24},
		\qquad
		\mathcal{D}_{23}\cong \mathcal{D}_{14},
	\end{equation}
	and
	\begin{equation}
		\mathcal{D}_{12}+\mathcal{D}_{23}+\mathcal{D}_{13}\cong -\sum_{i=1}^4 M_{\Delta_i,s_i}^2 .
	\end{equation}
	The important difference from flat space is that these are differential operators, so in general they do not commute. This noncommutativity is the main new ingredient in the AdS version of color-kinematics duality.
	
\begin{figure}[H]
		\centering
		\begin{tikzpicture}[scale=0.88, every node/.style={font=\scriptsize}]
			\def\contact#1{
				\begin{scope}[xshift=#1cm]
					\draw (-0.9,-0.85) -- (0.9,-0.85);
					\coordinate (v) at (0,0.15);
					\foreach \x/\lab in {-0.72/1,-0.24/2,0.24/3,0.72/4}{
						\draw (\x,-0.85) -- (v);
						\fill (\x,-0.85) circle (1.2pt);
						\node[below] at (\x,-0.85) {$P_{\lab}$};
					}
					\fill (v) circle (2pt);
				\end{scope}
			}
			\def\exchange#1#2#3#4#5#6{
				\begin{scope}[xshift=#1cm]
					\draw (-0.9,-0.85) -- (0.9,-0.85);
					\coordinate (vl) at (-0.34,0.12);
					\coordinate (vr) at (0.34,0.12);
					\draw (-0.72,-0.85) -- (vl);
					\draw (-0.24,-0.85) -- (vl);
					\draw (0.24,-0.85) -- (vr);
					\draw (0.72,-0.85) -- (vr);
					\draw[thick] (vl) -- node[above] {$\left(\mathcal D_{#6}+M^2\right)^{-1}$} (vr);
					\foreach \x/\lab in {-0.72/#2,-0.24/#3,0.24/#4,0.72/#5}{
						\fill (\x,-0.85) circle (1.2pt);
						\node[below] at (\x,-0.85) {$P_{\lab}$};
					}
					\fill (vl) circle (2pt);
					\fill (vr) circle (2pt);
				\end{scope}
			}
			\contact{-5.0}
			\node at (-5.0,-1.45) {$D_{\Delta_1,\Delta_2,\Delta_3,\Delta_4}$};
			\draw[->,thick] (-3.85,-0.15) -- (-2.95,-0.15) node[midway,above] {$\hat A_4$};
			\exchange{-1.8}{1}{2}{3}{4}{12}
			\node at (-1.8,-1.45) {$s$ channel};
			\node at (0,-0.15) {$+$};
			\exchange{1.8}{1}{4}{2}{3}{14}
			\node at (1.8,-1.45) {$t$ channel};
			\node at (3.6,-0.15) {$+$};
			\exchange{5.4}{1}{3}{2}{4}{13}
			\node at (5.4,-1.45) {$u$ channel};
			\node at (1.8,-2.05) {$C_g\left(\mathcal D_I+M^2\right)^{-1}\hat N_g\,D_{\Delta_1,\Delta_2,\Delta_3,\Delta_4}$};
		\end{tikzpicture}
		\caption{Schematic differential representation of a four-point boundary correlator. A single contact Witten diagram is acted on by local numerator operators and inverse quadratic-Casimir operators, one for each internal bulk line, to generate the cubic $s$-, $t$-, and $u$-channel contributions. The displayed external ordering in each channel indicates the corresponding partition; spin labels on the internal state and overall normalizations are suppressed.}
		\label{fig:differential_witten_representation}
	\end{figure}
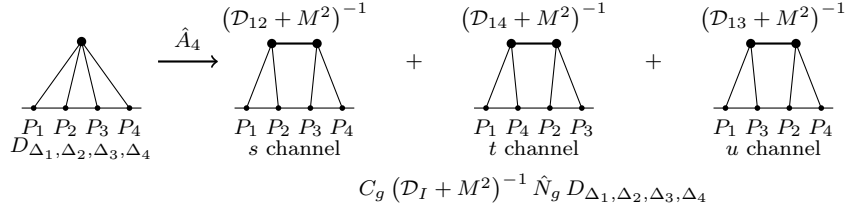

	The basic claim of the differential representation is that an $n$-point boundary correlator can be written as
	\begin{equation} \label{eq:DifRepCorrel}
		A_n = \hat{A}_n\, D_{\Delta_1,\ldots,\Delta_n},
		\qquad
		D_{\Delta_1,\ldots,\Delta_n}
		=
		\int_{\mathrm{AdS}} dX \prod_{i=1}^n E_{\Delta_i}(X,P_i),
	\end{equation}
	where
	\begin{equation}
		E_{\Delta_i}(X,P_i)=\frac{1}{(-2P_i\cdot X)^{\Delta_i}}
	\end{equation}
	is the scalar bulk-boundary propagator and $D_{\Delta_1,\ldots,\Delta_n}$ is the contact Witten diagram. In this language, the contact $D$-function plays a role analogous to the momentum-conserving delta function in flat space. For the tree-level correlators admitting this representation, the operator $\hat{A}_n$ can be organized as
	\begin{equation}
		\hat{A}_n
		=
		\sum_{g\in \text{cubic}}
		C_g
		\prod_{I\in g}
		\frac{1}{\mathcal{D}_I+M_{\Delta_I,s}^2}
		\,\hat{N}_g ,
	\end{equation}
	where the sum runs over cubic Witten-diagram topologies, $C_g$ is the color factor, the inverse operators
	\begin{equation}
		\frac{1}{\mathcal{D}_I+M_{\Delta_I,s}^2}
	\end{equation}
	are the AdS analogues of propagators, and $\hat{N}_g$ are local differential operators that play the role of kinematic numerators. This representation makes the analogy with flat-space amplitudes manifest: bulk-bulk propagators are replaced by inverse Casimir operators, while the kinematic dependence is encoded in local differential operators acting on a single contact diagram. The four-point version of this construction is illustrated in Fig.~\ref{fig:differential_witten_representation}.
	
	\subsubsection{BCJ relations, color-kinematics duality, and double copy}
	A useful consequence of this structure is that the derivation of BCJ relations parallels the flat-space one \cite{Diwakar:2021juk}. In the embedding-space formulation, one writes the vector of DDM-basis partial correlators as
	\begin{equation}
		A(1,\alpha,n)=\sum_{\beta\in S_{n-2}} \hat m(1,\alpha,n|1,\beta,n)\,\hat N_{1,\beta,n}\,D_{d,\dots,d}\ ,
	\end{equation}
	where $\hat m$ is the matrix of bi-adjoint scalar exchange operators obtained from its flat-space counterpart by the replacement $s_I\rightarrow \mathcal{D}_I$, and $\hat N_{1,\beta,n}$ are differential kinematic numerators. Exactly as in flat space, any left null vector of $\hat m$ gives a relation among partial correlators,
	\begin{equation}
		\sum_{\beta\in S_{n-2}} \hat v(\beta)\,\hat m(1,\beta,n|\alpha)=0
		\qquad \Longrightarrow \qquad
		\sum_{\beta\in S_{n-2}} \hat v(\beta)\,A(1,\beta,n)=0 \ .
	\end{equation}
	Assuming that the null vectors of the flat-space bi-adjoint matrix continue to null vectors of $\hat m$ after the replacement $s_I\rightarrow \mathcal{D}_I$, one is led to the AdS fundamental BCJ relations \cite{Diwakar:2021juk}
	\begin{equation}
		0=\mathcal{D}_{12}A(1,2,\dots,n)+\sum_{j=3}^{n-1}\left(\mathcal{D}_{12}+\sum_{k=3}^{j}\mathcal{D}_{2k}\right)A(1,3,\dots,j,2,j+1,\dots,n) \ .
	\end{equation}
	
	At four points this can be seen explicitly. The two DDM-basis partial correlators are
	\begin{equation}
		\begin{pmatrix}
			A(1,2,3,4)\\[2mm]
			A(1,3,2,4)
		\end{pmatrix}
		=
		\underbrace{
			\begin{pmatrix}
				\displaystyle \frac{1}{\mathcal{D}_{12}}+\frac{1}{\mathcal{D}_{23}} &
				\displaystyle -\,\frac{1}{\mathcal{D}_{23}} \\[3mm]
				\displaystyle -\,\frac{1}{\mathcal{D}_{23}} &
				\displaystyle \frac{1}{\mathcal{D}_{13}}+\frac{1}{\mathcal{D}_{23}}
		\end{pmatrix}}_{\hat m}
		\begin{pmatrix}
			\hat N_{1,2,3,4}\\[2mm]
			\hat N_{1,3,2,4}
		\end{pmatrix}
		D_{d,d,d,d} \ ,
	\end{equation}
	which is the embedding-space version of the flat space Eq.~\eqref{eq:DerivBCJ}. One then considers the differential row vector
	\begin{equation}
		\hat v=\begin{pmatrix}\mathcal{D}_{12}&-\mathcal{D}_{13}\end{pmatrix},
	\end{equation}
	and checks that it annihilates $\hat m$ from the left, namely
	\begin{equation}
		\mathcal{D}_{12}\,\hat m(1,2,3,4|\alpha)-\mathcal{D}_{13}\,\hat m(1,3,2,4|\alpha)\cong 0,
		\qquad
		\alpha=(1,2,3,4),(1,3,2,4)\ .
	\end{equation}
	Repeating the same computation for $\alpha=(1,3,2,4)$ and then multiplying the vector of partial correlators from the left by $\hat v$, one obtains
	\begin{equation}
		\mathcal{D}_{12}A(1,2,3,4)=\mathcal{D}_{13}A(1,3,2,4)\ , \label{eq:4pt_bcj_difop}
	\end{equation}
	which is the AdS analogue of the standard flat-space four-point BCJ relation.
	
	A first explicit example is the four-point adjoint scalar correlator with gluon exchange \cite{Herderschee:2022ntr} arising from the Lagrangian
	\be
	\mathcal{L}_{\text{gauge}}
	=-\frac{1}{4}\,F_{\mu\nu}^{a}F^{a,\mu\nu}
	-\frac{1}{2}\,D_{\mu}\phi^{a}D^{\mu}\phi^{a}
	-\frac{1}{2}\,D_{\mu}\tilde{\phi}^{a}D^{\mu}\tilde{\phi}^{a}
	-\frac{1}{4}\,f^{abc}f^{cde}\,\phi^{a}\tilde{\phi}^{b}\phi^{d}\tilde{\phi}^{e}\ .
	\ee
	For external scalars, the $s$-channel contribution takes the remarkably simple form
	\begin{equation}
		A_s(1,2,3,4)
		=
		\frac{1}{\mathcal{D}_{12}}(\mathcal{D}_{23}-\mathcal{D}_{13})\, D_{d,d,d,d} \ ,
	\end{equation}
	where $D_{d,d,d,d}$ was defined in Eq.~\eqref{eq:DifRepCorrel}, so that the full color-dressed correlator is
	\begin{equation}
		A(\phi_1\phi_2\phi_3\phi_4)
		=
		\left(
		C_s \frac{1}{\mathcal{D}_{12}}\hat{N}^{\phi\phi\phi\phi}_s
		+
		C_t \frac{1}{\mathcal{D}_{23}}\hat{N}^{\phi\phi\phi\phi}_t
		+
		C_u \frac{1}{\mathcal{D}_{13}}\hat{N}^{\phi\phi\phi\phi}_u
		\right)
		D_{d,d,d,d},
	\end{equation}
	with numerators obtained from the flat-space ones by the replacement $s_{ij}\rightarrow \mathcal{D}_{ij}$ \cite{Herderschee:2022ntr}:
	\be
	\begin{aligned}
		&\hat{N}_{s}^{\phi\phi\phi\phi}=\mathcal{D}_{23}-\mathcal{D}_{13}\ ,\qquad
		&\hat{N}_{t}^{\phi\phi\phi\phi}=\mathcal{D}_{13}-\mathcal{D}_{12}\ ,\qquad
		&\hat{N}_{u}^{\phi\phi\phi\phi}=\mathcal{D}_{12}-\mathcal{D}_{23}\ , \\
		&\hat{N}_{s}^{\phi\phi\tilde{\phi}\tilde{\phi}}=\mathcal{D}_{23}-\mathcal{D}_{13}\ ,\qquad
		&\hat{N}_{t}^{\phi\phi\tilde{\phi}\tilde{\phi}}=-\mathcal{D}_{23}\ ,\qquad
		&\hat{N}_{u}^{\phi\phi\tilde{\phi}\tilde{\phi}}=\mathcal{D}_{13}\ ,
	\end{aligned}
	\ee
	which satisfy an operator kinematic Jacobi identity. In particular, the four-point BCJ relation becomes
	\begin{equation}
		\mathcal{D}_{12} A(1,2,3,4)=\mathcal{D}_{13} A(1,3,2,4) , \label{eq:4ptBCJAds}
	\end{equation}
	exactly mirroring the flat-space relation. As in flat space, the representation is not unique: there is a generalized gauge freedom under shifts of the numerator operators by terms proportional to the corresponding channel operators, provided that the extra pieces either annihilate the common contact term or sum to zero in the Jacobi combination \cite{Diwakar:2021juk}. This freedom is useful for moving between equivalent differential representations without changing the correlator.
	
	The NLSM provides a particularly clean illustration of these ideas. It was shown in \cite{Diwakar:2021juk} that the AdS boundary correlators of the NLSM satisfy differential BCJ relations and admit kinematic numerators that are the direct embedding-space analogues of the flat-space ones under the replacement $s_{ij}\rightarrow \mathcal{D}_{ij}$. At four points, a convenient color/kinematics-satisfying choice is
	\be
	\hat{N}_s=\mathcal{D}_{12}\mathcal{D}_{13}\ ,\qquad
	\hat{N}_u=-\mathcal{D}_{13}\mathcal{D}_{12}\ ,\qquad
	\hat{N}_t=-[\mathcal{D}_{12},\mathcal{D}_{13}] \ ,
	\ee
	so that $\hat{N}_s+\hat{N}_t+\hat{N}_u=0$. The commutator term vanishes on the four-point contact diagram, and the resulting partial correlators obey the four-point BCJ relation above. The same strategy was carried out explicitly through six points, where after absorbing contact terms into cubic numerators and discarding commutators that annihilate the relevant $D$-functions, one again finds a differential realization of color-kinematics duality and the corresponding fundamental BCJ relation. Thus, for the NLSM, the embedding-space differential representation has been checked explicitly at both four and six points \cite{Diwakar:2021juk}. For completeness, the six-point scalar correlator was shown to obey the differential fundamental BCJ relation
	\begin{equation}
		\begin{aligned}
			0=
			\mathcal{D}_{12}A(1,2,3,4,5,6)
			+
			(\mathcal{D}_{12}+\mathcal{D}_{23})A(1,3,2,4,5,6) \\
			-
			(\mathcal{D}_{25}+\mathcal{D}_{26})A(1,3,4,2,5,6)
			-
			\mathcal{D}_{26}A(1,3,4,5,2,6) \ .
		\end{aligned}
	\end{equation}
	
	The construction of the YM four-point correlator proceeds somewhat differently. Starting from the embedding-space YM action, one first determines the three-point correlator and then builds the four-point exchange diagram in position space. The result is reduced to a linear combination of $D$-functions by using embedding-space identities and, in the case of a four-dimensional boundary, can be organized explicitly into color-ordered pieces. This made it possible to check directly that the resulting four-point YM correlator obeys the differential BCJ relation
	$
	\mathcal{D}_{12}A(1,2,3,4)=\mathcal{D}_{13}A(1,3,2,4)
	$
	for $d=4$, strongly supporting the existence of a differential color/kinematics representation for YM boundary correlators in AdS \cite{Diwakar:2021juk}.
	
	An example including gravitons is the scalar four-point correlator with graviton exchange \cite{Herderschee:2022ntr} arising from the Lagrangian
	\be
	\mathcal{L}_{\text{gravity}}
	=-\frac{\sqrt{-g}}{2}\left(g^{\mu\nu}\partial_{\mu}\varphi\,\partial_{\nu}\varphi+g^{\mu\nu}\partial_{\mu}\tilde{\varphi}\,\partial_{\nu}\tilde{\varphi}\right)\ .
	\ee
	Here the structure is again very close to the flat-space amplitude, but with an important correction. The full result can be organized as
	\begin{equation}
		\begin{aligned}
			M
			=
			\frac{1}{16}
			\left[
			\frac{1}{\mathcal{D}_{12}+2d}\left(\hat{N}_s^2+2d\,\mathcal{D}_{12}\right)
			+
			\frac{1}{\mathcal{D}_{23}+2d}\left(\hat{N}_t^2+2d\,\mathcal{D}_{23}\right) \right. \\
			\left.
			+
			\frac{1}{\mathcal{D}_{13}+2d}\left(\hat{N}_u^2+2d\,\mathcal{D}_{13}\right)
			\right]
			D_{d,d,d,d},
		\end{aligned} \label{eq:scalar_ex_DC_embedding}
	\end{equation}
	where $\hat{N}_{s,t,u}$ are the gauge-theory BCJ numerators with the same replacement $s_{ij}\rightarrow \mathcal{D}_{ij}$. Thus the naive double copy is corrected by terms proportional to the graviton Casimir eigenvalue $M^2_{\rm graviton}=2d$. These corrections are subleading in the large-$d$ limit since the action of conformal generators on the contact term yields a factor of $d$, so the usual flat-space double copy form is recovered asymptotically. The need for these corrections had already been realized in \cite{Diwakar:2021juk}, which suggested that the most naive prescription must be supplemented by shifts in the conformal weights of the external contact diagram. It was also realized in \cite{Diwakar:2021juk} that the embedding space three-point correlators satisfy the simple multiplicative double copy only in the large dimension limit. Thus, color-kinematics duality survives in AdS in a differential sense, but the straightforward flat-space double copy prescription at finite $d$ requires curvature-dependent modifications.

	\subsubsection{Alternative differential operators}
	A closely related differential representation was proposed in Ref.~\cite{Li:2022tby}. It uses embedding-space weight-shifting operators\footnote{For a review on weight shifting operators in embedding space and how to translate them to momentum space see Appendices A and B of \cite{Baumann:2019oyu}.} rather than the momentum-space operators in Eq.~\eqref{eq:weight_shifting_mom}. The basic idea is nevertheless similar to the construction of Ref.~\cite{Lee:2022fgr}: the flat-space polarization vectors and momenta are replaced by differential operators acting on scalar contact Witten diagrams, with a prescribed operator ordering.

	For a symmetric traceless primary of dimension $\Delta_i$ and spin $s_i$, the spin-up and dimension-up operators are~\cite{Costa:2011dw,Karateev:2017jgd}
	\begin{align}
		&D_i^{0+,\mu}
		=
		(s_i+\Delta_i)U_i^\mu
		+P_i^\mu U_i\cdot\partial_{P_i},
		\nonumber\\
		&D_i^{+0,\mu}
		=
		c_1\partial_{P_i}^{\mu}
		+c_2 P_i^\mu\partial_{P_i}^2
		+c_3 U_i^\mu\partial_{U_i}\cdot\partial_{P_i}
		+c_4 U_i\cdot\partial_{P_i}\,\partial_{U_i}^{\mu}
		\nonumber\\
		&\quad
		+c_5 P_i^\mu U_i\cdot\partial_{P_i}\,\partial_{U_i}\cdot\partial_{P_i}
		+c_6 U_i^\mu U_i\cdot\partial_{P_i}\,\partial_{U_i}^2
		+c_7 P_i^\mu (U_i\cdot\partial_{P_i})^2\partial_{U_i}^2 .
		\label{eq:spin_dimension_weight_shifts}
	\end{align}
	where the $c_i$ coefficients are given explicitly in \cite{Karateev:2017jgd}. These are then normalized as
	\begin{align}
		{\mathsf E}_i^\mu
		=&
		-\frac{P_i\cdot\partial_{P_i}+U_i\cdot\partial_{U_i}}{
			(P_i\cdot\partial_{P_i})(P_i\cdot\partial_{P_i}+1)}
		D_i^{0+,\mu},
		\\
		{\mathsf P}_i^\mu=&\frac{2}{
			(P_i\cdot\partial_{P_i}+1)(d+P_i\cdot\partial_{P_i}-2)
			(d+2P_i\cdot\partial_{P_i}-2)}
		D_i^{+0,\mu}.
		\label{eq:normalized_weight_shifts}
	\end{align}
	The operator $\mathsf{E}_{i}^{\mu}$ raises the spin of the $i$th operator, while $\mathsf{P}_{i}^{\mu}$ raises its conformal dimension and plays the role of a derivative or momentum insertion. This is the embedding-space analogue of the momentum-space construction in Section~\ref{sec:mom_dif_op}, but the operators act on boundary position-space contact diagrams rather than directly on reduced momentum-space correlators. The $\mathsf{P}_{i}$ do not obey ordinary momentum conservation as operator identities. Instead, flat-space-like relations only emerge after using conformal Ward identities, transversality, and a fixed ordering prescription.
	
	The three-point Yang--Mills correlator illustrates the construction most directly.  Denoting by ${\cal W}_{\Delta_1,\Delta_2,\Delta_3}$ the scalar contact Witten diagram, the cubic color-stripped Yang--Mills result is
	\begin{align}
		{\cal M}_{3,{\rm YM}}^{a_1a_2a_3}
		&=
		-\frac{d-1}{2}\,g_{\rm YM}
		\,\mathsf{T}_{\rm O}
		\Big[
		({\mathsf E}_2\cdot{\mathsf E}_3)({\mathsf E}_1\cdot{\mathsf P}_2)
		-(1\leftrightarrow 2)
		\nonumber\\
		&\hspace{4.2cm}
		+(2\to1,\,1\to3)
		\Big] {\cal W}_{\Delta_1,\Delta_2,\Delta_3} .
		\label{eq:three_point_yang_mills_weight_shift}
	\end{align}
	Thus the cubic Yang--Mills correlator is obtained from the flat-space cubic vertex by replacing the contact diagram with the momentum conserving delta function and performing the replacements
	\begin{equation}
		\epsilon_i^\mu\longrightarrow {\mathsf E}_i^\mu,
		\qquad
		p_i^\mu\longrightarrow {\mathsf P}_i^\mu \ ,
		\label{eq:flat_space_weight_shift_dictionary}
	\end{equation}
	together with the ordering prescription $\mathsf{T}_{\rm O}$, which places all ${\mathsf E}_i$'s to the left of the corresponding ${\mathsf P}_i$'s when they act on the same point.  The analogy with the momentum-space construction of Sec.~\ref{sec:mom_dif_op} is therefore quite direct, although here the operators are embedding-space weight-shifting operators rather than derivatives with respect to boundary momenta.
	
	The graviton case has the same structure and is given by the flat-space uplift
	\begin{equation}
		{\cal M}_{3,{\rm grav}}
		=
		\mathsf{T}_{\rm O}
		\left[
		{\cal M}_{3,{\rm grav}}^{\rm flat}
		\right]_{\epsilon\to{\mathsf E},\,p\to{\mathsf P},\,\delta\to{\cal W}}.
		\label{eq:three_point_gravity_weight_shift}
	\end{equation}
	At this level the differential double copy is immediate.  Although ${\mathsf P}_i$ is not a conserved momentum, effective three-point identities such as ${\mathsf E}_3\cdot {\mathsf P}_2\to -{\mathsf E}_3\cdot {\mathsf P}_1$ and ${\mathsf P}_1\cdot{\mathsf P}_2\to0$ hold inside the correlator.  Consequently
	\begin{equation}
		\widehat{{\cal M}}_{3,{\rm grav}}
		=
		\frac{32\sqrt{8\pi G}}{(d-1)^2g_{\rm YM}^2}
		\,\mathsf{T}_{\rm O}
		\left[
		\left(\widehat{{\cal M}}_{3,{\rm YM}}\right)^2
		\right],
		\label{eq:three_point_weight_shift_double_copy}
	\end{equation}
	where $\widehat{{\cal M}}$ is the correlator with the contact seed ${\cal W}$ stripped off. Thus, at three points the result is conceptually parallel to the differential double copy of Ref.~\cite{Lee:2022fgr}, but formulated with embedding-space weight-shifting operators rather than the momentum-space operators.
	
	At four points, exchange diagrams introduce inverse Casimir operators.  The relevant denominator is ${\cal D}_{ij}^{\Delta,s}=-\mathcal{D}_{ij}-M_{\Delta,s}^2$, where $M_{\Delta,s}$ was defined in Eq.~\eqref{eq:Casimir} and $\mathcal{D}_{ij}$ in Eq.~\eqref{eq:embedding_mandels}. For the exchanged gluon, $(\Delta,s)=(d-1,1)$, so $M_{\Delta,s}=0$ as in the discussion above. The four-point Yang--Mills answer can then be written schematically as
	\begin{equation}
		{\cal M}_{4,{\rm YM}}
		=
		\frac{(d-1)^2}{4}
		\,\mathsf{T}_{\rm O}
		\left[
		{\cal M}_{4,{\rm YM}}^{\rm flat}
		\right]_{
			\epsilon\to{\mathsf E},\,p\to{\mathsf P},\,\delta\to{\cal W},\,
			1/s_{ij}\to 1/(2{\cal D}_{ij}^{d-1,1})}.
		\label{eq:four_point_yang_mills_weight_shift}
	\end{equation}
	For gravity the same uplift captures the flat-space part, but an extra AdS contact contribution remains,
	\begin{equation}
		{\cal M}_{4,{\rm grav}}
		=
		\mathsf{T}_{\rm O}
		\left[
		{\cal M}_{4,{\rm grav}}^{\rm flat}
		\right]_{
			\epsilon\to{\mathsf E},\,p\to{\mathsf P},\,
			1/s_{ij}\to 1/(2{\cal D}_{ij}^{d,2})}
		+\frac{1}{R_{\rm AdS}^2}{\cal M}_{4,{\rm grav}}^{\rm AdS},
		\label{eq:four_point_gravity_weight_shift}
	\end{equation}
	with
	\begin{align}
		\widetilde{{\cal M}}_{4,{\rm grav}}^{\rm AdS}
		=
		4\pi Gd
		\big[
		&({\mathsf E}_1\cdot{\mathsf E}_2)^2({\mathsf E}_3\cdot{\mathsf E}_4)^2
		-4({\mathsf E}_1\cdot{\mathsf E}_2)({\mathsf E}_2\cdot{\mathsf E}_3)
		({\mathsf E}_3\cdot{\mathsf E}_4)({\mathsf E}_4\cdot{\mathsf E}_1) \nonumber \\
		&+{\rm permutations.}
		\big].
		\label{eq:adS_gravity_contact_term}
	\end{align}
	This additional term comes from the cosmological-constant part of the Einstein-Hilbert action and vanishes in the flat-space limit.  In AdS it is required by bulk gauge invariance, equivalently by conservation of the boundary stress tensor.
	
	The four-point BCJ relation is obtained after passing to a second, related differential representation.  The operators ${\mathsf P}_i$ do not give a simple momentum-conservation identity at four points.  Instead one rewrites contractions in terms of conformal generators,
	\begin{equation}
		({\mathsf E}_i\cdot{\mathsf P}_j)({\mathsf E}_k\cdot{\mathsf P}_l)
		\longrightarrow
		4{\mathsf E}_i^\mu {\mathsf E}_k^\nu
		D_{j\mu}{}^\rho D_{l\rho\nu},
		\qquad
		{\mathsf P}_i\cdot{\mathsf P}_j
		\longrightarrow
		-D_i\cdot D_j  \ ,
		\label{eq:momentum_to_conformal_generator_replacement}
	\end{equation}
	where $D_{i\mu\nu}$ denotes the coordinate-space projection of the embedding-space generator $D_i^{AB}$, with the embedding indices pulled back to boundary coordinate indices, and the conformal Ward identity, Eq.~\eqref{eq:CWI}, gives the replacement for momentum conservation.  In this representation the color-ordered current correlator takes the form
	\begin{equation}
		\widetilde{{\cal M}}[1234]
		=
		\frac{1}{{\cal D}_{12}^{d-1,1}}\mathcal{N}_s
		-\frac{1}{{\cal D}_{23}^{d-1,1}}\mathcal{N}_t \ ,
		\label{eq:color_ordered_yang_mills_denominators}
	\end{equation}
	where the differential numerators are obtained by uplifting flat-space BCJ numerators and then reordering, that is,
	\be
	\mathcal{N}_s
	=
	\mathsf{T}_{\rm O}\left(
	n_s^{\mathrm{flat}}\big|_{\epsilon\to\mathcal{E},\,p\to\mathcal{P}}
	\right) \, .
	\ee
	These differential kinematic numerators satisfy the Jacobi identities and the correlators satisfy the BCJ relations in Eq.~\eqref{eq:4pt_bcj_difop}. This suggests a four-point double copy of the form
	\begin{equation}
		\widetilde{{\cal M}}_{4,{\rm grav}}
		\sim
		\mathsf{T}_{\rm O}
		\left[
		\frac{\mathcal{N}_s^2}{{\cal D}_{12}^{d,2}}
		+\frac{\mathcal{N}_t^2}{{\cal D}_{23}^{d,2}}
		+\frac{\mathcal{N}_u^2}{{\cal D}_{24}^{d,2}}
		\right]
		+\widetilde{{\cal M}}_{\rm res} \ ,
		\label{eq:four_point_weight_shift_double_copy_problem}
	\end{equation}
	but the unresolved part, $\widetilde{{\cal M}}_{\rm res}$, is not easy to guess. This is another manifestation of the same phenomenon seen in Eq.~\eqref{eq:scalar_ex_DC_embedding}: in AdS the naive square of the gauge-theory expression misses curvature-dependent lower-derivative terms. The weight-shifting approach therefore provides a sharp three-point double copy and a promising four-point Yang--Mills BCJ representation, but a complete finite-radius four-point gravity double copy still requires additional AdS contact data.
	
	\subsection{Isometric space}
	The embedding-space description reviewed above makes the action of the conformal group on AdS boundary data manifest, and for maximally symmetric spaces it provides a very efficient language for writing correlators. The isometric-space formalism of \cite{Cheung:2022pdk} extends this idea to a broader class of curved backgrounds, namely symmetric spaces, by replacing the embedding-space realization of the conformal generators with the algebra of spacetime isometries. In this sense, isometric space is a direct generalization of embedding space. Flat space and AdS then arise as special cases, with translations and conformal generators playing the role of the corresponding isometric momenta.  Related covariant formulations of color-kinematics duality directly at the level of equations of motion were developed in \cite{Cheung:2021zvb}. In the following, we review the construction of \cite{Cheung:2022pdk} of BCJ relations in isometric space.
	
	For a symmetric space $\mathcal{M}\simeq G/H$, one starts from an overcomplete basis of Killing vectors
	\begin{equation}
		K_A = K_A{}^\mu \partial_\mu ,
	\end{equation}
	obeying the Killing algebra
	\begin{equation}
		[K_A,K_B] = F_{AB}{}^{C} K_C \ .
	\end{equation}
	It is convenient to define the bilinear form
	\begin{equation}
		g_{AB}=F_{AC}{}^{D}F_{BD}{}^{C} \ ,
	\end{equation}
	that raises and lowers isometric indices and the associated Lie-derivative operators
	\begin{equation}
		\mathbb{D}_A \equiv \mathcal{L}_{K_A}.
	\end{equation}
	For scalars on a symmetric space, the Laplacian is equal to the quadratic Casimir built from these operators,
	\begin{equation}
		\nabla^2 \phi = \mathbb{D}^2 \phi,
		\qquad
		\mathbb{D}^2 \equiv g^{AB}\mathbb{D}_A\mathbb{D}_B .
	\end{equation}
	This is the basic observation that makes the formalism useful: propagators become inverse Casimirs and interaction vertices can be written directly in terms of isometry generators \cite{Cheung:2022pdk}. The external states are encoded by wavefunctions $\psi(p_i,x)$ solving the free massless scalar equation
	\begin{equation}
		\mathbb{D}^2\psi(p_i,x)=0 .
	\end{equation}
	Here $p_i$ labels the kinematic quantum numbers of the solution. The key new ingredient is that the action of a spacetime isometry on the coordinate $x$ can be compensated by an action on the external data,
	\begin{equation}
		\left(\mathbb{D}+\mathbb{D}_i\right)\psi(p_i,x)=0 ,
	\end{equation}
	where $\mathbb{D}_i$ is the \emph{isometric momentum} of leg $i$. These operators satisfy the same algebra as the Killing vectors,
	\begin{equation}
		[\mathbb{D}_{iA},\mathbb{D}_{iB}] = F_{AB}{}^{C}\,\mathbb{D}_{iC},
	\end{equation}
	and the previous two equations imply the on-shell condition
	\begin{equation}
		\mathbb{D}_i^2 \psi(p_i,x)=0 .
	\end{equation}
	Thus the kinematics are formally close to both flat space and AdS: in flat space $\mathbb{D}$ reduces to translations and $\psi\sim e^{ip\cdot x}$ with $p_i^2=0$, while in AdS $\mathbb{D}$ becomes a conformal generator and $\psi$ is the bulk-boundary propagator for a massless scalar. The only essential novelty is that the isometric momenta are generically non-commutative.
	
	For several external legs one defines the product wavefunction
	\begin{equation}
		\psi(p_1,\ldots,p_n;x)=\prod_{i=1}^n \psi(p_i,x),
	\end{equation}
	which obeys the analogue of momentum conservation,
	\begin{equation}
		\left(\mathbb{D}+\sum_{i=1}^n \mathbb{D}_i\right)\psi(p_1,\ldots,p_n;x)=0 .
	\end{equation}
	It is particularly useful to define a generalization of the momentum-conserving delta function in flat space which is given by the spacetime integral,
	\begin{equation}
		\Delta_n \equiv \Delta(p_1,\ldots,p_n)=\int_x \psi(p_1,\ldots,p_n;x),
	\end{equation}
	and it is dubbed the n-point contact correlator. In flat space it indeed reduces to the momentum-conserving delta function, while in AdS it is the contact Witten diagram. It satisfies
	\begin{equation}
		\left(\sum_{i=1}^n \mathbb{D}_i\right)\Delta_n=0,
		\qquad
		\mathbb{D}_i^2 \Delta_n=0 .
	\end{equation}
	One also defines sums over subsets of legs,
	\begin{equation}
		\mathbb{D}_I \equiv \sum_{i\in I}\mathbb{D}_i ,
	\end{equation}
	whose squares are subset Casimirs. For later use it is important that
	\begin{equation}
		[\mathbb{D}_I^2,\mathbb{D}_J^2]=0
		\qquad\text{if}\qquad
		I\supseteq J,\;\; I\subseteq J,\;\;\text{or}\;\; I\cap J=\varnothing ,
	\end{equation}
	and similarly
	\begin{equation}
		[\mathbb{D}_I,\mathbb{D}_J^2]=0
		\qquad\text{if}\qquad
		I\supseteq J\;\;\text{or}\;\; I\cap J=\varnothing .
	\end{equation}
	These commutation properties are the curved-space replacement for the ordinary commutativity of Mandelstam invariants. 
	
	Now, we introduce the main observable which is the on-shell correlator, defined as the curved-space analogue of LSZ reduction,
	\begin{equation}
		A_n \equiv A(p_1,\ldots,p_n)
		=
		\left(\prod_{i=1}^{n}\int_{x_i}
		\psi(p_i,x_i)\,\nabla_{x_i}^{2}\right)\,
		\langle \phi(x_1)\cdots \phi(x_n)\rangle \ ,
	\end{equation}
	For symmetric spaces this can be rewritten in a differential form dubbed the \emph{isometric representation},
	\begin{equation}
		A(p_1,\ldots,p_n)=\mathbb{A}(\mathbb{D}_1,\ldots,\mathbb{D}_n)\,\Delta(p_1,\ldots,p_n),
		\label{eq:isometric_corr}
	\end{equation}
	which is the direct generalization of the differential representation of AdS boundary correlators \cite{Diwakar:2021juk,Herderschee:2022ntr} introduced above. The operator $\mathbb{A}$ is built from propagators $1/\mathbb{D}^2$ and vertices written in terms of the isometric momenta. In flat space this reproduces the ordinary scattering amplitude, while in AdS it reproduces the boundary correlator.

	A practical way to construct $\mathbb{A}$ is to solve the classical equations of motion perturbatively, equivalently using Berends--Giele recursion. The resulting diagrammatic prescription can be summarized as follows. One first enumerates the tree graphs contributing to the desired correlator, with external legs amputated. Then, one chooses an external leg as a root and the others as leaves. Next one evaluates each graph from the leaves towards the root, multiplying the corresponding vertices, propagators, and external wavefunctions in that order. Finally, each propagator or vertex is written in terms of the total isometric momentum flowing from above it, which is equivalent to eliminating the root momentum using total isometric momentum conservation. This last point is important: it implies that propagators can always be expressed in terms of nested or disjoint subset Casimirs, so they commute with one another, and moreover each propagator also commutes with all numerator factors lying below it in the graph. Hence local tree-level on-shell correlators admit a clean numerator/denominator split,
	\begin{equation}
		\mathbb{A}_n
		=
		\sum_{\alpha}
		\frac{1}{\mathbbm{d}_{\alpha}(\mathbb{D}_1,\ldots,\mathbb{D}_n)}
		\,\mathbbm{n}_{\alpha}(\mathbb{D}_1,\ldots,\mathbb{D}_n),
		\label{eq:num_denom}
	\end{equation}
	with mutually commuting denominators \cite{Cheung:2022pdk}. This structural fact is what makes a curved-space version of color-kinematics duality possible. Furthermore, because nonlinear field redefinitions only generate extra terms proportional to the on-shell condition of the root leg, the resulting on-shell correlators are invariant under changes of field basis, just as in flat space.
	
	\subsubsection{Biadjoint-scalar and NLSM on-shell correlators}
	The simplest example relevant for color-kinematics duality is the biadjoint scalar. Writing the field as $\phi^{\bar a} = \phi^{a\bar a} T_a$,  its curved-space equation of motion is
	\begin{equation}
		\nabla^2 \phi^{\bar c} + \frac{1}{2}\,f^{\bar c}{}_{\bar a\bar b}\,[\phi^{\bar a},\phi^{\bar b}] = 0 ,
	\end{equation}
	with dual-color current
	\begin{equation}
		\mathcal{K}^{\bar c}_{\mu}
		=
		f^{\bar c}{}_{\bar a\bar b}\,
		\mathrm{tr}\!\left(
		\phi^{\bar a}\overleftrightarrow{\nabla}_{\mu}\phi^{\bar b}
		\right),
	\end{equation}
	satisfying $\nabla^\mu \mathcal{K}^{\bar c}_{\mu}=0$.  In isometric frame this becomes
	\begin{equation}
		\mathbb{D}^2 \phi^{\bar c} + \frac{1}{2}\,f^{\bar c}{}_{\bar a\bar b}\,[\phi^{\bar a},\phi^{\bar b}] = 0 ,
	\end{equation}
	with
	\begin{equation}
		\mathcal{K}^{\bar c}_{\mu}
		=
		f^{\bar c}{}_{\bar a\bar b}\,
		\mathrm{tr}\!\left(
		\phi^{\bar a}\overleftrightarrow{\mathbb{D}}_{\mu}\phi^{\bar b}
		\right) \ ,
	\end{equation}
	where we defined the spacetime-indexed isometric derivative $\mathbb{D}_\mu\equiv K_\mu^A\mathbb{D}_A$. The BAS on-shell correlators are then exactly the cubic scalar tree expressions dressed by color and dual-color factors,
	\begin{align}
		\mathbb{A}(\mathbb{D}_1,\mathbb{D}_2,\mathbb{D}_3)
		&=
		f_{a_1a_2a_3}f_{\bar a_1\bar a_2\bar a_3},
		\\[3pt]
		\mathbb{A}(\mathbb{D}_1,\mathbb{D}_2,\mathbb{D}_3,\mathbb{D}_4)
		&=
		f_{a_1a_2}{}^{b}f_{ba_3a_4}\,
		f_{\bar a_1\bar a_2}{}^{\bar b}f_{\bar b\bar a_3\bar a_4}\,
		\frac{1}{\mathbb{D}_{12}^2}
		+\text{2 terms} \ ,
	\end{align}
	and so on. Since the denominators commute, these formulas are literally the flat-space BAS amplitudes with ordinary propagators replaced by inverse subset Casimirs.
	
	The corresponding NLSM is most conveniently described in terms of a chiral current $j^\mu = j^{a\mu} T_a$. The derived second-order current equation and the divergence constraint are
	\begin{equation}
		\nabla^2 j^\mu + \nabla^{[\mu}\nabla^{\nu]}j_\nu + [j^\nu,\nabla_\nu j^\mu]=0 ,
		\qquad
		\nabla_\mu j^\mu = 0  \ .
	\end{equation}
	Passing to isometric frame with $j^A \equiv j^\mu K^A_{\mu}$, one obtains
	\begin{equation} \label{eq:NLSM_iso}
		\mathbb{D}^2 j^{cC} + f_{ab}{}^{c}\,j^{aA}\mathbb{D}_A j^{bC}=0\ ,
		\qquad
		\mathbb{D}_A j^A =0 \ .
	\end{equation}
	To relate this formulation to the pion field, one introduces the group-valued field $\mathcal{U}$ and writes
	\begin{equation}
		j^A = i\,\mathcal{U}^{-1}\mathbb{D}^A\mathcal{U},
		\qquad
		\mathcal{U}=1-i\pi+\cdots ,
	\end{equation}
	so that
	\begin{equation}
		j^{aA}=\mathbb{D}^A\pi^a+\cdots .
	\end{equation}
	The scalar can then be extracted from the current by contracting with
	\begin{equation}
		\widetilde{\varepsilon}_A=\frac{q_A}{q\!\cdot\!\mathbb{D}},
		\qquad
		\pi^a=\widetilde{\varepsilon}_A j^{aA}+\cdots ,
	\end{equation}
	where $q_A$ is an arbitrary reference vector. The linearized seed for Berends--Giele recursion is then
	\begin{equation}
		\pi^a(p_1,x)=\psi^a(p_1,x),
		\qquad
		j^{aA}(p_1,x)=\mathbb{D}^A\psi^a(p_1,x).
	\end{equation}
	If one writes the classical pion solution as
	\begin{equation}
		\pi^a(p_1,\ldots,p_{n-1};x)
		=
		\Pi^{a_1\cdots a_{n-1}a}(\mathbb{D}_1,\ldots,\mathbb{D}_{n-1})\,
		\psi_{a_1\cdots a_{n-1}}(p_1,\ldots,p_{n-1};x),
	\end{equation}
	then the corresponding on-shell correlator is obtained by amputation,
	\begin{equation}
		\mathbb{A}_{a_1\cdots a_n}(\mathbb{D}_1,\ldots,\mathbb{D}_n)
		=
		\mathbb{D}_{1\cdots n}^2\,
		\Pi_{a_1\cdots a_n}(\mathbb{D}_1,\ldots,\mathbb{D}_{n-1}) .
	\end{equation}
	The three-point scalar correlator vanishes on-shell, while the four-point correlator can be organized as
	\begin{equation}
		\mathbb{A}_{a_1a_2a_3a_4}(\mathbb{D}_1,\mathbb{D}_2,\mathbb{D}_3,\mathbb{D}_4)
		=
		\frac{1}{2}\,f_{a_1a_2}{}^{b}f_{ba_3a_4}\,\mathbb{D}_1\!\cdot\!\mathbb{D}_3
		+
		\frac{1}{2}\,f_{a_1a_3}{}^{b}f_{ba_2a_4}\,\mathbb{D}_1\!\cdot\!\mathbb{D}_2 .
		\label{eq:NLSM4_isometric}
	\end{equation}
	The dependence on $q_A$ drops out, as it must. A particularly striking feature is that these NLSM correlators are literally the flat-space amplitudes with the replacement
	\begin{equation}
		p_I \;\longrightarrow\; \mathbb{D}_I \ ,
	\end{equation}
	similar to the AdS case. The reason is that in a two-derivative theory the interaction vertices only depend on invariants of the form
	\begin{equation}
		\mathbb{D}_I\!\cdot\!\mathbb{D}_J
		=
		\frac{1}{2}\Big(
		\mathbb{D}_{IJ}^2-\mathbb{D}_I^2-\mathbb{D}_J^2
		\Big),
	\end{equation}
	and these are built from mutually commuting subset Casimirs whenever the relevant sets are nested or disjoint. This explains why the non-commutativity of the individual isometric momenta does not obstruct the final scalar correlators.
	
	\subsubsection{Color-kinematics duality and BCJ relations}
	We are now ready to see how the color-kinematics duality and BCJ relations arise in the isometric representation of scalar field theories. The field-theoretic color-kinematics map is especially transparent. The dual-color generator is replaced by an isometry generator,
	\begin{equation}
		T_{\bar a}\quad\longrightarrow\quad \mathbb{D}_A ,
	\end{equation}
	and a dual-colored field ${\cal V}^{\bar a}$ is sent to a vector field ${\cal V}^A$. At the level of the structure constants, one has
	\begin{equation}
		f^{\bar c}{}_{\bar a\bar b}\,{\cal V}^{\bar a}{\cal W}^{\bar b}
		\quad\longrightarrow\quad
		{\cal V}^{A}\mathbb{D}_A{\cal W}^{C}
		-
		{\cal W}^{A}\mathbb{D}_A{\cal V}^{C},
	\end{equation}
	which is precisely the kinematic structure appearing in the NLSM equation of motion. This map is natural because the ordinary dual-color commutator
	\begin{equation}
		[{\cal V}^{\bar a}T_{\bar a},{\cal W}^{\bar b}T_{\bar b}]
		=
		{\cal V}^{\bar a}{\cal W}^{\bar b}f^{\bar c}{}_{\bar a\bar b}T_{\bar c}
	\end{equation}
	is mirrored by the commutator of field-dependent Lie derivatives,
	\begin{equation}
		[{\cal V}^{A}\mathbb{D}_A,{\cal W}^{B}\mathbb{D}_B]
		=
		\Big(
		{\cal V}^{A}\mathbb{D}_A{\cal W}^{C}
		-
		{\cal W}^{A}\mathbb{D}_A{\cal V}^{C}
		\Big)\mathbb{D}_C  \ ,
	\end{equation}
	which is thought of as acting on scalars. In this sense the BAS$\rightarrow$NLSM map sends the algebra of dual color to the algebra of gauged isometries \cite{Cheung:2022pdk}. Applying the same replacement to the BAS dual-color current yields a kinematic current whose conservation enforces the kinematic Jacobi identities.
	
	At four points one may therefore write the NLSM correlator in the cubic form
	\begin{equation}
		\mathbb{A}_4
		=
		\frac{c_s\,\mathbbm{n}_s}{\mathbb{D}_{12}^2}
		+
		\frac{c_t\,\mathbbm{n}_t}{\mathbb{D}_{23}^2}
		+
		\frac{c_u\,\mathbbm{n}_u}{\mathbb{D}_{31}^2},
	\end{equation}
	with the color factors given by Eq.~\eqref{eq:color_4pt} and the kinematic factors by 
	\begin{align}
		\mathbbm{n}_s &=
		\frac{1}{q\!\cdot\!\mathbb{D}_4}
		\Big[
		(\mathbb{D}_3\!\cdot\!\mathbb{D}_{12})(\mathbb{D}_1\!\cdot\!\mathbb{D}_2)
		(q\!\cdot\!\mathbb{D}_1-q\!\cdot\!\mathbb{D}_2) \nonumber\\
		&\qquad
		-(\mathbb{D}_1\!\cdot\!\mathbb{D}_2)
		(\mathbb{D}_1\!\cdot\!\mathbb{D}_3-\mathbb{D}_2\!\cdot\!\mathbb{D}_3)
		(q\!\cdot\!\mathbb{D}_3)
		\Big], \nonumber\\
		\mathbbm{n}_t &=
		\frac{1}{q\!\cdot\!\mathbb{D}_4}
		\Big[
		(\mathbb{D}_1\!\cdot\!\mathbb{D}_{23})(\mathbb{D}_2\!\cdot\!\mathbb{D}_3)
		(q\!\cdot\!\mathbb{D}_2-q\!\cdot\!\mathbb{D}_3) \nonumber\\
		&\qquad
		-(\mathbb{D}_2\!\cdot\!\mathbb{D}_3)
		(\mathbb{D}_2\!\cdot\!\mathbb{D}_1-\mathbb{D}_3\!\cdot\!\mathbb{D}_1)
		(q\!\cdot\!\mathbb{D}_1)
		\Big], \nonumber\\
		\mathbbm{n}_u &=
		\frac{1}{q\!\cdot\!\mathbb{D}_4}
		\Big[
		(\mathbb{D}_2\!\cdot\!\mathbb{D}_{31})(\mathbb{D}_3\!\cdot\!\mathbb{D}_1)
		(q\!\cdot\!\mathbb{D}_3-q\!\cdot\!\mathbb{D}_1) \nonumber\\
		&\qquad
		-(\mathbb{D}_3\!\cdot\!\mathbb{D}_1)
		(\mathbb{D}_3\!\cdot\!\mathbb{D}_2-\mathbb{D}_1\!\cdot\!\mathbb{D}_2)
		(q\!\cdot\!\mathbb{D}_2)
		\Big] \ ,
	\end{align}
	which satisfy the kinematic Jacobi identity
	$\mathbbm{n}_s+\mathbbm{n}_t+\mathbbm{n}_u=0$. Note that this is a different realization of the kinematic numerators than the AdS one described above where the numerators were defined as acting on the contact Witten diagram.
	
	Finally, for the color-ordered on-shell representatives constructed in this isometric frame, the fundamental BCJ relations follow from current conservation by means of the \emph{null color replacement}. The observation is that the divergence of the BAS dual-color current has the same structure as the BAS cubic interaction provided one makes the replacement
	\begin{equation}
		f^{a_n}{}_{a_I a_J}
		\quad\longrightarrow\quad
		\delta f^{\,a_n}{}_{a_I a_J}
		=
		\delta_{a_I a_J}\,
		\overleftrightarrow{\mathbb{D}}^{\,2},
	\end{equation}
	or, in the on-shell correlator attached to the root leg,
	\begin{equation}
		f^{a_n}{}_{a_I a_J}
		\quad\longrightarrow\quad
		\delta_{a_I a_J}\,
		\big(\mathbb{D}_J^2-\mathbb{D}_I^2\big) .
	\end{equation}
	Since the transformed expression computes the divergence of a conserved current, it must vanish. Rearranging the four-point BAS correlator then gives the simplest BCJ identity,
	\begin{equation}
		\mathbb{D}_{23}^2\,A(1,4,2,3)-\mathbb{D}_{12}^2\,A(1,2,4,3)=0 \  , 
	\end{equation}
	which is the generalization of Eq.~\eqref{eq:4ptBCJAds}. The same logic extends to arbitrary multiplicity and applies equally to BAS and the NLSM. In particular, the null-color replacement preserves the relevant Jacobi identities, so current conservation directly implies the full set of fundamental BCJ relations in curved symmetric spacetimes \cite{Cheung:2022pdk}.
	
	One might expect the existence of color-kinematics duality and BCJ relations to lead immediately to a physical double copy, but this is not straightforward. Replacing color with kinematics in the NLSM equations suggests a curved-space analogue of the special Galileon, but the resulting chiral tensor is not automatically conserved. As a result, the cancellation of the reference-vector dependence in the would-be external polarizations is not guaranteed, and the conjectured double copy equations appear inconsistent unless one imposes extra constraints by hand. Thus, it remains to be seen whether there is an appropriate prescription for a double copy in isometric space. Similarly, it remains an open question whether this formulation can exist and be useful for fields with spin.

	\subsection{Coordinate space for generic spacetimes}
	The previous sections focused mainly on representations of correlators in spaces with enough symmetry to make the action of isometries explicit, either through momentum-space wavefunction coefficients, Mellin amplitudes, embedding-space conformal generators, or weight-shifting operators.  The approach of \cite{Sivaramakrishnan:2021srm} is somewhat different.  Most of the construction is formulated directly in a generic curved spacetime, without assuming conformal symmetry, spatial translations, or any other isometry.  The basic idea is to replace the S-matrix by an object which keeps the external legs ``on-shell'' while still being defined in position space.
	
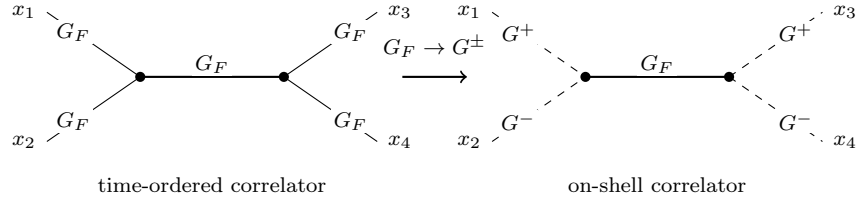
\begin{figure}[H]
		\centering
		\begin{tikzpicture}[scale=0.95, every node/.style={font=\scriptsize}]
			\coordinate (lL) at (-4.1,0);
			\coordinate (lR) at (-2.1,0);
			\coordinate (rL) at (2.1,0);
			\coordinate (rR) at (4.1,0);
			\draw (-5.4,0.9) node[left] {$x_1$} -- node[above left,fill=white,inner sep=1pt] {$G_F$} (lL);
			\draw (-5.4,-0.9) node[left] {$x_2$} -- node[below left,fill=white,inner sep=1pt] {$G_F$} (lL);
			\draw (lR) -- node[above right,fill=white,inner sep=1pt] {$G_F$} (-0.8,0.9) node[right] {$x_3$};
			\draw (lR) -- node[below right,fill=white,inner sep=1pt] {$G_F$} (-0.8,-0.9) node[right] {$x_4$};
			\draw[thick] (lL) -- node[above,fill=white,inner sep=1pt] {$G_F$} (lR);
			\fill (lL) circle (2pt);
			\fill (lR) circle (2pt);
			\node at (-3.1,-1.5) {time-ordered correlator};
			\draw[->,thick] (-0.45,0) -- (0.45,0) node[midway,above=3pt] {$G_F\to G^\pm$};
			\draw[dashed] (0.8,0.9) node[left] {$x_1$} -- node[above left,fill=white,inner sep=1pt] {$G^+$} (rL);
			\draw[dashed] (0.8,-0.9) node[left] {$x_2$} -- node[below left,fill=white,inner sep=1pt] {$G^-$} (rL);
			\draw[dashed] (rR) -- node[above right,fill=white,inner sep=1pt] {$G^+$} (5.4,0.9) node[right] {$x_3$};
			\draw[dashed] (rR) -- node[below right,fill=white,inner sep=1pt] {$G^-$} (5.4,-0.9) node[right] {$x_4$};
			\draw[thick] (rL) -- node[above,fill=white,inner sep=1pt] {$G_F$} (rR);
			\fill (rL) circle (2pt);
			\fill (rR) circle (2pt);
			\node at (3.1,-1.5) {on-shell correlator};
		\end{tikzpicture}
		\caption{A tree-level time-ordered correlator and the corresponding on-shell correlator in a generic curved spacetime. Only the external Feynman propagators are replaced by Wightman functions; the internal line remains a Feynman propagator. The assignments of $G^+$ and $G^-$ shown here are illustrative, with a nonvanishing correlator requiring both signs.}
		\label{fig:generic_spacetime_onshell_correlator}
	\end{figure}

	In an ordinary time-ordered correlator the external legs of a tree diagram are Feynman propagators, $G_F(x_i,x)$.  These are off shell Green functions: acting on one endpoint with the kinetic operator gives a delta-function source.  By contrast, a Wightman function $G^\pm(x_i,x)$ obeys the homogeneous free equation in each argument, and therefore provides an on-shell external leg. The ``on-shell correlator’’ of \cite{Sivaramakrishnan:2021srm} is defined diagrammatically by computing the same weakly-coupled tree diagrams as for a time-ordered correlator, but replacing the external Feynman propagators by Wightman functions.  In flat space this reproduces the usual S-matrix, up to the standard overall momentum-conserving delta function.  In Lorentzian AdS this gives the same transition amplitudes discussed above, now viewed as a special case of a more general curved-spacetime construction. This replacement is illustrated in Fig.~\ref{fig:generic_spacetime_onshell_correlator}.
	
	The central technical tool is a contact representation.  One introduces enumerated differential operators $d_i^\mu$, defined at the level of the integrand by letting $d_i$ differentiate only the $i$-th external Wightman function,
	\be
	d_i^\mu G^\pm(x_j,x)=\delta_{ij}\nabla_x^\mu G^\pm(x_i,x) .
	\ee
	Thus $d_i$ plays the role of the momentum of leg $i$, but only inside the contact integrand. A scalar on-shell correlator can then be written schematically as
	\be
	{\cal M}_n^{\rm on-shell}(x_i)
	=
	\int_{\cal M}\dd^{d+1}x\,\sqrt{-g}\,
	A\!\left(d_i,d_i^{-1}\right)
	\prod_{i=1}^n G^\pm(x_i,x) \ ,
	\label{eq:curved_contact_representation}
	\ee
	where $A$ is the contact representation ``amplitude'', which is now an operator built from the $d_i$’s, their inverses, and curvature terms in a specific order defined by the Feynman diagrams.  The outermost $d_i$ obeys $\sum_i d_i^\mu=0$ and the innermost $d_i^2=0$. This is the direct analogue of the contact Witten diagram representation used above in AdS: in AdS the contact seed is the integral of a product of bulk-to-boundary propagators, while here the same idea is written in a way that does not rely on conformal symmetry.  In flat space, $d_i^\mu$ becomes multiplication by the external momentum, $\sum_i d_i^\mu=0$ becomes momentum conservation, and Eq.~\eqref{eq:curved_contact_representation} reduces to the usual amplitude integrand.
	
	The important new feature is that this representation is not unique.  In flat space, different forms related by momentum conservation, integration by parts, equations of motion, or generalized gauge transformations give the same amplitude.  In curved space these operations can differ by commutators of covariant derivatives and inverse kinetic operators.  These commutators are controlled by the curvature.  Thus two contact representations which are equivalent in the flat-space limit can give different-looking curved-space expressions, differing by curvature couplings or terms which vanish on the external equations of motion.  This non-uniqueness is not merely cosmetic: it affects how one should square differential numerators in a double copy.
	
	The NLSM gives a simple four-point test of this construction.  The contact integrand can be written in a cubic form, but the inverse kinetic operators are not unique because, in a curved background, the two sides of a channel are no longer interchangeable before acting on the contact term.  One convenient representative is
	\begin{align}
		A_{4,{\rm NLSM}}&=c_s\left(\frac{1}{\Box_{12}}\Box_{12}+\frac{1}{\Box_{34}}\Box_{34}\right)K_s \nonumber \\
		&+c_t\left(\frac{1}{\Box_{14}}\Box_{14}+\frac{1}{\Box_{23}}\Box_{23}\right)K_t \nonumber \\
		&+c_u\left(\frac{1}{\Box_{13}}\Box_{13}+\frac{1}{\Box_{24}}\Box_{24}\right)K_u \ ,
		\label{eq:nlsm_curved_contact_integrand}  
	\end{align}
	where
	\begin{align}
		K_s=d_{14}-d_{24}-d_{13}+d_{23}\ ,  \nonumber \\
		K_t=-d_{12}+d_{24}+d_{13}-d_{34}\ , \nonumber \\
		K_u=-d_{14}+d_{34}+d_{12}-d_{23}\ .
		\label{eq:nlsm_curved_Ks}
	\end{align}
	
	The numerator identification is slightly different from the flat-space one.  The starting point is not a unique cubic denominator, but an operator representation of the contact vertex in which one has inserted a trivial factor $\Box^{-1}\Box$ before performing the Wick contractions.  After contracting the external legs, each channel therefore comes with two possible inverse kinetic operators.  For example, the $s$-channel term contains both $\Box_{12}^{-1}$ and $\Box_{34}^{-1}$, since neither side of the channel is preferred in a curved background.  The numerator is then defined as the sum of the operators multiplying these inverse boxes,
	\be
	n_s=(\Box_{12}+\Box_{34})K_s\ ,\qquad
	n_t=(\Box_{14}+\Box_{23})K_t\ ,\qquad
	n_u=(\Box_{13}+\Box_{24})K_u\ .
	\label{eq:nlsm_curved_numerators}
	\ee
	Thus the notation should not be read as a standard cubic representation with one common propagator per channel.  Rather, it is a compact way of packaging the permutation-symmetric contact representation.  In flat space the two inverse boxes in each channel become equivalent, so this reduces to the usual numerator-denominator form. These numerators obey the usual identities when evaluated in the contact representation.  Thus the four-point NLSM realizes color-kinematics duality in generic curved spacetimes.  We note that there remains some generalized gauge freedom in the choice of numerators.  As in flat space, this leaves the correlator unchanged, but in curved space different representatives can give different curvature terms after squaring.
	
	While color-kinematics can be generalized to generic spacetimes in the NLSM, the BCJ relations are more restrictive and are not automatic in a general curved background. At four points they can hold for suitable color-ordered representatives \cite{Sivaramakrishnan:2021srm}. The double copy is similarly subtle. The BAS$\rightarrow$NLSM color-kinematics replacement survives in this representation.  Starting from the BAS amplitude,
	\be
	A_{4,{\rm BAS}}
	=c_s\bar c_s\left(\frac{1}{\Box_{12}}+\frac{1}{\Box_{34}}\right)+c_t\bar c_t\left(\frac{1}{\Box_{14}}+\frac{1}{\Box_{23}}\right)+c_u\bar c_u\left(\frac{1}{\Box_{13}}+\frac{1}{\Box_{24}}\right)\ ,
	\label{eq:bas_curved_contact_integrand}
	\ee
	one replaces the second color factor by the corresponding operator numerator, with a specified ordering, to reproduce Eq.~\eqref{eq:nlsm_curved_contact_integrand}.  Thus the zeroth-copy-to-single-copy map works in generic curved backgrounds at this order. A complementary position-space construction was developed in \cite{Prabhu:2020avf}, relating perturbative solutions of the biadjoint scalar theory to Yang--Mills solutions on a fixed curved background. At linear order, an explicit differential map was found on maximally symmetric spacetimes, while, given such a linearized map, the correspondence can be extended to arbitrary order in perturbation theory on more general curved backgrounds.
	
	The NLSM-squared double copy is less canonical.  The replacement is made before performing the Wick contractions, at the level of the operator representation.  For example, take the NLSM numerator operator
	\be
	n_s=\Box_{12}d_{14}\ .
	\ee
	Squaring means replacing the color factor by a second copy of the numerator operator, but this can be done on different sides of the inverse box.  Two natural choices are
	\be
	\hat A^{(a)}{({\rm NLSM})^2}
	=\Box_{12}d_{14}\frac{1}{\Box_{12}}\Box_{12}d_{14}
	=\Box_{12}d_{14}^2\ ,
	\label{eq:nlsm_squared_a}
	\ee
	and
	\be
	\hat A^{(b)}{({\rm NLSM})^2}
	=\frac{1}{\Box_{12}}\Box_{12}d_{14}\Box_{12}d_{14}
	=d_{14}\Box_{12}d_{14}\ .
	\label{eq:nlsm_squared_b}
	\ee
	Thus the two squarings differ by
	\be
	\hat A^{(a)}{({\rm NLSM})^2}-\hat A^{(b)}{({\rm NLSM})^2}
	=[\Box_{12},d_{14}]d_{14}\ .
	\label{eq:nlsm_squared_ordering_ambiguity}
	\ee
	The commutator vanishes in flat space, but in a curved background it produces curvature-dependent terms.  Hence the result of the squaring depends on the ordering prescription, as well as on the generalized gauge representative chosen before taking the double copy.
	
	Since a unique special-Galileon theory on a completely general curved spacetime is not known, this ambiguity cannot be fixed by comparison with a universal target theory. In AdS, however, these on-shell correlators reduce to the AdS amplitudes discussed above, and the generalized gauge freedom is sufficient to match the known four-point AdS special-Galileon result \cite{Bonifacio:2018zex}. This parallels the earlier AdS examples: the flat-space double copy fixes the leading derivative structure, while finite-radius consistency requires additional curvature-dependent terms.
	
	It would be interesting to understand whether the generic-spacetime contact representation itself can be extended to Yang--Mills theory and to higher multiplicity.  Such an extension could lead to a broader formulation of color-kinematics duality and double copy in curved backgrounds.  However, the generic case faces several obstructions: the contact representation is not unique, different choices of representative can generate inequivalent curvature couplings after squaring, and there is no unique curved-space double copy target.
	
	\subsection{Twistor space} \label{sec:twistor_correl}
	
	In this section, we explore twistor space as an alternative representation of boundary correlation functions in (A)dS, in which kinematic constraints are trivialized. This approach has been developed in several recent works~\cite{Baumann:2024ttn,CarrilloGonzalez:2025qjk,Bala:2025gmz,Bala:2025jbh,Bala:2025qxr,Ansari:2025fvi,CarrilloGonzalez:2026eum}, with earlier related constructions appearing in~\cite{Adamo:2016rtr,Adamo:2017zkm,Roehrig:2020kck,Eberhardt:2020ewh}. Twistor space is especially well suited to boundary correlators of conserved currents because conformal symmetry and their associated conservation is incorporated automatically into the formalism. The simplification is particularly striking for spinning correlators, for which the tensor structures required in more conventional representations are encoded much more economically. As we will see, this provides a natural setting in which to formulate a simple double copy.
	
	Instead of introducing twistor space in the standard way, we will consider a spinorial formalism for embedding space which will be connected to twistors and dual twistors. Hence, we start by reviewing the results of \cite{Binder:2020raz}.
	
	\subsubsection{Bispinors}
	We have seen in Sec.~\ref{sec:spinorhel} that it is useful to trade tensorial indices for spinorial ones by using a different representation of the Lorentz group.  Meanwhile, in Sec.~\ref{sec:correll_embedding} we saw that embedding-space variables make the action of the conformal group linear. In the AdS$_4$/CFT$_3$ context, these two ideas combine naturally into a bispinor formalism \cite{Binder:2020raz}, in which both bulk and boundary points are encoded by spinors carrying an $Sp(4,\mathbb C)$ index. We denote the fundamental $Sp(4,\mathbb C)$ indices by $\mathcal{A},\mathcal{B},\ldots$, and raise and lower them with the symplectic form $\Omega_{\mathcal{A}\mathcal{B}}$.

	The starting point is the embedding coordinates $X$, and $P$ from Eq.~\eqref{eq:emb_bulk} and Eq.~\eqref{eq:emb_bdy} which in this case are five-dimensional and satisfy 
	\begin{equation*}
		X^2=-1 \ , \quad P^2=0 \ ,
	\end{equation*}
	that is, they can be decomposed similar to massive and massless momenta in 5d spinor-helicity variables \cite{Chiodaroli:2022ssi,Pokraka:2024fao}. Thus, the little group for the bulk is $SO(1,3)$ and for the boundary it is $SO(2,1)\times \mathbb{R}_{+}$.

	One can then reconstruct the bulk embedding-space point through the bilinears
	\begin{align}
		X^{\mathcal A\mathcal B}
		&=
		T^{\mathcal A\alpha}T^{\mathcal B\beta}\epsilon_{\alpha\beta}
		+i\,\Omega^{\mathcal A\mathcal B}
		\nonumber\\
		&=
		\Omega^{\mathcal A\mathcal A'}\Omega^{\mathcal B\mathcal B'}
		\bar T_{\mathcal A'}{}^{\dot\alpha}\bar T_{\mathcal B'}{}^{\dot\beta}
		\epsilon_{\dot\alpha\dot\beta}
		-i\,\Omega^{\mathcal A\mathcal B}\ .
		\label{eq:bispinorX}
	\end{align}
	where $\alpha,\dot\alpha$ are the four-dimensional $SL(2,\mathbb{C})$ indices already introduced in Sec.~\ref{sec:spinorhel}, which in this case encode the little group. In vector notation this is the same bulk point introduced earlier in Eq.~\eqref{eq:embedding_coord}. The pair of bulk bispinors $T^{\mathcal A}{}_{\alpha}\ , \  \bar T_{\mathcal A}{}^{\dot\alpha}$ are Killing spinors of AdS and obey the contracted constraints $T_{\mathcal A}{}_{\alpha}T^{\mathcal A}{}_{\beta}=2i\,\epsilon_{\alpha\beta}\ , \ \bar T^{\mathcal A}{}^{\dot\alpha}\bar T_{\mathcal A}{}^{\dot\beta}=2i\,\epsilon^{\dot\alpha\dot\beta} \ , \ T^{\mathcal A}{}_{\alpha}\bar T_{\mathcal A}{}^{\dot\beta}=0$.
	
	The corresponding boundary point is
	\begin{equation}
		P^{\mathcal A\mathcal B}=\Lambda^{\mathcal B a}\Lambda^{\mathcal A}{}_{a}\ ,
		\label{eq:P_bispinor}
	\end{equation}
	where $a,b,\ldots$ are the three-dimensional spinor indices. The boundary bispinors satisfy $\Lambda_{\mathcal A}^{ a}\Lambda^{\mathcal A}{}_{a}=0$ and can be obtained as the boundary limit of the bulk bispinors. Working in Poincare coordinates, this reads
	\be
	\Lambda^{\mathcal A}{}_{a}=\lim_{z\rightarrow0}\frac{\sqrt{z}}{2}\left(\delta^\alpha_a T^{\mathcal A}{}_{\alpha}+\delta^{\dot\alpha}_a\bar T^{\mathcal A}{}_{\dot\alpha}\right) \ .
	\ee
	This is the bispinor version of the projective null-cone description reviewed in Sec.~\ref{sec:correll_embedding}. In particular, both the bulk and the boundary points are quadratic in the corresponding bispinors, and all conformal invariants may be written in terms of their contractions.  Furthermore, the conservation operator, see Eq.~\eqref{eq:embedding_conserv}, is now encoded in
	\be \label{eq:bispinor-current-conservation}
	\nabla^{I}_{a_{1}a_{2}}\,J^{a_{1}\cdots a_{2S}}=0\ ,\qquad
	\text{with}\quad
	\nabla^{I}_{ab}\equiv-\Lambda^{\mathcal{A}}_{(a}(\Gamma^{I})_{\mathcal{A}}{}^{\mathcal{B}}\frac{\partial}{\partial\Lambda^{\mathcal B b)}}\ ,
	\ee
	where $(\Gamma^{I})_{\mathcal{A}}{}^{\mathcal{B}}$ are the five-dimensional gamma matrices.
	
	\subsubsection{Twistors}
	Twistors and dual twistors of a d-dimensional space can be thought of as the spinors in $d+2$ dimensions; for an introductory review of twistor space, see \cite{Adamo:2017qyl}. In the present case, the four-dimensional bulk twistor and dual twistor are six-dimensional spinors. We will generically work with twistors of complexified spacetimes and only impose reality conditions at the end when explicitly mentioned. A (dual) twistor, $Z^{\mathcal{M}}$ ($W_{\mathcal{M}}$), transforms in the (anti-) fundamental of $SL(4,\mathbb{C})$. Thus, just like the embedding space formalism, twistors are useful for writing conformally invariant structures since they carry a linear action of the conformal group. It will be useful to work with projective (dual) twistors which are related via the equivalence relation ($W_{\mathcal{M}}\sim r W_{\mathcal{M}}$ ) $Z^{\mathcal{M}}\sim r Z^{\mathcal{M}}$ where $r\in \mathbb{C}^*$. Thus, $Z^{\mathcal{M}},W_{\mathcal{M}} \in \mathbb{CP}^3$. 
	
	The relation with the bulk bispinors is obtained by using embedding the five-dimensional spinors in the six-dimensional ones \cite{CarrilloGonzalez:2025qjk}, that is, using $Sp(4,\mathbb{C})\subset SL(4,\mathbb{C})$. Explicitly, this relation is given by\footnote{The normalization in Eq.~\eqref{eq:T_Tbar_from_twistors} was chosen to match the Poincare coordinates bispinors of \cite{Binder:2020raz}.}
	\begin{equation}
		Z^\mathcal{M}=\sqrt z \  \delta_{\mathcal A}^{\mathcal{M}} T^{\mathcal A\alpha}\lambda_\alpha\ ,\qquad
		W_\mathcal{M}=\delta^{\mathcal A}_{\mathcal{M}}\sqrt z \ \bar T_{\mathcal A}{}^{\dot\alpha}\tilde\lambda_{\dot\alpha}\ ,
		\label{eq:T_Tbar_from_twistors}
	\end{equation}
	where $\lambda_\alpha$ and $\tilde\lambda_{\dot\alpha}$ are projective spinors with equivalence relations $\lambda_\alpha \sim r\lambda_\alpha $ and $\tilde\lambda_{\dot\alpha}\sim r \tilde\lambda_{\dot\alpha}$, that is, $\lambda^\alpha, \ \tilde\lambda^{\dot\alpha}\in \mathbb{CP}^1$.  To be more precise, the relation in Eq.~\eqref{eq:T_Tbar_from_twistors} is between bispinors and twistors valued in the projective spinor bundle, not in projective twistor space. The standard decomposition of twistors into Weyl spinors of opposite chirality is \begin{equation}
		Z^{\mathcal{M}}=(\lambda_\alpha,\mu^{\dot\alpha})\ ,\qquad
		W_{\mathcal{M}}=(\tilde\mu^\alpha,-\tilde\lambda_{\dot\alpha})\ ,
	\end{equation}
	and the maps from projective twistor and dual twistor space to the corresponding projective spinor bundles are given by the incidence relations 
	\be
	\mu^{\dot\alpha}=x^{\alpha\dot{\alpha}}\lambda_\alpha\ ,\qquad
	\tilde\mu^\alpha=x^{\alpha\dot{\alpha}}\tilde\lambda_{\dot\alpha}\ .
	\label{eq:twistor_incidence}
	\ee
	Taking the boundary limit gives the boundary twistor \cite{CarrilloGonzalez:2025qjk}
	\begin{equation}
		\boldsymbol{\Lambda}^{\mathcal A}\equiv\pi^a\Lambda^{\mathcal A}{}_{a}= \lim_{z\to 0}\delta^{\mathcal A}_{A}Z^A\  \ ,
	\end{equation}
	where $\pi^a \in \mathbb{CP}^1$ is a three-dimensional projective spinor with the equivalence relation $\pi^a\sim r \pi^a$. In this language, the conformal generators act linearly as
	\begin{equation}
		T^{3d}_{\mathcal A\mathcal B}
		=
		\boldsymbol{\Lambda}_{(\mathcal A}
		\frac{\partial}{\partial \boldsymbol{\Lambda}^{\mathcal B)}}\ ,
		\label{eq:3d_conf_generator_twistor}
	\end{equation}
	which is the three-dimensional analogue of the linear twistor action of the four-dimensional conformal generators.
	
	The boundary twistors were first introduced in \cite{Baumann:2024ttn} and have been used in \cite{Baumann:2024ttn,CarrilloGonzalez:2025qjk,Bala:2025gmz,Bala:2025jbh,Bala:2025qxr,Ansari:2025fvi} to construct boundary correlators.  Given the remarkable simplifications that occur when flat-space amplitudes are reformulated in twistor variables—where tree amplitudes acquire support on simple holomorphic configurations, MHV rules admit a natural twistor-action origin, and symmetries and recursion relations become more transparent \cite{Witten:2003nn,Roiban:2004yf,Mason:2009sa,Adamo:2011pv,Cachazo:2012kg,Arkani-Hamed:2009hub}, it is natural to ask whether analogous structures survive in curved spacetimes. In flat space, twistor methods turn locality and factorization into geometric statements in twistor space, making hidden kinematic organization manifest. It is therefore worthwhile to understand whether curved-space observables, such as boundary correlators or wavefunction coefficients, admit a similarly twistorial description in which at least part of this geometric and algebraic structure remains visible.
	
	As we will see now, twistor space is particularly useful to construct boundary correlators of conserved currents. In the bulk, there is an isomorphism (Penrose transform) between solutions of the zero-rest mass equation for a spin $s$  anti self-dual field,
	\be \label{eq:zrm}
	\nabla^{\alpha_1\dot{\alpha}_1}\phi_{\alpha_1 ... \alpha_{2s}}=0 \ .
	\ee
	and twistor space functions\footnote{This is an oversimplification. The Penrose transform is actually an isomorphism between cohomology classes $H^1 (\mathbb{PT}, \mathcal O (k))$ and solutions to the zero-rest-mass equation. The twistor correlators are usually written using \v{C}ech cohomology classes, but the Dolbeault perspective can also be useful to regulate divergences. $f(Z)$ is a \v{C}ech cohomology class representative (an explicit $p$-cocycle).} with homogeneity $-2s-2$,
	\be
	f(r Z)=r^{-2s-2} f(Z) \ . \label{eq:twistor_hom}
	\ee
	In the scalar case, the isomorphism is for conformally coupled scalars satisfying $(\nabla^2-R/6)\phi=0$. To obtain the corresponding coordinate space solution from the twistor space function, one can use the Penrose transform 
	\begin{equation}\label{PTlambda}
		\phi_{\alpha_1 ... \alpha_{2s}} = \oint D\lambda  \, \lambda_{\alpha_1}... \lambda_{\alpha_{2s}} f(Z) \ , 
	\end{equation}
	for twistors. The corresponding version for self-dual fields which are solutions to their corresponding zero-rest mass equation, $\nabla^{\alpha_1\dot{\alpha}_1}\phi_{\dot \alpha_1 ... \dot \alpha_{2s}}=0$ is given by the derivative Penrose transform,
	\begin{equation}\label{PTlambda+}
		\phi_{\dot \alpha_1 ... \dot \alpha_{2s}} = \oint D\tilde\lambda  \;\frac{\partial}{\partial \tilde\mu^{\dot \alpha_1}}... \frac{\partial}{\partial\tilde \mu^{\dot \alpha_{2s}}} f(W)\ .
	\end{equation}
	Sometimes, we will refer to this as the positive helicity version. Equivalently, one may use the Fourier-dual multiplication transform with a dual-twistor representative. Here, the measures are \cite{Penrose:1986ca} $D\lambda\equiv Z \cdot I \cdot dZ=z \braket{\lambda d \lambda}$ and $D\tilde \lambda\equiv W \cdot I \cdot dW=z [\tilde \lambda d \tilde \lambda]$, and the twistor and dual twistor are given by Eq.~\eqref{eq:T_Tbar_from_twistors}. We have also introduced $I^{\mathcal{MN}}$, the infinity twistor which breaks the 4d conformal symmetry down to the (A)dS isometries. In other words, the infinity twistor is what defines the conformal structure of the spacetime and defines points at infinity by $I^{\mathcal{MN}}X_{\mathcal{MN}}=0$ , where $X$ is the 6d embedding space coordinate. In the present case, the infinity twistor is simply given by the symplectic form as
	\be 
	I^{\mathcal{MN}}=\delta_{\mathcal A}^{\mathcal{M}}\delta_{\mathcal B}^{\mathcal{N}}\Omega^{{\mathcal AB}} \ .
	\ee
	In the boundary limit, Eq.~\eqref{eq:zrm} becomes 
	\be \label{eq:consev}
	\partial_{a_{1}a_{2}}\,J^{a_{1}\cdots a_{2S}}=0\ , 
	\ee
	where $\partial_{a_{1}a_{2}}\equiv \frac{\partial}{\partial x^{a_{1}a_{2}}}$ is the three-dimensional partial derivative in spinor notation. This is the the flat-frame projection of Eq.~\eqref{eq:bispinor-current-conservation}, since $\partial_{ab}=V_I\nabla^I_{ab}$ for an embedding-space vector $V^I$ satisfying $P_I V^I=-1$. In other words, the boundary twistor functions give conserved currents. Thus, we can use twistor space to write (A)dS propagators and boundary correlators. To do so, we use nested Penrose transforms. Below, we will give the examples at two and three-points.
	
	\subsubsection{Lower-point correlators and double copy}
	We start from the AdS propagators between self-dual and anti-self-dual fields. As in embedding space is useful to encode these fields in polynomials to avoid the index clutter; hence, we consider $\mathcal{F}^s= t^{\alpha_1}...t^{\alpha_{2s}} \mathcal{F}_{\alpha_1...\alpha_{2s}}$, and $\mathcal{\bar{F}}^s= \bar t^{\dot \alpha_1}...\bar t^{\dot \alpha_{2s}} \mathcal{\bar F}_{\dot \alpha_1...\dot \alpha_{2s}}$, where $t^{\alpha_i}$ and $\bar t^{\dot \alpha_i}$ are constant 4d spinors. In embedding space, these are encoded by \cite{Binder:2020raz}
	\be
	\tau^\mathcal{M} = t_{\alpha} T^{\mathcal{A} \alpha} \delta^\mathcal{M}_\mathcal{A} \ , \quad \bar\tau_\mathcal{M} = \bar t^{\dot{\alpha}} \bar T_{\mathcal{A} \dot \alpha} \delta_\mathcal{M}^\mathcal{A} \ .  \label{eq:polar_embed}
	\ee
	The simplest guess gives the coordinate space (A)dS propagator as 
	\begin{equation}\label{eq:AdsProp}
		\braket{\mathcal{F}^s_{1}(X_1) \bar{\mathcal{F}}^s_{2}(X_2)} \!\sim \oint\!\oint  (Z_1 \cdot I \cdot dZ_1) (W_2 \cdot I \cdot dW_2)\frac{ (\tau_1 \cdot I \cdot Z_1)^{2s} (\bar \tau_2 \cdot I\cdot W_2)^{2s}}{ (Z_1 \cdot W_2)^{2s+2}}  \ .
	\end{equation}
	where the naive bulk twistor representative is
	\be
	f_s^{\rm bulk}(Z_1,W_2)=\frac{1}{ (Z_1 \cdot W_2)^{2s+2}} \label{eq:twistor2pt}
	\ee
	and it can easily be seen to satisfy a double copy relation
	\be
	f_2^{\rm bulk}=\frac{(f_1^{\rm bulk})^2}{f_0^{\rm bulk}} \ .  \label{eq:twistorDC}
	\ee
	We note that Eq.~\eqref{eq:AdsProp} actually gives a vanishing result after performing the integral using the residue theorem. Instead, one has to consider using Dolbeault representatives as in \cite{CarrilloGonzalez:2025qjk}, but we will not show the details of this construction here. Considering a reality condition with real twistors as in \cite{Baumann:2024ttn}, would lead to a result that blows up, unless one works on non-projective twistor space. Accordingly, the boundary expression below should be understood as shorthand for a representative obtained after the required regularization. Taking the boundary limit of Eq.~\eqref{eq:twistor2pt} we obtain the boundary correlator of conserved currents in twistor space 
	\be
	f_s(\boldsymbol{\Lambda}_1,\boldsymbol{\Lambda}_2)=\frac{1}{ (\boldsymbol{\Lambda}_1 \cdot \boldsymbol{\Lambda}_2)^{2s+2}}  \ , \label{eq:bdytwistor2pt}
	\ee
	which satisfies the analogous relation $f_2=f_1^2/f_0$. This is transformed to position space via the boundary Penrose transform
	\be
	\braket{O^s_1(P_1)O^s_2(P_2)}=\oint\oint D\pi_1 D\pi_2\braket{l_1 \pi_1}^{2s} \braket{l_2 \pi_2}^{2s} f_s(\boldsymbol{\Lambda}_1,\boldsymbol{\Lambda}_2) \ ,
	\ee
	where $l^a$ are the boundary limit of the spinors $t^\alpha$, $\bar{t}^{\dot\alpha}$ obtained using $\delta^\alpha_a$ and  $\delta^{\dot\alpha}_a$. These are related to the bispinor polarizations of \cite{Binder:2020raz} as $\sigma^\mathcal{A}= l^a \Lambda^\mathcal{A}_a$. In \cite{CarrilloGonzalez:2025qjk,Bala:2025qxr}, it was shown how to extend this to non-conserved currents. 
	
	The three-point correlators do not suffer from the issues previously mentioned at two points and also have simple expressions. These can be guessed simply from the requirement of the homogeneity of the twistor function in Eq.~\eqref{eq:twistor_hom} for a given spin $s$ conserved current.  While in the boundary, twistors and dual twistors are isomorphic due to the existence of the symplectic form, the Penrose transform implemented via a boundary limit of Eq.~\eqref{PTlambda} or Eq.~\eqref{PTlambda+} will lead to different helicity configurations in momentum space as discussed below. Hence, correlators are constructed by mixing the boundary limits from twistor and dual twistors accordingly. For correlators involving operators with equal spin $s\geq1$, there are two parity even cases corresponding to different helicity configurations. The parity odd correlators will not be discussed here, and were constructed in \cite{Bala:2025gmz}.
	These correlators are given by \cite{CarrilloGonzalez:2025qjk}
	\begin{equation}\label{samehelicity}
		\braket{O_1^{s} O_2^{s} O_3^{s}}= \# \oint D \pi_{123} \frac{\braket{l \pi_1}^{2s} \braket{l \pi_2} ^{2s}\braket{l \pi_3}^{2s}} {(\mathbf{\Lambda}_1\cdot\mathbf{\Lambda}_2)^{s+1} (\mathbf{\Lambda}_2\cdot\mathbf{\Lambda}_3)^{s+1}  (\mathbf{\Lambda}_3\cdot\mathbf{\Lambda}_1)^{s+1}} \ .
	\end{equation}
	and
	\begin{equation}\label{oppositehelicity}
		\begin{aligned}
			\braket{\tilde O_1^{s} \tilde O_2^{s} \tilde O_3^{s}}= \# \oint D \pi_{123} 
			\frac{\braket{l \pi_1}^{2s} \braket{l  \pi_2}^{2s}}{(\mathbf{\Lambda}_1\cdot\mathbf{\Lambda}_2)^{3s+1}} \braket{l \frac{\partial}{\partial \mu_3}}^{2s}
			\frac{ \log(\mathbf{\Lambda}_2\cdot\mathbf{\Lambda}_3)\log(\mathbf{\Lambda}_3\cdot\mathbf{\Lambda}_1)}{ (\mathbf{\Lambda}_2\cdot\mathbf{\Lambda}_3)^{1-s}  (\mathbf{\Lambda}_3\cdot\mathbf{\Lambda}_1)^{1-s}} \ .
		\end{aligned}
	\end{equation}
	As expected from our discussion, the former,
	\be
	f_s^{(---)}(\mathbf{\Lambda}_1,\mathbf{\Lambda}_2,\mathbf{\Lambda}_3)=\frac{1} {(\mathbf{\Lambda}_1\cdot\mathbf{\Lambda}_2)^{s+1} (\mathbf{\Lambda}_2\cdot\mathbf{\Lambda}_3)^{s+1}  (\mathbf{\Lambda}_3\cdot\mathbf{\Lambda}_1)^{s+1}} \ ,\label{eq:3pt_twistor_hd}
	\ee
	gives the all minus helicity configurations, which are given by higher derivative interactions in the bulk. It also gives the conformally coupled scalar three-point when $s=0$. Meanwhile, the latter,  
	\be
	f_s^{(--+)}(\mathbf{\Lambda}_1,\mathbf{\Lambda}_2,\mathbf{\Lambda}_3)= \frac{\log(\mathbf{\Lambda}_2\cdot\mathbf{\Lambda}_3) \log(\mathbf{\Lambda}_3\cdot\mathbf{\Lambda}_1)}{(\mathbf{\Lambda}_1\cdot\mathbf{\Lambda}_2)^{3s+1} (\mathbf{\Lambda}_2\cdot\mathbf{\Lambda}_3)^{1-s}  (\mathbf{\Lambda}_3\cdot\mathbf{\Lambda}_1)^{1-s}} \ , \label{eq:3pt_twistor_leading}
	\ee
	gives the $(--+)$ configuration corresponding to leading interactions such as Yang-Mills for spin one and Einstein-Hilbert for spin two. Both cases satisfy the double copy relation in Eq.~\eqref{eq:twistorDC}, although for the $(--+)$ one has to remove the logarithm from the representative, that is,
	\begin{equation}
		f_2^{(---)}= \frac{
			\left( f_1^{(---)}
			\right)^2
		}{
			\frac{1}{
				(\boldsymbol{\Lambda}_1\!\cdot\!\boldsymbol{\Lambda}_2)\,
				(\boldsymbol{\Lambda}_2\!\cdot\!\boldsymbol{\Lambda}_3)\,
				(\boldsymbol{\Lambda}_3\!\cdot\!\boldsymbol{\Lambda}_1)
			}
		} \ , \quad 
		\frac{f_2^{(--+)}}{\log(\mathbf{\Lambda}_3\cdot\mathbf{\Lambda}_1)}=\frac{
			\left( f_1^{(--+)}/\log(\mathbf{\Lambda}_3\cdot\mathbf{\Lambda}_1)
			\right)^2
		}{
			\frac{1}{
				(\boldsymbol{\Lambda}_1\!\cdot\!\boldsymbol{\Lambda}_2)\,
				(\boldsymbol{\Lambda}_2\!\cdot\!\boldsymbol{\Lambda}_3)\,
				(\boldsymbol{\Lambda}_3\!\cdot\!\boldsymbol{\Lambda}_1)
			}
		} 
	\end{equation}
	We also note that these twistor correlators are related to those in \cite{Baumann:2024ttn} after imposing an appropriate reality condition as discussed in \cite{CarrilloGonzalez:2025qjk}. It is worth pausing to highlight the simplicity of the twistor space correlators in Eqs.~\eqref{eq:3pt_twistor_hd} and \eqref{eq:3pt_twistor_leading}. Compared to expressions in coordinate space, even in embedding space or using bispinors, these are remarkably compact and simple even for higher spins. Their twistor origin automatically gives the conserved correlator without having to impose additional constraints from differential operators. Furthermore, they can easily be guessed by requirements of their homogeneity, analytic structure, and parity.
	
	\subsubsection{Ward-Takahashi Identity}
	The correlators of conserved currents satisfy Ward-Takahashi identities (see Eq.~\eqref{eq:Ward_id}) that relate $n-$point correlators to $n-1$-point correlators.  For example, for the 3-point currents one has \cite{Baumann:2020dch}
	\begin{equation}
		\begin{aligned}
			\nabla^{x_3}_\mu \braket{\tilde J_1 \tilde J_2 \tilde J^\mu_3}\propto \delta^3(x_3 -x_1) \braket{(\delta \tilde J_1) \tilde J_2}+\delta^3(x_3 -x_2) \braket{\tilde J_1 (\delta \tilde J_2)}\ .
		\end{aligned}
	\end{equation}
	It is known that correlators arising from higher-derivative bulk interactions satisfy the homogeneous version where the RHS vanishes identically. That is the case for the correlator in Eq.~\eqref{eq:3pt_twistor_hd}. Meanwhile, for Eq.~\eqref{eq:3pt_twistor_leading} we obtain a non-trivial relation to Eq.~\eqref{eq:bdytwistor2pt}. In twistor space, the Ward-Takahashi identity can be rewritten by integrating over $x_3$ and applying the divergence theorem to a sphere at $x_2$ (excluding $x_1$). After performing the $x_3$ Penrose transform, the relation takes the form
	\be
	\int\braket{\lambda_3 d\lambda_3}\int dS^{ab}_{x_3}\frac{\partial} {\partial \mu^{a}_3} \frac{\partial}{\partial \mu^{b}_3} f^{\langle \tilde{O}_1^{s} \tilde{O}_2^{s} \tilde{O}_3^{s} \rangle}_s(\mathbf{\Lambda}_1,\mathbf{\Lambda}_2,\mathbf{\Lambda}_3)=f^{\langle \tilde{O}_1^{s} \tilde{O}_2^{s} \rangle}_s(\mathbf{\Lambda}_1,\mathbf{\Lambda}_2) \ . \label{eq:twistor_ward_id}
	\ee
	Here, we considered the derivative Penrose transform since it is the one relevant for the leading interactions. In the case of the higher-derivative interaction one can replace $\partial_\mu \rightarrow \lambda$ and find that the RHS vanishes identically. It was checked in \cite{CarrilloGonzalez:2025qjk} that \eqref{eq:twistor_ward_id} is satisfied for the YM correlator.
	
	\subsubsection{Four-points and beyond}
	One of the appeals of the twistor space representation is the possibility of reproducing the flat space story where recursion relations become highly transparent and allow one to construct higher-point amplitudes in a simple manner. These are still open directions in the twistor correlators program, but some observations at four points have already been made in \cite{Ansari:2025fvi}. The object considered there is not a generic four-point correlator. It is a stripped Lorentzian Wightman function, for a theory with cubic interactions, in the special kinematic region where the middle operators have spacelike momenta $p_2^2>0$ and $p_3^2>0$, which causes all but one contribution to the correlator to vanish.  In this regime, the exchange contribution factorizes into two three-point Wightman functions and can be interpreted as a Wightman conformal partial wave. The twistor correlators were obtained by translating these factorizations that arise in momentum space to twistor space via the inverse-half Fourier transform that will be introduced in the following section. For conformally coupled scalars,
	\be\label{eq:scalar4pttwistorCPW}
	f^{(\cO\cO|\cO|\cO\cO)}
	(\mathbf{\Lambda}_1,\mathbf{\Lambda}_2,\mathbf{\Lambda}_3,\mathbf{\Lambda}_4)
	=
	\frac{
		\delta(\mathbf{\Lambda}_1\!\cdot\!\mathbf{\Lambda}_2)
		\delta(\mathbf{\Lambda}_3\!\cdot\!\mathbf{\Lambda}_4)
		\delta(v-1)}
	{
		|\mathbf{\Lambda}_1\!\cdot\!\mathbf{\Lambda}_3|
		|\mathbf{\Lambda}_2\!\cdot\!\mathbf{\Lambda}_4|}  \ ,
	\ee
	where we have ignored numerical factors and defined the cross-ratio as
	\be
	v\equiv
	\frac{(\mathbf{\Lambda}_1\!\cdot\!\mathbf{\Lambda}_4)
		(\mathbf{\Lambda}_2\!\cdot\!\mathbf{\Lambda}_3)}
	{(\mathbf{\Lambda}_1\!\cdot\!\mathbf{\Lambda}_3)
		(\mathbf{\Lambda}_2\!\cdot\!\mathbf{\Lambda}_4)}  \ .
	\ee
	For the MHV Yang-Mills helicity configuration $(+,-,+,-)$, after stripping $f^{A_1A_2B}f^{A_3A_4B}$, the analogous spin-one partial wave is
	\begin{equation}\label{eq:YM4pttwistorCPW}
		\begin{aligned}
			f^{(JJ|J|JJ)}
			(\mathbf{\Lambda}_1,\mathbf{\Lambda}_2,\mathbf{\Lambda}_3,\mathbf{\Lambda}_4)
			={}&
			\Sgn(\mathbf{\Lambda}_1\!\cdot\!\mathbf{\Lambda}_2)
			\Sgn(\mathbf{\Lambda}_2\!\cdot\!\mathbf{\Lambda}_3)
			\Sgn(\mathbf{\Lambda}_3\!\cdot\!\mathbf{\Lambda}_4)
			\Sgn(\mathbf{\Lambda}_4\!\cdot\!\mathbf{\Lambda}_1) \\
			&\times
			\frac{1}{|\mathbf{\Lambda}_1\!\cdot\!\mathbf{\Lambda}_3|^4}
			\frac{\dd^3}{\dd v^3}
			\left[ v^3\Sgn\!\left(1-\frac{1}{v}\right)\right] .
		\end{aligned}
	\end{equation}
	For the corresponding graviton helicity configuration, the stress-tensor exchange is
	\begin{equation}\label{eq:grav4pttwistorCPW}
		\begin{aligned}
			f^{(s)}_{(TT|T|TT)}
			(\mathbf{\Lambda}_1,\mathbf{\Lambda}_2,\mathbf{\Lambda}_3,\mathbf{\Lambda}_4)
			={}&
			|\mathbf{\Lambda}_1\!\cdot\!\mathbf{\Lambda}_2|
			|\mathbf{\Lambda}_2\!\cdot\!\mathbf{\Lambda}_3|
			|\mathbf{\Lambda}_3\!\cdot\!\mathbf{\Lambda}_4|
			|\mathbf{\Lambda}_4\!\cdot\!\mathbf{\Lambda}_1| \\
			&\times
			\frac{1}{|\mathbf{\Lambda}_1\!\cdot\!\mathbf{\Lambda}_3|^8}
			\frac{1}{v}\frac{\dd^6}{\dd v^6}
			\left[ v^7\left|1-\frac{1}{v}\right|\right] .
		\end{aligned}
	\end{equation}
	Thus the compactness of the four-point result is tied to the special Wightman kinematics and to the factorized exchange contribution. Contact terms and the other channels are not included in Eqs.~\eqref{eq:scalar4pttwistorCPW}--\eqref{eq:grav4pttwistorCPW}, so these formulae should not be read as giving the full four-point correlator. In this kinematic regime the double copy is not manifest in the closed real-twistor functions, but it becomes a simple square in their Schwinger-parameterized twistor representation, as reviewed in the Grassmannian section below. We remark that these twistor correlators use real twistor variables, as opposed to the results above that showcase a double copy in complex projective twistor space. It might additionally be the case, as it is in amplitudes, that one requires the full correlator to observe the double copy relation.
	
	Finally, we note that it is also possible to generalize the twistor space construction to four-dimensional boundaries\footnote{In \cite{Bala:2026hag}, a proposal for this construction using real ambitwistors was put forward.} by using nested Penrose transforms in ambitwistor space \cite{CarrilloGonzalez:2026eum}. Here, a double copy relation naturally arises from the boundary point of view although it does not correspond to the expected relation for the dual bulk interpretation. We also note that the ambitwistor-string construction of Ref.~\cite{Eberhardt:2020ewh} gives an AdS analogue of the CHY formula for scalar correlators and indicates a potential obstruction to a naive AdS double copy, since the left and right integrand operators need not commute. 
	
	\subsubsection{A possible KLT double copy} \label{sec:twistor_klt}
	Before concluding the twistor-space section, we will briefly mention a realization of the KLT kernel in twistor space due to \cite{Adamo:2024hme}.  The flat-space KLT double copy was already reviewed in Sec.~\ref{sec:klt_and_strings}, where the gravity amplitude was written as two color-ordered Yang--Mills amplitudes glued by the KLT momentum kernel, see Eq.~\eqref{eq:KLT_gravity}. The result of \cite{Adamo:2024hme} is a twistorial analogue of this statement. Instead of applying the KLT kernel after evaluating the amplitudes in momentum space, one rewrites the twistor-space formula for gravity itself so that its integrand looks like two Yang--Mills twistor integrands multiplied by a new kernel.
	
	The flat-space starting point is the relation between two known formulas for tree amplitudes. The Yang--Mills side is the RSVW formula \cite{Roiban:2004yf,Witten:2003nn}, while the gravity side is the Cachazo--Skinner formula \cite{Cachazo:2012kg}. Both are written as integrals over holomorphic maps $Z^A(\sigma)=U^A_{a_1\cdots a_d}\sigma^{a_1}\cdots\sigma^{a_d}$, 
	from the Riemann sphere to twistor space. Here $\sigma$ denotes a point on the worldsheet sphere and its degree fixes the helicity sector: degree $d$ corresponds to an $\mathrm{N}^{d-1}\mathrm{MHV}$ amplitude. 
	
	More explicitly, the RSVW formula for the color-ordered Yang--Mills amplitude in the degree-$d$ sector is \cite{Roiban:2004yf,Witten:2003nn}
	\be \label{eq:RSVW}
	\mathcal{A}_{n,d}[\rho]=\int d\mu_d\,\widetilde{\mathcal I}^{\,\tilde h}_n[\rho]\prod_{i\in h}a_i\prod_{j\in\tilde h}b_j\ ,
	\ee
	where $d\mu_d$ denotes the standard measure over the moduli space of degree-$d$ holomorphic maps modulo $GL(2,\mathbb{C})$, $a_i$ and $b_j$ are the positive- and negative-helicity gluon wavefunctions\footnote{These are the same holomorphic functions in twistor space, or more precisely cohomology classes, that were described above.} whose dependence on $Z(\sigma_i)$ is suppressed in the notation here, and
	\be
	\widetilde{\mathcal I}^{\,\tilde h}_n[\rho]\equiv (\Delta_{\tilde h})^{\,4}\,{\rm PT}_n[\rho]\ ,\qquad
	\Delta_{\tilde h}\equiv\prod_{\substack{i,j\in\tilde h\\ i<j}}(ij)\ ,\qquad
	{\rm PT}_n[\rho]\equiv\prod_{i=1}^{n}\frac{D\sigma_{\rho(i)}}{\big(\rho(i)\rho(i+1)\big)}\ ,
	\ee
	where the product in the Parke--Taylor forms ${\rm PT}_n[\rho]$ is cyclic and $	\Delta_{\tilde h}$ is the Vandermonde determinant.
	
	The corresponding gravity amplitude is given by the Cachazo--Skinner formula \cite{Cachazo:2012kg},
	\be
	M_{n,d}=\int d\mu_d\,(\Delta_{\tilde h})^{\,8}\,\det{}'(\mathbb{H})\,\det{}'(\mathbb{H}^{\vee})\prod_{i\in h}{\rm h}_i\prod_{j\in\tilde h}\tilde {\rm h}_j\ .
	\label{eq:cachazo_skinner}
	\ee
	The functions ${\rm h}_i$ and $\tilde {\rm h}_j$ are the positive- and negative-helicity graviton wavefunctions on twistor space, of homogeneity $2$ and $-6$, respectively. The matrices $\mathbb{H}$ and $\mathbb{H}^{\vee}$ are the corank-one Hodges matrices associated with the positive- and negative-helicity sectors, respectively, and $\det{}'$ denotes their reduced determinants.
	
	The main observation in \cite{Adamo:2024hme} is that the Cachazo--Skinner gravity integrand can be rearranged into the form
	\be
	\mathcal{M}_{n,d}
	=
	\sum_{\substack{
			b\rho,b\omega\in\mathcal{S}(h)\\
			a\bar\rho,a\bar\omega\in\mathcal{S}(\tilde h)
	}}
	\int \dd\mu_{n,d}\,
	\mathcal{I}_{n}^{\tilde h}[a\bar\rho b\rho]\,
	S_{n,d}[\rho,\bar\rho|\omega,\bar\omega]\,
	\mathcal{I}_{n}^{\tilde h}[\bar\omega^{T}ab\omega] 
	\prod_{i\in h}{\rm h}_i\prod_{j\in \tilde h}\tilde {\rm h}_j\ ,
	\label{eq:twistor_KLT_flat}
	\ee
	This is the twistor-space analogue of Eq.~\eqref{eq:KLT_gravity}. Here, $\dd\mu_{n,d}$ integrates over the moduli, $U$, of the degree-$d$ holomorphic map modulo the $\mathrm{GL}(2,\mathbb C)$ redundancy, after extracting one factor of $D\sigma_i$ for each external insertion from the Parke--Taylor forms. The remaining $D\sigma_i$ factors are carried by the color-ordered RSVW Yang--Mills integrands $\mathcal{I}_{n}^{\tilde h}$, which therefore play the role of the two single-copy factors. The sets $h$ and $\tilde h$ split the external states according to helicity; and satisfy $|\tilde h|=d+1$ and $|h|=n-d-1$. The notation in the sum means that $a\in\tilde h$ and $b\in h$ are fixed reference elements: $b\rho,b\omega\in\mathcal{S}(h)$ denotes orderings of $h$ which start with $b$, while $a\bar\rho,a\bar\omega\in\mathcal{S}(\tilde h)$ denotes orderings of $\tilde h$ which start with $a$. Equivalently, $\rho,\omega$ run over orderings of $h\setminus\{b\}$, while $\bar\rho,\bar\omega$ run over orderings of $\tilde h\setminus\{a\}$. The barred orderings therefore refer to the $\tilde h$ set, and the unbarred orderings to the $h$ set. The new ingredient is the kernel $S_{n,d}$, which replaces the usual momentum-space KLT kernel, and whose inverse can be related to the integrand of BAS doubly color-ordered amplitudes. 
	
	This looks more complicated than the familiar KLT formula because the helicity grading\footnote{Here helicity-grading means that the amplitude is decomposed into sectors of fixed helicity configuration (equivalently, fixed MHV degree or fixed map degree), rather than summed over all helicity sectors.} is kept manifest before the moduli integral is done. A useful feature of the kernel is that it factorizes into the two helicity sectors,
	\be
	S_{n,d}[\rho,\bar\rho|\omega,\bar\omega]
	=
	\mathcal{D}(\omega,\bar\omega)\,
	\mathbb{S}_{h}[\rho|\omega]\,
	\mathbb{S}_{\tilde h}[\bar\rho|\bar\omega]
	\ .
	\label{eq:twistor_KLT_chiral_kernel}
	\ee
	The two factors $\mathbb{S}_{h}$ and $\mathbb{S}_{\tilde h}$ are built from sums over tree graphs on the corresponding ordered sets of external particles. In this way, the graph-theory identities used in \cite{Adamo:2024hme} rewrite the determinant factors in the Cachazo--Skinner formula into the kernel which glues the two RSVW Yang--Mills integrands. In the MHV sector, $d=1$, the moduli integral can be done explicitly and the result reproduces the usual field-theory KLT relation. Away from MHV, the formula keeps the map degree fixed, so it is closer to the helicity-graded RSVW and Cachazo--Skinner representations than to the standard momentum-space KLT basis.
	
	The same construction can be formally extended to AdS$_4$. One starts from the AdS version of the Cachazo--Skinner formula constructed in \cite{Adamo:2015ina}. The expression is very close to the flat-space one, but the matrices appearing in the gravity integrand are now built using the AdS infinity twistor, schematically
	\be
	S_{n,d}
	\quad\longrightarrow\quad
	S_{n,d}^{\Lambda}
	=
	\mathcal{D}(\omega,\bar\omega)\,
	\mathbb{S}_{h}^{\Lambda}[\rho|\omega]\,
	\mathbb{S}_{\tilde h}^{\Lambda}[\bar\rho|\bar\omega]
	\ .
	\label{eq:twistor_KLT_AdS_kernel}
	\ee
	The superscript on $\mathbb{S}^{\Lambda}$ indicates that the flat twistor contractions have been replaced by contractions defined with the AdS infinity twistor. With this replacement, the AdS gravity formula can again be written in the same schematic form as Eq.~\eqref{eq:twistor_KLT_flat}, with $S_{n,d}$ replaced by $S_{n,d}^{\Lambda}$. This provides an attractive candidate for a KLT double copy in AdS twistor space, but it should be interpreted with care. The formula is a statement about the AdS generalization of the Cachazo--Skinner integrand. Its direct relation to standard AdS boundary correlators, computed for example from Witten diagrams in position or momentum space, is not yet fully established. In particular, one should not conclude that Eq.~\eqref{eq:twistor_KLT_AdS_kernel} is already the same object as the Mellin-space or momentum-space AdS double copies discussed earlier. A matching calculation is still needed. The conservative lesson is that the twistor-space organization of the flat-space KLT double copy survives a natural AdS deformation, but its precise interpretation as a double copy of physical AdS boundary correlators remains open.

	\subsection{Grassmannian}
	We now turn to the final representation of (A)dS boundary correlators considered in this review: the Grassmannian. The Grassmannian $Gr(k,n)$ is the space of $k$-dimensional planes in an $n$-dimensional vector space~\cite{Griffiths:1994prl,harris1992algebraic}. The cosmological Grassmannian\footnote{More precisely, this is the orthogonal Grassmannian, $OGr(k,n)$, which is a restricted Grassmannian in which the planes satisfy an additional quadratic constraint.} construction introduced in~\cite{Arundine:2026fbr} provides a particularly useful representation of boundary correlators of conserved currents. As in twistor space, the kinematic constraints are naturally incorporated into the geometry of the construction. In addition, its direct connection to momentum space makes the factorization properties of the correlators particularly transparent and closely analogous to those of flat-space scattering amplitudes. These features make the Grassmannian especially useful for constructing higher-point correlators and for exposing possible double copy relations.
	
	We first review the relation between the Grassmannian and twistor space before introducing the construction directly from momentum space. In this section, we denote three-dimensional spinors by $\lambda^\alpha$, following the conventions of~\cite{Baumann:2024ttn,Arundine:2026fbr}. Since no bulk spinors appear in this discussion, this notation should not lead to ambiguity with the spinors introduced in the previous section.
	
	\subsubsection{From twistor to momentum space and back}
	One can recover the momentum space correlators from the twistor space ones by using the half-Fourier transform, first introduced by Witten in \cite{Witten:2003nn} for 4d amplitudes. To do so, one has to choose a reality condition where the momentum spinor-helicity variables and the twistors are real. In 4d this is split signature, and in 3d this corresponds to Lorentzian signature. The three-dimensional version of the half-Fourier transform is given by \cite{Baumann:2024ttn}
	\be
	g(\lambda_i,\bar{\lambda}_i)\equiv \int d^2\mu_i e^{i\bar{\lambda}_i\cdot\mu_i} f \left(\mathbf{\Lambda}_i\right) \ , \label{eq:hft}
	\ee
	When acting on the twistor correlator we take
	\be
	\braket{J_1^{s_1}\cdots J_n^{s_n}}= \int \prod_{i=1}^n d^2\mu_i \braket{l_i \lambda_i}^{2s_i} \braket{\lambda_i \bar\lambda_i} e^{i\sum_{i=1}^n\bar{\lambda}_i\cdot\mu_i}  f \left(\{\mathbf{\Lambda}_i\}\right)
	\ee
	where we are transforming not just the twistor correlator $f(\{\mathbf{\Lambda}_i\})$, but the full polarized version. The additional factors of $\braket{\lambda_i \bar\lambda_i}$ can be guessed from the dilatation weights of the twistor and momentum space correlators \cite{Bala:2025qxr}. The polarization spinors are taken to be $l_i^\alpha=\bar\lambda_i^{\alpha}/\sqrt{p_i}$, where $p_i\equiv|\mathbf p_i|$ and
	\be
	p_i^{\alpha\beta}=\lambda_i^{(\alpha}\bar\lambda_i^{\beta)} \ .
	\ee
	Here, we have focused on the negative helicity correlators, but the positive ones can be obtained analogously from the derivative Penrose transform which is the boundary limit of Eq.~\eqref{PTlambda+}. Using this, one finds
	\be
	\frac{\braket{J_1^{s_1}\cdots J_n^{s_n}}}{\prod_{i=1}^n p_i^{s_i-1}}=\int \prod_{i=1}^n d^2\mu_i\, e^{i\sum_{i=1}^n\bar{\lambda}_i\cdot\mu_i}  f \left(\{\mathbf{\Lambda}_i\}\right) \ .
	\ee
	This is precisely the version used in \cite{Baumann:2024ttn} which shows that the correlators in momentum space are those normalized by $p^{\Delta-2}$ (since for 3d conserved currents $\Delta=s+1$ ) and which matches the results in \cite{Baumann:2020dch}. One can consider the inverse of these transformations and obtain the twistor correlators as
	\be
	f \left(\{\mathbf{\Lambda}_i\}\right)
	=\int \prod_{i=1}^n d^2\bar{\lambda}_i\, e^{-i\sum_{i=1}^n\bar{\lambda}_i\cdot\mu_i}  \frac{\braket{J_1^{s_1}\cdots J_n^{s_n}}}{\prod_{i=1}^n p_i^{s_i-1}} \ .
	\ee
	
	\subsubsection{From twistors to Grassmannian}
	In \cite{Baumann:2024ttn}, it was noticed that simple expressions for the twistor correlators arise when one uses Schwinger parameters. To show this, we take the real twistor correlators, which are obtained by replacing $z^{-n-1}\rightarrow \delta^{[n]}(x)$ for $n\geq0$ and $z^{-n-1}\log(z)\rightarrow \delta^{[n]}(x) $ for $n\leq -1$ in Eqs. \eqref{eq:3pt_twistor_hd} and \eqref{eq:3pt_twistor_leading}, where the delta function is defined as 
	\be
	\delta^{[n]}(x)\equiv i^{-n}\int_{-\infty}^{\infty}\frac{\ud c}{2\pi}\,e^{-icx}c^n
	=
	\begin{cases}
		\dfrac{\ud^n}{\ud x^n}\delta(x)\,, & n\geq 0\ ,\\
		\dfrac{1}{2(|n|-1)!}\operatorname{sign}(x)x^{|n|-1}\,, & n\leq -1\  ,
	\end{cases}
	\ee
	which uses the principal value prescription for $c^n$.
	While in the boundary twistor correlators section we did not distinguish between $\boldsymbol{\Lambda}_\mathcal{A}$ and $ \boldsymbol{\Lambda}^\mathcal{A}$ since indices are raised and lowered with $\Omega^{\mathcal{A}\mathcal{B}}$, it will be useful to keep this distinction in the following since they will be related to different components of the three-momenta. Nevertheless, one should remember that these are boundary (dual) twistors, not the bulk ones from Section \ref{sec:twistor_correl}. To distinguish them, we will denote them in bold: $\mathbf{W}_\mathcal{A}\equiv\boldsymbol{\Lambda}_\mathcal{A}$ and $\mathbf{Z}^\mathcal{A}\equiv \boldsymbol{\Lambda}^\mathcal{A}$. Given this, the real versions of the three-point correlators in Eqs. \eqref{eq:3pt_twistor_hd} and \eqref{eq:3pt_twistor_leading} are 
	\be
	f_s^{\text{higher}}(\mathbf{Z}_1,\mathbf{Z}_2,\mathbf{Z}_3)=\delta^{[s]}(\mathbf{Z}_1\cdot\mathbf{Z}_2)\delta^{[s]}(\mathbf{Z}_2\cdot\mathbf{Z}_3)\delta^{[s]}(\mathbf{Z}_3\cdot\mathbf{Z}_1)\ ,\label{eq:3pt_twistor_hd_real}
	\ee
	and  
	\be
	\widetilde f_s^{\text{leading}}(\mathbf{W}_1,\mathbf{W}_2,\mathbf{W}_3)= \delta^{[-s]}(\mathbf{W}_1\cdot\mathbf{W}_2)\delta^{[-s]}(\mathbf{W}_2\cdot\mathbf{W}_3)\delta^{[-s]}(\mathbf{W}_3\cdot\mathbf{W}_1)\ \ , \label{eq:3pt_twistor_leading_real}
	\ee
	respectively \cite{Baumann:2024ttn}. As before, those involving dual twistors are transformed to position space via the derivative Penrose transform Eq.~\eqref{PTlambda+}. For $\boldsymbol{Z}_i\cdot \boldsymbol{Z}_j$ we introduce a corresponding Schwinger parameter $c_{ij}$, which allows us to write the twistor correlators as
	\be
	f(\{\boldsymbol{Z}_i\})=
	\int \prod_{(ij)=(12),(23),(31)}\frac{\ud c_{ij}}{2\pi}\,
	e^{-i\left(c_{12}\boldsymbol{Z}_1\cdot \boldsymbol{Z}_2+c_{23}\boldsymbol{Z}_2\cdot \boldsymbol{Z}_3+c_{31}\boldsymbol{Z}_3\cdot \boldsymbol{Z}_1\right)}
	A(\{c_{ij}\}) \, , \label{eq:Twistor_cs}
	\ee
	and similarly for dual twistors. The Grassmannian correlators, $A(\{c_{ij}\}) $,  corresponding to Eqs. \eqref{eq:3pt_twistor_hd_real} and \eqref{eq:3pt_twistor_leading_real} are
	\be
	A_s^{\text{higher}}(\{c_{ij}\})=c_{12}^{s}c_{23}^{s}c_{31}^{s}\,,
	\qquad
	A_s^{\text{leading}}(\{c_{ij}\})=\frac{1}{c_{12}^{s}c_{23}^{s}c_{31}^{s}} \, , \label{eq:3pt_grassm}
	\ee
	respectively. We can observe that a very simple double copy relation arises in both cases,
	\be
	A_2(\{c_{ij}\})=\frac{\left(A_1(\{c_{ij}\})\right)^2}{A_0(\{c_{ij}\})}
	\ .
	\ee
	\subsubsection{From momentum space to Grassmannian}
	We will now review the construction of \cite{Arundine:2026fbr} for the cosmological Grassmannian. The Grassmannian representation can be used either for de Sitter wavefunction coefficients or, after the usual AdS continuation and rescaling, for AdS boundary correlators. In the normalization used here, the object represented by the Grassmannian integral is the rescaled momentum-space correlator
	\begin{equation}
		\psi_n(\{\lambda_i,\bar\lambda_i\})\equiv
		\left(\prod_{i=1}^n k_i^{2-\Delta_i}\right)
		\langle J_1^{h_1}\cdots J_n^{h_n}\rangle \, ,
	\end{equation}
	with the overall momentum-conserving delta function included unless a prime is written. This rescaling follows from the arguments described above that relate Grassmannian to twistors as we will see in the following. Following \cite{Arundine:2026fbr}, the external kinematic data can be collected into a $2n\times2$ matrix
	\begin{equation}
		\Lambda=\begin{pmatrix}
			\lambda_1^1&\lambda_1^2\\[-1mm]
			\vdots&\vdots\\
			\lambda_n^1&\lambda_n^2\\[1mm]
			\bar\lambda_1^1&\bar\lambda_1^2\\[-1mm]
			\vdots&\vdots\\
			\bar\lambda_n^1&\bar\lambda_n^2
		\end{pmatrix},
		\qquad
		Q=\begin{pmatrix}0&1_n\\ 1_n&0\end{pmatrix} .
	\end{equation}
	The quadratic momentum-conservation constraint is simply
	\begin{equation}
		\Lambda^T Q\Lambda=0\, ,
	\end{equation}
	and the homogeneous special-conformal Ward identity for the discontinuity of the correlator can be written as
	\begin{equation}
		\frac{\partial}{\partial\Lambda^{I\alpha}}Q^{IJ}
		\frac{\partial}{\partial\Lambda^{J\beta}}\psi_n(\Lambda)=0\, .
	\end{equation}
	The basic idea of the cosmological Grassmannian is to solve these two quadratic constraints by introducing an auxiliary $n\times2n$ matrix $C_{aI}$ satisfying
	\begin{equation}
		CQC^T=0\, ,
		\qquad
		C\Lambda=0\, .
		\label{eq:review_C_constraints}
	\end{equation}
	We label the $2n$ columns of $C$ by $I=\bar{1},\bar{2},\ldots,\bar{n},1,2,\ldots,n$, so that $\bar{i}$ and $i$ denote the barred and unbarred columns associated with the $i$-th external leg, respectively. The first condition in Eq.~\eqref{eq:review_C_constraints} says that the rows of $C$ span a null $n$-plane in a $2n$-dimensional space with metric $Q$, so $C\in {\rm OGr}(n,2n)$, the orthogonal Grassmannian. The second condition linearizes momentum conservation: any solution of $C\Lambda=0$ may be written as $\Lambda=(CQ)^TP$ for some $n\times2$ matrix $P$, and then $\Lambda^TQ\Lambda=P^T(CQC^T)P=0$.
	
	The correlator is then written as an integral over this auxiliary space,
	\begin{equation}
		\psi_n(\Lambda)=\int \ud C\,\delta(C\Lambda)\,A_n(C)\, ,
		\qquad
		\ud C\equiv \frac{\ud^{n\times2n}C}{{\rm GL}(n)}\,
		\delta(CQC^T)\, .
		\label{eq:review_grassmannian_integral}
	\end{equation}
	The ${\rm GL}(n)$ quotient reflects the fact that $C$ and $RC$, with $R\in {\rm GL}(n)$, describe the same null plane. Thus $A_n(C)$ must be written in terms of the $n\times n$ minors
	\begin{equation}
		(I_1\cdots I_n)\equiv \epsilon^{a_1\cdots a_n}C_{a_1I_1}\cdots C_{a_nI_n}\, .
	\end{equation}
	Under $C\mapsto RC$ these minors scale as $(I_1\cdots I_n)\mapsto r(I_1\cdots I_n)$, where $r=\det R$. The integral is ${\rm GL}(n)$ invariant if
	\begin{equation}
		A_n(r(I_1\cdots I_n))=r^{-(n-3)}A_n((I_1\cdots I_n))\, .
		\label{eq:review_GL_scaling}
	\end{equation}
	Little group covariance imposes a second constraint. With
	\begin{equation}
		\lambda_i\mapsto \rho_i\lambda_i\, ,
		\qquad
		\bar\lambda_i\mapsto \rho_i^{-1}\bar\lambda_i\, ,
	\end{equation}
	the rescaled correlator obeys
	\begin{equation}
		\psi_n\mapsto \left(\prod_{i=1}^n\rho_i^{-2h_i}\right)\psi_n\, .
	\end{equation}
	Since $C\Lambda$ is kept invariant by $C\mapsto C\rho^{-1}$, a minor containing a barred column $\bar\imath$ carries weight $\rho_i^{-1}$ while a minor containing an unbarred column $i$ carries weight $\rho_i$. Hence
	\begin{equation}
		A_n((I_1\cdots I_n))\mapsto
		\left(\prod_{i=1}^n\rho_i^{-2h_i}\right)
		A_n((I_1\cdots I_n))\, .
		\label{eq:review_LG_scaling}
	\end{equation}
	Equations \eqref{eq:review_GL_scaling} and \eqref{eq:review_LG_scaling} are the kinematic bootstrap conditions on the Grassmannian integrand.
	
	We now describe how the map from Grassmannian to momentum space arises via twistor space. Taking a dual twistor $\mathbf{W}_i\equiv(\bar\mu_{i,\alpha},\bar\lambda_i^\alpha)$, one can perform the half-Fourier transform, Eq.~\eqref{eq:hft}, of Eq.~\eqref{eq:Twistor_cs} in the $\bar\mu_i$ variables and find
	\begin{equation}
		\psi_n(\Lambda)=\int \ud c_{ij}\,
		\delta(\lambda_i^\alpha-c_{ij}\bar\lambda_j^\alpha)\,A_n(c_{ij})\, ,
		\label{eq:review_gauge_fixed_integral}
	\end{equation}
	where repeated particle labels in $c_{ij}\bar\lambda_j^\alpha$ are summed and which, at first sight, does not seem to be equivalent to Eq.~\eqref{eq:review_grassmannian_integral}. To see the equivalence, one has to gauge-fix and pick a Grassmannian branch. The orthogonal Grassmannian has two disconnected components, called the right and left branches. The two branches are the two disconnected choices allowed by the quadratic constraints on the Grassmannian variables\footnote{More explicitly, the orthogonality condition on $C$ relates complementary maximal minors with a relative sign; the two possible signs define the two connected components \cite{ElMaazouz2024}.}. For example, the condition $CQC^T=0$ implies $(\bar1\bar2\cdots\bar n)(1\bar2\cdots\bar n)=0$, so one can be on a branch where $(1\bar2\cdots\bar n)=0$, called right branch, or on the complementary left branch where $(\bar1\bar2\cdots\bar n)=0$. On the right branch one can fix the ${\rm GL}(n)$ redundancy by choosing
	\begin{equation}
		C=(1_{n\times n},C_n)\, ,
		\qquad
		(C_n)_{ij}=-c_{ij}\, ,
		\qquad
		c_{ij}=-c_{ji}\, .
		\label{eq:review_right_branch_gauge}
	\end{equation}
	It is precisely the right-branch gauge-fixed Grassmannian integral which reduces Eq.~\eqref{eq:review_grassmannian_integral} to Eq.~\eqref{eq:review_gauge_fixed_integral}. The left branch can be obtained by exchanging an odd number of barred and unbarred columns. Similarly, replacing an odd number of dual twistors $\mathbf{W}_i\equiv(\bar\mu_{i,\alpha},\bar\lambda_i^\alpha)$ by twistors $\mathbf{Z}_i\equiv(\lambda_i^\alpha,\mu_i^\alpha)$ gives charts for the other branch. In practice these different charts are related by flips $i\leftrightarrow \bar\imath$ and correspond to different helicity configurations.
	
	For later use, it is useful to record how the delta functions in Eq.~\eqref{eq:review_gauge_fixed_integral} localize the integral. The delta functions include three-momentum conservation, which leaves $2n-3$ constraints on the $n(n-1)/2$ parameters $c_{ij}$, allowing $c_{ij}$ to be expressed in terms of $(n-2)(n-3)/2$ parameters. At three points, there is no parameter left and, on the right branch,
	\begin{equation}
		\psi_3'=\frac{1}{4}\,A_3(c_{ij})\bigg|_{c_{ij}=\langle ij\rangle/E}\, ,
		\qquad
		E=k_1+k_2+k_3\, .
		\label{eq:review_three_point_localization}
	\end{equation}
	where $\psi'$ is the delta stripped wavefunction/correlator. At four points one free parameter remains. A convenient parameterization is
	\begin{equation}
		c_{ij}(\tau)=\frac{\langle ij\rangle}{E}
		+\tau\,\frac{\epsilon_{ijkl}\langle\bar k\bar l\rangle}{2}\, ,
		\qquad
		E=\sum_{i=1}^4 k_i\, ,
		\label{eq:review_cij_tau}
	\end{equation}
	so that, up to an overall Jacobian convention,
	\begin{equation}
		\psi_4'=\int\frac{\ud\tau}{2\pi i}\,A_4(c_{ij}(\tau))\, .
		\label{eq:review_four_point_tau_integral}
	\end{equation}
	The choice of contour determines which discontinuity of the full correlator is produced.
	
	\subsubsection{Three- and four-point Grassmannian correlators}
	At three points little group covariance fixes all the possible answers and gives the results already found from twistor space. For all-plus helicities, on the right branch one has the RHS of Eq.~\eqref{eq:3pt_grassm}. On the left branch, the corresponding all-plus helicities Grassmannian correlator is the LHS of Eq.~\eqref{eq:3pt_grassm}. Instead of working with simple gauge fixed expressions, it can be insightful to look at the gauge-invariant three-point formula. The leading and higher-derivative all-plus correlators for equal spins are
	\begin{align}
		A_{3,+++}^{({\rm leading})}
		&=\left(\frac{(\bar1\bar2\bar3)^2}{{\cal K}}\right)^s \  ,
		\label{eq:review_A3_gi_leading}\\[1mm]
		A_{3,+++}^{({\rm higher})}
		&=\left(
		\frac{(1\bar{2}\bar{3})^2(\bar{1}2\bar{3})^2(\bar{1}\bar{2}3)^2}{\mathcal{K}^3}
		\right)^{s}\, .
		\label{eq:review_A3_gi_higher}
	\end{align}
	where the little group invariant $\mathcal{K}$ is defined as 
	\begin{equation}
		{\cal K}\equiv (i\bar\imath j)(\bar\jmath k\bar k)\, ,
		\label{eq:review_Kappa}
	\end{equation}
	which is independent of the cyclic choice of $\{i,j,k\}$ up to the identities of ${\rm OGr}(3,6)$. Other helicity configurations are obtained by flipping barred and unbarred columns on the corresponding external leg. For example, the $++-$ configuration follows from $3\leftrightarrow\bar3$. This is what helicity means in this formalism: the sign of the helicity is not encoded by a separate polarization tensor, but by which chiral spinor representative, barred or unbarred, appears in the minors associated with that leg. Thus changing helicity is implemented algebraically by exchanging the corresponding columns of the Grassmannian matrix. This also explains why the support can move between the two branches. For instance, for leading interactions the $+++$ correlator is supported on the right branch, while the $++-$ correlator is supported on the left branch. For higher-derivative interactions these assignments are reversed. The branch is therefore not itself the interaction type; the interaction type is instead detected by the order of the ${\cal K}$ pole and by the flat-space limit. The gauge-invariant three-point Grassmannian correlators also satisfy a trivial double copy.
	
	At four points, the correlator can be expressed as in  Eq.~\eqref{eq:review_four_point_tau_integral}. It can be bootstrapped by looking at the flat space limit and using unitarity. The building blocks for the correlator are the little group invariant minors 
	\begin{equation}
		S=(\bar1\bar2 12)\, ,
		\qquad
		T=(\bar1\bar4 14)\, ,
		\qquad
		U=(\bar1\bar3 13)\, .
		\label{eq:review_STU}
	\end{equation}
	In the right-branch gauge these are
	\begin{equation} \label{eq:grassm_mandelstams}
		S=c_{13}c_{24}-c_{14}c_{23}\, ,
		\qquad
		T=c_{13}c_{24}-c_{12}c_{34}\, ,
		\qquad
		U=-c_{14}c_{23}-c_{12}c_{34}\, .
	\end{equation}
	Using \eqref{eq:review_cij_tau}, they become quadratic functions of $\tau$, while their sum obeys
	\begin{equation}
		{\cal E}(\tau)\equiv -\frac{S+T+U}{2}=\tau E\, .
		\label{eq:review_calE}
	\end{equation}
	Thus the pole at $S+T+U=0$ is the Grassmannian avatar of the total-energy pole. If
	\begin{equation}
		A_4\supset \frac{1}{{\cal E}^{r+1}}\,{\cal A}_4((ij\bar k\bar l))\, ,
	\end{equation}
	then its flat-space singularity is extracted by the substitution rule
	\begin{equation}
		\frac{1}{{\cal E}^{r+1}}\,{\cal A}_4((ij\bar k\bar l))
		\mapsto
		\binom{2r}{r}\frac{1}{E^{2r+1}}\,{\cal A}_4(\langle ij\rangle[kl])
		\equiv \frac{(2r)!}{E^{2r+1}}M_4\, ,
		\label{eq:review_flat_space_substitution}
	\end{equation}
	where $M_4$ is the flat-space amplitude. This substitution assumes that ${\cal A}_4$ depends only on minors containing two barred and two unbarred columns; Ref.~\cite{Arundine:2026fbr} discusses exceptions to this assumption.
	
	To fix the correlator, it is also necessary to use unitarity. For the exchange of a massless particle in the s-channel one has the following cutting rule in momentum space \cite{Goodhew:2021oqg,Baumann:2021fxj}
	\be
	\mathrm{Disc}_{k_s^2}[\psi'_4]
	=
	-\frac{1}{2k_s}\sum_h
	\mathrm{Disc}_{k_s^2}[\psi_{3,L}^{\prime(-h)}]\,
	\mathrm{Disc}_{k_s^2}[\psi_{3,R}^{\prime(+h)}] \, ,
	\ee
	where $\mathrm{Disc}_{k_s^2}[\psi]=\psi(k_s)-\psi(-k_s)\,.$ Here $k_s$ is the internal energy. It has been shown in \cite{Arundine:2026fbr} that this implies the Grassmannian factorization
	\begin{equation}
		\mathop{\rm Res}_{S=0}A_4=
		\sum_h A_{3,L}^{(-h)}A_{3,R}^{(+h)}\, .
		\label{eq:review_factorization}
	\end{equation}
	This is the same algebraic form as on-shell factorization of flat-space amplitudes.  For four conformally coupled scalars exchanging a photon in the $s$ channel, factorization gives
	\begin{equation}
		A_{4,\gamma}=\frac{1}{S+T+U}\frac{T-U}{S}\, ,
		\label{eq:review_photon_exchange}
	\end{equation}
	while for the  graviton exchange one obtains
	\begin{equation}
		A_{4,g}=\frac{1}{(S+T+U)^2}
		\left(\frac{(U-T)^2}{S}-\frac{S}{3}\right)\, .
		\label{eq:review_graviton_exchange}
	\end{equation}
	These expressions are obtained by imposing the residues \eqref{eq:review_factorization}, completing the result by the minimal contact terms compatible with the expected spin-one or spin-two exchange structure, and checking the total-energy limit through \eqref{eq:review_flat_space_substitution}. Performing the $\tau$ integral then reproduces the corresponding momentum-space correlators. A related development is the formulation of the scalar four-point function of minimal Vasiliev higher-spin gravity in de Sitter space as an integral over ${\rm OGr}(4,8)$ \cite{De:2026shn}.
	
	While one would hope that these are related by a simple double copy prescription, there is a clear obstruction given by the contact term in the graviton exchange. These contact terms arise when fixing the correlator using the flat space limit. For spin one, the factorization residue already has the form of the first Legendre polynomial, so no extra term regular at the exchange pole is required. For spin two, the residue must be completed to the traceless spin-two structure, equivalently to the second Legendre polynomial. This completion adds a term regular at the exchange pole, which appears as a contact term in Grassmannian space. This mismatch of the naive double copy has been observed before \cite{Herderschee:2022ntr}. As reviewed in Section~\ref{sec:correll_embedding}, the naive double copy of the four-point scalar correlator with spin-one interactions does not give the one with graviton interactions unless the kinematic numerators are  corrected by terms proportional to the quadratic Casimir of intermediate states and lower-dimension differential operators, see Eq.~\eqref{eq:scalar_ex_DC_embedding}. It would be interesting to formulate the Grassmannian double copy in an analogous manner.
	
	For the pure Yang--Mills example, one first fixes the full color-ordered Grassmannian function $A(1^-2^-3^+4^+)$ by gluing the three-point correlators on the $S=0$ and $T=0$ factorization poles, and then extending the result away from these poles by allowing only simple poles at $S=0$, $T=0$, and $S\pm T\pm U=0$, while imposing $S\leftrightarrow T$ symmetry and the Kleiss--Kuijf relation. The result of this bootstrap is \cite{Arundine:2026fbr}
	\begin{equation}
		\begin{aligned}
			A(1^-2^-3^+4^+)
			&=\frac{(12\bar3\bar4)^2}{ST}
			\left(
			\frac{3}{S+T+U}
			+\frac{1}{S+T-U}
			\right.\\
			&\hspace{2.2cm}\left.
			-\frac{1}{-S+T+U}
			-\frac{1}{S-T+U}
			\right)\, .
		\end{aligned}
		\label{eq:review_YM_grassmannian_full}
	\end{equation}
    One can further consider a reduced correlator given by the part even under $U\mapsto -U$, $\hat A\equiv \frac12[A+A(U\mapsto -U)]$ which is related to a particularly simple external-energy discontinuity of the full color-ordered correlator. For pure Yang--Mills, the corresponding color-ordered $--++$ representative is
	\begin{equation}
		\hat A(1^-2^-3^+4^+)=
		2\left(\frac{1}{S+T+U}+\frac{1}{S+T-U}\right)
		\frac{(12\bar3\bar4)^2}{ST}\, .
		\label{eq:review_YM_grassmannian}
	\end{equation}
	The first term has the same form as the flat-space amplitude written in terms of the Grassmannian Mandelstams, while the second term is obtained by the sign flip $U\mapsto -U$. Its flat space limit, given by the $S+T+U=0$ singularity, reads
	\begin{equation}
		\hat A(1^-2^-3^+4^+)\longrightarrow
		-\frac{1}{E}\frac{\langle12\rangle^2[34]^2}{s\,t}\, ,
	\end{equation}
	which is the expected color-ordered Yang--Mills amplitude. After the $\tau$ integral this gives the discontinuity $\mathop{\rm Disc}_{k_1^2,k_3^2}\psi$ of the full correlator, rather than the full correlator itself. The full answer can be reconstructed from a dispersive integral over this discontinuity, but from the twistor-space perspective it would be more satisfying to have a Grassmannian object that directly obeys the inhomogeneous conformal Ward identities, as happens for the complex twistor-space correlators reviewed above.
	
	The corresponding four-graviton Grassmannian correlator first appeared in \cite{Bala:2026lvw} and subsequently \cite{Arundine:2026fbr} presented the same expression and checked it by reconstructing the associated momentum-space correlator. It reads
	\begin{equation}
		\begin{aligned}
			A_{4,\mathrm{GR}}^{--++}
			&=-2\frac{(12\bar3\bar4)^4}{STU}
			\Bigg[
			\frac{3}{(S+T+U)^2}
			-\frac{1}{(-S+T+U)^2}\\
			&\hspace{2.9cm}
			-\frac{1}{(S-T+U)^2}
			-\frac{1}{(S+T-U)^2}
			\Bigg]\, .
		\end{aligned}
		\label{eq:review_graviton_grassmannian}
	\end{equation}
	With the appropriate contour, the $\tau$ integral reproduces a triple discontinuity of the four-graviton wavefunction coefficient, while the additional dispersive integral needed to reconstruct the full correlator is still unknown. From this expression, it is not immediately clear if a double copy is satisfied. One would need to compute the 4-point biadjoint scalar correlator and use it to construct the KLT kernel or the invariant propagator analogue appearing in the BCJ double copy.
	
	A related investigation was performed in \cite{Ansari:2025fvi}, where the correlators were constructed in the special kinematic limit defined in Section \ref{sec:twistor_correl} above Eq.~\eqref{eq:scalar4pttwistorCPW}. The Schwinger-parameter functions corresponding to Eq.~\eqref{eq:YM4pttwistorCPW} and \eqref{eq:grav4pttwistorCPW} are
	\begin{align}
		\mathcal{M}_{4,\mathrm{YM}}
		&=
		\frac{c_{1}^3c_3^3}{c_{1}c_{3}c_2^2c_4^2} \, , \\
		\mathcal{M}_{4,\mathrm{GR}}
		&=
		\left(\frac{c_{1}^3c_3^3}{c_{1}c_{3}c_2^2c_4^2}\right)^2
		=
		\mathcal{M}_{4,\mathrm{YM}}^2 \, .
	\end{align}
	We note that this kinematic limit corresponds to taking $S\rightarrow 0$, with $S$ given in Eq.~\eqref{eq:grassm_mandelstams}, which leads to the four-point Grassmannian correlators being given by products of the three-point Grassmannian correlators, and hence trivially satisfying a double copy. While these correlators are related by a squaring procedure, this is too simple to describe a satisfactory four-point double copy, not only because they correspond to a special kinematic limit, but because there is no trace of color-kinematics duality. Nevertheless, it provides an interesting check that relates the graviton and gluon structures. Recent explorations have shown how one can obtain more general correlators when working with supersymmetric theories \cite{Bala:2026bdx,Bala:2026hdm,Huang:2026tsh} and \cite{Arundine:2026myr} further develops this Grassmannian approach for exchange correlators with generic mass and spin, but no double copy was constructed. Additionally, the discontinuities of four-dimensional correlators have been constructed from a Grassmannian formulation and shown to satisfy a double copy relation at three points~\cite{Bala:2026trw} and four points for an exchange channel~\cite{S:2026qwn}. Furthermore, a cosmological BCFW bridge gives algebraic recursion relations for orthogonal-Grassmannian four-point correlators and a positive-geometry interpretation of the four-gluon result \cite{Bala:2026lvw}. It remains an open question whether color-kinematics duality, BCJ relations, and the double copy can be formulated for full Grassmannian correlators instead of just discontinuities.

	\subsection{\texorpdfstring{Scattering on Nontrivial Backgrounds}{Scattering on Nontrivial Backgrounds}}
	
	The constructions reviewed above concern correlators whose external data are specified at a boundary. A conventional scattering problem on a curved spacetime instead requires sufficiently controlled asymptotic regions in which particle states can be defined. The two approaches reviewed here retain such an interpretation in complementary ways. Sandwich plane waves have flat in- and out-regions and therefore admit an S-matrix, while a coherent state of asymptotic on-shell quanta can be reinterpreted as a classical background field. In both cases the background changes the double copy itself: one must account for dressed external states, memory or tail terms, and a map between the gauge and gravitational backgrounds.
	
	\subsubsection{Plane-wave scattering and double copy}\label{sec:plane_wave_scattering_dc}
	
	\begin{figure}[!t]
		\centering
		\begin{tikzpicture}[x=1.18cm,y=1.12cm,every node/.style={font=\normalsize}]
			\fill[blue!8] (-0.9,-1.6) rectangle (0.9,1.6);
			\draw[blue!45,dashed] (-0.9,-1.6) -- (-0.9,1.6);
			\draw[blue!45,dashed] (0.9,-1.6) -- (0.9,1.6);
			\draw[->] (-3.35,0) -- (3.55,0) node[below right] {$u$};
			\draw[->] (0,-1.55) -- (0,1.85) node[above left] {$v$};
			\node[below] at (-0.9,-1.6) {$u_1$};
			\node[below] at (0.9,-1.6) {$u_2$};
			\node[align=center] at (-2.25,1.12) {flat in-region\\$H_{ab}=0$\\$\dot{\mathrm{A}}_a=0$};
			\node[align=center] at (2.25,1.12) {flat out-region\\$H_{ab}=0$\\$\dot{\mathrm{A}}_a=0$};
			\node[align=center] at (0,-1.02) {sandwich wave\\$H_{ab}(u)$ or $\dot{\mathrm{A}}_a(u)$};
		\end{tikzpicture}
		\caption{Schematic sandwich plane wave. The transverse directions $x^a$ are suppressed. The curvature profile $H_{ab}(u)$, or gauge field strength $\dot{\mathrm{A}}_a(u)$, has support only for $u_1<u<u_2$, leaving flat in- and out-regions where scattering states can be defined.}
		\label{fig:sandwich_plane_wave}
	\end{figure}
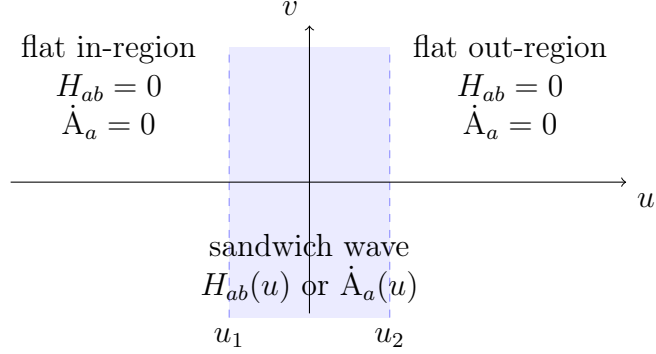
	
	We first consider the spacetime construction of Ref.~\cite{Adamo:2017nia}. The setting is a sandwich plane wave, as shown schematically in Fig.~\ref{fig:sandwich_plane_wave}, so the background is flat before and after a finite interval of retarded time. This gives a genuine scattering problem while keeping enough symmetry to write the external states explicitly. A gravitational plane wave can be written in Brinkmann coordinates as
	\begin{equation}
		ds^2=-2du\,dv+H_{ab}(u)x^a x^b\,du^2+\delta_{ab}dx^a dx^b\, ,
		\qquad \delta^{ab}H_{ab}=0\, ,
		\label{eq:pw_brinkmann_background}
	\end{equation}
	where $a,b=1,\ldots,d-2$. The trace condition is the vacuum Einstein equation. The relation to Einstein--Rosen coordinates, in which the metric is
	\begin{equation}
		ds^2 = -2\,dU\,dV + \gamma_{ij}(U)\,dy^i\,dy^j \, ,
		\label{eq:pw_EinsteinRosen} 
	\end{equation}
	is encoded by a Brinkmann-to-Rosen vielbein $E^a{}_i(u)$ and its inverse $E^i{}_a(u)$, with
	\begin{equation}
		\ddot E_a{}_i=H_{ab}E^b{}_i\, ,
		\qquad
		\gamma^{ij}=E^{(i}{}_aE^{j)}{}_b\delta^{ab}\, ,
		\qquad
		\sigma_{ab}=\dot E^i{}_aE_{bi}\, .
		\label{eq:pw_vielbein}
	\end{equation}
	Here $E_{bi}=\delta_{bc}E^c{}_i$, $\gamma^{ij}$ is the inverse transverse Einstein--Rosen metric, and $\sigma_{ab}$ contains the expansion and shear of the null congruence. It is also useful to define
	\begin{equation}
		F^{ij}(u)=\int^u ds\,\gamma^{ij}(s)\, ,
		\qquad
		\Omega(u)=\left|\gamma^{-1}(u)\right|^{1/4}=\left|\det E^a{}_i(u)\right|^{-1/2}\, .
		\label{eq:pw_F_Omega}
	\end{equation}
	For a sandwich plane wave, $H_{ab}$ has compact support in $u$. One may then choose an in-state or out-state vielbein which approaches the identity in the corresponding flat region. The two choices are inequivalent because the wave changes the transverse congruence, which is the origin of gravitational memory.
	
	The gauge-theory analogue is a plane wave valued in the adjoint of a given Lie algebra which lives in a Minkowski spacetime with metric given by Eq.~\eqref{eq:pw_brinkmann_background} with $H_{ab}=0$. Here we will take this plane wave background to be valued in the Cartan subalgebra to be able to define a double copy, see also the discussion in Section~\ref{sec:sd_twistor_klt}. Suppressing the algebra generator, its Rosen- and Brinkmann-gauge potentials are
	\begin{equation}
		A=-\mathrm{A}_a(u)dx^a
		\quad\longrightarrow\quad
		A=x^a\dot{\mathrm{A}}_a(u)\,du\, ,
		\qquad
		F=\dot{\mathrm{A}}_a(u)dx^a\wedge du\, .
		\label{eq:pw_gauge_background}
	\end{equation}
	The arrow indicates the gauge transformation generated by $x^a\mathrm{A}_a(u)$ and the fact that the gauge field is linear in $x$ is a first indication of the double copy structure. Taking $\dot{\mathrm{A}}_a$ to have compact support again defines a sandwich wave. The evolution problem is unitary and there is no particle creation on either background. Nevertheless, a momentum eigenstate in the in-region does not generically evolve into a momentum eigenstate in the out-region. The gauge and gravitational memory effects are therefore already visible in the definition of the asymptotic states.
	
	The external wavefunctions make the background dependence explicit. On the gravitational plane wave background, a solution of the Hamilton-Jacobi equation 
	\begin{equation}
		g^{\mu\nu}(\partial_\mu \phi)(\partial_\nu \phi)=0\, ,
	\end{equation}
	is given by 
	\begin{equation}
		\Phi= \Omega(u) e^{i \phi_k} \ , \quad \phi_k
		=k_0v+\frac{k_0}{2}\sigma_{ab}x^ax^b+k_iE^i{}_ax^a+\frac{k_ik_j}{2k_0}F^{ij}\, ,
		\label{eq:pw_graviton_phase}
	\end{equation}
	where $(k_0,k_i)$ are the $d-1$ conserved momenta associated with the plane-wave isometries. One can similarly find a gauge field solution on the gravitational plane wave background,
	\begin{equation}
		A=\Phi \epsilon\cdot dX \ , \quad \epsilon\cdot dX
		=\varepsilon_a dx^a+\varepsilon^a\left(\frac{k_i}{k_0}E^i{}_a+\sigma_{ab}x^b\right)du\, .
		\label{eq:pw_graviton_polarization}
	\end{equation}
	where $\varepsilon_a$ is a constant null transverse polarization. Finally, the corresponding linearized graviton is
	\begin{equation}
		h_{\mu\nu}dX^\mu dX^\nu
		=\Phi  \left[(\epsilon\cdot dX)^2-\frac{i}{k_0}\varepsilon^a\varepsilon^b\sigma_{ab}du^2\right] \, .
		\label{eq:pw_graviton_wavefunction}
	\end{equation}
	The second term is absent for a flat-space polarization square and is the first indication of a curvature tail. The local momentum is the one-form
	\begin{equation}
		K_\mu dX^\mu=d\phi_k\, ,
		\qquad
		g^{\mu\nu}K_\mu K_\nu=0\, ,
		\qquad
		g^{\mu\nu}K_\mu\epsilon_\nu=0\, .
		\label{eq:pw_graviton_momentum}
	\end{equation}
	Thus $k_0$ and $k_i$ label the state globally, while $K_\mu(u,x)$ and $\epsilon_\mu(u,x)$ are the background-dressed momentum and polarization entering the local interaction. It is clear that, similar to some examples shown above, the double copy at fixed background requires an additional term to give the gravitational solution. For the present case, one can avoid this by double copying the background itself. We therefore turn to solutions on gauge backgrounds.
	
	For a gluon on a non-trivial (Cartan-valued) gauge background, the analogous phase and polarization are
	\begin{align}
		\widetilde\phi_k
		&=k_0v+\left(k_a+e\mathrm{A}_a\right)x^a+\frac{f(u)}{2k_0}\, ,
		\qquad
		f(u)=\int^u ds\,\left(k_a+e\mathrm{A}_a(s)\right)^2\, ,
		\label{eq:pw_gluon_phase}\\
		\widetilde\epsilon\cdot dX
		&=\widetilde\varepsilon_a dx^a
		+\frac{\widetilde\varepsilon^a\left(k_a+e\mathrm{A}_a\right)}{k_0}du\, ,
		\label{eq:pw_gluon_polarization}
	\end{align}
	with $\widetilde\varepsilon_a$ constant. The linearized gluon and its gauge-covariant local momentum are
	\begin{equation}
		a_\mu^A=T^A\widetilde\epsilon_\mu e^{i\widetilde\phi_k}\, ,
		\qquad
		\widetilde K_\mu dX^\mu=-ie^{-i\widetilde\phi_k}D_\mu e^{i\widetilde\phi_k}dX^\mu\, ,
		\qquad
		\widetilde K^2=\widetilde K\cdot\widetilde\epsilon=0\, .
		\label{eq:pw_gluon_wavefunction}
	\end{equation}
	Equations~\eqref{eq:pw_graviton_wavefunction} and \eqref{eq:pw_gluon_wavefunction} are the curved-background replacements of the plane-wave external states used in the flat-space amplitudes reviewed in Section~\ref{sec:flatDC}.
	
	\paragraph{Three-point amplitudes and the double copy}
	
	The full three-point amplitudes contain a $d-1$-conserved momentum delta function, a remaining $u$ integral, and background-dependent phases. For the double copy, the relevant objects are the stripped tree-level integrands. On the gravitational plane wave background, the gluon integrand is the simple final answer
	\begin{equation}
		\mathcal M_3(A_1,A_2,A_3)
		=\epsilon_1\cdot\epsilon_3\,K_1\cdot\epsilon_2
		+\epsilon_1\cdot\epsilon_2\,K_2\cdot\epsilon_3
		+\epsilon_2\cdot\epsilon_3\,K_3\cdot\epsilon_1\, .
		\label{eq:pw_gluon_amplitude}
	\end{equation}
	Here $K_r=d\phi_{k_r}$ is the dressed local momentum, and $\epsilon_r$ is the dressed gluon polarization defined in Eq.~\eqref{eq:pw_graviton_polarization} for the state labelled by $r$. The corresponding graviton integrand is
	\begin{equation}
		\mathcal M_3(h_1,h_2,h_3)
		=\left(\mathcal M_3(A_1,A_2,A_3)\right)^2
		-i k_{10}k_{20}k_{30}\,\sigma^{ab}\mathcal C_a\mathcal C_b\, ,
		\label{eq:pw_graviton_integrand}
	\end{equation}
	with
	\begin{equation}
		\mathcal C_a
		=\frac{\epsilon_2\cdot\epsilon_3}{k_{10}}\varepsilon_{1a}
		+\frac{\epsilon_1\cdot\epsilon_3}{k_{20}}\varepsilon_{2a}
		+\frac{\epsilon_1\cdot\epsilon_2}{k_{30}}\varepsilon_{3a}\, .
		\label{eq:pw_tail_vector}
	\end{equation}
	The first term in Eq.~\eqref{eq:pw_graviton_integrand} is the expected square of the gluon integrand on the same gravitational background. The second term is the curvature tail from the graviton wavefunction in Eq.~\eqref{eq:pw_graviton_wavefunction}. Thus the most naive curved-space double copy seems to fail: holding the gravitational background fixed and simply squaring the gluon result misses the shear-dependent contribution.
	
	The square root of this tail becomes visible only after comparing with Yang--Mills theory on a gauge plane wave background. Taking all three gluon momenta to be outgoing, the stripped gauge-background integrand can be written as
	\begin{equation}
		\mathcal A_3(a_1,a_2,a_3)
		=F(\{k_{r0},k_{ra},\widetilde\varepsilon_r\})
		+\mathsf C(\{k_{r0},k_{ra},\widetilde\varepsilon_r\}|\mathrm{A})\, ,
		\label{eq:pw_gauge_split}
	\end{equation}
	where, using transverse contractions,
	\begin{align}
		F
		&=-\frac{\widetilde\varepsilon_1\cdot\widetilde\varepsilon_3}{k_{20}}
		\left(k_{10}\,k_2\cdot\widetilde\varepsilon_2
		-k_{20}\,k_1\cdot\widetilde\varepsilon_2\right)
		+\mathrm{cyclic}\, ,
		\label{eq:pw_gauge_flat_part}\\
		\mathsf C
		&=-\frac{\widetilde\varepsilon_1\cdot\widetilde\varepsilon_3}{k_{20}}\,
		\mathrm{A}\cdot\widetilde\varepsilon_2\,
		\left(k_{10}e_2-k_{20}e_1\right)
		+\mathrm{cyclic}\, .
		\label{eq:pw_gauge_memory_part}
	\end{align}
	The term $F$ is the flat three-gluon numerator in the lightfront parametrization, while $\mathsf C$ is the part linear in the background gauge potential. This is the gauge-theory memory contribution, analogous to the shear term in the gravitational plane wave.
	
	The curved-background double copy is then a three-step prescription:
	\begin{enumerate}
		\item Flip the background charge (the sign of the background color factor in the gauge-plane-wave integrand) in one copy, so that $\widetilde{\mathcal A}_3=F-\mathsf C$ and
		\begin{equation}
			\mathcal A_3\,\widetilde{\mathcal A}_3=F^2-\mathsf C^2\, .
			\label{eq:pw_gauge_product}
		\end{equation}
		
		\item Replace the transverse gauge-background data by the gravitationally dressed data,
		\begin{equation}
			k_{ra}\longrightarrow k_{ri}E^i{}_a\, ,
			\qquad
			\widetilde\varepsilon_{ra}\longrightarrow\varepsilon_{ra}\, .
			\label{eq:pw_kinematic_replacement}
		\end{equation}
		
		\item Replace the quadratic gauge-background memory by the gravitational shear,
		\begin{equation}
			e_r e_s\mathrm{A}_a\mathrm{A}_b\longrightarrow
			\begin{cases}
				i k_{r0}\sigma_{ab}\, , & r=s\, ,\\
				i(k_{r0}+k_{s0})\sigma_{ab}\, , & r\neq s\, .
			\end{cases}
			\label{eq:pw_replacement_map}
		\end{equation}
	\end{enumerate}
	Calling the combined replacement map $\rho$, the first two steps map $F^2$ to the square in Eq.~\eqref{eq:pw_graviton_integrand}, while the third maps $\mathsf C^2$ to the graviton tail,
	\begin{equation}
		\rho\!\left(\mathsf C^2\right)
		=i k_{10}k_{20}k_{30}\sigma^{ab}\mathcal C_a\mathcal C_b\, .
		\label{eq:pw_tail_double_copy}
	\end{equation}
	Therefore
	\begin{equation}
		\mathcal M_3(h_1,h_2,h_3)
		=\rho\!\left(\mathcal A_3\,\widetilde{\mathcal A}_3\right)\, .
		\label{eq:pw_integrand_double_copy}
	\end{equation}
	This is the precise sense in which Ref.~\cite{Adamo:2017nia} finds a plane-wave double copy: the graviton amplitude is not obtained by squaring gluons on the same fixed gravitational background, but by double copying gluons on the corresponding gauge plane wave and then mapping the memory data. We note that Ref.~\cite{Adamo:2017sze} reproduced the same three-point amplitudes from type-II and heterotic ambitwistor strings on plane-wave backgrounds. However, that formulation did not make an ambitwistor space double copy manifest. It would be interesting to revisit this question in light of recent ambitwistor space constructions for four-dimensional boundary correlators \cite{CarrilloGonzalez:2026eum}.
	
	At four points, color--kinematics duality on a fixed Yang--Mills plane-wave background was studied in \cite{Adamo:2018mpq}. The amplitude can still be organized into the usual $s$-, $t$-, and $u$-channel color structures, but the corresponding kinematic numerators do not satisfy an exact Jacobi identity because the propagators carry channel-dependent information about the background-dressed exchanged momenta. After mapping the three channels to a common propagator structure, it can be seen that the Jacobi identity for the kinematic numerators is not satisfied. However, the obstruction is highly constrained: every term violating the Jacobi identity is proportional to a background-deformed analogue of the usual transversality condition $k_i\cdot\epsilon_i=0$. This leads to a weaker form of the kinematic Jacobi relation in which equality holds modulo these deformed gauge-condition terms. The four-point double copy was conjectured, but it has not been checked to correspond to the expected gravitational amplitude.
	
	The plane-wave double copy was further extended in \cite{Adamo:2020qru} to the radiation generated by massive particles propagating through strong gauge-theory and gravitational plane-wave backgrounds, which are treated exactly. The leading back-reaction was computed both classically, from background-coupled worldline dynamics, and from tree-level three-point non-linear Compton amplitudes, with the classical limit of the latter reproducing the corresponding radiation fields. The gluon-emission amplitude was shown to double copy to the graviton-emission amplitude through a background-dependent extension of the plane-wave replacement rules, which in the classical limit gives a direct map between the gluon and graviton radiation fields. There is also a direct relation to the classical double copy reviewed in Sec.~\ref{sec:kerr-schild}. At the level of the background, the linear gauge potential $x^a\dot{\mathrm{A}}_a(u)du$ is replaced by the quadratic Brinkmann profile $H_{ab}(u)x^ax^bdu^2$; at the level of three-point gluon integrands, this appears as the substitution $e_r\mathrm{A}_a\to k_{r0}\sigma_{ab}x^b$, which maps gluons on the gauge background to gluons on the gravitational background.
	
	\subsubsection{Coherent-state backgrounds}\label{sec:coherent_state_backgrounds}
	
	An earlier perturbative approach considered massive scalar pair production in general background gauge fields and constructed its double copy to particle production in corresponding gravitational and axio-dilaton backgrounds through next-to-leading order \cite{Ilderton:2024oly}. This construction also gives FRW spacetimes from a class of time-dependent gauge backgrounds and led to a conjectured all-orders map. Reference~\cite{Ilderton:2025gug} gives a more general route from vacuum amplitudes to amplitudes on backgrounds. Instead of starting with a curved spacetime, one starts with ordinary flat-space asymptotic states in which some of the gluons, or gravitons, are placed in coherent states. The associated c-number background is built from the coherent-state profiles; when the bra and ket profiles coincide it reduces to the usual field expectation value.
	
	\paragraph{Coherent states and amplitudes.}
	
	The starting point is a matrix element in which the asymptotic states are coherent states of gluons,
	\begin{equation}
		\mathcal A[\alpha',\alpha]
		=\langle\alpha',\mathrm{out}|T_{\rm YM}|\alpha,\mathrm{in}\rangle\, .
		\label{eq:coherent_matrix_element}
	\end{equation}
	Here $T_{\rm YM}$ denotes the Yang--Mills transition matrix in vacuum, and $|\mathrm{in}\rangle$, $|\mathrm{out}\rangle$ may also contain ordinary number states. The coherent states are built by acting on the vacuum with the displacement operator:
	\begin{align}
		|\alpha\rangle&=D(\alpha)|0\rangle\, ,\nonumber\\
		D(\alpha)
		&=e^{\int_k
			\left(\widehat a_s^{a\dagger}(k)\alpha_s^a(k)
			-\widehat a_s^a(k)\bar\alpha_s^a(k)\right)}=e^{\int_k\widehat a_s^{a\dagger}\alpha_s^a}
		e^{-\int_k\widehat a_s^a\bar\alpha_s^a}
		e^{-\frac{1}{2}\int_k|\alpha_s^a|^2} \ ,
		\label{eq:coherent_displacement}
	\end{align}
	where $a$ is a color index, $s$ labels the polarization, and repeated indices imply summation. The normal-ordered form is useful because the Gaussian factor in the last line is part of the coherent-state overlap. A basic property of a single coherent state is that it has a nonzero field expectation value for the asymptotic gauge field
	\begin{equation}
		\langle\alpha|	\widehat A_\mu^a(x)|\alpha \rangle
		=\sum_s\int_k\left[
		\alpha_s^a(k)\epsilon_\mu^s(k)e^{-ik\cdot x}
		+\bar{\alpha}_s^{a}(k)\bar\epsilon_\mu^s(k)e^{ik\cdot x}
		\right],
		\label{eq:coherent_operator_field}
	\end{equation}
	where $\int_k\equiv\int \frac{d^3 {\bf k}}{(2\pi)^3\,2|{\bf k}|}$. This gives the real classical free field associated with $\alpha$.

	One can then write the coherent states in Eq.~\eqref{eq:coherent_matrix_element} explicitly:
	\begin{align}
		\mathcal A[\alpha',\alpha]
		&=e^{-\frac{1}{2}\int_k(|\alpha_s^a|^2+|\alpha_s^{\prime a}|^2)}
		\langle\mathrm{out}|
		e^{\int_k\widehat a_s^a(k)\bar\alpha_s^{\prime a}(k)}
		T_{\rm YM}e^{\int_k\widehat a_s^{a\dagger}(k)\alpha_s^a(k)}
		|\mathrm{in}\rangle\, .
		\label{eq:coherent_unexpanded}
	\end{align}
	This can be expanded in the total number $n$ of gluons supplied by the coherent states as
	\begin{equation}
		\mathcal A[\alpha',\alpha]=e^{-\frac{1}{2}\int_k(|\alpha_s^a|^2+|\alpha_s^{\prime a}|^2)}\sum_n\mathcal A_{(n)} \ .
		\label{eq:coherent_generic_expansion}
	\end{equation}
	The first two coefficients are
	\begin{align}
		\mathcal A_{(0)}
		&=\langle\mathrm{out}|T_{\rm YM}|\mathrm{in}\rangle\, ,\nonumber\\
		\mathcal A_{(1)}
		&=\int_k\left[
		\bar\alpha_s^{\prime a}(k)\langle\mathrm{out},k^{a,s}|T_{\rm YM}|\mathrm{in}\rangle
		+\alpha_s^a(k)\langle\mathrm{out}|T_{\rm YM}|k^{a,s},\mathrm{in}\rangle
		\right].
		\label{eq:coherent_low_orders}
	\end{align}
	Thus the coherent profiles act as wavepackets multiplying ordinary vacuum amplitudes with extra gluons inserted in the initial or final state. We denote such a vacuum amplitude by $\mathcal A(\ldots\bar i\ldots j\ldots)$, with barred labels indicating final-state coherent gluons and unbarred labels indicating initial-state coherent gluons. The order-$n$ term, with $r$ final coherent-state gluons, is
	\begin{align}
		\mathcal A_{(n)}
		&=\sum_{r=0}^n\frac{1}{r!(n-r)!}
		\sum_{\{a_i,s_i\}}\int_{k_1}\cdots\int_{k_n}
		\nonumber\\
		&\hspace{0.5cm}\times
		\bar\alpha_{s_1}^{\prime a_1}(k_1)\cdots
		\bar\alpha_{s_r}^{\prime a_r}(k_r)
		\alpha_{s_{r+1}}^{a_{r+1}}(k_{r+1})\cdots
		\alpha_{s_n}^{a_n}(k_n)
		\nonumber\\
		&\hspace{0.5cm}\times
		\mathcal A(\bar1,\ldots,\bar r,r+1,\ldots,n)\ ,
		\label{eq:coherent_expansion}
	\end{align}
	where the indices of the external state have been absorbed in $\mathcal A$. These vacuum amplitudes must be understood with the disconnected pieces that are normally discarded. The reason is simple: an initial coherent-state gluon may propagate freely into a final coherent state gluon.
	
	The matrix element in Eq.~\eqref{eq:coherent_unexpanded} can then be simplified by assuming that ordinary number states have negligible phase-space overlap with the opposite coherent profiles\footnote{If the phase-space separation assumption fails, the leftover exponentials can remove gluons from the number states, and the corresponding lower-multiplicity terms must also be kept.}, using the identity $D^\dagger(\alpha)\widehat a_s^a(k)D(\alpha)=\widehat a_s^a(k)+\alpha_s^a(k)$ and the analogous shift of creation operators by $\bar\alpha'_s$, extracting the disconnected free coherent-to-coherent propagators into a prefactor $e^W$, and commuting exponentials, one finds
	\begin{equation}
		\mathcal A[\alpha',\alpha]
		=e^{W[\alpha',\alpha]}
		\langle\mathrm{out}|T_{\rm YM}[A(a+\alpha,a^\dagger+\bar{\alpha}')]|\mathrm{in}\rangle\,  ,
		\label{eq:coherent_background_equivalence}
	\end{equation}
	where
	\begin{equation}
		W[\alpha',\alpha]
		=-\frac{1}{2}\int_k|\alpha_s^a|^2
		-\frac{1}{2}\int_k|\alpha_s^{\prime a}|^2
		+\int_k\bar\alpha_s^{\prime a}\alpha_s^a\, .
		\label{eq:coherent_overlap}
	\end{equation}
	Thanks to the shift of the mode operators, the transition matrix now corresponds to scattering on the background field given by
	\begin{align}
		A_\mu^a(x)
		&=\sum_s\int_k\left[
		\alpha_s^a(k)\epsilon_\mu^s(k)e^{-ik\cdot x}
		+\bar\alpha_s^{\prime a}(k)\bar\epsilon_\mu^s(k)e^{ik\cdot x}
		\right] \\
		&\equiv\int d^4k\,A_\mu^a(k)e^{-ik\cdot x} \delta(k^2) \theta(k^0)\, .
		\label{eq:coherent_background}
	\end{align}
	Since all modes are on shell, Eq.~\eqref{eq:coherent_background} solves the linearized Yang--Mills equations. 
	
	One can further make the connection between coherent state amplitudes and background fields more explicit by defining a spin-stripped amplitude where the polarization vectors and momentum-conserving delta functions are factored out. For example, for the one-gluon vacuum  $\mathcal A(1)=\epsilon_s^\mu(k)\delta^{(4)}(\Delta Q-k)\mathcal A_\mu^a(-k)$.
	Associating the stripped polarizations with the background profiles rather than with the vacuum replaces the exposed external gluon lines by insertions of $A_\mu^a(k)$. Permutation symmetry then gives the order-$n$ contribution
	\begin{align}
		\mathcal A_{(n)}[A]
		&=\frac{1}{n!}\int d^4k_1\cdots d^4k_n\,
		\delta^{(4)}\!\left(\Delta Q-\sum_{i=1}^nk_i\right)
		\nonumber\\
		&\hspace{0.5cm}\times
		\prod_{i=1}^n A_{\mu_i}^{a_i}(k_i)\,
		\mathcal A_{a_1\cdots a_n}^{\mu_1\cdots\mu_n}
		(-k_1,\ldots,-k_n)\, ,
		\label{eq:coherent_background_expansion}
	\end{align}
	where $\Delta Q$ is the difference between the total momenta of the states $|\mathrm{out}\rangle$ and $|\mathrm{in}\rangle$, and $\mathcal A_{a_1\cdots a_n}^{\mu_1\cdots\mu_n}$ is a spin-stripped vacuum amplitude.
	
	The gravitational construction is parallel. The matrix element $\mathcal M[\beta',\beta]$ is defined analogously, with $T_{\rm YM}$ replaced by $T_{\rm grav}$. Anticipating the double copy, the coherent state is allowed to contain graviton, two-form, and dilaton modes. With $P$ collecting species and polarization labels for $h_{\mu\nu}$, $B_{\mu\nu}$, and $\varphi$, write
	\begin{equation}
		|\beta\rangle
		=e^{\sum_P\int_k\beta_P(k)\widehat a_P^\dagger(k)}e^{-\frac{1}{2}\sum_P\int_k|\beta_P(k)|^2}|0\rangle\, ,
		\label{eq:coherent_gravity_state}
	\end{equation}
	and similarly for $|\beta'\rangle$. All other steps follow analogously to the gauge theory. The corresponding background is the ``fat graviton''
	\begin{align}
		\mathcal H_{\mu\nu}(x)
		&\equiv h_{\mu\nu}(x)+B_{\mu\nu}(x)+\varphi_{\mu\nu}(x)\nonumber\\
		&=\sum_P\int_k\left[
		\beta_P(k)\epsilon_{\mu\nu}^P(k)e^{-ik\cdot x}
		+\bar\beta'_P(k)\bar\epsilon_{\mu\nu}^P(k)e^{ik\cdot x}
		\right].
		\label{eq:coherent_fat_graviton}
	\end{align}
	The trace polarization $\varphi_{\mu\nu}$ represents the dilaton contribution, up to the usual gauge terms. The graviton part defines a perturbative metric $\eta_{\mu\nu}+ h_{\mu\nu}$ obeying the linearized Einstein equations, while $B_{\mu\nu}$ and $\varphi$ obey the corresponding free equations of a two-form field. In direct analogy with Eq.~\eqref{eq:coherent_background_expansion},
	\begin{align}
		\mathcal M_{(n)}[\mathcal H]
		&=\frac{1}{n!}\int d^4k_1\cdots d^4k_n\,
		\delta^{(4)}\!\left(\Delta Q-\sum_{i=1}^nk_i\right)
		\nonumber\\
		&\hspace{0.5cm}\times
		\prod_{i=1}^n\mathcal H_{\mu_i\nu_i}(k_i)\,
		\mathcal M^{\mu_1\nu_1\cdots\mu_n\nu_n}(-k_1,\ldots,-k_n)\, .
		\label{eq:coherent_gravity_background_expansion}
	\end{align}
	Thus the coherent-state construction separates the double copy structure into two pieces: the usual map between spin-stripped vacuum amplitudes, and the induced map between the background wavepackets that multiply them.
	
	\paragraph{Double copy and classical backgrounds.}
	
	The double copy for the vacuum spin-stripped amplitudes follows as in Section~\ref{sec:flatDC}. For the purposes of this discussion, we do not need to specify the realization of the double copy. For each gluon polarization $s$, choose the second polarization vector $\widetilde\epsilon_\nu^s$ used in the double copy and decompose
	\begin{equation}
		\epsilon_\mu^s(k)\widetilde\epsilon_\nu^s(k)
		=\sum_PC_P^s\epsilon_{\mu\nu}^P(k)\, ,
		\qquad
		P\in\{h,B,\varphi\}\, .
		\label{eq:coherent_polarization_dc}
	\end{equation}
	The coefficients $C_P^s$ must be chosen consistently for every amplitude in the coherent-state expansion so that all multiplicities land in the same gravitational target theory. With this choice, the required vacuum map is
	\begin{align}
		\mathcal A(\bar1,\ldots,\bar r,r+1,\ldots,n)
		\longrightarrow{}&
		\sum_{P_1,\ldots,P_n}
		\bar C_{P_1}^{s_1}\cdots\bar C_{P_r}^{s_r}
		C_{P_{r+1}}^{s_{r+1}}\cdots C_{P_n}^{s_n}
		\nonumber\\
		&\times
		\mathcal M(\bar1,\ldots,\bar r,r+1,\ldots,n)\, .
		\label{eq:coherent_vacuum_dc}
	\end{align}
	
	Applying Eq.~\eqref{eq:coherent_vacuum_dc} to the one-particle term fixes the gravity coherent profiles:
	\begin{equation}
		\beta_P(k)=\sum_{a,s}C_P^s\alpha_s^a(k)\, ,
		\qquad
		\beta'_P(k)=\sum_{a,s}C_P^s\alpha_s^{\prime a}(k)\, .
		\label{eq:coherent_profile_dc}
	\end{equation}
	The explicit sum over color reflects the fact that every color component is copied in the same way. It also means that the resulting explicit double copy expression depends on the chosen color and polarization basis. Using the same relation at every multiplicity turns the whole gauge-theory expansion into the gravitational one,
	\begin{equation}
		\mathcal A_{(n)}
		\longrightarrow
		\mathcal M_{(n)}\, .
		\label{eq:coherent_order_by_order_dc}
	\end{equation}
	There is a small but important normalization subtlety with this procedure. If one double copies only the vacuum amplitudes in the expansion, the leading Gaussian normalization in Eq.~\eqref{eq:coherent_generic_expansion} still contains $\alpha,\alpha'$, not $\beta,\beta'$. One can in principle argue that this is a property of the asymptotic coherent state so it can be set by hand and does not satisfy a double copy. However, one can define a non-perturbative operator statement that fixes this automatically:
	\begin{equation}
		T_{\rm YM}\longrightarrow T_{\rm grav}\, ,
		\qquad
		\widehat a_s^{a\dagger}(k)
		\longrightarrow\sum_PC_P^s\widehat a_P^\dagger(k)\, ,
		\qquad
		\widehat a_s^a(k)
		\longrightarrow\sum_P\bar C_P^s\widehat a_P(k)\, .
		\label{eq:coherent_operator_dc}
	\end{equation}
	This replacement maps the full displacement operators, and hence their normalization factors, from the gauge coherent states to the gravity coherent states. Consequently,
	\begin{equation}
		\langle\alpha',\mathrm{out}|T_{\rm YM}|\alpha,\mathrm{in}\rangle
		\longrightarrow
		\langle\beta',\mathrm{out}|T_{\rm grav}|\beta,\mathrm{in}\rangle\, .
		\label{eq:coherent_amplitude_dc}
	\end{equation}
	In the background-field language, after choosing a vacuum double copy representation for the spin-stripped amplitudes, the same statement is
	\begin{equation}
		\mathcal A^{a_1\cdots a_n}_{\mu_1\cdots\mu_n}
		\longrightarrow
		\mathcal M_{\mu_1\nu_1\cdots\mu_n\nu_n}\, ,
		\qquad
		A_\mu^a(x)
		\longrightarrow
		\mathcal H_{\mu\nu}(x)\, ,
		\label{eq:coherent_background_dc}
	\end{equation}
	with $\mathcal H_{\mu\nu}$ given by Eqs.~\eqref{eq:coherent_fat_graviton} and \eqref{eq:coherent_profile_dc}. The first arrow is schematic: it is the chosen amplitudes double copy for spin-stripped vacuum amplitudes, with the same polarization coefficients $C_P^s$ used at all multiplicities. The second arrow is the induced classical map between backgrounds.
	
	A key consequence is that the coherent profile is not squared. The color generator is removed and the polarization vector is doubled, while $\alpha_s^a(k)$ is carried linearly into $\beta_P(k)$. A rule such as $\beta\sim\alpha^a\alpha^a$ would require extra dimensionful data, often supplied in other classical double copies by a spectator or biadjoint scalar. That structure is not generated by the coherent-state derivation. This is also why the construction does not immediately include Coulomb, Schwarzschild, or other sourced fields: the coherent profiles here describe asymptotically free radiation.
	
	\paragraph{Example: plane waves}
	Plane waves provide the simplest example and a useful comparison with the plane-wave scattering construction above. In four dimensions a Cartan-valued gauge plane wave in Rosen gauge can be written as
	\begin{equation}
		A_\mu(x)=T\,\delta_\mu^i a_i(u)\, ,
		\qquad i=1,2\, ,
		\label{eq:coherent_plane_wave_gauge}
	\end{equation}
	where $T$ is a fixed Cartan generator and $a_i(u)$ are real functions of lightfront time. Its Fourier representation has support at zero transverse momentum; in the lightfront helicity basis
	\begin{equation}
		\epsilon_\mu^s(k)
		=\frac{1}{\sqrt2}\left(\frac{k_1+is k_2}{k_u},0,1,is\right),
		\qquad
		a_s(u)=\frac{a_1(u)-is a_2(u)}{\sqrt2}\, ,
		\label{eq:coherent_plane_wave_helicity}
	\end{equation}
	the polarization vectors become constant on the support of the profile. A symmetric polarization double copy produces only gravitons and gives
	\begin{equation}
		h_{\mu\nu}(x)=\sum_{s=\pm}\epsilon_{\mu\nu}^s a_s(u)\, ,
		\qquad
		ds^2=-2du\,dv+\gamma_{ij}(u)dx^idx^j\, ,
		\label{eq:coherent_plane_wave_gravity}
	\end{equation}
	with
	\begin{equation}
		\gamma_{ij}
		=\delta_{ij}-\sqrt2
		\begin{pmatrix}
			a_1(u)&a_2(u)\\
			a_2(u)&-a_1(u)
		\end{pmatrix}\, .
		\label{eq:coherent_plane_wave_metric}
	\end{equation}
	The factor of $\sqrt2$ follows from the normalization of the doubled polarization tensors. This is a plane-wave metric, but for arbitrary $a_i(u)$ it is guaranteed only to solve the linearized vacuum Einstein equation. This differs from the exact sandwich-wave double copy in Sec.~\ref{sec:plane_wave_scattering_dc}: that construction is formulated in Brinkmann gauge, relates three-point amplitudes on exact gauge and gravitational plane waves, and includes memory through the replacement rule \eqref{eq:pw_replacement_map}. The coherent-state map instead doubles an on-shell Rosen-gauge radiation profile. The Rosen and Brinkmann gauge potentials are related by a large gauge transformation, and the corresponding relation between gauge transformations and diffeomorphisms remains part of the open problem.
	
	This illustrates the general status of the construction. Coherent states give an amplitudes-based double copy for scattering on any background admitting an on-shell coherent-state description, and they induce a classical map between the associated backgrounds. The output background is automatically a solution at linear order; it becomes an exact nonlinear solution only in special sectors. This is complementary to the Kerr--Schild maps reviewed in Sec.~\ref{sec:kerr-schild} and to the background-and-perturbation viewpoint reviewed in Sec.~\ref{sec:backgrounds_perturbations}. In the self-dual sector, the coherent-state prescription overlaps with the operator Kerr--Schild construction and can produce exact vacuum solutions; we return to that case in Subsection~\ref{subsec:coherent_sd_backgrounds}.
	
	\subsubsection{Hawking radiation from scattering amplitudes}\label{sec:hawking_radiation_dc}
	
	In this subsection we restore the gravitational and gauge couplings to display the physical temperature and charge dependence.
	
	The examples above use backgrounds whose scattering interpretation is controlled either by flat asymptotic regions or by coherent-state data. Hawking radiation provides a different kind of time-dependent problem: particle creation associated with the formation of a black hole. A simple model of this process is an ingoing Vaidya spacetime in which a spherical shell of null matter falls in and the exterior geometry changes from flat space to Schwarzschild. In the sharp-shell approximation used in \cite{Aoude:2025jvt}, the mass function is proportional to $\Theta(v)$, with $v=t+r$, so the geometry can be written in Kerr--Schild form. Applying the Kerr--Schild prescription reviewed in Sec.~\ref{sec:kerr-schild} gives:
	\begin{equation}
		g_{\mu\nu}=\eta_{\mu\nu}+\frac{2G M\,\Theta(v)}{r}k_\mu k_\nu\, ,
		\qquad
		A_\mu=\frac{g Q_B\,\Theta(v)}{4\pi r}k_\mu\, ,
		\qquad k_\mu dx^\mu=dv\, .
		\label{eq:vaidya_hawking_single_copy}
	\end{equation}
	The single-copy background is therefore a null shell of background charge $Q_B$ together with the Coulomb field produced after the shell has passed.
	
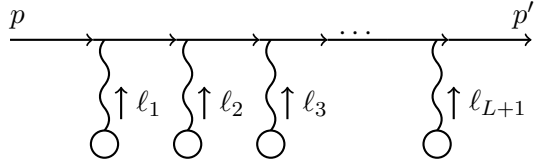
\begin{figure}[!t]
		\centering
		\begin{tikzpicture}[x=1cm,y=1cm,every node/.style={font=\small}]
			\draw[thick,->] (-3.45,0) -- (-2.35,0);
			\draw[thick,->] (-2.35,0) -- (-1.25,0);
			\draw[thick,->] (-1.25,0) -- (-0.15,0);
			\draw[thick,->] (-0.15,0) -- (0.75,0);
			\draw[thick] (0.75,0) -- (1.55,0);
			\node at (1.15,0.08) {$\cdots$};
			\draw[thick,->] (1.55,0) -- (2.35,0);
			\draw[thick,->] (2.35,0) -- (3.45,0);
			\node[above] at (-3.35,0) {$p$};
			\node[above] at (3.35,0) {$p'$};
			\foreach \x/\lab in {-2.2/{\ell_1},-1.1/{\ell_2},0/{\ell_3},2.2/{\ell_{L+1}}}
			{
				\draw[thick]
				(\x,0)
				.. controls (\x-0.08,-0.10) and (\x-0.08,-0.20) .. (\x,-0.30)
				.. controls (\x+0.08,-0.40) and (\x+0.08,-0.50) .. (\x,-0.60)
				.. controls (\x-0.08,-0.70) and (\x-0.08,-0.80) .. (\x,-0.90)
				.. controls (\x+0.08,-1.00) and (\x+0.08,-1.10) .. (\x,-1.20);
				\draw[thick,->] (\x+0.22,-1.05) -- (\x+0.22,-0.62);
				\node[right] at (\x+0.28,-0.84) {$\lab$};
				\draw[thick,fill=white] (\x,-1.38) circle (0.18);
			}
		\end{tikzpicture}
		\caption{Eikonal comb diagrams entering the Hawking radiation amplitude. The scalar probe propagates from momentum $p$ to $p'$, while each circular background insertion transfers momentum $\ell_i$ through the same three-point scalar--background coupling. In the leading geometric-optics limit these insertions factorize in impact parameter space and exponentiate into the eikonal phase \cite{aoude2024amplitudeshawkingradiation}.}
		\label{fig:hawking_eikonal_ladder}
	\end{figure}

	We proceed to analyze the scattering around these backgrounds. Consider a massless scalar of charge $Q$ and gauge coupling $g$ propagating in the external gauge field $A_\mu$ of Eq.~\eqref{eq:vaidya_hawking_single_copy} in the geometric-optics limit. Following the KMOC approach \cite{Kosower:2018adc,Kosower:2022yvp} and its extension to curved backgrounds \cite{Adamo:2022rmp}, let $|\psi\rangle$ be the incoming one-particle wavepacket, with position-space profile $\varphi(v)$ supported at $v<0$, and let $|p\rangle$ be an outgoing on-shell one-particle state of energy $E=p^0$. Writing $b^\mu(v)=(v,\mathbf{0})$, the matrix element $\langle p|S|\psi\rangle$ is the outgoing momentum-space wavefunction, obtained by evolving $|\psi\rangle$ with the scalar $S$-matrix in this external background and projecting onto $|p\rangle$. Keeping the leading geometric-optics/eikonal term and fixing the eikonal trajectory at $b^\mu=b_0^\mu$,
	\begin{align}
		\langle p'|\hat S-1|\psi\rangle_{\mathrm{LO}}&\propto
		\int d^4q\,\delta(2p'\cdot q)\,e^{-iq\cdot b_0}\,
		i\mathcal{A}_3(p\to p')\, ,
	\end{align}
	where $q=p'-p\ll p$ is the momentum transfer. The tree-level background insertion is
	\begin{equation}
		\mathcal{A}_3(p\to p')=-gQ \widetilde A^\mu(q)(2p_\mu+q_\mu)\, .
	\end{equation}
	Here $\widetilde A^\mu(q)$ is the Fourier transform of Eq.~\eqref{eq:vaidya_hawking_single_copy}. $\mathcal{A}_3(p\to p')$ can be thought of as the scattering amplitude between two scalars and a gauge field, or equivalently, as a one-to-one amplitude for the massless probe in the gauge field background in Eq.~\eqref{eq:vaidya_hawking_single_copy}. The eikonal comb diagrams, shown schematically in Fig.~\ref{fig:hawking_eikonal_ladder}, are then built from repeated insertions of the same three-point coupling to the background.
	Their resummation in impact parameter space gives
	\begin{align}
		A(E)\equiv \langle p|S|\psi\rangle
		&\sim \int_{-\infty}^0 dv\,\varphi(v)e^{ip\cdot b(v)}
		e^{i\chi(v)},\nonumber\\
		\chi(v)
		&=2\alpha\log\left(\frac{-v}{\mu}\right),
		\qquad
		\alpha=\frac{g^2Q Q_B}{4\pi}\, ,
		\label{eq:em_hawking_eikonal}
	\end{align}
	up to a $v$-independent infrared phase, with $\mu$ an IR scale. Thus $\chi(v)$ is the eikonal phase accumulated by the charged probe in the Coulomb region produced by the shell.
	
	The gravitational calculation follows in an analogous manner using the Vaidya metric in Eq.~\eqref{eq:vaidya_hawking_single_copy}. The only difference is the coefficient of the logarithm. With the sign choice in which the probe is attracted to the background charge, the double copy replaces the factor $-\alpha$ by $2GM E$, which is encoded by $g\rightarrow \sqrt{8\pi G}$, $Q_B\rightarrow M$, and $-Q\rightarrow E$. The gravitational phase is then $\chi_{\rm GR}(v)=-4GM E\log(-v/\mu)$.
	In the gravitational case, it is known that the logarithmic phase is the universal near-horizon behavior of the relevant modes in a black-hole formation geometry and encodes particle creation. 
	
	Take $\varphi(v)=(2\pi/E_0)e^{-iE_0v}$ for the incoming spherical mode. $A(E,E_0)$ and $B(E,E_0)$ are the Bogoliubov coefficients which relate a future outgoing mode of energy $E$ to the positive- and negative-frequency parts of a past mode of energy $E_0$. Following \cite{aoude2024amplitudeshawkingradiation}, they can be computed as generalized amplitudes, $A(k,p)=\langle\Omega|S^\dagger a(k)S a^\dagger(p)|\Omega\rangle$ and $B(k,p)=\langle\Omega|a(p)S^\dagger a(k)S|\Omega\rangle$.
	The Bogoliubov coefficient $A(E,E_0)$ is the Hawking amplitude obtained from the resummed eikonal phase in Eq.~\eqref{eq:em_hawking_eikonal} using the spherical incoming state, and $B(E,E_0)$ is obtained by analytic continuation as $B(E,E_0)=A(E,-E_0)$. Suppressing wavepacket and normalization factors, the differential number spectrum is then
	\begin{align}
		\frac{dn}{dE}
		&\propto
		\left(\frac{E_0}{E}\right)^2 |B(E,E_0)|^2 \nonumber\\
		&\propto e^{-4\pi GME}
		\frac{4\pi GM E}{\sinh(4\pi GM E)} \nonumber\\
		&\propto \frac{1}{e^{8\pi GM E}-1}\, .
	\end{align}
	The energy dependence in the gravitational amplitude is what turns this into a Planck distribution with Hawking temperature $T_{\rm H}=1/(8\pi G M)$ for Schwarzschild, agreeing with the original computation in \cite{Hawking:1975vcx} up to greybody factors. From the double copy perspective, this naturally arises from the squaring of kinematic numerators that lead to a quadratic momentum dependence of the amplitude instead of a linear one.
	
	By contrast, the abelian single-copy spectrum does not seem to be thermal in energy $E$. The interpretation proposed in \cite{Aoude:2025jvt} is instead chemical-potential-like: the exponential factor depends on the charge product $\alpha$, so it acts as a fugacity for the probe charge in the background rather than as an energy-dependent Boltzmann weight.
	
	Another way to see the same point is through a worldline computation \cite{Ilderton:2025aql}. In this case, the double copy sends the charge coefficient in the single-copy logarithmic phase to the energy-dependent coefficient in the near-horizon Vaidya mode as before. At the level of wavefunctions, this is accompanied by $v\rightarrow v-v_0=v+4GM$, where $v_0=-4GM$ is the last escaping ray in the Vaidya geometry and the corresponding single-copy ray is $v=0$. This shift changes the phase but not the modulus of the pair-creation amplitude. It is also shown that the mid-time pair-creation amplitude double copies to the Hawking amplitude, while the abelian gauge-theory spectrum remains non-thermal.
	
	Retaining the non-abelian nature of the single copy, rather than reducing immediately to the abelian version, reveals the thermal structure of the gauge-theory root \cite{Carrasco:2025bgu}. The background has a fixed color orientation, so along an eikonal trajectory the Wilson line is generated by a single operator $c^aT^a$. For a probe color state satisfying $c^aT^a|\lambda\rangle=\lambda|\lambda\rangle$, path ordering becomes trivial and the logarithmic phase is labeled by the eigenvalue $\lambda$. The non-abelian logarithm prefactor becomes $gQ_B \lambda$ and the emission probability reads
	\begin{equation}
		|A(E,\lambda)|^2
		\propto \frac{1}{E^2}
		\frac{\pi gQ_B\lambda}{\sinh(\pi gQ_B\lambda)}\, .
		\label{eq:color_thermal_hawking}
	\end{equation}
	The prefactor is the usual soft-radiation energy dependence, while the thermal factor is now a function of the color eigenvalue. To form an inclusive color spectrum one must also include the density of available eigenvalues; in the large-$N_c$ limit, with $N_c$ the number of colors, this can be modeled by random matrix theory, which gives a Wigner semicircle law \cite{Anninos:2020ccj,Wigner1993,Carrasco:2025bgu}. Thus, the spectrum of radiated color charge is given by
	\begin{equation}
		\frac{dn}{d\lambda}
		\propto \frac{\pi gQ_B\lambda}{\sinh(\pi gQ_B\lambda)}\,
		\sqrt{R^2-\lambda^2}\, .
	\end{equation}
	For $gQ_B\gg R$, and in particular for $gQ_B\lambda\gg1$ over the relevant support, the hyperbolic-sine factor dominates over the semicircle density and the spectrum becomes Planck-like in the color variable, up to the cutoff $|\lambda|\le R$. This is a regime of large eikonal phase, equivalently a semiclassical regime, not a breakdown of perturbation theory. The central point is independent of the choice of density of states: the thermal variable on the gauge-theory side is color, not energy. In this sense the gravitational Planck factor is the double copy image of color thermality, because the color factor in the gauge-theory root is replaced by a second kinematic numerator on the gravity side.
	
	A complementary nonperturbative perspective comes from a worldline instanton treatment of the same color thermal structure \cite{Carrasco:2026ijt}. The color thermal factor is a winding contribution to the worldline action, with effective temperature set by the finite source charge, $T_c\propto (gQ_B)^{-1}$. For a macroscopic source in a color representation $\mathcal R$, the charge magnitude is measured by the quadratic Casimir, $C_2(\mathcal R)\simeq Q_B^2$. If the emitted pair carries color eigenvalue $\lambda$, color conservation changes the effective source charge to $Q_B-|\lambda|$, so the relevant Casimir change is $\Delta C_2\simeq Q_B^2-(Q_B-|\lambda|)^2=2|\lambda|(Q_B-|\lambda|/2)$. The pair-production probability $P_{\rm pair}$ is the modulus squared of the pair-creation amplitude, and its exponent is controlled by this Casimir change, $P_{\rm pair}\propto |B(E,E_0)/A(E,E_0)|^2\propto e^{-2\pi g\Delta C_2}=e^{-4\pi g|\lambda|(Q_B-|\lambda|/2)}$, valid for $|\lambda|\ll Q_B$. The double copy sends the source charge to the black-hole mass and the emitted color eigenvalue to the emitted energy, $Q_B\rightarrow M$ and $\lambda\rightarrow E$. Thus, the color Casimir change double copies to the entropy change as $2\pi g\Delta C_2 \rightarrow -\Delta S_{\rm BH}$, where $\Delta S_{\rm BH}=S_{\rm BH}(M-E)-S_{\rm BH}(M)=-8\pi G E(M-E/2)$ and $S_{\rm BH}=A_{\rm H}/(4G)=4\pi G M^2$. Thus the full exponent $-2\pi g\Delta C_2$ maps to $\Delta S_{\rm BH}$, or equivalently to the corresponding change in horizon area, giving the Parikh--Wilczek tunneling factor $\Gamma\propto e^{\Delta S_{\rm BH}}$ \cite{Parikh:1999mf}. This provides evidence that the double copy can capture not only the leading thermal factor but also the first backreaction correction.
	
	These examples suggest that radiation observables give a useful testing ground for double copy structures beyond stationary classical solutions. In particular, they isolate which parts of the Hawking spectrum are controlled by long-distance logarithmic phases, color spectra, and source depletion, and which parts are expected to be sensitive to dynamical background data.
	\section{Classical and Field-Level Double Copy}\label{sec:classical_field_level}
	Classical double copy constructions relate gravitational solutions directly to gauge and biadjoint-scalar solutions, rather than through scattering amplitudes or correlators. There is not, however, a single universal field-level prescription, although different versions agree in most cases. The Kerr--Schild construction acts on a special metric decomposition, the Weyl and Cotton constructions factorize gauge-invariant curvatures, and the twistor-space construction relates representatives whose Penrose transforms solve the corresponding field equations. The convolutional double copy instead works off shell at linear order and organizes the map through the BRST complexes of the two gauge theories. We review these prescriptions below.

	\subsection{\texorpdfstring{Kerr--Schild Double Copy}{Kerr-Schild Double Copy}} \label{sec:kerr-schild}
	We begin with the Kerr--Schild double copy, the first classical prescription formulated directly for exact gravitational solutions\footnote{An earlier realization of this relationship in a different context was found in \cite{Didenko:2008va,Didenko:2009td}.} \cite{Monteiro:2014cda}. Its scope is restricted by the Kerr--Schild ansatz, but within that class it covers many physically important geometries, including black holes and gravitational waves. The construction has been extended to a broad range of backgrounds, matter sources, and spacetime dimensions \cite{Monteiro:2014cda,Luna:2015paa,Luna:2016due,Luna:2016hge,Ilderton:2018ldc,Gurses:2018ckx,Bah:2019sda,CarrilloGonzalez:2019pxo,BahjatAbbas:2020mrf,Gumus:2020has,Alkac:2021kut,Alkac:2022wkv,Dempsey:2023gza,Easson:2023ldu,Ortaggio:2024abc,Ridgway:2015fdl,Ceresole:2023wxg,Easson:2020esh,Chawla:2023bsu,He:2023iew}. Self-dual Kerr--Schild metrics \cite{Monteiro:2014cda,Chacon:2020fmr,Chacon:2025hh,Alkac:2021bav,Andrzejewski:2019hub} will be discussed separately in Subsection~\ref{sec:self_dual}. Here, we mainly review the construction in \cite{Bahjat-Abbas:2017htu,Carrillo-Gonzalez:2017iyj}.
	
	A Kerr--Schild spacetime is an exact metric that can be written as
	\begin{equation}
		g_{\mu\nu}=\bar{g}_{\mu\nu}+\phi k_\mu k_\nu,
		\label{eq:KSmetric}
	\end{equation}
	where $\bar{g}_{\mu\nu}$ is the base or background metric, $\phi$ is a scalar profile, and the Kerr--Schild vector $k^\mu$ is null and geodesic with respect to both metrics,
	\begin{equation}
		g^{\mu\nu}k_\mu k_\nu=\bar{g}^{\mu\nu}k_\mu k_\nu=0,\quad\quad k^\mu\nabla_\mu k^\nu=k^\mu \bar{\nabla}_\mu k^\nu=0 \  ,
		\label{eq:ksvec}
	\end{equation}
	where $\nabla^\mu$ is the covariant derivative of $g_{\mu\nu}$ and $\bar{\nabla}^\mu$ is the covariant derivative of $\bar{g}_{\mu\nu}$. The null condition makes the inverse metric terminate at first order in $\phi$, while the geodesic condition removes the remaining nonlinear terms from the mixed-index Ricci tensor. This is the mechanism behind the exact linearization: although $h_{\mu\nu}=\phi k_\mu k_\nu$ need not be a small perturbation, the relevant Einstein equations are linear in the Kerr--Schild profile. The associated spacetimes are algebraically special and are generically of Petrov type~II or more special \cite{Stephani:2003tm}; see~\ref{app:spinorial_gr_petrov}. Explicitly,
	\begin{equation} \label{R curved}
		R^\mu{}_\nu = \bar R^\mu{}_\nu -\phi k^\mu k^\lambda \bar R_{\lambda \nu} +\tfrac{1}{2} \left[ \bar \nabla^\lambda \bar \nabla^\mu (\phi k_\lambda k_\nu ) +   \bar \nabla^\lambda \bar \nabla_\nu (\phi k^\mu k_\lambda ) -  \bar \nabla^2 (\phi k^\mu k_\nu )  \right] \ ,
	\end{equation}
	where $ \bar R^\mu{}_\nu$ is the Ricci tensor of the background metric. 
	
	The Kerr--Schild double copy states that, given an exact gravitational solution in Kerr--Schild form, the single copy is a gauge field solving the linearized Yang--Mills equations, while the zeroth copy is a scalar field solving the linearized biadjoint scalar equations: 
	
	\begin{mybox}[Kerr--Schild Double Copy]
		\begin{center}
			\renewcommand{\arraystretch}{1.8}
			\begin{tabularx}{\textwidth}{l X}
				\textbf{Gravity} &   \vspace{-1cm} 
				\begin{equation}
					h_{\mu\nu}= \phi \, k_\mu k_\nu 
					\label{eq:KS_gravity}
				\end{equation} \vspace{-1cm}  \\  
				\textbf{Yang--Mills} &   \vspace{-1cm} 
				\begin{equation}
					A_\mu^a = c^a k_\mu \phi 
					\label{eq:KS_gauge}
				\end{equation}  \vspace{-1cm}  \\ 
				\textbf{Biadjoint Scalar} &   \vspace{-1cm} 
				\begin{equation}
					\phi^{a\, b'} = c^a c^{b'}\phi 
					\label{eq:KS_bas}
				\end{equation}  \vspace{-0.5cm} 
			\end{tabularx}
		\end{center}
	\end{mybox}
	where $c^a$ are constant color charges in the adjoint. In other words, the color-kinematics replacements are given by
	\begin{equation}
		c^a \leftrightarrow k^\mu \ .
	\end{equation}
	That is, one copy of the null vector is replaced by a constant color vector to obtain the gauge field, and the second copy is replaced to obtain the biadjoint scalar. Because the color direction is constant, the commutator term in the Yang--Mills field strength vanishes and the single copy is effectively Abelian. The gauge and scalar equations are therefore linear even though the gravitational metric is exact. This construction resembles the BCJ double copy reviewed in Section~\ref{sec:bcj}.
	
	On a maximally symmetric background, this relation follows directly by rewriting the trace-reversed Einstein equations for the Kerr--Schild metric in $d$ dimensions \cite{Carrillo-Gonzalez:2017iyj}:
	\begin{equation}
		-2 \left( T^\mu_{\ \nu} - \delta^\mu_{\ \nu} \frac{T}{d - 2} \right)=2(\bar R^\mu{}_\nu - R^\mu{}_\nu)= \left( \bar \nabla_\lambda F^{\lambda\mu} + \tfrac{(d-2)}{d(d-1)} \bar R A^\mu \right) k_\nu  + X^\mu{}_\nu \ , \label{eq:EinstEqKS}
	\end{equation}
	where, $F^{\lambda\mu}$ is the field strength for the Abelian gauge field $A^\mu=k^\mu \phi$. The Lorentz indices are raised using the base metric unless stated otherwise and the additional tensor $X^\mu{}_\nu$ is given by
	\begin{align}
		X^\mu{}_\nu\equiv& - \bar\nabla_\nu\left[A^\mu\left(\bar\nabla_\lambda k^\lambda+\frac{k^\lambda \bar\nabla_\lambda \phi}{\phi}\right)\right]  \nonumber \\
		+& \, F^{\rho\mu}\bar{\nabla}_\rho k_\nu-\bar{\nabla}_\rho\left(A^\rho \bar{\nabla}^\mu k_\nu-A^\mu \bar{\nabla}_\rho k_\nu \right) \ .
	\end{align} 
	To extract the linearized Yang--Mills, that is, Maxwell's equations, we contract Eq.~\eqref{eq:EinstEqKS} with a Killing vector $V^\nu$ of both the background and full metric. The result depends on whether the Killing vector is timelike or null. For stationary solutions (timelike $V^\nu$) one finds 
	\begin{equation}
		{\bar\nabla}_\lambda F^{\lambda\mu} =J^\mu \ , \label{Aeom}
	\end{equation}
	while for wave solutions (null $V^\nu$)
	\begin{equation}
		{\bar\nabla}_\lambda F^{\lambda\mu} + \tfrac{(d-2)}{d(d-1)} \bar R A^\mu  =J^\mu \ , \label{eq:KS_gauge_eom}
	\end{equation}
	where in both cases
	\begin{equation}
		J^\mu\equiv-\tfrac{2 V^\nu}{V^\rho k_\rho}\left({T^\mu}_\nu-\delta^\mu_\nu \tfrac{T}{d-2}\right). \label{eq:KS_current}
	\end{equation}
	The contraction with $V^\nu$ projects the gravitational source onto the current seen by the single copy. 
	
	The distinction between stationary and wave solutions is important: a timelike Killing vector removes the curvature-dependent term in Eq.~\eqref{eq:EinstEqKS}, whereas a null Killing vector leaves the effective mass term displayed in Eq.~\eqref{eq:KS_gauge_eom}. Thus the curved-background single copy is not obtained by merely replacing partial derivatives with covariant derivatives.
	Similarly, the zeroth copy equation is obtained by further contracting Eq. \eqref{eq:EinstEqKS} with the same Killing vector which for stationary solutions it gives
	\begin{equation}
		\bar \nabla^2\phi-\tfrac{2(d-3)}{d(d-1)} \bar R \phi=j \ , \label{eq:KSphi_stat}
	\end{equation}
	and for waves
	\begin{equation}
		\bar \nabla^2\phi-\tfrac{(d-4)}{d(d-1)} \bar R \phi=j \ , \label{eq:KSphi_waves}
	\end{equation}
	where the source is now
	\begin{equation}
		j=\frac{V_\nu J^\nu}{V^\rho k_\rho} \ . \label{eq:KS_phi_source}
	\end{equation}
	The curvature-dependent terms may be interpreted as non-minimal couplings to the background. For the vector, such a term is not gauge invariant by itself, which limits a direct interpretation as an ordinary mass; massive extensions of the Kerr--Schild double copy are discussed in \cite{Garcia-Compean:2024uie,Garcia-Compean:2024zze,Gonzalez:2021ztm}. For the scalar, Eq.~\eqref{eq:KSphi_stat} becomes the conformally coupled scalar equation in four dimensions, while Eq.~\eqref{eq:KSphi_waves} becomes the massless scalar equation. In fact, the value of the scalar mass is not arbitrary. The value of the curvature-dependent mass term is precisely the $s=0$ continuation of the curvature-dependent mass term in the AdS Fronsdal equations \cite{Fronsdal:1978rb}. This realization allows the fields $\phi k_{\mu_1}\cdots k_{\mu_s}$ to solve the free Fronsdal equations for every integer spin, thereby organizing the Kerr--Schild multicopies into a higher-spin tower \cite{Didenko:2022qxq}.
	
	Another relevant point is that the Kerr--Schild metric decomposition has a rescaling ambiguity:
	\begin{equation}
		k_\mu \to f k_\mu, \qquad \qquad \quad \phi \to \phi / f^2  \ , \label{eq:KS_scaling}
	\end{equation}
	which leads to different single and zeroth copies depending on the choice of $f$. For flat space, vacuum solutions, it is straightforward to fix this scaling by requiring that $\phi$ satisfies $\nabla^2\phi=0$. In other cases, this can be more involved and require intuition of what is expected from the double copy. For example, it has been proposed that one can require that the Lie derivative of the curvature tensors and scalar vanish through the double copy prescription \cite{Beetar:2024ptv}. A related subtlety is that the residual symmetries preserving the Kerr--Schild ansatz need not be mapped directly by the double copy, as shown for the Schwarzschild solution in \cite{Holton:2026eha}.
	
	\subsubsection{\texorpdfstring{Example: Kundt--AdS waves}{Example: Kundt--AdS waves}}\label{subsec:ks_kundt_ads}
	As a radiative example, consider the four-dimensional Kundt waves on an AdS background, which belong to the non-expanding Petrov type~N sector \cite{Carrillo-Gonzalez:2017iyj,Ortaggio:2024abc}. In coordinates $(u,v,x,y)$, a convenient background metric is
	\be
	\dd\bar s^2=\frac{1}{P^2}\left(-4x^2\,\dd u\,\dd v-v^2\,\dd u^2+\dd x^2+\dd y^2\right),
	\qquad P=1+\frac{\Lambda}{12}(x^2+y^2)\ ,
	\label{eq:ks_kundt_ads_background}
	\ee
	where $\Lambda<0$ for AdS. The Kerr--Schild vector and scalar profile can be chosen as
	\be
	k=\frac{x}{P}\,\dd u\ ,\qquad\phi=\frac{P}{x}H(u,x,y)\ ,
	\label{eq:ks_kundt_ads_data}
	\ee
	so that the full metric is
	\be
	\dd s^2=\dd\bar s^2+\frac{x}{P}H(u,x,y)\,\dd u^2\ .
	\label{eq:ks_kundt_ads_metric}
	\ee
	This is an exact vacuum solution provided the wave profile satisfies
	\be
	\left(\partial_x^2+\partial_y^2+\frac{2\Lambda}{3P^2}\right)H(u,x,y)=0\ .
	\label{eq:ks_kundt_ads_profile}
	\ee
	The Kerr--Schild replacement then gives an especially simple null gauge potential and zeroth copy,
	\be
	A=\phi k=H(u,x,y)\,\dd u\ ,\qquad \phi^{a\,b'}=c^a c^{b'}\phi\ .
	\label{eq:ks_kundt_ads_copies}
	\ee
	Using Eq.~\eqref{eq:ks_kundt_ads_profile}, the color-stripped fields obey
	\be
	\bar\nabla_\mu F^{\mu\nu}+\frac{\bar R}{6}A^\nu=0\ ,\qquad
	\bar\nabla^2\phi=0\ ,
	\label{eq:ks_kundt_ads_copy_eoms}
	\ee
	with $\bar R=4\Lambda$. This realizes explicitly the wave equations in Eqs.~\eqref{eq:KS_gauge_eom} and \eqref{eq:KSphi_waves}: unlike the stationary examples, the vector copy retains a curvature-dependent term, whereas the scalar copy is minimally coupled. The gravitational, gauge, and scalar profiles share the same null wavefronts.
	
	\subsubsection{\texorpdfstring{Example: Kerr--(A)dS black hole}{Example: Kerr-(A)dS black hole}}
	As a simple yet non-trivial example, let us examine a rotating black hole in (anti-)de Sitter space.  We will obtain the single and zeroth copies of the four-dimensional Kerr--(A)\,dS geometry by expressing the full metric in Kerr--Schild form.  It is convenient to write the background (A)\,dS metric in oblate spheroidal coordinates~\cite{Gibbons:2004uw}:
	\begin{align}
		\bar g_{\mu\nu}\, \mathrm d x^\mu \mathrm d x^\nu
		\;=\;
		-\frac{\Delta}{\Omega}\,f\,\mathrm d t^{2}
		+\frac{\rho^{2}\,\mathrm d r^{2}}{(r^{2}+a^{2})f}
		+\frac{\rho^{2}\,\mathrm d\theta^{2}}{\Delta}
		+\frac{(r^{2}+a^{2})\sin^{2}\theta}{\Omega}\,\mathrm d\varphi^{2}. \label{eq:KerrAds_mzero}
	\end{align}
	Here $\Lambda$ is the (A)dS cosmological constant, $a$ is the Kerr rotation parameter, $M$ denotes the mass parameter, and we have introduced the shorthand notation
	\begin{equation}
		f      \equiv 1-\frac{\Lambda r^{2}}{3},
		\qquad
		\rho^{2}\equiv r^{2}+a^{2}\cos^{2}\theta,
		\qquad
		\Delta \equiv 1+\frac{\Lambda a^{2}}{3}\cos^{2}\theta,
		\qquad
		\Omega \equiv 1+\frac{\Lambda a^{2}}{3}.
	\end{equation}
	The associated Kerr--Schild null vector and scalar profile read
	\begin{align}
		k_\mu\,\mathrm d x^\mu
		&= \frac{\Delta}{\Omega}\,\mathrm d t
		+\frac{\rho^{2}}{(r^{2}+a^{2})f}\,\mathrm d r
		-\frac{a\sin^{2}\theta}{\Omega}\,\mathrm d\varphi,
		&
		\qquad
		\phi &= \frac{2GM\,r}{\rho^{2}}. \label{eq:kphi_ks_kerrads}
	\end{align}
	In the non-rotating limit $a\to0$, these expressions reduce to the (A)\,dS-Schwarzschild solution. The Kerr--(A)\,dS geometry is sourced by a negative surface-mass density distributed on a disk of radius $a$ located at $r=0$.  
	The disk rotates rigidly about the $z$-axis with superluminal tangential velocity and is stabilized by a compensating radial pressure.  
	Its stress--energy tensor takes the Israel form generalized to (A)\,dS~\cite{Israel:1970kp}:
	\begin{align}
		T^{\mu}{}_{\nu}
		\;=\;
		-j\,\cos^{2}\theta\,
		\bigl(\xi^{\mu}\xi_{\nu}+u^{\mu}u_{\nu}\bigr),
		\qquad\qquad
		j
		\;=\;
		\frac{M}{4\pi a^{2}}\,
		\sec^{3}\theta\,
		\delta(r),
		\label{eq:jkerr}
	\end{align}
	where the orthonormal vectors
	\begin{equation}
		\xi_{\mu}
		= \frac{a\cos\theta}{\sqrt{\Delta}}\,(0,0,1,0),
		\qquad\qquad
		u_{\mu}
		= \frac{\sqrt{\Delta}\,\tan\theta}{\Omega}\,\!(1,0,0,-a)
	\end{equation}
	characterize, respectively, the azimuthal and temporal directions on the disk.  
	For $\Lambda=0$ these expressions reduce to the familiar flat-space Kerr source.
	
	Employing the Kerr--Schild double copy prescription, the corresponding Abelian gauge field is
	\begin{equation}
		A_{\mu}\;=\;k_{\mu}\,\phi,
	\end{equation}
	with $k_{\mu}$ and $\phi$ as in Eq.~\eqref{eq:kphi_ks_kerrads}. It satisfies the Maxwell equations in an (A)dS background with source
	\begin{equation}
		J^{\mu}=j\,\zeta^{\mu},
		\qquad\qquad
		\zeta_{\mu}=(2,0,0,2/a).
	\end{equation}
	Thus the single copy describes the electromagnetic field generated by a charged, rigidly rotating disk in (A)dS.  The field possesses both electric and magnetic components, the latter being proportional to the angular momentum of the source—mirroring the way angular momentum appears in the gravitational solution.\\
	
	\subsubsection{Double Copy of Backgrounds and Perturbations}\label{sec:backgrounds_perturbations}
	The construction reviewed above keeps the background geometry fixed and double copy fields defined on it. Schematically, the gravitational metric is split as
	\begin{equation}
		g_{\mu\nu}=\bar{g}_{\mu\nu}+h_{\mu\nu} 
		\label{eq:curvedbckgnd}
	\end{equation}
	where $\bar{g}_{\mu\nu}$ is the metric on which the gauge and BAS solutions are defined. This split is additional data: it may be selected by asymptotic boundary conditions or, for a family of black holes, by an $M\rightarrow0$ limit, but it is not determined uniquely by the full metric. A different construction copies both the background and the field placed on it. The gauge and gravity fields then take the form
	\be
	A^\mu=\bar{A}^\mu+a^\mu\ , \quad g_{\mu\nu}= \bar{g}_{\mu\nu}+h_{\mu\nu} \ . \label{eq:backgr_pert}
	\ee
	Here $\bar A^\mu$ is mapped to $\bar g_{\mu\nu}$, while $a^\mu$ is mapped to $h_{\mu\nu}$. As usual, the terminology ``perturbation'' refers to this decomposition and need not imply a small-field expansion: for Kerr--Schild metrics the dependence on the profile is exactly linear. The nontrivial requirement is that the two maps remain compatible when the equations for $a^\mu$ and $h_{\mu\nu}$ are evaluated on their respective copied backgrounds.
	
	In \cite{Bahjat-Abbas:2017htu} a Kerr--Schild double copy where both background and perturbations are doubled copied was proposed and dubbed type A. It was shown that when the background metric has a Kerr--Schild form, the double copy will hold in the expected way. In the following we summarize their construction. To start, we consider metrics that can be cast in a double Kerr--Schild form
	\be
	g_{\mu\nu} \;=\; \eta_{\mu\nu} + \phi_1\,k_\mu k_\nu + \phi_2\,k_\mu k_\nu \ .
	\ee
	An elementary example is the Schwarzschild--(A)dS family, with $\phi_1\propto r^2$ for the cosmological term and $\phi_2\propto M/r$ for the Schwarzschild term. Since both terms use the same Kerr--Schild vector, this is a split of a single Kerr--Schild metric rather than a genuine double Kerr--Schild geometry with two distinct null directions.
	Grouping the first term into the background and the second into the perturbation yields an explicit double copy map:
	\begin{align*}
		\bar g_{\mu\nu} &= \eta_{\mu\nu} + \phi_1\,k_\mu k_\nu , &
		h_{\mu\nu} &= \phi_2\,k_\mu k_\nu ,\\[4pt]
		\bar A_\mu^a     &= c^a\,\phi_1\,k_\mu , &
		a_\mu^a    &= c^a\,\phi_2\,k_\mu ,
	\end{align*}
	where $c$ is a color factor.  The recipe extends to multi-Kerr--Schild metrics by partitioning the sum over distinct Kerr--Schild vectors into two disjoint subsets, one forming the background and the other the perturbation.  Each subset separately solves its field equations, so the correspondence holds term-by-term. The zeroth copy can be constructed analogously by replacing the remaining Kerr--Schild vector in each term by a second color vector, thus producing a biadjoint scalar
	\be
	\Phi \;=\; \bar\Phi + \tilde\Phi , 
	\quad\text{with}\quad
	\bar\Phi = c^a c^{a'}\phi_1,\;
	\tilde\Phi = c^a c^{a'}\phi_2 ,
	\ee
	that obeys the linearized biadjoint equation on the background $\bar\Phi$. Note that in these linearized cases, the equations of motion on the non-trivial backgrounds are the same as the equations of motion on the trivial background since all the background contributions appear non-linearly. 
	This term-by-term prescription depends on the multi-Kerr--Schild decomposition and is therefore more restrictive than an arbitrary background-field split. Its value is that it makes the two layers of the map explicit: one subset of Kerr--Schild terms defines the copied background, while the remaining subset defines a solution propagating on that background.\\
	
	In flat space, the relation between the Kerr--Schild prescription and the amplitudes double copy can be traced to on-shell three-point data \cite{Arkani-Hamed:2019ymq}. For the relevant classical solutions, the same special structure that linearizes the field equations allows the spacetime field to be reconstructed without introducing independent higher-point information. This helps explain why a local replacement $c^a\leftrightarrow k^\mu$ can reproduce an exact gravitational solution. An analogous derivation on curved backgrounds is not yet available: it would require identifying how the Kerr--Schild map is encoded in three-point correlators or other curved-space observables.
	
	The prescription is not limited to the original spin two black-hole and wave examples. It has been extended to higher-spin fields \cite{Didenko:2022qxq}, as well as to more general backgrounds and higher-dimensional solutions \cite{Ortaggio:2023cdz,Ortaggio:2023rzp,Hassaine:2024mfs,Srinivasan:2025hro,Alencar:2026zdz,Rodriguez:2026zxm}. In these generalized constructions the full metric may have weaker algebraic speciality than the standard Kerr--Schild class, and the chosen background need not be the $M\rightarrow0$ member of the solution family. It can instead be an extremal geometry obtained in the limit $M\rightarrow M^{\mathrm{ext.}}$, where $M^{\mathrm{ext.}}$ is the extremal mass.
	
	In some cases, this change of background can recast a double Kerr--Schild perturbation of flat space as a single Kerr--Schild perturbation of an extremal curved geometry \cite{Hassaine:2024mfs}. In a less customary approach, it also allows us to construct a double copy in fixed charged backgrounds. While a standard interpretation as a double copy of the latter situation may seem contrived, this technique can still be useful for deriving new exact charged gravitational solutions. 
	
	The Weyl double copy, to which we now turn, avoids the field-level dependence by formulating the relation directly in terms of curvatures.

	\subsection{Weyl Double Copy} \label{sec:classical_Weyl}
	The Weyl double copy \cite{Luna:2018dpt,Godazgar:2020zbv} is a gauge-invariant, curvature-level relation. Rather than decomposing the metric, it expresses the Weyl curvature in terms of two Maxwell field strengths and a scalar. This makes the prescription insensitive to gauge transformations of the gravitational and gauge potentials, but it also means that reconstructing those potentials is a separate problem. The factorization is most transparent in spinorial variables, although tensor formulations are also available \cite{Alawadhi:2019urr,Alawadhi:2020jrv}. Section~\ref{sec:spinorhel} introduced the relation between Lorentz tensors and $SL(2,\mathbb{C})$ spinors for amplitudes. In curved spacetime the conversion is performed locally, using an orthonormal frame,
	\be
	g_{\mu\nu}={e_\mu}^\mathrm{a}{e_\nu}^\mathrm{b}\eta_{\mathrm{ab}} \ ,
	\ee
	where $\eta_{\mathrm{ab}}$ is the Minkowski metric. The transformation from Lorentz to spinor ($SL(2,\mathbb{C})$) indices is now given by
	\be
	v_{\alpha\dot{\alpha}}
	\;=\;
	v_\mu{e_\mathrm{a}}^\mu(\sigma^\mathrm{a})_{\alpha\dot{\alpha}} \ , 
	\qquad
	\sigma^\mathrm{a}=\{\mathrm{I},\sigma^i\} \ .
	\ee
	
	We can now write the curvature tensor of the gauge and gravitational theories in spinor notation. The YM field strength is given by
	\be
	F_{\alpha\dot{\alpha}\beta\dot{\beta}}=f_{\alpha\beta} \, \varepsilon_{\dot{\alpha} \dot{\beta}} + \bar{f}_{\dot{\alpha} \dot{\beta}} \, \varepsilon_{\alpha\beta} \ ,
	\ee
	where $f_{\alpha\beta}$ and $\bar{f}_{\dot{\alpha} \dot{\beta}}$ are symmetric bispinors, and if the field strength is real, they are the complex conjugate of each other. The bispinor $\bar{f}_{\dot{\alpha} \dot{\beta}}$ encodes the self-dual component of the field strength, that is, the components satisfying $F=i\star F$, or in components, $F_{\mu\nu}=i(\sqrt{|g|}/2)\epsilon_{\mu\nu\rho\lambda}F^{\rho\lambda}$, and $f_{\alpha\beta}$ encodes the anti-self-dual part satisfying $F=-i\star F$. The YM equations of motion now read
	\be
	{D^\alpha}_{\dot{\alpha}}f_{\alpha\beta}=0\ ,
	\ee
	where ${D^\alpha}_{\dot{\alpha}}$ is the spinor of the gauge connection $\nabla^\mu+i A^\mu$. On the gravitational side, we will consider the Weyl tensor, which is given by
	\be
	C_{\alpha\dot{\alpha}\beta\dot{\beta}\gamma\dot{\gamma}\sigma\dot{\sigma}}=\Psi_{\alpha\beta\gamma\sigma} \, \varepsilon_{\dot{\alpha} \dot{\beta}} \varepsilon_{\dot{\gamma} \dot{\sigma}}+ \bar{\Psi}_{\dot{\alpha} \dot{\beta}\dot{\gamma} \dot{\sigma}} \, \varepsilon_{\alpha\beta}\varepsilon_{\gamma\sigma} \ ,
	\ee
	where again $\Psi_{\alpha\beta\gamma\sigma}$ and $\bar{\Psi}_{\dot{\alpha} \dot{\beta}\dot{\gamma} \dot{\sigma}}$ are fully symmetric, encode the anti-self-dual and self-dual Weyl tensors respectively, and are related by complex conjugation when working with real metrics. 
	
	For vacuum spacetimes with or without a cosmological constant, the Einstein equations imply that the traceless Ricci tensor vanishes $S_{\mu\nu}=R_{\mu\nu}-R g_{\mu\nu}/4=0$. In this case, the Bianchi identities in spinorial form reduce to \cite{Stephani:2003tm}
	\be
	{\nabla^\alpha}_{\dot{\alpha}}\Psi_{\alpha\beta\gamma\sigma}=0 \ . \label{eq:Bianchi_spinor}
	\ee
	Exact gravitational solutions satisfying the vacuum Einstein equations can be fully classified by the principal null directions of the Weyl tensor. In four dimensions the Weyl spinor can always be factorized into four spinors,
	\be
	\Psi_{\alpha\beta\gamma\delta}
	= o_{(\alpha}\,\iota_{\beta}\,\kappa_{\gamma}\,\lambda_{\delta)} ,
	\ee
	and each principal spinor (say $o_\alpha$) defines a null vector $k^{\alpha\dot{\alpha}} = o^{\alpha}\,\bar{o}^{\dot{\alpha}}$; these null directions are the \emph{principal null directions} (PNDs) of the Weyl tensor. Algebraically, a PND is characterized by the cubic Petrov equation
	\be
	\Psi_{\alpha\beta\gamma\delta}\,o^{\beta}o^{\gamma}o^{\delta}=0 \ .
	\ee
	The multiplicities with which the four principal spinors coincide determine the Petrov type:
	four distinct PNDs (type~I), one double plus two simple (II), two double (D),
	one triple plus one simple (III) or a quadruple PND (N), while a vanishing Weyl spinor gives the conformally flat type~O.
	
	We only need the type~D and type~N cases, which are the cases constructed by the Weyl double copy. They can be written as
	\begin{align}
		\text{type N} &: \quad \Psi_{\alpha\beta\gamma\delta} = \Psi_4 o_\alpha o_\beta o_\gamma o_\delta \ , \label{eq:typeN}\\
		\text{type D} &: \quad \Psi_{\alpha\beta\gamma\delta} = \Psi_2 o_{(\alpha} o_\beta \iota_\gamma \iota_{\delta)} \ , \label{eq:typeD}
	\end{align}
	where $\Psi_4$ and $\Psi_2$ are the relevant Newman--Penrose scalars. Petrov type~D solutions correspond to the gravitational fields of isolated objects, such as black holes, while Petrov type~N solutions give transverse wave solutions. In these two cases the double copy is constructed as:
	
	\begin{mybox}[Weyl Double Copy]
		\begin{center}
			\renewcommand{\arraystretch}{1.8}
			\begin{tabularx}{\textwidth}{l X}
				\textbf{Gravity} &   \vspace{-1cm} 
				\begin{equation}
					\Psi_{\alpha\beta\gamma\delta} =\frac{f_{(\alpha \beta}f_{\gamma \delta)}}{S}
					\label{eq:Weyl_gravity}
				\end{equation} \vspace{-1cm}  \\  
				\textbf{Yang--Mills} &   \vspace{-1cm} 
				\begin{equation}
					f_{\alpha \beta}^a = c^a f_{\alpha \beta} 
					\label{eq:Weyl_gauge}
				\end{equation}  \vspace{-1cm}  \\ 
				\textbf{Biadjoint Scalar} &   \vspace{-1cm} 
				\begin{equation}
					S^{a\, b'} = c^a c^{b'}S
					\label{eq:Weyl_bas}
				\end{equation}  \vspace{-0.5cm} 
			\end{tabularx}
		\end{center}
	\end{mybox}
	where the color-stripped zeroth-copy scalar $S$ satisfies
	\be
	\label{eq:ScalarConfCoupl}
	\left(\bar{\nabla}^2-\frac{\bar R}{6}\right)S=0 \ ,
	\ee
	which is the equation of motion for a conformally coupled scalar in a background metric $\bar{g}_{\mu\nu}$ with covariant derivative $\bar{\nabla}_{\alpha\dot{\alpha}}$.  The gauge field satisfies the linearized Yang--Mills, i.e., Maxwell's equations
	\be
	{\bar{\nabla}^\alpha}\, _{\dot{\alpha}}f_{\alpha\beta}=0 \ ,
	\ee
	and the Weyl spinor satisfies
	\be\
	{\bar{\nabla}^\alpha}\, _{\dot{\alpha}}\Psi_{\alpha\beta\gamma\sigma}=0 \ .
	\ee
	
	For type~D and type~N solutions, additional symmetries allow the nonlinear gravitational curvature to obey Eq.~\eqref{eq:Bianchi_spinor}. In the type~N case, this is realized since the Goldberg--Sachs theorem ensures that the repeated principal null direction is geodesic and shear-free, and additional first-order differential relations fixing $S$ are imposed \cite{Godazgar:2020zbv}. For type~D, this follows from the valence-two Killing spinor aligned with the two repeated principal spinors \cite{Luna:2018dpt}. The scalar is therefore essential: it compensates the spacetime dependence and scaling weight of the product $f_{(\alpha\beta}f_{\gamma\delta)}$ and ensures its differential compatibility with the full geometry. Consequently, an algebraic factorization alone is not sufficient; the three fields must also satisfy their respective differential equations on a common background. This classical double copy resembles the KLT double copy reviewed in Sec.~\ref{sec:klt_and_strings}.
	
	The Weyl double copy has been studied extensively in Ricci-flat spacetimes \cite{Luna:2018dpt,Godazgar:2020zbv,ArmstrongWilliams:2025src,Godazgar:2021iae,Chawla:2024mse,Liu:2024byr,Alawadhi:2019urr,Sabharwal:2019ngs,Monteiro:2020plf,Easson:2021asd,Chacon:2025hh,Keeler:2020rcv,Chawla:2025uwu,Keeler:2024bdt,Monteiro:2021ztt} and on curved backgrounds \cite{Han:2022ubu,Han:2022wdc,Alkac:2024rwdc,Zhao:2025fiveD,Zhao:2025typeD,Chacon:2025hh,Kent:2025godel,Chawla:2023aligned,Alkac:2025iyw}. Generalizations to five dimensions appear in \cite{Zhao:2024wtn,Zhao:2024ljb}; the three-dimensional analogue is the Cotton double copy reviewed below.
	
	\bigskip
	\subsubsection{Example: Kundt--AdS waves}\label{subsec:weyl_kundt_ads}
	A simple Type N example is given by the AdS Kundt$(\Lambda)$ class. In particular, we focus on generalized pp-waves (plane-fronted waves with parallel propagation) whose metric is given by
	\begin{equation}
		ds^2=-F\,du^2+2\frac{q^2}{p^2}\,du\,dv-\frac{2}{p^2}\,dz\,d\bar z,
	\end{equation}
	where the functions $p$ and $q$ encode the AdS curvature dependence,
	\begin{equation}
		p=1+\frac{\Lambda}{6}z\bar z,\qquad
		q=\alpha(u)\!\left(1-\frac{\Lambda}{6}z\bar z\right),
	\end{equation}
	where $\alpha(u)$ is an arbitrary function. Meanwhile the wave profile is contained in $F$,
	\begin{equation}
		F=\frac{\Lambda}{3}\alpha^2\frac{q^2}{p^2}v^2-\frac{(q^2)_{,u}}{p^2}v-\frac{q}{p}\left(f_{,z}+\bar f_{,\bar z}-\frac{\Lambda}{3p}\,(z\bar f+\bar z f)\right) \ ,
	\end{equation}
	with $f=f(u,z)$ determining the wave profile. The Weyl tensor of this spacetime is captured by a single non-vanishing Newman--Penrose scalar, see Eq.~\eqref{eq:typeN}, which reads
	\begin{equation}
		\Psi_4=\frac{1}{2}pq\,\partial_{\bar z}^{\,3}\bar f \ .
	\end{equation}
	The corresponding Weyl double copy \cite{Han:2022wdc} is specified by a zeroth copy scalar,
	\begin{equation}
		S=C(u,\bar z)\frac{p}{q},
	\end{equation}
	together with a single copy gauge field of the form 
	\begin{equation}
		f_{\alpha\beta}=\Phi_2 i_{(\alpha}i_{\beta)} \ , \quad \Phi_2=\frac{p}{\sqrt{2}}\sqrt{C(u,\bar z)\,\partial_{\bar z}^{\,3}\bar f} \ .
	\end{equation}
	
	In this way, the Type N gravitational wave is related to a null Maxwell field on the same AdS background, while the scalar $S$ is the zeroth copy satisfying Eq.~\eqref{eq:ScalarConfCoupl}.
	Indeed, substituting the two expressions gives 
	\begin{equation}
		\Psi_4=\frac{\Phi_2^2}{S} \ . \label{eq:typeN_dc}
	\end{equation}
	The arbitrary factor $C(u,\bar z)$ therefore redistributes the profile between the Maxwell and scalar fields without changing the gravitational curvature. This is the curvature-level counterpart of the rescaling freedom encountered in the Kerr--Schild prescription.
	
	This is the same type~N Kundt--AdS sector illustrated with the Kerr--Schild map in Subsection~\ref{subsec:ks_kundt_ads}. The Kerr--Schild construction acts on the metric profile and directly produces a gauge potential, whereas the Weyl construction factorizes the gauge-invariant curvature. The two descriptions preserve the same repeated null direction, but their Maxwell representatives and auxiliary scalars need not agree term by term because their rescaling freedoms and coordinate normalizations differ.

	\subsubsection{\texorpdfstring{Example: Kerr--(A)dS black hole}{Example: Kerr-(A)dS black hole}}
	A simple example is the Kerr--(A)dS black hole. To construct the gravitational solution, start with the Maxwell field strength of a charged rotating disk of radius $a$ in an (A)dS background. Its field-strength spinor outside the disk is
	\be
	f_{\alpha\beta}=\Phi_1 o_{(\alpha}i_{\beta)} \ , \quad \Phi_1=\frac{1}{(r + i a \cos\theta)^2} \ ,
	\ee
	where the solution is written in the coordinate system of Eq.~\eqref{eq:KerrAds_mzero}. Similarly, the conformally coupled scalar solution in these coordinates is given by
	\be
	S=\frac{1}{(r + i a \cos\theta)} \ .
	\ee
	and thus the double copy is a Type D solution as in Eq.~\eqref{eq:typeD} with
	\be
	\Psi_2=\frac{1}{(r + i a \cos\theta)^3} \ ,
	\ee
	which in fact corresponds to the only non-zero Newman-Penrose scalar of the Kerr-(A)dS black hole written in Boyer-Lindquist coordinates. Here, we have set the mass of the black hole and the charge of the disk to one. For type D, we see that the double copy simplifies in terms of the Newman-Penrose scalars as 
	\begin{equation}
		\Psi_2=\frac{\Phi_1^2}{S}   \ ,
	\end{equation}
	which is analogous to Eq.~\eqref{eq:typeN_dc}. 
	
	Unsurprisingly, these solutions look the same in (A)dS and Minkowski since the equations of motion of the curvatures involved in the double copy are all invariant under conformal transformations. This version of the double copy is equivalent to what was found in the Kerr--Schild case in Sec.~\ref{sec:kerr-schild}. To make the connection explicit for the linearized biadjoint scalar, one has to take $\phi=S+\bar{S}$. The agreement for the Kerr--(A)dS double copies relies on the common alignment of the Kerr--Schild null congruence with the repeated principal null directions of the type~D curvature.
	
	For the flat-space double copy, the split-signature formulation makes the connection to scattering amplitudes particularly direct: using the KMOC formalism \cite{Kosower:2018adc}, the classical Weyl spinor can be reconstructed from the corresponding three-point amplitude and therefore inherits its double copy structure \cite{Monteiro:2020plf,Emond:2020lwi,Monteiro:2021ztt}. It remains an open question whether a similar relation can be made in the curved space version.

	\subsection{Cotton Double Copy}
	The Cotton double copy is the three-dimensional analogue of the Weyl double copy \cite{CarrilloGonzalez:2022mxx,Emond:2022uaf}. In three dimensions the Weyl tensor vanishes identically, and the Cotton tensor instead measures the local conformal curvature. Moreover, massless Einstein gravity has no propagating bulk graviton, so the natural comparison is between topologically massive Yang--Mills (TMYM) and topologically massive gravity (TMG), each of which carries one parity-violating local degree of freedom. The resulting relation expresses the linearized Cotton curvature as a product of TMYM field strengths. Its connection to scattering amplitudes was studied in \cite{Emond:2022uaf}, the curved-space wave construction in \cite{CarrilloGonzalez:2022mxx}, its minitwistor realization in \cite{CarrilloGonzalez:2022ggn}, and its extension to AdS black holes in \cite{Beetar:2024ptv}.
	
	In curved spacetime, topologically massive Yang--Mills is described by
	\begin{align}\label{eq:tpYM}
		S_{\text{TMYM}}
		&=\int d^3x\,\sqrt{-g}\Bigg[
		-\frac{1}{4}F^{a\mu\nu}F^{\ }_{a\mu\nu}
		+\frac{\mathrm g}{\sqrt{2}}\,A^{\mu a}J_{\mu a} \\
		&+\varepsilon_{\mu\nu\rho}\,\frac{m}{12}\Big(
		6A^{a\mu}\partial^{\nu}A^{\rho}_{a}
		+\,f_{abc}A^{a\mu}A^{b\nu}A^{c\rho}
		\Big)
		\Bigg],
	\end{align}
	where $m$ is the gauge-field mass and
	$\varepsilon_{\mu\nu\rho}=\sqrt{-g}\,\epsilon_{\mu\nu\rho}$ is the Levi--Civita tensor (with $\epsilon_{\mu\nu\rho}$ the Levi--Civita symbol). This theory propagates a single, parity-violating, massive spin one local degree of freedom \cite{Deser:1981wh} whose helicity is given by the sign of the mass. Restricting to color configurations
	$A^{\mu a}=c^a A^\mu$ and $J^{\mu a}=c^a J^\mu$, with constant color charge $c^a$ and adjoint $U(N)$ index $a$, the dynamics becomes linear in the spacetime fields. The resulting equations of motion take the form
	\begin{equation}\label{eq:TMYM}
		\nabla_\mu F^{\mu\nu}
		+\frac{m}{2}\,\varepsilon^{\nu\rho\gamma}F_{\rho\gamma}
		=J^\nu,
	\end{equation}
	where $F_{\mu\nu}$ is the linearized Yang--Mills field strength. 
	
	The double copy partner of TMYM is topologically massive gravity, whose action can be written as
	\begin{align}\label{eq:tpmGR}
		S_{\text{TMG}}
		&=\frac{1}{2}\int d^3x\,\sqrt{-g}\Bigg[
		-R+2\Lambda+\mathcal L_{\text{Matter}}
		\\ &-\frac{1}{2m}\,\varepsilon^{\mu\nu\rho}\Big(
		\Gamma^{\alpha}_{\mu\sigma}\partial_{\nu}\Gamma^{\sigma}_{\alpha\rho}
		+\frac{2}{3}\Gamma^{\alpha}_{\mu\sigma}\Gamma^{\sigma}_{\nu\beta}\Gamma^{\beta}_{\rho\alpha}
		\Big)
		\Bigg],
	\end{align}
	where $m$ is now the mass of the graviton. Like TMYM, this theory propagates a single, parity-violating, massive spin 2 bulk degree of freedom \cite{Deser:1981wh} and the sign of the mass gives the helicity of the massive graviton. Varying with respect to the metric yields the field equations
	\begin{equation}\label{eq:tmg}
		G_{\mu\nu}+\frac{1}{m}C_{\mu\nu}
		=-\frac{T_{\mu\nu}}{2}-\Lambda g_{\mu\nu},
	\end{equation}
	where $G_{\mu\nu}$ is the Einstein tensor, $T_{\mu\nu}$ is the stress-energy tensor, and the Cotton tensor is
	\begin{equation}\label{eq:cotton}
		C^{\mu\nu}
		=\varepsilon^{\mu\alpha\beta}\nabla_{\alpha}
		\left(R^{\nu}_{\ \beta}-\frac{1}{4}\delta^{\nu}_{\ \beta}R\right).
	\end{equation}
	
	The Cotton double copy, just like the Weyl one, is better described using spinorial variables. The three-dimensional version has been developed in \cite{Milson:2012ry,castillo20033,doi:10.1063/1.1592611}. It relies on the isomorphism between the 3d Lorentz group and the spin group, $SO(1,2)\simeq SL(2,\mathrm{R})/\mathbb{Z}_2,$ which lets one recast tangent-space Lorentz transformations in terms of $SL(2,\mathrm{R})$ spinors. In this language a tangent-space vector can be represented as a symmetric bispinor,
	\be
	v_{\mathrm{a}}=-(\sigma_{\mathrm{a}})_{AB}v^{AB} \ ,
	\ee
	where the matrices $\sigma_{\mathrm{a}}$ provide a basis of $SL(2,\mathrm{R})$ and obey the 3d Clifford algebra. As before, the conversion between tangent and coordinate indices is achieved via the frame field $e^\mu_{\mathrm{a}}$ as
	$v^\mu=-{e^\mu}_{\mathrm{a}}(\sigma^{\mathrm{a}})_{AB}v^{AB}$. Spinor indices $A,B\in\{1,2\}$ are raised and lowered using the two-dimensional Levi--Civita symbol $\epsilon_{AB}$, with conventions
	$\psi^A=\psi_B\epsilon^{BA}\ , \  \psi_A=\epsilon_{AB}\psi^B$ . 
	
	The Cotton double copy relates the curvatures of the topologically massive theories as\footnote{Note that the conventions here follow \cite{CarrilloGonzalez:2022ggn} which differ by factors of $\sqrt{2}$ in various places with respect to \cite{CarrilloGonzalez:2022mxx}.} 
	\begin{mybox}[Cotton Double Copy]
		\begin{center}
			\renewcommand{\arraystretch}{1.8}
			\begin{tabularx}{\textwidth}{l X}
				\textbf{Gravity} &   \vspace{-1cm} 
				\begin{equation}
					C_{ABCD}=m \frac{f_{(AB}f_{CD)}}{S}
					\label{eq:Cotton_gravity}
				\end{equation} \vspace{-1cm}  \\  
				\textbf{Yang--Mills} &   \vspace{-1cm} 
				\begin{equation}
					f_{AB}^a = c^a f_{AB} 
					\label{eq:Cotton_gauge}
				\end{equation}  \vspace{-1cm}  \\ 
				\textbf{Biadjoint Scalar} &   \vspace{-1cm} 
				\begin{equation}
					S^{a\, b'} = c^a c^{b'}S
					\label{eq:Cotton_bas}
				\end{equation}  \vspace{-0.5cm} 
			\end{tabularx}
		\end{center}
	\end{mybox}
	where the curvature spinors satisfy the equations of motion of TMG and TMYM which in spinor notation read
	\begin{align}
		\nabla^{EA}C_{BECD}&=m{C^A}_{BCD} \ ,\\
		\nabla^{CA}f_{BC}&=m{f^A}_B \ , 
	\end{align}
	where $C^{\mu\nu}=\sigma^\mu_{AB}\sigma^\nu_{CD}C_{ABCD}$ and the dual field strength is given by $f^\mu=-\sigma^\mu_{AB}f^{AB}$. Meanwhile the scalar field satisfies
	\be
	\left(\bar{\nabla}^2-m^2-\frac{R}{6}\right)S=0 \ .
	\label{eq:ZerothCopy_m_3d}
	\ee
	The explicit factor of $m$ in Eq.~\eqref{eq:Cotton_gravity} is chosen by convention and reflects the first-order massive equations obeyed by both curvature spinors. The scalar equation contains the same mass scale, but its curvature coupling is not the three-dimensional conformal coupling. This is consistent with the fact that the topological mass already breaks scale invariance. We next illustrate the construction for type~N waves and type~D warped geometries.
	
	\subsubsection{Example: AdS pp-waves}\label{subsec:cotton_ads_pp_waves} We proceed to describe the double copy relation for pp-waves. For TMG, any solution that admits a null Killing vector, well-defined through all space, is a pp-wave solution
	\cite{Gibbons:2008vi}. In AdS the metric of pp-waves can always be written as \cite{Chow:2009km}
	\begin{equation}
		\mathrm{d} s^{2}=\mathrm{d} y^{2}-2 \mathrm{e}^{2 \frac{y}{L}} \mathrm{d} u \mathrm{~d} v+\mathrm{e}^{\frac{(1-mL)}{L} y} f(u) \mathrm{d} u^{2} ,
	\end{equation}
	where $u,v$ are lightcone coordinates, and $L=1/\sqrt{-\Lambda}$ is the AdS radius. The dS solution is obtained by taking $L\rightarrow i L$. On the TMYM side, we can write the pp-wave solution as 
	\be
	A^a=c^a e^{-m y} g(u) du \ .
	\ee
	For Type N solutions, it is simpler to express the double copy in terms of the Newman-Penrose scalars as
	\begin{equation}
		\Psi_4=m\frac{\Phi_2^2}{S} \ . \label{eq:CottonDCNP}
	\end{equation}
	For the AdS pp-waves one has that the Newman-Penrose scalars of the double, single, and zeroth copy are
	\begin{align}
		\Psi_4&=m^3e^{-(\frac{3}{L}+m)y}f(u) \ , \\
		\Phi_2&=-me^{-(\frac{2}{L}+m)y} g(u) \ , \\
		S&=\frac{g(u)^2}{f(u)}e^{-(\frac{1}{L}-m) y} \ .
	\end{align}
	which clearly satisfies Eq.~\eqref{eq:CottonDCNP} and the scalar field is a solution of Eq.~\eqref{eq:ZerothCopy_m_3d}. The functions $f(u)$ and $g(u)$ encode independent profile choices before the scalar is fixed. Their ratio in $S$ is precisely what makes the spacetime dependence and normalization match in Eq.~\eqref{eq:CottonDCNP}.
	
	\subsubsection{Example: Warped black holes}\label{subsec:cotton_warped_black_holes}
	Type D solutions in TMG are locally warped AdS spacetimes, which are homogeneous deformations of AdS$_3$ in which one fiber direction over an AdS$_2$ base is stretched or squashed \cite{Chow:2009km}. Similar to the BTZ black hole, in TMG one can obtain black hole solutions as discrete quotients of warped AdS \cite{Anninos:2008fx}. In \cite{CarrilloGonzalez:2022ggn}, it was shown that one can construct a Cotton double copy for these solutions. 
	
	For simplicity, take the spacelike-squashed AdS$_3$ metric in Poincar\'e coordinates \cite{Chow:2009km}: 
	\begin{equation}
		\text{d}s^2=\frac{9}{m^2-27\Lambda}\Bigg[ \frac{-\text{d}t^2+\text{d}x^2}{x^2}+\frac{4m^2}{m^2-27\Lambda}\Bigg( \text{d}z+\frac{\text{d}t}{x} \Bigg)^2 \Bigg] \ ,
		\label{eq:ssads}
	\end{equation}
	where the factor
	\begin{equation}
		\lambda\equiv\frac{4m^2}{m^2-27\Lambda} \ ,
	\end{equation}
	is the squashing parameter: for $\lambda>1$ we have stretching, for $\lambda<1$ squashing, and when $\lambda=1$ the metric is AdS. To interpret this solution as a double copy, one can think of the warped AdS with cosmological constant $\Lambda$ as a deformation of an AdS spacetime with cosmological constant $\bar{\Lambda}=-m^2/9$, since $\Lambda=\bar{\Lambda}$ gives $\lambda=1$, so the squashing deformation vanishes. The Cotton tensor of this metric reads
	\begin{equation} \label{eq:tracelesRicciBH}
		C^{\mu\nu}=m\left(\Lambda-\bar{\Lambda } \right)\left(g^{\mu\nu}-3l^\mu l^\nu\right) \ ,
	\end{equation}
	where we used the spacelike Killing vector $l^\mu = ((m^2-27\Lambda)/(6m))(0,0,1)^\mu$. The corresponding Cotton spinor can be written as
	\begin{equation}
		\Psi_{ABCD}\propto \frac{r_{(A}r_{B}s_{C}s_{D)}}{\braket{rs}^5} \ ,  \quad  \text{with} \quad \braket{rs} =\left[6 (-\bar{\Lambda})^{1/2}(\Lambda-\bar{\Lambda})\right]^{-\frac13} \ .
	\end{equation}
	The single copy of the squashed AdS has a dual field strength
	\begin{equation} \label{eq:typeDstarF}
		\star F^\mu = 6\left(\frac{\Lambda-\bar{\Lambda }}{x}\right)l^\mu.
	\end{equation}
	with a field strength spinor
	\be
	\phi_{AB}\propto\frac{r_{(A}s_{B)}}{\braket{rs}^3}
	\ee
	In the TMG double copy, the Cotton tensor obeys
	\begin{equation} \label{eq:LieCotton}
		\mathcal{L}_l C^{\mu\nu}\propto \mathcal{L}_l g^{\mu\nu} -3 \mathcal{L}_l l^\mu l^\nu=0  \ .
	\end{equation}
	Similarly, for the TMYM single copy the dual field strength satisfies
	\begin{equation} \label{eq:LieCotton2}
		\mathcal{L}_l \star F^{\mu}=0  \ .
	\end{equation}
	This shows how the double copy relation preserves symmetries. We also note that the electric field of the single copy can be shown to be orthogonal to the spacelike Killing vector, and, as expected, there is no magnetic field.
	
	Finally, the zeroth copy is the scalar $S=1/\braket{rs}$, which satisfies $\bar{\nabla}^2S=0$ because it lives on an AdS background with $\bar{\Lambda}=-m^2/9$. 
	
	\subsection{Double Copy in Twistor Space}\label{sec:field_twistor_double_copy}
	Twistor space provides a common origin for several curvature-level double copies. In four dimensions, the Penrose transform explains the Weyl double copy and extends the construction beyond type~D and type~N solutions at linear order \cite{White:2020sfn,Chacon:2021wbr,Chacon:2021hfe,Chacon:2021lox,Adamo:2021dfg,Armstrong-Williams:2023ssz}. In flat space, twistor methods also provide a bridge to the amplitudes picture: classical spacetimes can be reconstructed from S-matrix data, while the relation between the twistor-space construction and the local position-space Weyl double copy can be understood through the corresponding integral transforms \cite{Guevara:2021yud,Luna:2022dxo}.
	
	In three dimensions, the analogous minitwistor construction yields the Cotton double copy in flat space \cite{CarrilloGonzalez:2022ggn} and in AdS \cite{Beetar:2024ptv}. The essential simplification is that on-shell spacetime fields are represented by cohomology classes with definite homogeneity, so multiplication and division of representatives can implement the spin and mass assignments of a double copy before the Penrose transform is performed. We focus here on the curved three-dimensional construction and refer to \cite{White:2024pve} for a review of the flat-space literature.
	
	We first need to introduce the three-dimensional version of twistor space, which is referred to as minitwistor space. We will work in Euclidean space following \cite{Beetar:2024ptv}. The minitwistor space of $\mathbb{H}$, denoted $\MT$, is \cite{tsai}\footnote{For the hyperbolic construction used here, the relation to four-dimensional twistor space is a scaling reduction. Viewing four-dimensional Minkowski space as the embedding space of hyperbolic slices, projectivizing the radial $\mathbb C^*$ action produces the projective spinor and minitwistor data on each slice \cite{Seet:2024vmh}.}
	\begin{equation} \label{eq:MT}
		\MT =\{(\mu^{\dot\alpha},\lambda_{\alpha}), \mu^{\dot\alpha} \in \cp^1, \lambda_{\alpha} \in \cp^1 \} \,.
	\end{equation}
	Unlike four-dimensional twistor space, the pair $(\mu^{\dot\alpha},\lambda_{\alpha})$ has two independent projective scalings. Explicitly, $([\mu^{\dot\alpha}],\allowbreak[\lambda_{\alpha}])\allowbreak\sim\allowbreak([u\mu^{\dot\alpha}],\allowbreak[v\lambda_{\alpha}])$, with independent $u,v\in\C^*$. Thus, functions on minitwistor space have independent homogeneities (scaling weight) with respect to either pair of coordinates: 
	\begin{equation}\label{scaling}
		f(u\mu^{\dot\alpha},v\lambda_\alpha) = u^p v^q f(\mu^{\dot\alpha},\lambda_\alpha)\,.
	\end{equation}
	where $f$ is a homogeneous function of weight $(p,q)$. The incidence relation connecting minitwistor space to hyperbolic space is
	\begin{equation}
		\mu^{\dot\alpha}=X^{\alpha\dot\alpha}\lambda_{\alpha}\ .
		\label{eq:incidence}
	\end{equation}
	Here $X^{\alpha\dot\alpha}$ is the bispinor corresponding to the projective coordinate $X^\mu$ in four-dimensional (complex) Minkowski space satisfying $X\cdot X\neq0$, since hyperbolic space can be described as $\mathbb{H}=\mathbb{CP}^3\setminus\{X\cdot X=0\}$. A non-projective coordinate in hyperbolic space can be obtained by choosing the normalized representative
	\begin{equation} \label{eq:CoordHyperbolic}
		x^\mu=\frac{L X^\mu}{\sqrt{-X\cdot X}}\ .
	\end{equation}
	which satisfies $x\cdot x=-L^2$. In the rest of this subsection, we set $L=1$. This is gives the usual embedding space coordinates. We write $\mu_X^{\dot\alpha}$ for the incidence value before normalization and $\mu_x^{\dot\alpha}$ after restricting to the representative in Eq.~\eqref{eq:CoordHyperbolic}.
	
	The three-dimensional metric is the pullback of the embedding-space metric to this normalized hyperboloid,
	\begin{equation}
		\dd s_{\mathbb H}^2
		=\eta_{\mu\nu}\,\dd\!\left(\frac{X^\mu}{\sqrt{-X\cdot X}}\right)
		\dd\!\left(\frac{X^\nu}{\sqrt{-X\cdot X}}\right)\ .
		\label{eq:minitwistor_induced_metric}
	\end{equation}
	To obtain the real Euclidean slice, impose the Hermitian reality condition $\widehat X^{\alpha\dot\alpha}=(X^{\alpha\dot\alpha})^\dagger$ and take $y_0,y_1,y_2$ to be real. A parametrization of the embedding coordinate is given by Poincar\'e coordinates and reads
	\begin{equation}
		X^{\alpha\dot\alpha}
		=\begin{pmatrix}
			1 & y_1-i y_2\\
			y_1+i y_2 & y_0^2+y_1^2+y_2^2
		\end{pmatrix}\ ,
		\label{eq:minitwistor_poincare_embedding}
	\end{equation}
	where $\det(X^{\alpha\dot\alpha})=-X\cdot X=y_0^2$. Substitution into Eq.~\eqref{eq:minitwistor_induced_metric} therefore gives
	\begin{equation}
		\dd s_{\mathbb H}^2=\frac{1}{y_0^2}
		\left(\dd y_0^2+\dd y_1^2+\dd y_2^2\right)\ ,
		\label{eq:minitwistor_h3_metric}
	\end{equation}
	where the boundary lies at $y_0\to0$. See Sec.~\ref{sec:correll_embedding} for the general embedding-space formalism and its use for AdS bulk and boundary points.
	
	To formulate the double copy, we use the minitwistor Penrose transform. This relates holomorphic functions on minitwistor space to solutions of field equations on the corresponding spacetime\footnote{More precisely, for an open subset $U\subset\MT$ the minitwsitor Penrose transform gives an isomorphism between cohomology classes and solutions to equations of motion. For $s\geq1/2$ it gives
		\begin{equation*}
			H^{1}\!\left(U,\cO(-s-1-m\,|\,-s-1+m)\right)
			\cong
			\left\{\nabla_{\h\,\alpha_1}^{\phantom{\h\alpha}\beta}\Psi_{\beta\alpha_2\ldots\alpha_{2s}}
			=-m\Psi_{\alpha_1\ldots\alpha_{2s}}\right\} \ ,
		\end{equation*}
		and, for $s=0$,
		\begin{equation*}
			H^1\!\left(U,\cO(-1-m\,|\,-1+m)\right)
			\cong
			\left\{\nabla_{\h\,\alpha\beta}\nabla_{\h}^{\alpha\beta}\Psi=-2(m^2-1)\Psi\right\} \ ,
		\end{equation*}
		where $H^1\!\left(U,\cO(p,q)\right)$ denotes the first Čech cohomology group of co-chains that can be thought of as holomorphic functions of scaling weight $(p,q)$. This identifies functions that differ by a coboundary. In the contour formula this is the statement that adding terms holomorphic on only one side of $\gamma$ gives no residue and therefore does not change the spacetime field.}  \cite{alma990162587480107026,tsai}. The required functions have the scaling property
	\begin{equation}
		f_s(u\mu^{\dot\alpha},v\lambda_\alpha)
		=u^{-s-1-m}v^{-s-1+m}f_s(\mu^{\dot\alpha},\lambda_\alpha),
		\qquad s\geq0\ .
		\label{eq:minitwistor_spin_scaling}
	\end{equation}
	For $s\geq\frac12$, holomorphic functions with this scaling are isomorphic to solutions of the massive spin-$s$ equation on hyperbolic space,
	\begin{equation}
		\nabla_{\h\,\alpha_1}^{\phantom{\h\alpha}\beta}\Psi_{\beta\alpha_2\ldots\alpha_{2s}}
		=-m\Psi_{\alpha_1\ldots\alpha_{2s}}\ ,
		\label{eq:minitwistor_spin_eom}
	\end{equation}
	where $	\nabla_{\h\,\alpha_1}^{\phantom{\h\alpha}\beta}$ is the covariant derivative of the hyperbolic space. For $s=0$, the holomorphic function is isomorphic to solutions of the scalar equation
	\begin{equation}
		\nabla_{\h\,\alpha\beta}\nabla_{\h}^{\alpha\beta}\Psi=-2\left(m^2-1\right)\Psi\ ,
		\label{eq:minitwistor_scalar_eom}
	\end{equation}
	which is equivalent to Eq.~\eqref{eq:ZerothCopy_m_3d}.
	
	In practice, we only use the Penrose transform in one direction, from the minitwistor function to the spacetime field. After identifying the undotted $SL(2,\mathbb{C})$ indices of the four-dimensional embedding space with those of the double cover of the complex three-dimensional Lorentz group, the resulting fields obey the equations entering the Cotton double copy of topologically massive theories. The integral form of the Penrose transform is
	\begin{equation}\label{penrosetransform}
		\Psi_{\alpha_1\ldots\alpha_{2s}}(X) = \cint\la\lambda\rd\lambda\ra\,\lambda_{\alpha_1}\ldots\lambda_{\alpha_{2s}}\,f_s(\mu^{\dot\alpha}_{X},\lambda_\alpha)\,,
	\end{equation}
	where $\la\lambda\rd\lambda\ra\equiv\lambda_{\alpha}\rd\lambda^{\alpha}=\epsilon^{\alpha\beta}\lambda_{\beta}\rd\lambda_{\alpha}$. All dependence on $X$ enters through $\mu_X^{\dot\alpha}$ in Eq.~\eqref{eq:incidence}; evaluating the result on the normalized representative $x^\mu$ in Eq.~\eqref{eq:CoordHyperbolic} gives the field on hyperbolic space.
	
	The minitwistor double copy is then a product of these holomorphic functions, directly parallel to the twistor-space correlator relations above. The spin-two function is written as
	\begin{equation}
		f_{s=2,  m_G}( \mu^{\dot\alpha},\lambda_{\alpha})
		=\frac{f_{s=1,  m_V}(\mu^{\dot\alpha},\lambda_\alpha) f'_{s=1,  m_V}(\mu^{\dot\alpha},\lambda_\alpha)}{f_{s=0,  m_\phi}( \mu^{\dot\alpha},\lambda_\alpha)} \ ,
		\label{eq:DCminitwistors}
	\end{equation}
	and the weights match provided
	\begin{equation}
		m_G= 2 m_V- m_\phi \ .
	\end{equation}
	Indeed, using Eq.~\eqref{eq:minitwistor_spin_scaling}, the right-hand side has the spin-two homogeneity
	\begin{align}
		f_{s=2,  m_G}(u \mu^{\dot\alpha},v\lambda_{\alpha})&=
		v^{ 2m_V- m_\phi-3} u^{ -( 2m_V- m_\phi)-3} f_{s=2,  2m_V- m_\phi}(\mu^{\dot\alpha},\lambda_\alpha) \ .
	\end{align}
	As in the correlator examples, this simple multiplication rule is a statement before the Penrose transform. The contour integral does not necessarily turn products of holomorphic minitwistor functions into spacetime fields satisfying a Cotton double copy. This only happens for type~N and type~D spacetimes. There is also a residual choice of function: adding terms with poles only on one side of the contour $\gamma$ leaves the transformed field unchanged, but can obscure the quotient form in Eq.~\eqref{eq:DCminitwistors}. This is because we are strictly working with cohomology class representatives as explained in the footnote above. Fixing this freedom intrinsically is still not understood in curved space. A second difference from the flat-background twistor double copy concerns the masses. In flat space the gluon and graviton masses are matched, consistently with the BCJ picture in which the propagator denominator is unchanged. In the AdS construction the homogeneity condition instead requires
	\begin{equation}\label{correctMass}
		m_G= m_V+\text{sign}( m_{G})= m_\phi + 2 \ \text{sign}( m_{G}) \ ,
	\end{equation}
	or equivalently in terms of the conformal dimensions\footnote{Irreducible representations of EAds$_3$ can be labeled in the $\mathfrak{sl}(2,\mathbb{C})$ basis by their holomorphic and anti-holomorphic conformal weights $h,\bar{h}$. The irreducible representations in the $\mathfrak{so}(1,3)$ and the $\mathfrak{sl}(2,\mathbb{C})$ basis are related as \cite{Kessel:2018zqm}
		$D(\Delta,s)=(\mathcal{D}_+(h), \mathcal{D}_-(\bar{h}))$
		where $\Delta=h+\bar{h}$, $s =\vert h-\bar{h} \vert$ and the helicity is given by
		$\eta=h-\bar{h}$.}
	\begin{equation}\label{correct}
		\Delta_{\pm,G}= \Delta_{\pm,V}\pm \text{sign}(\eta_{G})= \Delta_{\pm,\phi}\pm 2\  \text{sign}(\eta_{G})\ , 
	\end{equation}
	where $\eta_G$ is the helicity of the massive graviton whose sign is the same as that of the mass $m_G$. The conformal dimensions are related to the mass by
	\begin{equation}\label{deltam}
		\Delta_\pm= 1 \pm m_{(s)} \ .
	\end{equation}
	
	Under the AdS/CFT correspondence, the bulk AdS fields are dual to boundary operators. Given the prescription above, conserved stress energy tensors are the double copy of conserved currents, as expected. With this prescription, we have that for positive helicity modes the double copy relates solutions as  
	\begin{mybox}[Minitwistor Double Copy (AdS)]
		\begin{center}
			\renewcommand{\arraystretch}{1.8}
			\begin{tabularx}{\textwidth}{l X}
				\textbf{Gravity} &   \vspace{-1cm} 
				\begin{equation}
					f_{s=2,  {m}_\phi+2}=\frac{f_{s=1,  {m}_\phi+1} f'_{s=1,  {m}_\phi+1}}{f_{s=0,  {m}_\phi}} 
					\label{eq:twistor_gravity}
				\end{equation} \vspace{-1cm}  \\  
				\textbf{Yang--Mills} &   \vspace{-1cm} 
				\begin{equation}
					f_{s=1,  {m}_\phi+1}^a = c^a f_{s=1,  {m}_\phi+1}
					\label{eq:twistor_gauge}
				\end{equation}  \vspace{-1cm}  \\ 
				\textbf{Biadjoint Scalar} &   \vspace{-1cm} 
				\begin{equation}
					f_{s=0, {m}_\phi}^{a\, b} = c^a c'^{b}f_{s=0,  {m}_\phi}
					\label{eq:twistor_bas}
				\end{equation}  \vspace{-0.5cm} 
			\end{tabularx}
		\end{center}
	\end{mybox}
	Note that now we have allowed for the possibility of an asymmetric double copy involving two different spin one representatives $f$ and $f'$. This is equally valid thanks to the homogeneity of the representatives.
	
	\paragraph{Type~N waves}
	The AdS pp-waves reviewed in Subsection~\ref{subsec:cotton_ads_pp_waves} have a direct minitwistor realization \cite{Beetar:2024ptv}. A representative with a simple pole in $\la\lambda A\ra$ produces, through the Penrose transform, a field whose principal spinor $A_\alpha$ has multiplicity $2s$. For the mass branch $m_V=m_\phi-1$ and $m_G=m_\phi-2$, one may choose
	\be
	f_{s,m_\phi-s}
	=\frac{\la\lambda B\ra^{m_\phi-2s}}
	{\la\lambda A\ra[\mu C]^{m_\phi+1}}\ ,
	\qquad s=0,1,2\ ,
	\label{eq:minitwistor_typeN_representatives}
	\ee
	where $A_\alpha$, $B_\alpha$, and $C_{\dot\alpha}$ are constant spinors. The pole structure makes the double copy relation manifest,
	\be
	f_{2,m_\phi-2}=\frac{f_{1,m_\phi-1}^{2}}{f_{0,m_\phi}}\ .
	\label{eq:minitwistor_typeN_dc}
	\ee
	The corresponding spacetime fields have one repeated principal spinor and are therefore type~N. To make contact with the pp-wave of Subsection~\ref{subsec:cotton_ads_pp_waves}, introduce null Poincar\'e coordinates through the embedding-space matrix \cite{Beetar:2024ptv}
	\be
	X^{\alpha\dot\alpha}
	=\begin{pmatrix}
		1 & u\\
		v & y_0^2+uv
	\end{pmatrix}\ .
	\label{eq:minitwistor_ppwave_null_embedding}
	\ee
	Here $y_0$ is the radial coordinate. On the Euclidean real slice $v=\bar u$, whereas after analytic continuation $u$ and $v$ become independent Lorentzian null coordinates. In these coordinates, choose
	\be
	\la AB\ra=1\ ,\qquad A_0=1\ ,\quad A_1=0\ ,\qquad
	C_{\dot 0}=1\ ,\quad C_{\dot 1}=0\ .
	\label{eq:minitwistor_ppwave_spinor_choice}
	\ee
	Thus $A_\alpha$ is aligned with the repeated principal null direction, while $B_\alpha$ fixes its normalization through $\la AB\ra=1$. The Penrose transforms of Eq.~\eqref{eq:minitwistor_typeN_representatives} then reduce to
	\be
	\phi_0=-y_0^{m_\phi+1}\ ,\qquad
	f_{\alpha\beta}=-A_\alpha A_\beta y_0^{m_\phi+1}\ ,\qquad
	C_{\alpha\beta\gamma\delta}=-A_\alpha A_\beta A_\gamma A_\delta y_0^{m_\phi+1}\ .
	\label{eq:minitwistor_ppwave_fields}
	\ee
	Up to the overall normalizations used in Eq.~\eqref{eq:CottonDCNP}, these are the scalar, TMYM field strength, and Cotton spinor of the massive AdS pp-wave. A non-constant regular factor in the representative restores the arbitrary profile in the null coordinate $u$.
	
	\paragraph{Warped black holes}
	The warped black holes of Subsection~\ref{subsec:cotton_warped_black_holes} also have a direct minitwistor realization. Type~D requires two distinct repeated principal spinors, generated by a pair of third-order poles in the spin-2 representative. For the branch $(m_G,m_V,m_\phi)=(3,2,1)$ in AdS units, a convenient set is
	\be
	f_{2,3}=\frac{1}{[\mu A]^3[\mu B]^3}\ ,\qquad
	f_{1,2}=\frac{1}{[\mu A]^2[\mu B]^2}\ ,\qquad
	f_{0,1}=\frac{1}{[\mu A][\mu B]}\ ,
	\label{eq:minitwistor_typeD_representatives}
	\ee
	where $A_{\dot\alpha}$ and $B_{\dot\alpha}$ are constant and linearly independent. Using the spin-frame freedom, we may choose
	\be
	\begin{gathered}
		A_{\dot 0}=1\ ,\quad A_{\dot 1}=0\ ,\qquad
		B_{\dot 0}=0\ ,\quad B_{\dot 1}=\mathcal N\ ,\\
		\mathcal N\equiv[AB]
		=\left[6(-\bar\Lambda)^{1/2}(\Lambda-\bar\Lambda)\right]^{-1/3}\ .
	\end{gathered}
	\label{eq:minitwistor_typeD_spinor_frame}
	\ee
	Here $[AB]\equiv\epsilon^{\dot\alpha\dot\beta}A_{\dot\alpha}B_{\dot\beta}$. Other choices related by an $SL(2,\mathbb C)$ transformation give the same spacetime fields. The representatives satisfy
	\be
	f_{2,3}=\frac{f_{1,2}^{2}}{f_{0,1}}\ .
	\label{eq:minitwistor_typeD_dc}
	\ee
	Their Penrose transforms yield a type~D Cotton spinor, a TMYM field strength with the same two principal spinors, and the scalar zeroth copy. After matching the overall normalization, these fields reproduce the warped-AdS Cotton double copy in Eqs.~\eqref{eq:tracelesRicciBH}--\eqref{eq:typeDstarF}. The homogeneities implement the required relation between the distinct spin-2, spin-1, and scalar masses; they do not preserve a common mass. In this example, the two-pole structure also reproduces the type~D principal-spinor pattern, and the transformed fields are invariant under the spacelike Killing vector $l^\mu$, as in Eqs.~\eqref{eq:LieCotton} and \eqref{eq:LieCotton2}.
	
	\subsection{Convolutional Double Copy}
	
	The convolutional double copy \cite{Anastasiou:2018rdx,Anastasiou:2014qba,Godazgar:2022gfw,Beneke:2021ilf,Ferrero:2020vww,Borsten:2020xbt,Luna:2020adi,LopesCardoso:2018xes} is an off-shell, field-level realization of the schematic relation ``gravity = gauge theory squared.'' Rather than acting on a particular exact solution, it maps two gauge-fixed Yang--Mills BRST complexes to linearized gravitational fields and their gauge structure on a fixed background. Its central ingredient is a convolution product dressed by an inverse spectator biadjoint scalar, together with a derivative rule that allows Yang--Mills gauge transformations to descend to linearized diffeomorphisms. We first review the translation-invariant flat-space construction, and then explain how group convolution on homogeneous spaces and a mixed Mellin--spatial convolution in (A)dS replace it when ordinary spacetime translations are unavailable.
	
	\subsubsection{Flat-Space Convolution and the BRST Dictionary}\label{subsec:convolution_flat}
	
	In flat space, the convolution is the position-space counterpart of multiplication in momentum space, while the spectator biadjoint scalar contracts the two color algebras and supplies the inverse propagator needed for the correct mass dimension. The resulting dictionary includes the ghost and gauge-fixing sectors as well as the propagating gluon modes. Consequently, the linearized diffeomorphism symmetry and, in suitable gauges, the gravitational gauge condition are inherited from the Yang--Mills BRST transformations.
	
	Consider the YM BRST complex $(A_\mu^a,c^a,\bar c^a)$, where $A_\mu^a$ and $\widetilde A_\mu^{a'}$ denote two linearized Yang--Mills gauge fields, with ghosts and antighosts $(c^a,\bar c^a)$ and $(\widetilde c^{a'},\widetilde{\bar c}^{a'})$. The basic flat-space convolution of two fields is
	\be
	(f\star g)(x)=\int \dd^D y\,f(y)g(x-y)\ .
	\label{eq:flat_convolution_star}
	\ee
	It satisfies the important derivative property
	\be
	\partial_\rho(f\star g)=(\partial_\rho f)\star g=f\star(\partial_\rho g)\ ,
	\label{eq:flat_convolution_derivative_rule}
	\ee
	where the second equality follows by integrating by parts and discarding boundary terms. This is precisely the property that allows gauge transformations in either Yang--Mills factor to become diffeomorphisms in the double copy. To formulate a double copy, the color indices are removed by inserting a biadjoint scalar $\Phi^{aa'}$. In practice, one uses a convolution defined as
	\be
	X\circ Y\equiv X^a\star(\Phi^{-1})_{aa'}\star Y^{a'}\ ,
	\label{eq:circ_product_flat}
	\ee
	which in momentum space becomes an ordinary product,
	\be
	\widetilde{X\circ Y}(p)=\widetilde X^a(p)\,\widetilde{\Phi}^{-1}_{aa'}(p)\,\widetilde Y^{a'}(p)\ .
	\label{eq:circ_product_momentum}
	\ee

	At the linearized level the Yang--Mills BRST transformations take the form
	\be
	Q A_\mu=\partial_\mu c\ ,\qquad Qc=0\ ,\qquad Q\bar c=\frac{1}{\xi}G[A]\ ,
	\label{eq:YM_BRST_flat}
	\ee
	where $\xi$ is the gauge fixing parameter. The operator $Q$ is the nilpotent BRST differential which implements the linearized gauge transformations and $G[A]$ is the gauge-fixing functional. On the gravity side, the corresponding linearized fields are the graviton $h_{\mu\nu}$, the Kalb--Ramond two-form $B_{\mu\nu}$ and the dilaton $\varphi$, together with the diffeomorphism ghost $c_\mu$, the two-form ghost $d_\mu$, and the corresponding antighosts $\bar c_\mu$ and $\bar d_\mu$. Their linearized BRST transformations take the form
	\begin{align}
		Qh_{\mu\nu}&=2\partial_{(\mu}c_{\nu)}\ ,&
		QB_{\mu\nu}&=2\partial_{[\mu}d_{\nu]}\ ,&
		Q\varphi&=0\ ,\nonumber\\
		Qc_\mu&=0\ ,&
		Qd_\mu&=0\ ,\nonumber\\
		Q\bar c_\mu&=\frac{1}{\xi}G_\mu[h,\varphi]\ ,&
		Q\bar d_\mu&=\frac{1}{\xi}G_\mu[B]\ .
		\label{eq:gravity_BRST_flat}
	\end{align}
	The convolutional dictionary is arranged so that these transformations follow from Eq.~\eqref{eq:YM_BRST_flat} and the derivative rule in Eq.~\eqref{eq:flat_convolution_derivative_rule}.
	
	With the Lorenz gauge functional $G[A]=\partial^\nu A_\nu$, the full BRST dictionary in the one-parameter linear covariant gauge family is \cite{Anastasiou:2018rdx}
	\be
	\begin{aligned}
		h_{\mu\nu}={}&A_{(\mu}\circ\widetilde A_{\nu)}
		+a_1\frac{\partial_\mu\partial_\nu}{\Box}A^\rho\circ\widetilde A_\rho
		+a_2\frac{\partial_\mu\partial_\nu}{\Box}c^\alpha\circ\widetilde c_\alpha \\
		&+\frac{a_3}{\Box}\left(\partial A\circ\partial_{(\mu}\widetilde A_{\nu)}
		+\partial_{(\mu}A_{\nu)}\circ\partial\widetilde A\right)\\
		&+\eta_{\mu\nu}\left(b_1 A^\rho\circ\widetilde A_\rho
		+b_2 c^\alpha\circ\widetilde c_\alpha
		+\frac{b_3}{\Box}\partial A\circ\partial\widetilde A\right)\ ,\\
		B_{\mu\nu}={}&A_{[\mu}\circ\widetilde A_{\nu]}
		-\frac{1}{2\Box}\left(\partial A\circ\partial_{[\mu}\widetilde A_{\nu]}
		-\partial_{[\mu}A_{\nu]}\circ\partial\widetilde A\right)\ ,\\
		\varphi={}&A^\rho\circ\widetilde A_\rho+\frac{1}{\xi}c^\alpha\circ\widetilde c_\alpha
		+\left(\frac{1}{\xi^2}-1\right)\frac{1}{\Box}\partial A\circ\partial\widetilde A\ .
	\end{aligned}
	\label{eq:flat_bosonic_dictionary}
	\ee
	where $\partial A\equiv\partial^\rho A_\rho$, $\partial\widetilde A\equiv\partial^\rho\widetilde A_\rho$, and
	\be
	\begin{aligned}
		a_1&=\frac{1}{1-\xi}\ ,&
		b_1&=\frac{\xi}{(2-D)(\xi-1)}\ ,\\
		a_2&=\frac{1+\xi}{2(1-\xi)}\ ,&
		b_2&=\frac{b_1}{\xi}\ ,\\
		a_3&=-\frac{1}{2}\ ,&
		b_3&=\left(\frac{1}{\xi^2}-1\right)b_1\ .
	\end{aligned}
	\label{eq:flat_bosonic_dictionary_coefficients}
	\ee
	Here $c^\alpha=(c,\bar c)$ and
	\be
	c^\alpha\circ\widetilde c_\alpha=c\circ\widetilde{\bar c}-\bar c\circ\widetilde c
	\label{eq:ghost_singlet}
	\ee
	is the ghost singlet. The diffeomorphism and two-form ghosts are obtained by applying $Q$ to Eq.~\eqref{eq:flat_bosonic_dictionary}. At the same order,
	\be
	\begin{aligned}
		c_\mu={}&\frac{1}{4}\left[c\circ\widetilde A_\mu+A_\mu\circ\widetilde c
		-\frac{\xi+1}{\xi}\frac{\partial_\mu}{\Box}
		\left(c\circ\partial\widetilde A+\partial A\circ\widetilde c\right)\right]\ ,\\
		d_\mu={}&\frac{1}{4}\left[c\circ\widetilde A_\mu-A_\mu\circ\widetilde c
		-\frac{\xi+1}{\xi}\frac{\partial_\mu}{\Box}
		\left(c\circ\partial\widetilde A-\partial A\circ\widetilde c\right)\right]\ .
	\end{aligned}
	\label{eq:flat_ghost_dictionary}
	\ee
	Using Eq.~\eqref{eq:flat_convolution_derivative_rule} and the fact that the scalar dictionary is BRST closed, $Q\varphi=0$, one obtains the expected BRST transformations in Eq.~\eqref{eq:gravity_BRST_flat}. Thus the tensor product of the two Yang--Mills BRST systems reproduces the BRST complex of the NS--NS sector at linear order.
	
	The same construction also determines the gravitational gauge fixing. Applying $Q$ to the antighost dictionary, one reads off the gravity gauge functional from the Yang--Mills one. Choosing Lorenz gauge maps to de Donder gauge, $G_\mu[h]=\partial^\nu h_{\nu\mu}-\frac{1}{2}\partial_\mu h$.
	
	For applications to purely metric solutions it is useful to remove the two-form by taking the two Yang--Mills factors to be identical. A convenient Lorenz-gauge dictionary for the trace-reversed metric perturbation is then
	\be
	\bar h_{\mu\nu}=2A_\mu\circ A_\nu-\frac{2}{\Box}\partial_{(\mu}A_{\nu)}\circ\partial A\ ,\qquad \varphi=2A^\rho\circ A_\rho+4\xi c\circ\bar c\ ,
	\label{eq:lorenz_trace_reversed_dictionary}
	\ee
	where $\bar h_{\mu\nu}=h_{\mu\nu}-\frac{1}{2}\eta_{\mu\nu}h$. The ghost contribution is not optional if one wants a clean separation between the graviton and dilaton. For a point charge, the convolutional dictionary naturally gives the Janis--Newman--Winicour family; an appropriate choice of ghost data can set the dilaton charge to zero and recover the Schwarzschild perturbation in de Donder gauge \cite{Luna:2020adi}.
	
	A more general flat-space dictionary can be written for gauge-fixing functionals of the form
	\be
	G[A]=n^\mu A_\mu\ ,
	\label{eq:general_gauge_fixing_flat}
	\ee
	where $n^\mu$ can be a differential operator or a constant vector, as in axial or temporal gauge. The dictionary can then contain dimensionless combinations built from $A_\mu$, $nA$, $\partial A$, ghosts, and inverse operators such as $(n\partial)^{-1}$ and $\Box^{-1}$, whose role is to ensure the BRST transformation on the gravitational side, Eq.~\eqref{eq:gravity_BRST_flat}. Temporal gauge, $n^\mu=(1,0,0,0)$, is especially important in curved-space extension because it separates the time direction from the spatial convolution and is compatible with the conformal slicing used for de Sitter space \cite{Liang:2023zxo}.
	
	\subsubsection{Convolutional double copy on homogeneous spaces}\label{subsec:convolution_homogeneous_spaces}
	
	The flat-space construction relies on two related facts: the convolution product is defined by the translation group, and derivatives can be moved from the convolution to either factor. On a generic curved space there is no global translation group, and there is also no canonical way to multiply tensor fields at different points. In \cite{Borsten:2019prq,Borsten:2021zir}, it was shown that homogeneous spaces provide a controlled replacement for translations which gives rise to a well-defined convolutional double copy.
	
	Let $M=G/H$ be a compact Riemannian homogeneous space, where the compact Lie group $G$ acts transitively on $M$ and $H$ is the stabilizer of a reference point $\eta\in M$. The projection
	\be
	\pi:G\rightarrow M\ ,\qquad \pi(g)=g\eta
	\label{eq:homogeneous_projection}
	\ee
	realizes $G$ as a principal $H$-bundle over $M$. Thus points of $M$ are equivalence classes $gH$. Fields on $M$ can be pulled back to $G$, convolved using the group structure, and then projected back to $M$. The projection back to $M$ removes the dependence on the fiber $H$.
	
	The relevant geometric point is that tensor fields at different points cannot be multiplied before their indices are compared. On the group manifold this comparison is provided by left and right translations. For scalar functions on $G$ there are two natural group convolutions,
	\be
	(f\ast_{RL}\tilde f)(g')=\int_G\dd g\, f(g)\tilde f(g^{-1}g')\ ,\qquad
	(f\ast_{LR}\tilde f)(g')=\int_G\dd g\, f(g)\tilde f(g'g^{-1})\ ,
	\label{eq:group_function_convolution_L_R}
	\ee
	where $\dd g$ is the normalized Haar measure. 
	
	For tensor fields one must also specify how indices at different group elements are compared before the fields are multiplied. Let $S$ be a covariant rank-$p$ tensor and $T$ a covariant rank-$q$ tensor on $G$, so that in components they have $p$ and $q$ lower indices. The mixed tensor convolutions are
	\be
	S\otimes_{RL}T=\int_G\dd g\,\mathfrak r_g(S)\otimes L_{g^{-1}}^*T\ ,\qquad
	S\otimes_{LR}T=\int_G\dd g\,\mathfrak l_g(S)\otimes R_{g^{-1}}^*T\ .
	\label{eq:group_tensor_convolution_RL_LR}
	\ee
	Here $L_g$ and $R_g$ denote left and right multiplication on $G$, while $L_g^*$ and $R_g^*$ are the corresponding pullbacks acting on tensor indices. The operations $\mathfrak r_g$ and $\mathfrak l_g$ move the first tensor using right- and left-invariant frames, respectively. Thus the formula is the group-theoretic version of transporting tensor indices to the same point before multiplying the component functions.
	
	For one-forms this becomes particularly explicit. If $r^a$ and $l^a$ are right- and left-invariant one-form bases, write $\omega=(\omega_R)_a r^a=(\omega_L)_a l^a$ and similarly for $\tilde\omega$, then
	\be
	\omega\otimes_{RL}\tilde\omega=
	\big((\omega_R)_a\ast_{RL}(\tilde\omega_L)_b\big)r^a\otimes l^b\ ,\qquad
	\omega\otimes_{LR}\tilde\omega=
	\big((\omega_L)_a\ast_{LR}(\tilde\omega_R)_b\big)l^a\otimes r^b\ .
	\label{eq:one_form_tensor_convolution_components}
	\ee
	Thus the tensor convolution reduces to an ordinary scalar convolution of components, provided the components are taken in the invariant frames appropriate to the $RL$ or $LR$ prescription. The symmetric convolution, denoted by $\vee_{AB}$, is obtained by replacing the tensor product in Eq.~\eqref{eq:group_tensor_convolution_RL_LR} by the symmetric tensor product.
	
	The convolution on the homogeneous space $M=G/H$ is defined by pulling fields back to $G$, applying the group convolution, and then projecting back to $M$. The only subtlety is that the pullback to $G$ contains extra directions along the fiber $H$, which do not correspond to directions on $M$. These are removed by projecting onto the horizontal directions and averaging over the fiber. Denoting this projection by $\pi_H$, the homogeneous-space convolution is
	\be
	S\otimes_{AB}T=\pi_H\!\left(\pi^*S\otimes_{AB}\pi^*T\right)\ ,\qquad AB=RL,LR\ .
	\label{eq:homogeneous_tensor_convolution}
	\ee
	Here $\pi^*$ lifts the tensors from $M$ to $G$, $\otimes_{AB}$ performs the group convolution with the appropriate transport of tensor indices, and $\pi_H$ projects the result back to a tensor on $M$.
	
	The derivative property needed for the double copy involves the symmetrized covariant derivative rather than the exterior derivative. For a function $f$, it is simply
	\be
	\nabla f=df\ ,
	\label{eq:sym_cov_deriv_function}
	\ee
	while for a one-form $\omega$ it is
	\be
	(\nabla\omega)(X,Y)=(\nabla_X\omega)(Y)+(\nabla_Y\omega)(X)\ ,
	\label{eq:sym_cov_deriv_one_form}
	\ee
	which is the coordinate-free version of $2\nabla_{(\mu}\omega_{\nu)}$. For functions and one-forms, the homogeneous-space convolution satisfies the required non-Leibniz derivative rules
	\be
	\nabla(f\vee_{AB}\tilde f)=\nabla f\vee_{AB}\tilde f=f\vee_{AB}\nabla\tilde f\ .
	\label{eq:homogeneous_derivative_rule_functions}
	\ee
	These identities are the homogeneous-space analogue of Eq.~\eqref{eq:flat_convolution_derivative_rule}. They are special to the convolution built from the group action and to the symmetrized covariant derivative. The exterior derivative does not satisfy an equally useful rule for higher-degree forms on a general group manifold, which is why the generic homogeneous-space construction naturally reproduces the spin-two diffeomorphism sector, while the full Kalb--Ramond sector requires additional structure.
	
	The double copy can now be formulated on an ultrastatic spacetime
	\be
	\widehat M=\mathbb R\times M\ ,\qquad \dd s^2=-\dd t^2+g_{ij}(x)\dd x^i\dd x^j\ ,
	\label{eq:ultrastatic_homogeneous_background}
	\ee
	where $M=G/H$. A spacetime one-form is decomposed as
	\be
	\widehat A=A_0\,\dd t+A_1\ ,
	\label{eq:spacetime_one_form_split}
	\ee
	where, at fixed time, $A_0[t]\in\Omega^0(M)$ is a scalar on the spatial manifold and $A_1[t]\in\Omega^1(M)$ is a spatial one-form. The spacetime convolution is ordinary convolution in time combined with the homogeneous-space convolution in space. For two functions,
	\be
	(\widehat f\,\widehat\vee_{AB}\widehat g)[t]=\int_{-\infty}^{\infty}\dd t'\, f[t']\vee_{AB}g[t-t']\ .
	\label{eq:spacetime_homogeneous_convolution_functions}
	\ee
	and for two one-forms
	\be
	(\widehat\omega\,\widehat\vee_{AB}\widehat{\tilde\omega})_{i+j}[t]
	=\int_{-\infty}^{\infty}\dd t'\,\omega_{(i}[t']\vee_{AB}\tilde\omega_{j)}[t-t']\  . 
	\label{eq:spacetime_homogeneous_convolution_one_forms}
	\ee
	Here $i,j=0,1$, with $0$ denoting the time component and $1$ denoting the spatial component. The derivative rule on $\widehat M$ follows from the spatial identities in Eq.~\eqref{eq:homogeneous_derivative_rule_functions} together with integration by parts in the time convolution.
	
	At the linearized level the Yang--Mills BRST transformations on the ultrastatic background are
	\be
	Q\widehat A=\widehat\nabla c\ ,\qquad Qc=0\ ,
	\label{eq:homogeneous_YM_BRST}
	\ee
	where $\widehat\nabla c=\dd c$. The spin-two double copy dictionary is then
	\be
	h=\widehat A\ \widehat\vee_{AB}\ \widehat{\widetilde A}\ ,\qquad
	c_{\rm grav}=c\widehat\vee_{AB}\widehat{\widetilde A}
	+\widehat A\widehat\vee_{AB}\widetilde c\ ,
	\label{eq:homogeneous_graviton_ghost_dictionary}
	\ee
	where $c_{\rm grav}$ is the linearized diffeomorphism ghost. Applying $Q$ and using the homogeneous-space derivative rule gives
	\be
	Qh=\widehat\nabla\!\left(c\widehat\vee_{AB}\widehat{\widetilde A}
	+\widehat A\widehat\vee_{AB}\widetilde c\right)
	=\widehat\nabla c_{\rm grav}\ .
	\label{eq:homogeneous_Qh_result}
	\ee
	Similarly,
	\be
	Q\xi=Q\!\left(c\widehat\vee_{AB}\widehat{\widetilde A}
	+\widehat A\widehat\vee_{AB}\widetilde c\right)
	=-c\widehat\vee_{AB}\widehat\nabla\widetilde c+\widehat\nabla c\widehat\vee_{AB}\widetilde c=0\ ,
	\label{eq:homogeneous_Qxi_zero}
	\ee
	again by the same derivative rule. Thus the symmetric convolution of the two Yang--Mills gauge fields produces the correct linearized diffeomorphism BRST transformation of the graviton on the fixed homogeneous background.
	
	This construction should therefore not be viewed as a naive covariantization of the flat dictionary. The squaring procedure itself changes: it depends on the homogeneous-space structure $M=G/H$. The construction applies naturally to ultrastatic spacetimes with compact homogeneous spatial slices, such as the Einstein static universe, and gives a field-level double copy for linearized gravity on these fixed curved backgrounds.
	
	\paragraph{Sphere case}\label{subsubsec:sphere_convolution_case}
	
	The simplest non-trivial example is the two-sphere \cite{Borsten:2019prq}, $S^2\simeq SO(3)/SO(2)$. Scalar functions are expanded in spherical harmonics,
	\be
	f(\theta,\phi)=\sum_{\ell=0}^{\infty}\sum_{m=-\ell}^{\ell}f_\ell^mY_\ell^m(\theta,\phi)\ ,\qquad f_\ell^m=\int_{S^2}\dd\Omega\,f\,\overline{Y_\ell^m}\ .
	\label{eq:sphere_harmonic_transform}
	\ee
	The spherical convolution is inherited from the group convolution on $SO(3)$, using the fact that the spherical harmonics are matrix elements of Wigner $D$-matrices. For one-forms one uses a spin-weighted harmonic basis, so that the components of the gauge potential are convolved mode by mode with the correct transformation under the stabiliser $SO(2)$. In this language the convolution becomes algebraic in harmonic space, just as the flat convolution becomes algebraic in momentum space. For example, the scalar convolution factorizes as
	\be
	(k\cdot f)_\ell^m
	=
	2\pi\sqrt{\frac{4\pi}{2\ell+1}}\,
	k_\ell^m f_\ell^0\ .
	\label{eq:sphere_scalar_convolution_factorization}
	\ee
	This asymmetry arises from choosing a left convolution on the group.
	
	Adding a trivial time direction gives the $D=1+2$ Einstein static universe. The double copy dictionary then maps two Yang--Mills BRST systems on $\mathbb{R}\times S^2$ to the linearized diffeomorphism BRST transformation of the graviton on the same fixed background.
	
	\subsubsection{Convolutional double copy in (A)dS}\label{subsec:convolution_ads_ds}
	De Sitter and anti-de Sitter require a different treatment than the homogeneous spaces described above, which was formulated in \cite{Liang:2023zxo}. In the conformally flat slicing of de Sitter,
	\be
	\dd s^2=a(\eta)^2\left(-\dd\eta^2+\dd\mathbf{x}^2\right)\ ,\qquad a(\eta)=-\frac{1}{H\eta}\ ,
	\label{eq:dS_conformal_metric}
	\ee
	spatial translations remain manifest, but time translations are replaced by the scaling symmetry of the conformal-time coordinate. This motivates a mixed transform: Fourier transform in the spatial directions and Mellin transform in conformal time, following the same scale-adapted logic as in the Mellin and Mellin--Barnes representations reviewed above.
	
	For a field $f(\eta,\mathbf{x})$, one considers the Mellin-Fourier transform
	\be
	f(\eta,\mathbf{x})=\int_{\mathcal C}\frac{\dd s}{2\pi i}\int\frac{\dd^3\mathbf{k}}{(2\pi)^3}\,\eta^{-2s+3/2}e^{i\mathbf{k}\cdot\mathbf{x}}\widetilde f(s,\mathbf{k})\ ,
	\label{eq:mellin_fourier_transform}
	\ee
	where $\mathcal C$ is the Mellin contour. The convolution used in \cite{Liang:2023zxo} is the ordinary spatial convolution combined with the Mellin convolution in the scale variable,
	\be
	(f\star g)(\eta,\mathbf{x})=
	\int_0^\infty\frac{\dd\tau}{\tau}\int\dd^3\mathbf{y}\,
	f(\tau,\mathbf{y})\,g\!\left(\frac{\eta}{\tau},\mathbf{x}-\mathbf{y}\right).
	\label{eq:dS_mellin_spatial_convolution}
	\ee
	With the transform in Eq.~\eqref{eq:mellin_fourier_transform}, this definition implies
	\be
	\widetilde{f\star g}(s,\mathbf{k})=\widetilde f(s,\mathbf{k})\,\widetilde g(s,\mathbf{k})\ .
	\label{eq:dS_convolution_mellin_fourier_product}
	\ee
	The spectator-dressed product is then defined as in Eq.~\eqref{eq:circ_product_flat}, but with the convolution in Eq.~\eqref{eq:dS_mellin_spatial_convolution}, such that
	\be
	\widetilde{X\circ Y}(s,\mathbf{k})=\widetilde X^a(s,\mathbf{k})\,\widetilde\Phi^{-1}_{aa'}(s,\mathbf{k})\,\widetilde Y^{a'}(s,\mathbf{k})\ .
	\label{eq:dS_circ_product}
	\ee
	
	A useful gauge choice is temporal gauge, $A_0=0$ and $h_{0\mu}=0$. In de Sitter the covariant derivative of a spatial vector mixes temporal and spatial components, while imposing temporal gauge and the residual condition $c_0=0$ reduces the spatial BRST transformations to the same form as the flat temporal-gauge transformations discussed above. The flat temporal-gauge dictionary can therefore be extended by replacing the flat convolution with Eq.~\eqref{eq:dS_mellin_spatial_convolution}. With identical Yang--Mills factors one obtains
	\be
	\begin{aligned}
		h_{ij}={}&
		\mathfrak{a}_{1}A_i\circ A_j
		+2\mathfrak{a}_{123}\xi\frac{\partial_i\partial_j}{\partial_0}c\circ\bar c
		-2\mathfrak{a}_{123}\delta_{ij}\xi\,\partial_0c\circ\bar c\ ,\\
		\varphi={}&
		\mathfrak{b}_{1}H^2\eta^2
		\left(\delta_{ij}\left[A_i\circ A_j\right]
		-\frac{1}{\partial_i\partial_i}
		\left[\partial_kA_k\circ\partial_mA_m\right]\right)
		-2\mathfrak{b}_{2}\xi\,\partial_0c\circ\bar c\ ,
	\end{aligned}
	\label{eq:dS_temporal_dictionary}
	\ee
	where $\partial_0\equiv\partial_\eta$, $\partial_i\partial_i$ is the spatial Laplacian, $\mathfrak{a}_{123}\equiv\mathfrak{a}_{1}+\mathfrak{a}_{2}+\mathfrak{a}_{3}$, and the constants $\mathfrak{a}_{1}$, $\mathfrak{a}_{123}$, $\mathfrak{b}_{1}$, and $\mathfrak{b}_{2}$ parametrize the remaining freedom of the temporal-gauge dictionary. 
	
	In this gauge the de Sitter Schwarzschild perturbation is obtained from the convolutional double copy of a point-charge Maxwell solution on the de Sitter background. In Mellin--Fourier space the temporal-gauge gauge field and the corresponding biadjoint scalar are
	\be
	\widetilde A_i^a(s,\mathbf{k})
	=4\pi \,\frac{k_i}{\mathbf{k}^2}\,
	2\pi\delta\!\left(2s-\frac{1}{2}\right)\alpha^a\ ,\qquad
	\widetilde\Phi^{aa'}(s,\mathbf{k})
	=\frac{4\pi}{\mathbf{k}^2}\,
	2\pi i\left(2s-\frac{1}{2}\right)\delta^{aa'}\ ,
	\label{eq:dS_schwarzschild_mellin_fourier_gauge_scalar}
	\ee
	where $\alpha^a$ is a normalized color vector and the source is a unit charge. Choosing $\mathfrak{a}_{123}=0$, the double copy gives
	\be
	\widetilde h_{ij}(s,\mathbf{k})
	=\mathfrak{a}_{1}\widetilde{A_i\circ A_j}(s,\mathbf{k})
	=-\mathfrak{a}_{1}2\pi i\,\delta\!\left(2s-\frac{1}{2}\right)\,4\pi\frac{k_i k_j}{\mathbf{k}^2}\ ,
	\label{eq:dS_schwarzschild_mellin_fourier_convolution}
	\ee
	which is equal to the dS--Schwarzschild result
	\be
	\widetilde h_{ij}(s,\mathbf{k})
	=-\frac{r_s}{3H}\,2\pi i\,\delta\!\left(2s-\frac{1}{2}\right)
	4\pi\frac{k_i k_j}{\mathbf{k}^2}\ ,
	\label{eq:dS_schwarzschild_mellin_fourier_hij}
	\ee
	where $r_s$ is the Schwarzschild radius, if we take $\mathfrak{a}_{1}=\frac{r_s}{3H}$. Similarly, for the longitudinal point-charge field $A_i\propto k_i/\mathbf{k}^2$, the bracket in the dilaton dictionary vanishes in Mellin--Fourier space. Choosing $\mathfrak{b}_{2}=0$ then removes the ghost term and sets the dilaton to zero \cite{Liang:2023zxo}.
	
	The AdS construction is analogous in the Poincar\'e patch: one Mellin transforms in the radial coordinate $z$, works in the axial gauge $A_z=0$ and $h_{z\mu}=0$, and obtains the same temporal-gauge dictionary with $\partial_0$ replaced by $\partial_z$ and the spatial Kronecker delta replaced by the flat metric on the remaining coordinates \cite{Liang:2023zxo}.

	The main lesson is that the convolutional double copy is most robust when the background supplies enough symmetry to replace the flat translation group. Flat space uses Fourier modes, compact homogeneous spatial slices use group or harmonic modes, and the conformally flat (A)dS construction uses Mellin--Fourier modes. In all cases the BRST complex is the organizing principle: the gravitational gauge symmetry is not imposed after the fact, but is generated from the two Yang--Mills gauge symmetries through the convolution product.
	
	\subsection{Other Classical Double Copies and Related Extensions}
	
	Several extensions of the constructions above have a less direct relation to the standard amplitudes double copy. One group consists of solution-generating maps for modified gravity theories, including Chern--Simons gravity and bigravity \cite{Garcia-Compean:2024uie,Garcia-Compean:2024zze,Liu:2024byr}, and for theories with nonlinear or higher-derivative electrodynamic corrections \cite{Pasarin:2020qoa,Mkrtchyan:2022ulc,Lescano:2021ooe}. A related non-perturbative construction maps restricted ansatz sectors of quartic BAS theory, complexified $SU(2)$ Yang--Mills theory, and conformally flat gravity \cite{Armstrong-Williams:2026wzi}. These examples provide solution-generating relations, but a general color--kinematics interpretation has not yet been established. 
	
	Kerr--Schild methods have also been adapted to Double Field Theory and Exceptional Field Theory \cite{Lee:2018gxc,Angus:2021zhy,Cho:2021nim,Berman:2020xvs,Kim:2019jwm,Cho:2019ype}, where the enlarged field content and duality symmetries change the natural notion of a gravitational field. In Double Field Theory, the	generalized Kerr--Schild ansatz has further been extended to $\mathcal{N}=1$ supersymmetry and to higher-derivative heterotic theories, including corresponding higher-derivative corrections to the classical double copy, as well as supersymmetric $\alpha'$ corrections \cite{Lescano:2020nve,Rodriguez:2025hiu,Lescano:2021ooe}.
	
	\subsubsection{Background ambiguity and generalized classical maps}
	For Kerr--Schild metrics, the decomposition into a background and a Kerr--Schild profile usually supplies a natural spacetime on which to interpret the single copy. Outside this class, however, a gauge field constructed from the gravitational data may admit inequivalent curved- and flat-background interpretations. This distinction was investigated for the G\"odel universe in \cite{Kent:2025godel}. The G\"odel metric has no geodesic Kerr--Schild representation, but it is of Petrov type~D and therefore admits a Weyl single copy field. Contrary to standard double copy interpretations, \cite{Kent:2025godel} highlights that the derived single copy gauge fields can be naturally interpreted as living in the original curved gravitational background, see also Sec.~\ref{sec:unfolded}. The resulting G\"odel single copy gauge field naturally lives on the curved G\"odel geometry, inherits its characteristic homogeneity and rotation, and satisfies sourced Maxwell equations on that curved background, with the source related to the Ricci tensor. To interpret the single copy as living in a flat background, given the lack of a Kerr--Schild split, the proposal is instead to take the flat space limit of the metric in the single copy. In this flat-limit interpretation, the gauge field instead obeys vacuum Maxwell equations. 
	
	A separate flat-background single copy candidate can be proposed using the spacetime symmetries, but it neither follows from a controlled flat limit nor preserves all of the defining properties captured by the curved-space field. This example shows that the background for the single copy can change its interpretation. For the purposes of defining a double copy, in the spirit of the rest of the review, this gives an example where a classical double copy prescription does not seem to hold. While this focused on a flat background limit, similar issues may arise for sourced type-D solutions that are asymptotically (A)dS but do not admit a geodesic Kerr--Schild decomposition over the corresponding maximally symmetric background, as occurs, for example, for scalar-hairy solutions such as the Martínez--Troncoso--Zanelli black hole \cite{Martinez:2004nb}. In such cases, it is likely that an appropriate single copy in the (A)dS background cannot be found.
	
	A related proposal \cite{Kent:2025pvu} seeks to organize known metric-level maps and extend them beyond algebraically special spacetimes. It begins with a metric perturbation $h_{\mu\nu}$ about $\bar g_{\mu\nu}$ and a covariantly constant Killing vector $\xi^\mu$ of the background whose dual one-form is exact, $\bar g_{\mu\nu}\xi^\nu=\partial_\mu\lambda$. The contractions
	\be
	A_\mu = h_{\mu\nu}\,\xi^\nu  \ , \quad  \Phi=h_{\mu\nu}\xi^\mu\xi^\nu \ .
	\ee
	produce an Abelian gauge field $A_\mu$ satisfying Maxwell equations on both $g_{\mu\nu}$ and the base metric $\bar g_{\mu\nu}$ and a scalar $\Phi$ satisfying the corresponding scalar wave equation, interpreted as the linearized zeroth-copy equation, on $\bar g_{\mu\nu}$. For a Kerr--Schild metric these contractions reduce to the usual prescription. As a proof of concept, the construction gives a flat-space single copy for the Kasner metric and a corresponding curvature-level factorization, thereby extending this type of relation to an algebraically general Petrov type~I spacetime. The result provides evidence that exact classical gauge--gravity maps need not be restricted to algebraically special geometries. The construction depends on the chosen background and Killing vector and has not been connected to a scattering amplitudes double copy or shown to arise from a realization of color-kinematics replacements. It is therefore best regarded as a proposed organizing and solution-generating framework rather than as a universal classical double copy.
	
	It is also worth mentioning that electric--magnetic duality \cite{Huang:2019cja,Banerjee:2019saj} and the construction of Einstein--Maxwell or Einstein--Yang--Mills solutions have been studied in \cite{Easson:2022zoh,Alawadhi:2019urr}. These cases are more than a direct vacuum replacement because gravitational and gauge fields appear simultaneously on the gravity side.
	
	\subsubsection{Unfolded AdS Type D Construction} \label{sec:unfolded}
	
	Note that an earlier realization of the Kerr--Schild and Weyl double copy in AdS for Type D solutions was constructed in a different context in \cite{Didenko:2008va}. The authors of \cite{Didenko:2008va} construct a first-order differential system (unfolded) of equations for AdS black holes which makes symmetries manifest and integrability automatic. This generalizes the Cartan formalism by including $p$-form gauge fields. In tensor language the minimal unfolded system for a Killing vector $V^{a}$ in $\mathrm{AdS}_{4}$ reads
	\begin{align}
		D V^{a}&=\kappa^{a}{}_{b}\,e^{b},
		&
		D\kappa^{ab}&=-\frac{2}{3}\,\Lambda\,V^{[a}e^{b]},
		&
		R^{ab}&=\frac{\Lambda}{3}\,e^{a}\wedge e^{b},
		\label{eq:Killing}
	\end{align}
	where the two-form
	$\kappa_{ab}:=\nabla_{a}V_{b}-\nabla_{b}V_{a}$ is the Papapetrou/Killing two-form~\cite{Papapetrou:1966zz}, $e^{a}$ is the vierbein one form, $R^{ab}=\dd\Omega^{ab} +\Omega^{a}{}_{c}\wedge\Omega^{cb}$, and
	$D$ is the AdS covariant exterior derivative. These equations can be rewritten in spinor language as we will do below. In \cite{Didenko:2008va}, the authors introduced a mass deformation $M$ to describe Petrov type D spacetimes with a Weyl spinor
	\begin{equation}
		C_{\alpha\beta\gamma\delta}=f\,\kappa_{\alpha\beta}\kappa_{\gamma\delta} \ ,
	\end{equation}
    where $f$ depends on $M$. The resulting black-hole unfolded system for such spacetimes in spinor notation becomes
	\begin{align}
		\begin{aligned}
			\mathcal{D}V_{\alpha\dot\alpha}
			&=\tfrac12\,e^{\gamma}{}_{\dot\alpha}\,\kappa_{\gamma\alpha}
			+\tfrac12\,e_{\alpha}{}^{\dot\gamma}\,\bar\kappa_{\dot\gamma\dot\alpha},
			\\[6pt]
			\mathcal{D}\kappa_{\alpha\alpha}
			&=-\frac{\Lambda}{3}\,e_{\alpha}{}^{\dot\gamma}V_{\alpha\dot\gamma}
			+\frac{f}{4}\,e^{\beta\dot\beta}
			V^{\beta}{}_{\dot\beta}\,\kappa_{\beta\beta\alpha\alpha},
			\\[6pt]
			\mathcal{D}\bar\kappa_{\dot\alpha\dot\alpha}
			&=-\frac{\Lambda}{3}\,e^{\gamma}{}_{\dot\alpha}V_{\gamma\dot\alpha}
			+\frac{\bar f}{4}\,e^{\beta\dot\beta}
			V_{\beta}{}^{\dot\beta}\,\bar\kappa_{\dot\beta\dot\beta\dot\alpha\dot\alpha},
			\\[6pt]
			R_{\alpha\alpha}
			&=-\frac{\Lambda}{6}\,H_{\alpha\alpha}
			+\frac{f}{8}\,H^{\beta\beta}\,\kappa_{\beta\beta\alpha\alpha},
			\\[6pt]
			\bar R_{\dot\alpha\dot\alpha}
			&=-\frac{\Lambda}{6}\,\bar H_{\dot\alpha\dot\alpha}
			+\frac{\bar f}{8}\,\bar H^{\dot\beta\dot\beta}\,
			\bar\kappa_{\dot\beta\dot\beta\dot\alpha\dot\alpha},
			\\[6pt]
			\mathcal{D}e_{\alpha\dot\alpha}&=0,
			\\[6pt]
			\dd f
			&=-\Bigl(\tfrac{1}{12}f^{2}
			-\tfrac{5f\Lambda}{6\kappa^{2}}\Bigr)\,
			e^{\alpha\dot\gamma}V^{\alpha}{}_{\dot\gamma}\kappa_{\alpha\alpha},
			\\[6pt]
			\dd\bar f
			&=-\Bigl(\tfrac{1}{12}\bar f^{2}
			-\tfrac{5\bar f\Lambda}{6\bar\kappa^{2}}\Bigr)\,
			e^{\gamma\dot\alpha}V_{\gamma}{}^{\dot\alpha}\bar\kappa_{\dot\alpha\dot\alpha} \ ,
		\end{aligned}
		\label{eq:BHUS}
	\end{align}
	where $\mathcal{D}$ is the full covariant exterior derivative, not just the AdS covariant exterior derivative. Here, repeated same-letter spinor pairs use symmetric-spinor shorthand, e.g. $\kappa_{\alpha\alpha}\equiv\kappa_{(\alpha_1\alpha_2)}$, not a contraction; the four-index shorthand such as $\kappa_{\beta\beta\alpha\alpha}$ denotes the corresponding symmetric product.  The latter equations are consequences of the Bianchi identities for $R_{\alpha\alpha}$ and $\bar R_{\dot\alpha\dot\alpha}$ and have a general solution of the form
	\begin{equation}
		f \;=\; 6M\,\mathcal G^{3}/\kappa^{2} \ ,
	\end{equation}
	where $\mathcal G$ is the scalar entering this unfolded system; it is defined implicitly by $M\mathcal G^3-\mathcal G\sqrt{-\kappa^{2}}=-\Lambda/3$ and satisfies 
	\be
	d\mathcal G = -\frac{\mathcal G^2}{2 \sqrt{-\kappa^2}} \, e^{\alpha \dot{\alpha}} \, {V^{\alpha}}_{\dot{\alpha}} \, \kappa_{\alpha \alpha} 
	\ee
	These spacetimes can be written in Kerr--Schild form as 
	\begin{equation}
		g_{mn}=\bar g_{mn}+M(\mathcal G+\bar{\mathcal G})k_{m}k_{n}
	\end{equation}
	with $\bar g_{mn}$ the AdS background and $k_m$ the null geodesic Kerr--Schild vector obtained from projecting the Killing vector using 
	$\Pi^{\pm}_{\alpha}{}^{\beta}=
	\tfrac12(\delta_{\alpha}{}^{\beta}\pm\kappa_{\alpha}{}^{\beta}/\sqrt{-\kappa^{2}})$ as
	\[
	k_{\alpha\dot\alpha}=
	\frac{2V^{-}_{\alpha\dot\alpha}}
	{V^{-}\!\cdot V^{+}} \ ,
	\]
	where $ V^{-}\!\cdot V^{+}:=
	V^{-}_{\gamma\dot\gamma}V^{+\gamma\dot\gamma}$ and $V^{\pm}_{\alpha\dot\alpha}\equiv \Pi^{\pm}_{\alpha\beta}V^{\beta}{}_{\dot\alpha}$.
	
	In this context, the single copy arises by defining the anti-self-dual two-form
	\begin{equation}
		F_{\alpha\beta}
		:=\frac{\mathcal G^2}{\sqrt{-\kappa^{2}}}\,\kappa_{\alpha\beta} \ ,
		\label{eq:Fdef}
	\end{equation}
	with $F=\dd A$, and $A$ a potential aligned with the null congruence,
	\begin{equation}
		A_{m}=\tfrac12(\mathcal G+\bar{\mathcal G})k_{m} \ ,
	\end{equation}
	which resembles the single copy with $\phi\propto\mathcal G+\bar{\mathcal G}$. This two-form is related to the Killing--Yano tensor \cite{KYano}, with spinor components $Y_{\alpha \alpha}$, by
	\be
	Y_{\alpha \alpha} = \frac{i}{\mathcal G^3} F_{\alpha \alpha} \ .
	\ee
	Physically, a Killing--Yano tensor encodes a hidden symmetry of the spacetime that generates extra conserved quantities. Thus, the relation above identifies the Maxwell single copy with the same geometric two-form that encodes the hidden symmetry.
	
	Eqs.~\eqref{eq:BHUS} show that this two-form obeys the source-free Maxwell equations 
	\begin{equation}
		\mathcal{D}_{\gamma \dot{\alpha}} F_{\alpha}^{\ \gamma} = 0, \qquad \mathcal{D}_{\alpha \dot{\gamma}} \bar{F}^{\dot{\alpha} \dot{\gamma}} = 0 \ .
	\end{equation}
	Equivalently, in tensor notation the same field obeys $D_mF^m{}_{n}=\mathcal{D}_mF^m{}_{n}=0$, with $D$ the AdS-background derivative and $\mathcal{D}$ the full black-hole derivative.
	The electromagnetic field also shares the principal null direction of the Weyl tensor, and thus
	\begin{equation}
		C_{\alpha\beta\gamma\delta}
		\;=\;
		\frac{-6 M}{\mathcal G}\,
		F_{(\alpha\beta}F_{\gamma\delta)}\ ,
		\label{eq:weyl-vs-F}
	\end{equation}
	which resembles the Weyl double copy with $S\propto\mathcal G$. The difference with respect to the double copy scenario above is that the single copy, given by Maxwell equations, is considered as living in the full metric, not just the background. Nevertheless, as shown above, the equations of motion of the single copy will also be satisfied in the background, which allows us to interpret this as a double copy realization in fixed curved backgrounds.

    More recently, this construction was used to classify the Weyl double copies associated with the different adjoint orbits of $\mathfrak{so}(2,3)$, together with their Kerr--Schild representatives when they exist, and to extend the classification to linearized higher-spin multicopies \cite{Didenko:2026nip}
	
	\subsubsection{Weyl Doubling}\label{sec:weyl_doubling}
	
	The unfolded construction above already illustrates that a curvature factorization need not involve a Maxwell field defined only on an auxiliary background. A broader algebraic relation of this kind is known as Weyl doubling \cite{Alawadhi:2019urr,Alawadhi:2020jrv}. In contrast to the Weyl double copy of Sec.~\ref{sec:classical_Weyl}, the Abelian field strength is defined on the full gravitational spacetime, and the construction does not require an independent zeroth-copy field satisfying a scalar equation on a common background.
	
	The basic quadratic tensor entering this relation is obtained by projecting the product of two field strengths onto the algebraic symmetries of the Weyl tensor. In $D$ dimensions, it is
	\begin{align}
		C^{(D)}_{\mu\nu\rho\sigma}[F]={}&\Bigg(F_{\mu\nu}F_{\rho\sigma}-F_{\rho\mu}F_{\nu\sigma}-\frac{6}{D-2}g_{\mu\rho}{F_{\nu}}^{\lambda}F_{\sigma\lambda}\nonumber\\
		&\qquad+\frac{3}{(D-1)(D-2)}g_{\mu\rho}g_{\nu\sigma}F^2\Bigg)\Bigg|_{s}\ ,
		\label{eq:weyl_doubling_tensor}
	\end{align}
	where $F^2=F_{\lambda\tau}F^{\lambda\tau}$ and $\left.\right|_{s}$ denotes antisymmetrization separately in $\mu,\nu$ and in $\rho,\sigma$, with unit weight. All contractions in Eq.~\eqref{eq:weyl_doubling_tensor} are performed using the full spacetime metric $g_{\mu\nu}$. The relative coefficients subtract the traces of the quadratic Maxwell tensor so that $C^{(D)}_{\mu\nu\rho\sigma}[F]$ has the algebraic symmetries of a Weyl tensor.
	
	Weyl doubling holds when the spacetime curvature can be written as
	\begin{equation}
		C_{\mu\nu\rho\sigma}[g]=\frac{1}{S}C^{(D)}_{\mu\nu\rho\sigma}[F]\ .
		\label{eq:weyl_doubling_relation}
	\end{equation}
	The function $S$ compensates the relative spacetime dependence and normalization of the two sides. Although it plays the same algebraic role as the scalar in Eq.~\eqref{eq:Weyl_gravity}, it is not generally interpreted as an independent zeroth copy and is not required to satisfy Eq.~\eqref{eq:ScalarConfCoupl}. Thus the scalar prefactor is retained, but its zeroth-copy dynamics are not part of the Weyl-doubling prescription. In the vacuum examples motivating the construction, the field strength is additionally taken to satisfy the source-free Maxwell equations with respect to the full metric,
	\begin{equation}
		\nabla_{[\mu}F_{\nu\rho]}=0\ ,\qquad \nabla^\mu F_{\mu\nu}=0\ .
		\label{eq:weyl_doubling_maxwell}
	\end{equation}
	
	In four dimensions, the chiral projection of Eq.~\eqref{eq:weyl_doubling_tensor} reduces to the quadratic Maxwell-spinor numerator $f_{(\alpha\beta}f_{\gamma\delta)}$ appearing in Eq.~\eqref{eq:Weyl_gravity}. For a general real two-form, however, there is a second independent tensor with the algebraic symmetries of the Weyl tensor, obtained by replacing one field strength by its Hodge dual. Equivalently, the relation may be formulated separately for the self-dual and anti-self-dual sectors. This additional structure is special to four dimensions, where the Hodge dual of a two-form is again a two-form, and makes the relation between electric--magnetic duality and gravitational solution-generating transformations manifest \cite{Alawadhi:2019urr}.
	
	The factorization also imposes a strong algebraic alignment between the gauge and gravitational fields. If $k_\mu$ is an eigenvector of the field strength, then Eq.~\eqref{eq:weyl_doubling_relation} implies
	\begin{equation}
		{F_\mu}^{\nu}k_\nu=\lambda_k k_\mu\qquad\Longrightarrow\qquad C_{\mu\nu\rho\sigma}k^\nu k^\sigma=\Lambda_k k_\mu k_\rho\ ,
		\label{eq:weyl_doubling_alignment}
	\end{equation}
	for some scalar $\Lambda_k$. The null eigenvectors of $F_{\mu\nu}$ therefore determine principal null directions of the Weyl tensor. Generically, the two distinguished null directions place the resulting spacetime in the type~D class of the higher-dimensional alignment classification. In four dimensions, this is the tensorial counterpart of the shared principal spinors in the Weyl double copy.
	
	For Ricci-flat geometries, a useful field strength can often be constructed from a Killing vector, and the doubling relation has been verified for the Taub--NUT and Plebański--Demiański families as well as for the singly rotating Myers--Perry solution in five dimensions \cite{Alawadhi:2020jrv}. In four dimensions, an electric--magnetic duality rotation of this field maps through Eq.~\eqref{eq:weyl_doubling_relation} to the Ehlers transformation relating mass and NUT charge \cite{Alawadhi:2019urr}. The same algebraic idea also extends to situations in which $F_{\mu\nu}$ is a physical matter field already present in the gravitational theory, including Reissner--Nordström and Born--Infeld solutions, as well as supergravity branes whose curvature is related to the corresponding form-field strength \cite{Alawadhi:2020jrv}.
	
	The purely quadratic relation is nevertheless not universal. For generic multi-center Gibbons--Hawking geometries, the Weyl tensor also contains terms linear in derivatives of the Abelian field strength, while other higher-dimensional examples require additional geometric structures. Weyl doubling should therefore be regarded as an algebraic organization of particular gravitational curvatures rather than as a complete double-copy prescription. It retains the characteristic squaring and alignment of gauge and gravitational fields, but does not by itself provide a common background, a dynamical zeroth copy, or a color--kinematics interpretation.
	
	\subsubsection{Newman--Penrose Map}\label{sec:newman_penrose_map}
	
	The Newman--Penrose map relates a class of exact gravitational spacetimes to complex self-dual solutions of the vacuum Maxwell equations \cite{Elor:2020nqe}. Its input is not the complete metric profile alone, but the preferred null congruence. This distinguishes it from the Kerr--Schild prescription, which replaces one copy of the Kerr--Schild vector by color, and from the Weyl double copy, which factorizes the curvature. Where the constructions overlap, the real part of the Newman--Penrose field agrees with the usual single copy up to gauge transformations. The map nevertheless contains different information and is generally many-to-one, so it is most appropriately regarded as a closely related classical map rather than as an independent derivation of a double copy.
	
	In the flat-space map, the preferred congruence is the starting point. In a patch adapted to it, the flat background and the null one-form tangent to the congruence take the canonical form
	\be
	\dd s_0^2=2(\dd u\,\dd v-\dd\zeta\,\dd\bar\zeta)\ ,\qquad
	\ell=\dd u+\bar\Phi\,\dd\zeta+\Phi\,\dd\bar\zeta+\Phi\bar\Phi\,\dd v\ .
	\label{eq:np_null_congruence}
	\ee
	Here $\Phi$ is the local function that records the choice of null direction at each spacetime point. For a single Kerr--Schild metric $g_{\mu\nu}=\eta_{\mu\nu}+\ell_\mu \ell_\nu \phi$ that solves the vacuum Einstein equations, the function $\phi$ can be solved for in terms of $\Phi$ up to integration constants. Hence, the same $\Phi$ can correspond to solutions with different matter sources. The shear-free and geodesic conditions on $\ell$ are equivalent to
	\be
	(\Phi\partial_\zeta-\partial_v)\Phi=0\ ,\qquad
	(\Phi\partial_u-\partial_{\bar\zeta})\Phi=0\ ,
	\label{eq:np_shearfree_equations}
	\ee
	which in turn imply the linear wave equation
	\be
	\Box_0\Phi=2(\partial_u\partial_v-\partial_\zeta\partial_{\bar\zeta})\Phi=0\ .
	\label{eq:np_harmonic_scalar}
	\ee
	
	Thus the nonlinear congruence equations select a particular harmonic scalar. The map then applies a spin-raising operator $\widehat{k}_\mu$ to this scalar, in close analogy with the self-dual Kerr--Schild double copy reviewed in Section~\ref{sec:self_dual}, but without requiring $\widehat{k}_\mu\widehat{k}_\nu\Phi$ to solve the self-dual Einstein equations. Similar spin-raising operators also appear in the plane-wave scattering construction of Ref.~\cite{Adamo:2017nia}, reviewed in Sec.~\ref{sec:plane_wave_scattering_dc}, where Eq.~\eqref{eq:pw_graviton_polarization} dresses the scalar wave of Eq.~\eqref{eq:pw_graviton_phase} by a background-dependent polarization one-form. Up to an overall charge normalization,
	\be
	\widehat{k}=\dd v\,\partial_\zeta+\dd\bar\zeta\,\partial_u\ ,\qquad
	A=\widehat{k}\Phi\ .
	\label{eq:np_spin_raising}
	\ee
	Harmonicity of $\Phi$ then makes the resulting field strength obey
	\be
	F_{\mu\nu}=\frac{i}{2}\epsilon^{(0)}_{\mu\nu\rho\sigma}F^{\rho\sigma}\ ,\qquad
	\nabla^{(0)\mu}F_{\mu\nu}=0\ ,
	\label{eq:np_selfdual_maxwell}
	\ee
	where $\nabla^{(0)}$ and $\epsilon^{(0)}_{\mu\nu\rho\sigma}$ are the Levi--Civita connection and volume tensor of the flat background. This is the Lorentzian complexified version of Eq.~\eqref{eq:sd_definition} in Sec.~\ref{sec:self_dual}.
	
	The twistorial reformulation explains why the operator in Eq.~\eqref{eq:np_spin_raising} is the natural one. The Kerr theorem supplies the twistor origin of this step: with the dual-twistor convention used here, the same expanding shear-free null geodesic congruence is encoded by a holomorphic surface in projective dual twistor space. For the real congruence, the relevant representative is a null dual twistor,\footnote{The twistor-space Newman--Penrose map paper uses Penrose--Rindler conventions in which the object denoted $Z=(\omega^A,\pi_{A'})$ obeys $\omega^A=i x^{AA'}\pi_{A'}$ \cite{Farnsworth:2021wvs}. In the conventions of Sec.~\ref{sec:twistor_correl}, summarized in Eq.~\eqref{eq:twistor_incidence}, this has the chirality and incidence relation of the dual twistor $W_{\mathcal M}$; we translate their notation accordingly, with no raising or lowering of twistor indices implied.}
	\be
	W_{\mathcal M}=(\tilde\mu^\alpha,-\tilde\lambda_{\dot\alpha})\ ,\qquad
	\tilde\mu^\alpha=x^{\alpha\dot\alpha}\tilde\lambda_{\dot\alpha}\ ,
	\ee
	where $x^{\alpha\dot\alpha}$ is a position in Minkowski space, and null means $W_{\mathcal M}\bar{W}^{\mathcal M}=0$, with $\bar{W}^{\mathcal M}$ the complex-conjugate twistor \cite{Adamo:2017qyl}. The associated $\beta$-plane\footnote{This is the same two-plane called an $\alpha$-plane in Ref.~\cite{Farnsworth:2021wvs}. With the conventions used here, points of projective twistor space determine $\alpha$-planes, while points of projective dual twistor space determine $\beta$-planes; the difference is a chirality and orientation convention.} is generically a two-complex-dimensional totally null plane
	\be
	x^{\alpha\dot\alpha}=x_0^{\alpha\dot\alpha}+\rho^\alpha\tilde\lambda^{\dot\alpha}\ ,
	\label{eq:np_beta_plane}
	\ee
	with $\rho^\alpha$ arbitrary. When the dual twistor is null, this complex $\beta$-plane meets the Lorentzian real slice in a real null geodesic, whose tangent direction is $\bar{\tilde\lambda}^{\alpha}\tilde\lambda^{\dot\alpha}$; the Newman--Penrose map normalization then fixes its scale by requiring this real tangent direction to be transverse to the congruence in Eq.~\eqref{eq:np_null_congruence},
	\be
	\ell_{\alpha\dot\alpha}\bar{\tilde\lambda}^{\alpha}\tilde\lambda^{\dot\alpha}=1\ .    
	\label{eq:np_twistor_normalization}
	\ee
	The tangent vectors to the $\beta$-plane may be chosen as
	\be
	p^{\alpha\dot\alpha}=\bar{\tilde\lambda}^{\alpha}\tilde\lambda^{\dot\alpha}\ ,\qquad
	q^{\alpha\dot\alpha}=\tilde\mu^\alpha\tilde\lambda^{\dot\alpha}\ .
	\ee
	For $\tilde\mu^\alpha\bar{\tilde\lambda}_{\alpha}\neq0$, these determine a projectively normalized tangent bivector,
	\be
	\tau\equiv\frac{1}{\tilde\mu^\alpha\bar{\tilde\lambda}_{\alpha}}p\wedge q
	=-\delta^2\partial_u\wedge\partial_\zeta\ .
	\label{eq:np_twistor_bivector}
	\ee
	where $\delta$ is fixed by choosing the normalized dual-spinor representative $\tilde\lambda_{\dot\alpha}=\delta\bar\iota_{\dot\alpha}$ with $|\delta|=1$, where $(o^\alpha,\iota^\alpha)$ is the spin dyad adapted to the flat tetrad. Lowering one index with the flat metric in Eq.~\eqref{eq:np_null_congruence} turns this $\beta$-plane area bivector into the mixed one-form/vector operator
	\be
	\tau^{\mu\rho}\eta_{\rho\nu}\dd x^\nu\partial_\mu
	=\delta^2(\dd v\,\partial_\zeta+\dd\bar\zeta\,\partial_u)
	=\delta^2\widehat{k}\ .
	\label{eq:np_twistor_khat}
	\ee
	The scalar input also has a twistor origin. Writing $\tilde\mu^\alpha=-i\delta a\,o^\alpha+\beta\iota^\alpha$, one defines
	\be
	\Psi\equiv\frac{1}{\tilde\mu^\alpha\bar{\tilde\lambda}_{\alpha}}\ell_{\beta\dot\beta}\bar q^{\beta\dot\beta}
	=\bar\delta^2\Phi+\frac{\bar\delta\bar\beta}{i a}\ .
	\label{eq:np_twistor_scalar}
	\ee
	The second term is constant and is annihilated by the raising operator, while $|\delta|^2=1$ makes the phases in Eqs.~\eqref{eq:np_twistor_khat} and \eqref{eq:np_twistor_scalar} cancel. Thus the single copy can be written as
	\be
	A=\widehat{\kappa}\Psi\ ,\qquad \widehat{\kappa}\equiv\delta^2\widehat{k}\ ,
	\label{eq:np_twistor_single_copy}
	\ee
	which is equivalent to Eq.~\eqref{eq:np_spin_raising}. The spin-raising operator is therefore not chosen by hand: both it and the scalar on which it acts are built from the dual twistor and the congruence.
	
	The standard stationary examples then have a clear interpretation. For Schwarzschild, the real part of $A$ is gauge-equivalent to the Coulomb potential, while the imaginary part is its magnetic dual. For Kerr, the real part agrees with the rotating-disk field obtained from the Kerr--Schild single copy. Kinnersley's photon rocket illustrates the same point in a non-stationary setting: its retarded-cone congruence maps to the Li\'enard--Wiechert potential of the source worldline, while the Bondi mass aspect and possible electromagnetic charge change the gravitational profile without changing the congruence. Thus the map projects out data not encoded in the shear-free null congruence, which makes it useful but many-to-one.
	
	For curved backgrounds, the extension has been tested on a double Kerr--Schild form with a maximally symmetric base metric \cite{Farnsworth:2023mff},
	\be
	g_{\mu\nu}=\bar g_{\mu\nu}+\phi k_\mu k_\nu+\psi\ell_\mu\ell_\nu\ ,
	\label{eq:np_double_ks_metric}
	\ee
	where $k_\mu$ and $\ell_\mu$ are null, geodesic, shear-free, and mutually orthogonal. The orthogonality condition is naturally implemented after continuation to Kleinian signature, or in the complexified spacetime. Since the base is conformally flat, one may use the same adapted null coordinates as in Eq.~\eqref{eq:np_null_congruence}, with $\dd\bar s^2=\Omega^{-2}\dd s_0^2$ and $\Omega$ fixed by the constant curvature of the maximally symmetric background. In this patch the two Kerr--Schild directions can be parametrized by the same congruence data used above,
	\be
	k=\dd u+\bar\Phi_k\dd\zeta+\Phi_k\dd\bar\zeta+\Phi_k\bar\Phi_k\dd v\ ,\qquad
	\ell=\dd u+\bar\Phi_\ell\dd\zeta+\Phi_\ell\dd\bar\zeta+\Phi_\ell\bar\Phi_\ell\dd v\ .
	\label{eq:np_double_ks_one_forms}
	\ee
	In the complexified or Kleinian description the barred quantities are independent until a real slice is chosen. Each congruence determines a scalar, $\Phi_k$ or $\Phi_\ell$, and the curved-space map uses the same spin-raising operator associated with the conformally related flat metric, $\widehat{k}_0=\dd v\,\partial_\zeta+\dd\bar\zeta\,\partial_u$. Up to the conventional electric and magnetic normalizations, which are absorbed into $Q_e$ and $Q_m$ as in the other classical examples, the gauge potential is
	\be
	A=\widehat{k}_0\left(Q_e\Phi_\ell+Q_m\Phi_k\right)\ .
	\label{eq:np_double_ks_potential}
	\ee
	If both congruence scalars are harmonic with respect to the flat metric conformally related to $\bar g_{\mu\nu}$, Eq.~\eqref{eq:np_double_ks_potential} gives a self-dual Maxwell solution. Four-dimensional Maxwell theory is conformally invariant, so the same field also solves Maxwell equations on the maximally symmetric background. The scalar wave equation is not conformally invariant, however, so $\Phi_k$ and $\Phi_\ell$ should not be read as minimally coupled scalar fields on (A)dS.
	
	For the Kerr--Taub--NUT--(A)dS family, the required harmonicity can be checked explicitly \cite{Farnsworth:2023mff}. In the adapted conformally flat coordinates, each selected congruence scalar $\Phi_I$, with $I=k,\ell$, satisfies the same shear-free system as Eq.~\eqref{eq:np_shearfree_equations},
	\be
	(\Phi_I\partial_\zeta-\partial_v)\Phi_I=0\ ,\qquad
	(\Phi_I\partial_u-\partial_{\bar\zeta})\Phi_I=0\ ,
	\qquad I=k,\ell\ ,
	\label{eq:np_double_ks_shearfree_scalars}
	\ee
	and hence obeys $\Box_0\Phi_I=0$. The real part of Eq.~\eqref{eq:np_double_ks_potential} agrees, up to normalization and pure gauge terms, with the Kerr--Schild single copy in the Taub--NUT limit \cite{Luna:2015paa} and with the aligned-field construction for the full Kerr--NUT--(A)dS family \cite{Chawla:2023aligned}. The mass and NUT parameters become electric and magnetic charges, while rotation gives the corresponding rotating dyon. The full Newman--Penrose single copy can be written, up to pure gauge terms, as
	\be
	A\simeq
	\frac{Q_e+iQ_m}{4\pi(r+i a\cos\theta)}
	\left[
	\left(1+\frac{i\Lambda a r\cos\theta}{3}\right)\dd t+\dd r
	+i(r\cos\theta+i a)\dd\varphi
	\right]\ ,
	\label{eq:np_kerr_taub_nut_ads_gauge}
	\ee
	where $a$ is the rotation parameter. The physical real Maxwell field may be obtained by taking the real part of this complex self-dual representative. 
	
	This extension should be interpreted cautiously. The original Newman--Penrose map is defined for single Kerr--Schild geometries with an expanding shear-free null geodesic congruence. Its double Kerr--Schild version relies on conformal flatness and on explicit harmonicity checks for the two congruence scalars, and no general proof is known for arbitrary double Kerr--Schild spacetimes. The Kerr--Taub--NUT--(A)dS family therefore provides evidence for a curved-space extension and a useful bridge between the Kerr--Schild, Weyl, and twistor prescriptions, but not a universal Newman--Penrose double copy.

	\section{Self-Dual Double Copy}\label{sec:self_dual}
	
	The previous sections organized curved-space double copy constructions by the observable or by the type of classical map. There are many self-dual examples of such double copies, but self-dual theories have a specially clear double copy with a simple kinematic algebra and allow for many more insights and descriptions and hence we focus on these examples in this section. 
	
	In four dimensions, self-duality projects onto one chiral half of the field strength or Weyl tensor. This projection is a consistent truncation of Yang--Mills theory and gravity, and it makes the algebraic origin of the double copy unusually explicit: the relevant interactions can be written as products of structure constants. Curvature then enters through deformations of the kinematic algebra, through the choice of background, or through the observable used to probe the theory. This section follows that thread from flat space to AdS, cosmological backgrounds, self-dual curved backgrounds, radiative twistor data, exact solutions, boundary correlators, and chiral higher-spin extensions.
	
	Self-dual theories play a larger role than a single helicity truncation might suggest. Their equations are integrable, their perturbative self-dual S-matrix is one-loop exact with nonvanishing rational amplitudes, and the theories have natural twistor descriptions. One can also describe the full theory as a perturbation around the self-dual sector; interactions involving the opposite helicity are restored as perturbations around the self-dual action. This formulation is usually referred to as the Chalmers--Siegel action \cite{Chalmers:1996rq}, and its lightcone gauge is particularly enlightening since the tree-level MHV amplitudes can be constructed purely from three-point vertices, showcasing the kinematic algebra \cite{Monteiro:2011pc}. 
	
	\subsection{Flat-space self-dual double copy and heavenly equations}
	
	We first review the flat-space construction of Ref.~\cite{Monteiro:2011pc} and work in Euclidean signature. It is convenient to use complex double lightcone coordinates
	\eqs{
		u&=it+z \ , \quad v=it-z \ ,\\
		w&=x+iy \ , \quad \bar{w}=x-iy \ .
		\label{lccoords}
	}
	In these coordinates the flat background line element is $dw\,d\bar w-du\,dv$, and we write a self-dual gravitational line element as
	\be
	ds^2=dw\,d\bar w-du\,dv+h_{\mu\nu}\,dx^\mu dx^\nu\ .
	\label{eq:SDflatmetric}
	\ee
	The index split
	\be\label{coords_def}
	x^i=(u,w),\qquad y^\alpha=(v,\bar{w})\ ,
	\ee
	separates the two coordinates on which the kinematic Poisson bracket acts from their conjugate partners. This split may be viewed as separating holomorphic and antiholomorphic coordinates.
	
	In four dimensions, two-forms and Weyl tensors split into self-dual and anti-self-dual parts. The self-dual truncation keeps one of these chiral pieces by equating the curvature to its Hodge dual. For spin one and spin two in Euclidean signature this is
	\begin{equation}
		F_{\mu\nu}=\frac{1}{2}\epsilon_{\mu\nu\rho\lambda}F^{\rho\lambda} \ , \qquad
		C_{\mnrs}=\frac{1}{2}{\epsilon_{\mu\nu}}^{\eta\lambda}C_{\eta\lambda\rho\sigma} \ ,
		\label{eq:sd_definition}
	\end{equation}
	where $F_{\mu\nu}$ is the Yang--Mills field strength and $C_{\mnrs}$ is the Weyl tensor. In Lorentzian signature a real, nonvanishing two-form cannot obey this equation, so in that case the self-dual sector is usually treated after complexification.
	
	In lightcone gauge, $A_u=0$, the self-duality constraints of Yang--Mills theory can be solved in terms of a single adjoint scalar potential,
	\be
	\label{eq:sdYM}
	A_i=0,\qquad A_\alpha=\Pi_\alpha \Phi,\qquad
	\Pi_\alpha=\left(\partial_w,\partial_u\right)\ ,
	\ee
	with $\Phi=\Phi^aT^a$. The remaining equation of motion is
	\eqs{
		\Box_{\mathbb{R}^{4}}\Phi-\frac{i}{2}[\{\Phi,\Phi\}]=0 \ , \label{eq:sdYMflat}\\
		[\{f,g\}]=\varepsilon^{\alpha\beta}\left[\Pi_\alpha f,\Pi_\beta g\right]\ ,
		\label{eq:sdym_commutator_bracket}
	}
	where $[\ ,\ ]$ is the color Lie bracket, $\varepsilon^{\alpha\beta}$ is the antisymmetric tensor in the two-dimensional $\alpha$ index space, and the Poisson bracket is defined as
	\be
	\{f,g\}=\varepsilon^{\alpha\beta}\Pi_{\alpha }f\Pi_{\beta }g=\partial_w f\,\partial_u g-\partial_u f\,\partial_w g \ .
	\ee
	This equation follows from the cubic action
	\be
	S_{\rm SDYM}=\int\dd^4x\,\text{Tr}\left[\bar\Phi\left(\Box_{\mathbb{R}^{4}}\Phi-\frac{i}{2}[\{\Phi,\Phi\}]\right)\right]\ .
	\label{eq:sdym_action}
	\ee
	The field $\bar\Phi$ is a Lagrange multiplier carrying the opposite helicity and varying it imposes Eq.~\eqref{eq:sdYMflat}. The nonlinear term is explicitly $i\bar\Phi[\partial_u\Phi,\partial_w\Phi]$, so the self-dual action has only one cubic interaction, with one $\bar\Phi$ field and two $\Phi$ fields \cite{Monteiro:2011pc}. This is the first reason the self-dual sector is such a clean laboratory: the cubic vertex already contains all the nonlinearity.
	
	The second reason is that the kinematic part of the vertex is itself a Lie algebra. The action of the Poisson bracket on external states given by plane waves is given by
	\be
	\begin{aligned}
		\{e^{-ip_1\cdot x},e^{-ip_2\cdot x}\}&=-X(p_1,p_2)e^{-i(p_1+p_2)\cdot x} \ , \\
		X(p_1,p_2)&=p_{1w}p_{2u}-p_{1u}p_{2w}\ .
	\end{aligned}
	\label{eq:flat_poisson}
	\ee
	For any function $f(u,w)$, this bracket defines a Hamiltonian vector field
	\be
	L_f=\{f,\cdot\}=\partial_wf\,\partial_u-\partial_uf\,\partial_w,\qquad
	[L_f,L_g]=L_{\{f,g\}}\ .
	\label{eq:hamiltonian_vector_fields}
	\ee
	Its divergence in the $(u,w)$ plane vanishes, $\partial_uL_f^u+\partial_wL_f^w=0$, so the flow preserves the area form $du\wedge dw$. Conversely, locally every area-preserving vector field on the $x^i$ plane can be written in Hamiltonian form. Plane waves therefore give a convenient basis for the Lie algebra of area-preserving diffeomorphisms, with $X(p_1,p_2)$ playing the role of its structure constants. Their Jacobi identity is
	\eqs{
		&X(p_1,p_2)X(p_1+p_2,p_3)+X(p_2,p_3)X(p_2+p_3,p_1) \nonumber\\
		&\hspace{2.6cm}+X(p_3,p_1)X(p_3+p_1,p_2)=0\ .
		\label{eq:X_jacobi}
	}
	This is the kinematic algebra encoding the color-kinematics duality. In lightcone gauge, then, the SDYM vertex is automatically the product of a color structure constant and a kinematic structure constant. The color Jacobi identity and the kinematic Jacobi identity are visible off shell in the same representation, without first performing a generalized gauge transformation. This visibility is gauge dependent: the same algebra need not be manifest in an arbitrary set of variables \cite{Monteiro:2011pc,Chacon:2020fmr}.
	
	The gravitational counterpart follows by solving the self-dual metric constraints in the corresponding lightcone gauge, $h_{\mu u}=0$. The exact metric deformation is encoded in a scalar potential $\phi$ through
	\be
	h_{i\mu}=0,\qquad h_{\alpha\beta}=\Pi_\alpha\Pi_\beta\phi\ .
	\label{eq:sdg_flat_metric}
	\ee
	The scalar equation is Plebanski's second-heavenly equation,
	\eqs{
		\Box_{\mathbb{R}^{4}}\phi-\{\{\phi,\phi\}\}=0 \ , \label{eq:sdg_flat_eom}\\
		\{\{f,g\}\}=\frac{1}{2}\varepsilon^{\alpha\beta}\{\Pi_\alpha f,\Pi_\beta g\}\ ,
		\label{eq:sdg_double_bracket}
	}
	It follows from the action
	\be
	S_{\rm SDG}=\int\dd^4x\,\bar\phi\left(\Box_{\mathbb{R}^{4}}\phi-\{\{\phi,\phi\}\}\right)\ ,
	\label{eq:sdg_action}
	\ee
	where $\bar\phi$ is the opposite-helicity multiplier field. The double copy is now manifest at the level of the action:
	\begin{equation}
		\Phi\rightarrow\phi,\qquad \dfrac{i}{2}[\{\ ,\ \}]\rightarrow\{\{\ ,\ \}\}\ .
		\label{eq:self_dual_flat_double_copy}
	\end{equation} 
	Similarly, for the three-point vertices the color structure constant in SDYM is replaced by a second copy of the kinematic structure constant of the area-preserving diffeomorphism algebra,
	\begin{mybox}[Self-Dual Double Copy]
		\begin{center}
			\renewcommand{\arraystretch}{1.8}
			\begin{tabularx}{\textwidth}{l X}
				\textbf{Gravity} & \vspace{-1cm}
				\begin{equation}
					V_{\rm SDG}=\frac{1}{2}X(k_1,k_2)^2
					\label{eq:sdg_vertex}
				\end{equation} \\
				\textbf{Yang--Mills} & \vspace{-1cm}
				\begin{equation}
					V_{\rm SDYM}=\frac{1}{2}X(k_1,k_2)f^{a_1a_2a_3}
					\label{eq:sdym_vertex}
				\end{equation} 
				\vspace{-0.5cm}
			\end{tabularx}
		\end{center}
	\end{mybox}
	
	One can construct other interesting theories by considering deformations of the kinematic algebra. In the spirit of the BCJ double copy, this is similar to considering a squaring with different kinematic factors. One can therefore ask which other theories arise by replacing one or both factors by another antisymmetric bracket satisfying the appropriate Jacobi identity. A basic example is the Moyal deformation of the Poisson bracket,
	\be
	\{f,g\}^{M}=\frac{1}{i\hbar}(f\star g-g\star f),\qquad
	f\star g=f\exp\left[\frac{i\hbar}{2}
	\left(\overleftarrow{\partial_w}\overrightarrow{\partial_u}
	-\overleftarrow{\partial_u}\overrightarrow{\partial_w}\right)\right]g
	\label{eq:moyal_bracket}
	\ee
	which defines new structure constants $X^M$ on plane waves and reduces to the Poisson bracket as $\hbar\to0$. For example, replacing the color factor in SDYM by $X^M$ gives a Moyal-deformed second-heavenly equation with cubic kernel $X^M X$~\cite{Chacon:2020fmr}. These Moyal deformations are related to chiral higher-spin theories and define a $w_{1+\infty}$ algebra \cite{Monteiro:2022xwq,Bu:2022iak,Mason:2025pbz}. Other flat-space deformations were considered in \cite{Chacon:2020fmr}. Below, we will see how the self-dual theories living in curved space can also be viewed as deformations of the kinematic algebra, albeit of a different type.
	
	\subsection{AdS and conformally flat cosmological backgrounds} \label{sec:sd_ads_cosmo}
	
	We now consider the case where the background metric is not flat. We assume that both the self-dual gauge field and the self-dual graviton live on a fixed curved background. The simplest curved backgrounds to treat are conformally flat. We therefore restore a conformal factor and use
	\be
	ds^2=a(x^\mu)^2\left(dw\,d\bar w-du\,dv+h_{\mu\nu}\,dx^\mu dx^\nu\right)\ .
	\label{eq:SDfrw}
	\ee
	The metric with $h_{\mu\nu}=0$ is the fixed background, while $h_{\mu\nu}$ carries the self-dual gravitational field. For cosmological applications one takes $a=a(\tau)$ with $\tau=(u+v)/2$ together with an analytic continuation to Lorentzian signature, while Euclidean AdS$_4$ in the Poincar\'e patch corresponds to $a=2/(u-v)$.
	
	Yang--Mills theory is classically Weyl invariant in four dimensions. As a result, the local SDYM equation \eqref{eq:sdYMflat} continues to describe self-dual Yang--Mills fields after the conformal rescaling. Any nontrivial changes appear in boundary conditions and in the observables constructed from the solution. Gravity is more constrained. The conformal factor changes the self-duality condition on the curvature and deforms the scalar interaction that replaces the color Lie bracket \cite{Lipstein:2023pih}. 
	
	In AdS$_4$ with unit radius, the relevant self-duality condition is not imposed directly on the Riemann tensor. One first subtracts the constant-curvature part, so that pure AdS carries no self-dual excitation. This gives the shifted tensor
	\be
	T_{\mu\nu\rho\sigma}=R_{\mu\nu\rho\sigma}
	-\frac{\Lambda}{3}\left(g_{\mu\rho}g_{\nu\sigma}-g_{\nu\rho}g_{\mu\sigma}\right)\ .
	\label{eq:ads_self_dual_tensor}
	\ee
	which is imposed to be self-dual
	\be
	T_{\mu\nu\rho\sigma}=\frac{1}{2}\sqrt{g}\,\epsilon_{\mu\nu}{}^{\eta\lambda}T_{\eta\lambda\rho\sigma}\, .
	\ee
	Solving the self-duality condition in lightcone gauge gives
	\be
	h_{i\mu}=0,\qquad h_{\alpha\beta}=\Pi_{(\alpha}\tilde{\Pi}_{\beta)}\phi,\qquad
	\tilde{\Pi}_\alpha=\left(\partial_w,\partial_u-\frac{4}{u-v}\right)\ .
	\label{eq:ads_sdg_metric}
	\ee
	The scalar $\phi$ is conformally coupled on AdS,
	\be
	\sqrt{g}\left(-\Box_{\rm AdS}+m^2\right)\phi
	+4\left\{\left\{\frac{\phi}{u-v},\frac{\phi}{u-v}\right\}\right\}_{*}=0,\qquad m^2=-2\ .
	\label{eq:ads_sdg_eom}
	\ee
	The interaction is built from a deformed bracket,
	\eqs{
		\{f,g\}_{*}=\{f,g\}+\frac{2}{u-v}\left(f\partial_wg-g\partial_wf\right)\ ,
		\label{eq:ads_star_bracket}\\
		\{\{f,g\}\}_{*}=\frac{1}{2}\varepsilon^{\alpha\beta}\{\Pi_\alpha f,\Pi_\beta g\}_{*}\ .
		\label{eq:ads_double_bracket}
	}
	This is a Jacobi bracket: it obeys the Jacobi identity, but its Leibniz rule is modified by the explicit $u-v$ dependence. Acting on plane waves, it defines the deformed kinematic factor
	\eqs{
		\{e^{ik_1\cdot x},e^{ik_2\cdot x}\}_{*}
		=\tilde X(k_1,k_2)e^{i(k_1+k_2)\cdot x}\ ,\\
		\tilde X(k_1,k_2)=X(k_1,k_2)-\frac{2i}{u-v}(k_1-k_2)_w\ .
		\label{eq:ads_xtilde}
	}
	Both $X$ and $\tilde X$ obey Jacobi identities. The AdS three-point vertices are therefore still products of two structure constants,
	\be
	V_{\rm SDYM}=\frac{1}{2}X(k_1,k_2)f^{a_1a_2a_3},\qquad
	V_{\rm SDG}=\frac{1}{2}X(k_1,k_2)\tilde X(k_1,k_2)\ .
	\label{eq:ads_vertices}
	\ee
	although the AdS self-dual sector is now given by an asymmetric double copy. One factor remains the flat area-preserving-diffeomorphism structure constant, while the second factor is deformed by the AdS conformal factor:
	\be
	f^{a_1a_2a_3}\rightarrow\tilde X(k_1,k_2)\ .
	\label{eq:ads_dc_replacement}
	\ee
	As in flat space, one can perform a Moyal deformation which leads to a deformed $w_{1+\infty}$ algebra \cite{Lipstein:2023pih}. 
	
	The cosmological construction~\cite{CarrilloGonzalez:2024sto} uses the same lightcone framework but allows the conformal factor in Eq.~\eqref{eq:SDfrw} to have an arbitrary dependence on $\tau=(u+v)/2$. If the full Ricci tensor is required to be that of an FLRW stress tensor while the metric is on shell and self-dual, no genuinely new examples appear beyond the flat and (A)dS cases. A larger class is obtained by imposing self-duality of the Weyl tensor off shell,
	\be
	C_{\mu\nu\rho\sigma}=\frac{1}{2}\sqrt{g}\,\epsilon_{\mu\nu}{}^{\eta\lambda}C_{\eta\lambda\rho\sigma}\, .
	\ee
	The lightcone solution can then be written as
	\be
	h_{i\mu}=0,\qquad
	h_{\alpha\beta}=\frac{1}{4}\Pi_{(\alpha}{\Pi^\zeta}_{\beta)}\phi,\qquad
	\Pi^\zeta_\alpha=\left(\partial_w,\partial_u+2\zeta(u+v)\right)\ .
	\label{eq:cosmo_metric_ansatz}
	\ee
	The function $\zeta$ encodes the allowed Weyl frame. Self-duality requires
	\be
	\partial_u\left(\frac{\partial_u\zeta}{\zeta^2}\right)=0\ ,
	\label{eq:zeta_condition}
	\ee
	so $\zeta=-2/(u+v)$ or $\zeta$ is constant, with $\zeta=0$ included as the flat limit. The former case is in fact the (A)dS example described above and the tensor $T_{\mu\nu\rho\sigma}$ is nothing but the on-shell Weyl tensor. After a Weyl rescaling $\phi \rightarrow \phi/(L\zeta)$, the scalar equation becomes
	\be
	\Box_{\mathbb{R}^{4}}\phi-\frac{1}{L\zeta}\left\{\left\{\phi,\phi\right\}\right\}_{\zeta}=0\ ,
	\label{eq:cosmological_sdg_eom}
	\ee
	where the new bracket is a Jacobi bracket defined as
	\eqs{
		\{f,g\}_{\zeta}&=\{f,g\}+c_\zeta\,\zeta(u+v)\left(f\partial_wg-g\partial_wf\right)\ , \\
		c_\zeta&=2\left(\frac{\partial_u\zeta}{\zeta^2}-1\right)=\text{constant}\ .
		\label{eq:cosmological_jacobi}
	}
	This bracket has the same structural role as the AdS bracket: it satisfies the Jacobi identity but has a deformed Leibniz rule. At the level of the action, the cosmological double copy is implemented by the replacements
	\begin{equation}
		\Phi\rightarrow\phi,\qquad \dfrac{i}{2}[\{\ ,\ \}]\rightarrow\frac{1}{L\zeta}\{\{\ ,\ \}\}_{\zeta}\ .
		\label{eq:cosmo_dc_action}
	\end{equation}
	Acting on plane waves, the bracket defines the deformed kinematic factor
	\be
	X^\zeta(k_1,k_2)=X(k_1,k_2)-ic_\zeta\zeta(k_1-k_2)_w\ .
	\ee
	The corresponding three-point vertices are
	\begin{mybox}[Cosmological Self-Dual Double Copy]
		\begin{center}
			\renewcommand{\arraystretch}{1.8}
			\begin{tabularx}{\textwidth}{l X}
				\textbf{Gravity} & \vspace{-1cm}
				\begin{equation}
					V_{\rm SDG}=\frac{1}{2}\frac{1}{L\zeta}X(k_1,k_2)X^\zeta(k_1,k_2)
					\label{eq:cosmo_vertex}
				\end{equation} \\
				\textbf{Yang--Mills} & \vspace{-1cm}
				\begin{equation}
					V_{\rm SDYM}=\frac{1}{2}X(k_1,k_2)f^{a_1a_2a_3}
				\end{equation}
				\vspace{-0.5cm}
			\end{tabularx}
		\end{center}
	\end{mybox}
	This is again an asymmetric double copy, now adapted to cosmological conformal classes. 
	
	If only the Ricci scalar is fixed to match FLRW, the special cases include flat space, de Sitter, radiation domination with $a=\tau/L$, and coasting non-accelerating FLRW with $a=e^{H\tau}$. More general representatives of the same conformal classes have stress tensors that approach FLRW form when the additional $\partial_w^2\phi$ contribution is small \cite{CarrilloGonzalez:2024sto}. The important point is that the kinematic deformation is not chosen freely. It is fixed by the Jacobi bracket required by Weyl self-duality, so the double copy knows about the allowed curved conformal class.
	
	\subsection{Self-dual fields on self-dual backgrounds}
	
	The conformally flat examples above were fixed-background constructions: both the self-dual gauge field and self-dual graviton propagated on the same geometry. The construction of \cite{Brown:2023zxm} asks a different question. It starts with a nontrivial self-dual gauge background and asks whether the background itself can be included in the color-kinematics replacements. Thus the background is no longer a spectator. One may copy only the background, leaving a Yang--Mills fluctuation on the copied gravitational geometry, or copy both the background and the fluctuation.
	
	In the YM case, we keep an ansatz of the form Eq.~\eqref{eq:sdYM}, with the Yang--Mills scalar $\Phi\to\Phi_{\rm tot}=\Phi_0+\Phi_1$, where the adjoint-valued background $\Phi_0$ satisfies Eq.~\eqref{eq:sdYMflat} and $\Phi_1$ is the self-dual field on that background. Subtracting the background equation gives
	\be
	\Box^{\rm YM}_{\Phi_0}\Phi_1-\frac{i}{2}[\{\Phi_1,\Phi_1\}]=0\ ,
	\qquad
	\Box^{\rm YM}_{\Phi_0}f\equiv\Box_{\mathbb{R}^{4}}f-i[\{\Phi_0,f\}]\ .
	\label{eq:sdym_on_sdym_background}
	\ee
	The notation $\Box^{\rm YM}_{\Phi_0}$ is a shorthand for the kinetic operator around the gauge background. 
	
	On the gravitational side, let $\phi_0$ be a Ricci-flat self-dual background, so that it satisfies Eq.~\eqref{eq:sdg_flat_eom} with $\phi\to\phi_0$. We denote by $\Box_{\phi_0}$ the scalar wave operator on that background,
	\be
	\Box_{\phi_0}f\equiv\Box_{\mathbb{R}^{4}}f-2\{\{\phi_0,f\}\}\ ,
	\label{eq:sdg_background_wave_operator}
	\ee
	where we used the symmetry of the double bracket in Eq.~\eqref{eq:sdg_double_bracket}.
	
	The first replacement copies only the background,
	\be
	\Box^{\rm YM}_{\Phi_0}\longrightarrow\Box_{\phi_0}\ .
	\label{eq:sd_background_operator_replacement}
	\ee
	This gives Yang--Mills theory on the self-dual gravitational background obtained by the double copy,
	\be
	\Box_{\phi_0}\Phi_1-\frac{i}{2}[\{\Phi_1,\Phi_1\}]=0\ .
	\label{eq:sdym_on_sdg_background}
	\ee	
	One can then also copy the fluctuation by applying the same replacement as in flat space, $\Phi_1\to\phi_1$ and $\frac{i}{2}[\{\ ,\ \}]\to\{\{\ ,\ \}\}$. This gives the self-dual gravitational fluctuation equation on the self-dual gravitational background,
	\be
	\Box_{\phi_0}\phi_1-\{\{\phi_1,\phi_1\}\}=0\ .
	\label{eq:sdg_on_sdg_background}
	\ee
	Equivalently, the total scalar $\phi_0+\phi_1$ is again a solution of Eq.~\eqref{eq:sdg_flat_eom}. This is the so-called ``matched gauge'' of \cite{Brown:2023zxm}: the background and the fluctuation are treated by the same bracket structure, while the effect of the background sits in the kinetic operator.
	
	In this gauge the double copy is summarized as follows:
	\begin{mybox}[Self-Dual Background Double Copy]
		\begin{center}
			\renewcommand{\arraystretch}{1.8}
			\begin{tabularx}{\textwidth}{l X}
				\textbf{SDYM on SDYM} & \vspace{-1cm}
				\begin{equation}
					\Box^{\rm YM}_{\Phi_0}\Phi_1-\frac{i}{2}[\{\Phi_1,\Phi_1\}]=0
				\end{equation} \\
				\textbf{SDYM on SDG} & \vspace{-1cm}
				\begin{equation}
					\Box_{\phi_0}\Phi_1-\frac{i}{2}[\{\Phi_1,\Phi_1\}]=0
				\end{equation} \\
				\textbf{SDG on SDG} & \vspace{-1cm}
				\begin{equation}
					\Box_{\phi_0}\phi_1-\{\{\phi_1,\phi_1\}\}=0
				\end{equation}
				\vspace{-0.5cm}
			\end{tabularx}
		\end{center}
	\end{mybox}
	
	The matched gauge is useful for displaying the double copy of the equations, but it hides the kinematic algebra: the bracket is still the flat one and the background appears in $\Box_{\phi_0}$. There is another gauge, referred to as ``flipped gauge'', that is useful when one wants the background to deform the bracket itself. In flat space, the flipped and matched gauges are related by exchanging the lightcone directions. It works in a simple form only for backgrounds that can be written in Kerr--Schild form, which requires
	\be
	\phi_{0,uw}^2-\phi_{0,ww}\phi_{0,uu}=0\ .
	\label{eq:curved_ks_condition}
	\ee
	In that case the relevant derivatives are replaced by the background-dressed operators
	\eqs{
		\hat\Pi_\alpha^{\phi_0}=\left(\hat k_v(\phi_0),\hat k_{\bar w}(\phi_0)\right)\ , \\
		\hat k_v(\phi_0)=\partial_v-\frac{1}{2}\left(\phi_{0,ww}\partial_u-\phi_{0,uw}\partial_w\right)\ ,\\
		\hat k_{\bar w}(\phi_0)=\partial_{\bar w}-\frac{1}{2}\left(\phi_{0,uw}\partial_u-\phi_{0,uu}\partial_w\right)\ .
		\label{eq:curved_khat}
	}
	These define a background-dependent bracket,
	\be
	\{f,g\}_{\phi_0}
	=\varepsilon^{\alpha\beta}\hat\Pi_\alpha^{\phi_0}f\,\hat\Pi_\beta^{\phi_0}g\ .
	\label{eq:curved_poisson}
	\ee
	The self-duality equation for $\phi_0$ together with Eq.~\eqref{eq:curved_ks_condition} ensures the Jacobi identity. The corresponding color and gravitational brackets are
	\eqs{
		[\{f,g\}]_{\phi_0}
		&=\varepsilon^{\alpha\beta}\left[\hat\Pi_\alpha^{\phi_0}f,\hat\Pi_\beta^{\phi_0}g\right]\ ,\\
		\{\{f,g\}\}_{\phi_0}
		&=\frac{1}{2}\varepsilon^{\alpha\beta}
		\{\hat\Pi_\alpha^{\phi_0}f,\hat\Pi_\beta^{\phi_0}g\}_{\phi_0}\ .
		\label{eq:flipped_sdym}
	}
	The equations take the same schematic form as before, but the nonlinear brackets are replaced by $[\{\ ,\ \}]_{\phi_0}$ and $\{\{\ ,\ \}\}_{\phi_0}$. This exposes the deformed kinematic algebra. This algebraic structure is connected to the fact that one can define a Poisson bracket for self-dual spacetimes \cite{Dunajski:2000iq}.
	
	The curved kinematic algebra can be displayed explicitly in examples. In the present conventions, the self-dual plane-wave background is
	\be
	ds^2_{\rm PW}=dw\,d\bar w-du\,dv+2F(u)d\bar w^2,\qquad
	\phi_{0,uu}=2F(u)\ ,
	\label{eq:sd_pw_metric}
	\ee
	where $F(u)$ is the wave profile. This is a Kerr--Schild background, so the flipped gauge bracket above applies; in particular $\hat\Pi_\alpha^{\phi_0}=(\partial_v,\partial_{\bar w}+F(u)\partial_w)$ for this metric. We now want to find the new structure constants for the kinematic algebra in this example. First, we look for the analogue of a flat plane wave, which should solve the scalar wave equation on Eq.~\eqref{eq:sd_pw_metric} and reduce to $e^{-ik\cdot x}$ when $F\to0$. Considering a null vector $k^\mu$, one can parametrize the null condition by
	\be
	k_v=\rho k_w,\qquad k_{\bar w}=\rho k_u\ .
	\label{eq:sd_pw_null_parametrization}
	\ee
	Choose $G'(u)=F(u)$, and define
	\be
	Q_k=k \cdot x+\frac{G(u)}{\rho}k_w\ .
	\label{eq:pw_qk}
	\ee
	The covector $\partial_\mu Q_k$ is null, divergence-free, and geodesic with respect to Eq.~\eqref{eq:sd_pw_metric}. Hence every function of $Q_k$ is annihilated by $\Box_{\phi_0}$, and in particular $\Box_{\phi_0}e^{-iQ_k}=0$. The undeformed bracket of two such modes defines a background-dependent kinematic factor,
	\eqs{
		\{e^{-iQ_1},e^{-iQ_2}\}=-e^{-i(Q_1+Q_2)}X_{\rm PW}(k_1,k_2)\ ,\\
		X_{\rm PW}(k_1,k_2)=X(k_1,k_2)+
		k_{1w}k_{2w}\frac{\rho_1-\rho_2}{\rho_1\rho_2}F(u)\ .
		\label{eq:pw_structure}
	}
	In the holomorphic collinear limit $\rho_1=\rho_2$, the deformation vanishes. Consequently, the lifted $w$-algebra is undeformed \cite{Brown:2023zxm}, contrary to the conformally flat cases described above.
	
	The matched gauge gravitational double bracket does not simply square Eq.~\eqref{eq:pw_structure}; it contains an additional curvature term proportional to $F'(u)$. The simple double copy instead appears in the flipped gauge brackets introduced above. For the plane-wave background they give
	\be
	\{e^{-iQ_1},e^{-iQ_2}\}_{\phi_0}
	=-e^{-i(Q_1+Q_2)}\rho_1\rho_2X_{\rm PW}(k_1,k_2)\ ,
	\label{eq:pw_flipped_single_bracket}
	\ee
	and the corresponding double bracket squares this flipped gauge kinematic factor:
	\be
	\{\{e^{-iQ_1},e^{-iQ_2}\}\}_{\phi_0}
	=\frac{1}{2}e^{-i(Q_1+Q_2)}\left(\rho_1\rho_2X_{\rm PW}(k_1,k_2)\right)^2\ .
	\label{eq:pw_double_bracket}
	\ee
	
	The Eguchi--Hanson solution gives a more rigid instanton example. The gravitational background is encoded in the scalar
	\be
	\phi_0^{\rm EH}=\frac{m u^2}{2\bar w^2(uv-w\bar w)}\ ,
	\label{eq:eh_background_scalar}
	\ee
	Here $m$ is the Eguchi--Hanson parameter. This scalar solves Eq.~\eqref{eq:sdg_flat_eom} and obeys Eq.~\eqref{eq:curved_ks_condition}.
	
	The analogue of a plane wave, with the same null parametrization as Eq.~\eqref{eq:sd_pw_null_parametrization}, is now $e^{i\sqrt{R_k}}$ with the phase factor defined as
	\be
	R_k=\left(k \cdot x \right)^2
	-\frac{m(uk_u+wk_w)^2}{2(uv-w\bar w)^2}\ .
	\label{eq:eh_rk}
	\ee
	Then $e^{i\sqrt{R_k}}$ solves the scalar wave equation on the Eguchi--Hanson background and reduces to the ordinary plane wave in the limit $m\to0$. In the holomorphic collinear limit $\rho_1=\rho_2$, the bracket of two such modes defines
	\eqs{
		\{e^{i\sqrt{R_1}},e^{i\sqrt{R_2}}\}
		&=e^{i(\sqrt{R_1}+\sqrt{R_2})}X_{\rm EH}(k_1,k_2)\ ,\\
		X_{\rm EH}(k_1,k_2)&=\frac{X(k_1,k_2)}{\sqrt{R_1R_2}}
		\Bigg[(k_1\cdot x)(k_2\cdot x)\nonumber\\
		&\quad-\frac{m(uk_{1u}+wk_{1w})(uk_{2u}+wk_{2w})}{2(uv-w\bar w)^2}\Bigg]\ .
		\label{eq:eh_structure}
	}
	This has a deformation that does not vanish when uplifting to the w-algebra and gives a genuinely deformed version \cite{Brown:2023zxm} similar to the cosmological cases \cite{CarrilloGonzalez:2024sto}. In this case, both the matched and flipped gauge double copy give structure constants that are not simple squares. For example, in the Eguchi--Hanson background with flipped gauge we have
	\be
	\begin{aligned}
		\{\{e^{i\sqrt{R_1}},e^{i\sqrt{R_2}}\}\}_{\phi_0^{\rm EH}}
		={}&\frac{1}{2}e^{i(\sqrt{R_1}+\sqrt{R_2})}X_{\rm EH}(k_1,k_2)^2\\
		&+\text{curvature terms}\ .
	\end{aligned}
	\label{eq:eh_double_bracket}
	\ee
	The omitted terms are of order $m$ or higher and vanish in the flat limit, but they are not removed by going to the flipped gauge as in the waves example. Thus the contrast with Eq.~\eqref{eq:pw_double_bracket} is the main lesson of the two examples: self-dual curved backgrounds can support a color-to-kinematics replacement, and the flipped gauge brackets make this replacement especially sharp for self-dual plane waves, while less degenerate backgrounds such as Eguchi--Hanson still retain curvature corrections \cite{Brown:2023zxm}.
	
	\subsection{Twistor KLT kernel on self-dual radiative spacetimes} \label{sec:sd_twistor_klt}
	
	The twistor-space KLT construction reviewed in Sec.~\ref{sec:twistor_klt} has a self-dual curved-space extension for radiative spacetimes \cite{Adamo:2024hme}. These are complex asymptotically flat vacuum solutions whose free data are specified at complexified null infinity. The point of the construction is to ask whether the twistor KLT kernel can absorb the effect of a nontrivial self-dual background while preserving the two color-ordered factors. The extra background datum is the asymptotic shear $\tilde\sigma^0([\mu\,\bar{\lambda}],\,\lambda_{\alpha},\,\bar{\lambda}_{\dot{\alpha}})$, which deforms the complex structure on twistor space as
	\be
	\bar\nabla=\bar\partial+\tilde\sigma^0 D\bar\lambda\,\bar\lambda^{\dot\alpha}\frac{\partial}{\partial\mu^{\dot\alpha}},
	\qquad \bar\nabla^2=0\ .
	\label{eq:sd_radiative_dolbeault}
	\ee
	The condition $\bar\nabla^2=0$ is the integrability condition ensuring that the deformed twistor space remains a consistent complex manifold.
	
	The first object to keep track of is not yet a double copy formula, but the conjectural all-multiplicity graviton amplitude on the self-dual radiative background. With the same helicity split as in Sec.~\ref{sec:twistor_klt}, degree $d$ describes an $\mathrm{N}^{d-1}\mathrm{MHV}$ sector, with $\#\tilde h=d+1$ and $\#h=n-d-1$. Writing $N_m^{(k)}\equiv N^{(k)}(\sigma_m)$ and suppressing the arguments of the external twistor wavefunctions, the formula is \cite{Adamo:2022mev,Adamo:2024hme}
	\begin{multline}
		\mathcal{M}_{n,d}=
		\sum_{t=0}^{n-d-3}\sum_{p_1,\ldots,p_t}
		\int\dd\mu_d\,\Delta_{\tilde h}^8\,\mathrm{det}'(\mathbb{H}^{\vee})
		\prod_{i\in h}h_i\prod_{j\in\tilde h}\tilde h_j\\
		\times
		\left.
		\left(\prod_{m=1}^{t}
		D\sigma_m\wedge D\bar\lambda(\sigma_m)\,
		\frac{N_m^{(p_m-2)}}{p_m!}
		\frac{\partial^{p_m}}{\partial\varepsilon_m^{p_m}}
		\right)\mathrm{det}'(\mathcal{H})
		\right|_{\varepsilon=0}\ .
		\label{eq:sd_radiative_graviton_formula}
	\end{multline}
	The determinant $\mathrm{det}'(\mathbb{H}^{\vee})$ is unchanged from its flat-space counterpart in Eq.~\eqref{eq:cachazo_skinner}, while $\mathcal{H}$ replaces the positive-helicity matrix by an enlarged matrix that also knows about the background.
	
	The new ingredients in Eq.~\eqref{eq:sd_radiative_graviton_formula} have a simple interpretation. In curved spacetime, radiation can develop tails: an external graviton can interact with the background curvature before reaching null infinity. The integer $t$ counts how many such background insertions occur. For each insertion $m$, the integer $p_m\geq3$ records the order of that insertion in the expanded determinant. The auxiliary parameter $\varepsilon_m$ is only a bookkeeping variable; differentiating $p_m$ times and then setting all $\varepsilon_m$ to zero extracts the desired term.
	
	The background insertion itself is built from the Bondi news, $N^{(0)}=-\partial_u\tilde\sigma^0$, and its derivatives
	\be
	N^{(k)}(\sigma_m)=-\left.\frac{\partial^{k+1}}{\partial u^{k+1}}
	\tilde\sigma^0\big(u,\lambda(\sigma_m),\bar\lambda(\sigma_m)\big)
	\right|_{u=[\mu(\sigma_m)\bar\lambda(\sigma_m)]}\ .
	\label{eq:sd_radiative_news}
	\ee
	Thus $N_m^{(p_m-2)}$ inserts the appropriate $u$-derivative of the radiative data at the worldsheet point $\sigma_m$, evaluated using the twistor incidence relation $u=[\mu\,\bar\lambda]$.
	
	In \cite{Adamo:2024hme}, it was shown that this amplitude takes the KLT-like form
	\be
	\mathcal{M}_{n,d}=
	\sum_{\substack{b\rho,b\omega\in\mathcal{S}(h)\\
			a\bar\rho,a\bar\omega\in\mathcal{S}(\tilde h)}}
	\int\dd\mu_{n,d}\,
	\mathcal{I}_{n}^{\tilde h}[a\bar\rho b\rho]\,
	S^{N}_{n,d}[\rho,\bar\rho|\omega,\bar\omega]\,
	\mathcal{I}_{n}^{\tilde h}[\bar\omega^{T}ab\omega]
	\prod_{i\in h}h_i\prod_{j\in\tilde h}\tilde h_j\ .
	\label{eq:sd_radiative_twistor_klt}
	\ee
	The appearance of two color-ordered factors glued by a kernel is suggestive, but Eq.~\eqref{eq:sd_radiative_twistor_klt} is not automatically a double copy for scattering on a generic self-dual gauge background. The reason is that the known radiative-background RSVW formula in gauge theory contains non-abelian color transport through the background \cite{Adamo:2020yzi}. That transport is carried by holomorphic frame factors, whereas Eq.~\eqref{eq:sd_radiative_twistor_klt} keeps the two RSVW-like factors in the flat-space ordering basis and places all dependence on the gravitational radiative data in $S^N_{n,d}$.
	
	The kernel still factorizes chirally,
	\be
	S^{N}_{n,d}[\rho,\bar\rho|\omega,\bar\omega]
	=\mathcal{D}(\omega,\bar\omega)\,
	\mathbb{S}_{\tilde h}[\bar\rho|\bar\omega]\,
	\mathbb{S}^{N}_{h}[\rho|\omega]\ ,
	\label{eq:sd_radiative_chiral_kernel}
	\ee
	and only the positive-helicity factor $\mathbb{S}^{N}_{h}$ is deformed. This reflects the self-duality of the background: the background curvature couples to the same chiral half of the twistor formula, while the opposite-helicity determinant remains the flat-space one.
	
	More concretely, the analogue of Eq.~\eqref{eq:RSVW} in a self-dual radiative gauge background is given by Eq.~\eqref{eq:RSVW} with the Parke--Taylor factor replaced by \cite{Adamo:2020yzi}
	\be
	\prod_{i=1}^{n}\frac{H^{-1}(U,\sigma_i)T^{a_i}H(U,\sigma_i) D\sigma_{\rho(i)}}{\big(\rho(i)\rho(i+1)\big)}\ , \label{eq:sd_radiative_gauge_frame}
	\ee
	Here $H$ is the holomorphic trivialization of the background bundle, so the color generators are conjugated by the background frame.
	\be
	\bar\partial H(U,\sigma)=-\tilde A^0(\sigma)H(U,\sigma)D\bar\lambda(\sigma)\ ,
	\ee
	where $\tilde A^0$ is the characteristic radiative gauge-field data pulled back to twistor space. These non-abelian holomorphic frames do not have a known counterpart in Eq.~\eqref{eq:sd_radiative_twistor_klt}. For a Cartan-valued background, however, the conjugation is diagonal:
	\be
	H^{-1}(U,\sigma_i)T^{a_i}H(U,\sigma_i)
	=e^{e_i g(U,\sigma_i)}T^{a_i},
	\qquad
	\bar\partial g(U,\sigma)=\tilde A^0(\sigma)D\bar\lambda(\sigma) ,
	\label{eq:sd_radiative_cartan_frame}
	\ee
	with $e_i$ the charge of $T^{a_i}$ under the Cartan subalgebra.
	
	Denoting the resulting color-ordered integrand of the self-dual radiative RSVW formula by $\mathcal{I}^{\tilde h,e}_n$, the gravity formula can then be written as
	\be
	\mathcal{M}_{n,d}=
	\sum_{\substack{b\rho,b\omega\in\mathcal{S}(h)\\
			a\bar\rho,a\bar\omega\in\mathcal{S}(\tilde h)}}
	\int\dd\mu_{n,d}\,
	\mathcal{I}_{n}^{\tilde h,e}[a\bar\rho b\rho]\,
	S^{N}_{n,d}[\rho,\bar\rho|\omega,\bar\omega]\,
	\mathcal{I}_{n}^{\tilde h,-e}[\bar\omega^{T}ab\omega]
	\prod_{i\in h}h_i\prod_{j\in\tilde h}\tilde h_j\ .
	\label{eq:sd_radiative_charge_reversed_dc}
	\ee
	The two single copies carry opposite background charges, so their exponential frame factors cancel and one is left with the double copy formula from Eq.~\eqref{eq:sd_radiative_twistor_klt}. Equation~\eqref{eq:sd_radiative_charge_reversed_dc} therefore gives a charge-reversed double copy interpretation for Cartan-valued self-dual radiative gauge backgrounds. This is similar to the restriction to the Cartan subalgebra in Section~\ref{sec:plane_wave_scattering_dc} that allowed the double copy construction. No analogue of charge reversal is known for fully non-abelian radiative backgrounds, and the deformed positive-helicity kernel $\mathbb{S}^{N}_{h}$ has not been inverted on the space of color orderings as in \cite{Mizera:2016jhj}. Consequently, there is not yet a biadjoint-scalar interpretation of the inverse kernel. Nevertheless, Eq.~\eqref{eq:sd_radiative_charge_reversed_dc} provides a double copy for self-dual radiative amplitude integrands restricted to Cartan-valued background gauge fields.
	\vspace{0.2cm}
	\begin{mybox}[Radiative Self-Dual Double Copy]
		\begin{center}
			\renewcommand{\arraystretch}{1.8}
			\begin{tabularx}{\textwidth}{l X}
				\textbf{SD Radiative Gravity} &   \vspace{-1cm} 
				\begin{equation}
					\begin{aligned}
						&\mathcal{I}_{n}^{\tilde h,e}[a\bar\rho b\rho]\,
						S^{N}_{n,d}[\rho,\bar\rho|\omega,\bar\omega]
						\mathcal{I}_{n}^{\tilde h,-e}[\bar\omega^{T}ab\omega]\ ,
					\end{aligned}
					\label{eq:sd_rad_gravity}
				\end{equation} \vspace{-0.5cm}  \\ 
				\textbf{SD Radiative Yang--Mills} &   \vspace{-1cm} 
				\begin{equation}
					\mathcal{I}_{n}^{\tilde h,e}[a\bar\rho b\rho]
					\label{eq:sd_rad}
				\end{equation} 
			\end{tabularx}
		\end{center}
	\end{mybox}
	\vspace{0.2cm}

	\subsection{Non-perturbative self-dual solutions} \label{sec:sd_nonpert}
	
	Although not yet formulated as a double copy in curved spacetimes as defined in Section~\ref{sec:intro}, we briefly review some exact self-dual double copy examples which showcase the power of the self-dual sector. The preceding constructions were formulated either at the level of equations of motion or through twistor-space scattering data. There is also an exact-solution version given in terms of the Kerr--Schild double copy reviewed in Section~\ref{sec:kerr-schild}. A much earlier relation between the gauge and gravitational self-dual theories was observed in~\cite{Tod:1982} where self-dual Kerr--Schild solutions of the vacuum Einstein equations were constructed from null Maxwell fields in Minkowski space. For a broad class of self-dual vacuum classical solutions, the twistor-space and Kerr--Schild double copies from Section~\ref{sec:classical_field_level} are equivalent \cite{Albertini:2025ptks}. In the self-dual setting, Ref.~\cite{Monteiro:2014cda} recasts the Kerr--Schild vector as a commuting spin-raising differential operator $\hat k_\mu$. The operator version of the single, double, and zeroth copies is
	\begin{mybox}[Self-Dual Kerr-Schild Double Copy]
		\begin{center}
			\renewcommand{\arraystretch}{1.8}
			\begin{tabularx}{\textwidth}{l X}
				\textbf{Gravity} &   \vspace{-1cm}
				\begin{equation}
					h_{\mu\nu}=\hat k_\mu\hat k_\nu\phi
					\label{eq:classical_sd_operator_dc}
				\end{equation} \vspace{-1cm}  \\
				\textbf{Yang--Mills} &   \vspace{-1cm}
				\begin{equation}
					A_\mu^a=\hat k_\mu\Phi^a
					\label{eq:sd_operator_gauge}
				\end{equation}    \vspace{-0.5cm}
			\end{tabularx}
		\end{center}
	\end{mybox}
	where $\Phi^a$ is the self-dual Yang--Mills scalar potential, $\phi$ is the self-dual gravitational potential, and $c^{b'}$ is a constant color vector for the second algebra. The operator obeys the analogues of the Kerr--Schild null and geodesic conditions, including $\hat k^2=0$ and $\hat k\cdot\partial=0$. In the double lightcone coordinates of Eq.~\eqref{lccoords}, it can be chosen as
	\be
	\hat{k}_u=0,\qquad \hat{k}_v=\frac{1}{4}\partial_w,\qquad \hat{k}_w=0,\qquad \hat{k}_{\bar{w}}=\frac{1}{4}\partial_u\,.
	\ee
	Since $\hat k_i=0$ and $\hat k_\alpha=\frac{1}{4}\Pi_\alpha$, the ansatz is already in the gauges used in Eqs.~\eqref{eq:sdYM} and \eqref{eq:sdg_flat_metric}, so the equation of motion for the gauge field reduces precisely to the self-dual Yang--Mills equation introduced in Eq.~\eqref{eq:sdYMflat} up to numerical factors absorbed in the definition of the scalar. If the color direction is fixed, $\Phi^a=c^a\Phi_0$, the commutator term vanishes, the single copy is abelian, and the biadjoint scalar reduces to $\Phi^{a\,b'}=c^a c^{b'}\Phi_0$.
	
	The gravitational equation follows similarly. The Ricci-flat condition is equivalent to the scalar condition
	\be
	\Box_{\mathbb{R}^{4}}\phi-\frac{\kappa}{2}
	\left(\hat k^\mu\hat k^\nu\phi\right)\partial_\mu\partial_\nu\phi=0\ .
	\label{eq:sd_operator_gravity_reduction}
	\ee
	Using the component form of $\hat k_\mu$, the nonlinear term is the same object as the double bracket in Eq.~\eqref{eq:sdg_double_bracket}, up to the conventional normalization absorbed into $\phi$. Consequently Eq.~\eqref{eq:sd_operator_gravity_reduction} is the Plebanski equation reviewed in Eq.~\eqref{eq:sdg_flat_eom}. This makes the replacement of color by a second kinematic operator explicit at the level of the self-dual equations, not only at the level of the fields. Applied to the Eguchi--Hanson instanton, the map gives an abelian-like single copy whose field is dipole-like at large distance and carries neither electric nor magnetic charge. Thus the nontrivial topology of the gravitational instanton is not represented by an ordinary gauge charge \cite{Berman:2018hfs}.
	
	In Euclidean non-perturbative analyses of self-dual double copies, simple spherically symmetric power-law ansatze for the biadjoint scalar do not produce genuinely nonlinear solutions in exactly four dimensions \cite{ArmstrongWilliams:2022npd}. The four-dimensional solution is harmonic, so it solves only the linearized biadjoint equation and is naturally identified with the Eguchi--Hanson zeroth copy. More general instanton constructions use 't Hooft symbols\footnote{The 't Hooft symbols form an $\mathfrak{su}(2)$ subalgebra of $\mathfrak{so}(4)$. The anti-self-dual 't Hooft symbols $\bar\eta^i_{\mu\nu}$ form a constant basis of anti-self-dual two-forms on flat Euclidean $\mathbb R^4$. With $i,j,k=1,2,3$ and the fourth direction denoted by $4$, a common convention is $\bar\eta^i_{jk}=\epsilon_{ijk}$ and $\bar\eta^i_{j4}=-\delta^i_j$, with antisymmetry fixing the remaining components. The self-dual symbols $\eta^i_{\mu\nu}$ have the opposite sign in the $j4$ component.}, $\bar\eta^i_{\mu\nu}$, and operators of the form
	\be
	A_\mu=\hat k_\mu\phi,\qquad
	\hat k_\mu=\left(B_i\bar\eta^i_{\mu\nu}\right)\partial^\nu\ ,
	\label{eq:nonpert_khat}
	\ee
	with $B_iB^i=0$. This form was obtained after imposing that $\phi$ is harmonic, which implies the self-duality of the gauge field, and that the double copy is self-dual. One may also dress the abelian solution to obtain a non-abelian SU(2) instanton and its double copy
	\be
	A_\mu^a=-\bar\eta^a_{\mu\nu}\hat k^\nu\phi=-k_\mu^a\phi ,\qquad
	h_{\mu\nu}=-\hat k_\mu^a\hat k_\nu^a\phi\ .
	\label{eq:nonabelian_instanton_dc}
	\ee
	Surprisingly, the double copy of both abelian and non-abelian gauge fields coincides \cite{ArmstrongWilliams:2022npd}. This gives a constrained map between sectors of gauge instantons and self-dual gravitational instantons. Whether every non-abelian instanton has a gravitational double copy remains an open question.
	
	\subsection{Coherent-state self-dual backgrounds and focus waves}\label{subsec:coherent_sd_backgrounds}
	
	The coherent-state and self-dual Kerr--Schild double copies have different starting points. The coherent-state construction reviewed in Sec.~\ref{sec:coherent_state_backgrounds} is an amplitudes-based prescription: it builds a classical background from on-shell coherent profiles and doubles the polarization vector of each mode. By contrast, the operator Kerr--Schild construction reviewed in Sec.~\ref{sec:sd_nonpert} is a field-level ansatz for self-dual gauge and gravitational solutions, with exactness on the gravity side controlled by the nonlinear heavenly equation. The two constructions overlap when the coherent background can be written using an operator with the Kerr--Schild properties and its scalar profile obeys this additional equation.
	
	To display this overlap, consider a coherent background built from on-shell plane waves. When all modes have one helicity and one color direction, the non-abelian commutator vanishes and the background becomes an exact self-dual solution \cite{Ilderton:2025gug}. We use the Euclidean self-dual conventions of Eq.~\eqref{lccoords}, with complexified coordinates $x^\mu=(u,v,x^1,x^2)$ and flat metric $ds^2=dx^idx^i-du\,dv$. Choosing coherent profiles so that the gauge field contains only the positive-helicity polarization vector gives
	\begin{equation}
		A_\mu(x)
		=T\int_k\left[
		\epsilon_\mu^+(k)\bar\alpha'(k)e^{ik\cdot x}
		+\epsilon_\mu^+(k)\alpha(k)e^{-ik\cdot x}
		\right],
		\label{eq:coherent_sd_profile}
	\end{equation}
	where $T$ is a fixed Cartan generator. Since the color commutator is absent and each Fourier mode is on shell, Eq.~\eqref{eq:coherent_sd_profile} is an exact self-dual vacuum solution of the Yang--Mills equations.
	
	The same field can be packaged as
	\begin{equation}
		A_\mu=T\widehat V_\mu\phi\, ,
		\label{eq:coherent_sd_gauge}
	\end{equation}
	in terms of a scalar wave,
	\begin{equation}
		\phi(x)=\int_k\left[\bar\alpha'(k)e^{ik\cdot x}+\alpha(k)e^{-ik\cdot x}\right],
		\qquad
		\Box\phi=0\, ,
		\label{eq:coherent_sd_scalar}
	\end{equation}
	together with the non-local spin-raising operator
	\begin{equation}
		\widehat V_\mu
		=\frac{1}{\sqrt2}
		\left(\frac{\partial_1+i\partial_2}{\partial_u},0,1,i\right) \ .
		\label{eq:coherent_sd_v_operator}
	\end{equation}
	The operator obeys
	\begin{equation}
		[\widehat V_\mu,\widehat V_\nu]=0\, ,
		\qquad
		\eta^{\mu\nu}\widehat V_\mu\widehat V_\nu=0\, ,
		\qquad
		\widehat V^\mu\partial_\mu=0\, .
		\label{eq:coherent_sd_v_conditions}
	\end{equation}
	
	In Fourier space, $\widehat V_\mu$ inserts precisely the positive-helicity polarization vector. The coherent-state replacement that removes the color generator and doubles this polarization can therefore be written
	\begin{equation}
		g_{\mu\nu}
		=\eta_{\mu\nu}+\widehat V_\mu\widehat V_\nu\phi\, .
		\label{eq:coherent_sd_gravity}
	\end{equation}
	Because $\widehat V_\mu$ obeys Eq.~\eqref{eq:coherent_sd_v_conditions}, this field has the operator Kerr--Schild form of Eq.~\eqref{eq:classical_sd_operator_dc}. Thus polarization doubling in the coherent-state construction and replacing color by a second spin-raising operator in the self-dual Kerr--Schild construction give the same metric field. This identifies their common field configuration, but does not yet guarantee that it is an exact gravitational solution.
	
	For a generic scalar wave $\phi$, Eq.~\eqref{eq:coherent_sd_gravity} solves only the linearized self-dual Einstein equations. It becomes an exact self-dual vacuum metric when $\phi$ additionally satisfies
	\begin{equation}
		\phi\,\partial_u^{-2}\partial_z^2\phi
		=\left(\partial_u^{-1}\partial_z\phi\right)^2\, ,
		\qquad
		z=\frac{x^1+ix^2}{\sqrt2}\, .
		\label{eq:coherent_plebanski}
	\end{equation}
	Together with the wave equation in Eq.~\eqref{eq:coherent_sd_scalar}, this is the Plebanski equation already encountered in Eq.~\eqref{eq:sdg_flat_eom}, expressed in variables adapted to the non-local operator $\widehat V_\mu$. Indeed, it takes the same form of Eq.~\eqref{eq:sdg_flat_eom} if we introduce the local heavenly potential $\phi_{\rm H}=\partial_u^{-2}\phi$. On this subclass, the background map induced by the coherent-state prescription coincides with the self-dual Kerr--Schild map, while the coherent-state construction additionally relates any-multiplicity amplitudes on the exact backgrounds. Without Eq.~\eqref{eq:coherent_plebanski}, it still supplies an amplitudes double copy, but its gravitational background is only a linearized solution.
	
	\paragraph{Example: Focus-waves}
	A concrete exact example is supplied by focus-wave profiles. In Eq.~\eqref{eq:coherent_sd_profile}, set $\alpha'=0$ and take
	\begin{equation}
		\alpha(k)=\frac{4\pi}{\ell}f(k_u)
		\exp\left[-\frac{k_\perp^2}{2\ell k_u}\right],
		\qquad
		k_\perp^2=k_1^2+k_2^2\, ,
		\label{eq:coherent_focus_profile}
	\end{equation}
	with $\ell>0$. The scalar integral can then be performed explicitly:
	\begin{equation}
		\phi(x)
		=\frac{1}{1+i\ell v}\,
		f\!\left(u-\frac{\ell x_\perp^2}{2(1+i\ell v)}\right),
		\qquad
		x_\perp^2=(x^1)^2+(x^2)^2\, .
		\label{eq:coherent_focus_scalar}
	\end{equation}
	For example, $f(q)=e^{-i\omega q}$ gives a flying-focus wavepacket. Acting with $\widehat V_\mu$ gives
	\begin{equation}
		A_\mu(x)=T E_\mu(x)\phi(x)\, ,
		\qquad
		E_\mu(x)=\frac{1}{\sqrt2}
		\left(-i\frac{\ell(x^1+ix^2)}{1+i\ell v},0,1,i\right).
		\label{eq:coherent_focus_gauge}
	\end{equation}
	This scalar obeys Eq.~\eqref{eq:coherent_plebanski}, so the double copy is an exact self-dual metric,
	\begin{equation}
		g_{\mu\nu}
		=\eta_{\mu\nu}+E_\mu(x)E_\nu(x)\phi(x)\, .
		\label{eq:coherent_focus_gravity}
	\end{equation}
	The final expression is local in position space even though $\widehat V_\mu$ contains $\partial_u^{-1}$. This locality is special to the wavepacket: $\phi$ is an eigenfunction of the non-local part of $\widehat V_\mu$, and $[\widehat V_\mu,E_\nu]=0$. The example therefore explains why the coherent-state prescription can coincide with a simple Kerr--Schild-like position-space double copy in special self-dual cases, while one should not expect such locality for an arbitrary coherent background.
	
	\subsection{Boundary correlators in \texorpdfstring{AdS$_4$}{AdS4}}
	
	The scalar actions for self-dual YM and gravity on AdS$_4$ reviewed in Section~\ref{sec:sd_ads_cosmo} allow the double copy to be tested directly on boundary correlators \cite{Chowdhury:2024dcy}. In the lightcone variables used there, the self-dual interaction is cubic and linear in the negative-helicity field. A tree-level correlator in this sector therefore has one negative-helicity external field and all remaining external fields of positive helicity, and it is built entirely from cubic self-dual vertices. 
	
	A scalar field in AdS admits a useful representation in terms of a massless cubic theory on the flat half-space $z>0$, with its boundary at $z=0$.  After an appropriate Weyl rescaling, the kinetic term of the conformally coupled AdS scalar becomes that of a massless flat-space scalar, so the corresponding bulk-to-boundary and bulk-to-bulk propagators reduce to those of the half-space problem. The three-point scalar contact diagram and the four-point $s$-channel exchange diagram are then
	\be
	A_3^{\varphi^3}=\frac{1}{k_{123}},\qquad
	A_4^{\varphi^3,(s)}=\frac{1}{E E_L^{(s)}E_R^{(s)}}\ ,
	\label{eq:sd_ads_scalar_seeds}
	\ee
	where $k_i=|\vec k_i|$, $k_{ij}=k_i+k_j$, $E=k_{1234}$, $k_I^{(s)}=|\vec k_1+\vec k_2|$, and
	\be
	E_L^{(s)}=k_{12}+k_I^{(s)},\qquad E_R^{(s)}=k_{34}+k_I^{(s)}\ .
	\ee
	The pole at $E=0$ is reached by analytic continuation, and its residue gives the corresponding flat-space exchange diagram.
	
	The SDYM correlators are obtained by dressing each cubic vertex of the scalar diagram with one kinematic structure constant $X$ from Eq.~\eqref{eq:flat_poisson}. At three points,
	\be
	\mathcal{A}_3(1^+,2^+,3^-)=\frac{X_{1,2}}{k_{123}}\ ,
	\label{eq:sdym_ads_corr3}
	\ee
	while the $s$-channel exchange contribution is
	\be
	\mathcal{A}_4^{(s)}(1^+,2^+,3^+,4^-)
	=\frac{X_{1,2}X_{3,4}}{E E_L^{(s)}E_R^{(s)}}\ .
	\label{eq:sdym_ads_corr4s}
	\ee
	Thus each exchange diagram is the scalar diagram in Eq.~\eqref{eq:sd_ads_scalar_seeds} dressed by the two SDYM numerators associated with its vertices. Thus, the full four-point  color-ordered correlator reads,
	\be
	\mathcal{A}_4(1^+,2^+,3^+,4^-)
	=\frac{1}{E}\left(
	\frac{X_{1,2}X_{3,4}}{E_L^{(s)}E_R^{(s)}}
	+\frac{X_{1,4}X_{3,2}}{E_L^{(t)}E_R^{(t)}}
	\right)\ ,
	\label{eq:sdym_ads_corr4}
	\ee
	where the $t$-channel energies are obtained by the corresponding relabeling. For a single channel,
	\be
	\lim_{E\to0}E\mathcal{A}_4^{(s)}
	=\frac{X_{1,2}X_{3,4}}{(k_I^{(s)})^2-k_{12}^2}\ ,
	\ee
	which is the expected flat-space exchange diagram. After the $s$- and $t$-channels are combined, however, their residues cancel. The full AdS correlator need not vanish; it simply has no total-energy pole, in agreement with the vanishing four-point tree amplitude of SDYM as required from integrability of the theory.
	
	For self-dual gravity, the AdS deformation replaces the second factor of $X$ at each vertex by a differential operator acting on the scalar diagram. At three points this gives
	\be
	\mathcal{M}_3(1^+,2^+,3^-)
	=-X_{1,2}\hat{\mathcal D}_{1,2}\left(\frac{1}{k_{123}}\right),
	\qquad
	\hat{\mathcal D}_{1,2}=X_{1,2}\partial_{k_{123}}-i(k_{1w}-k_{2w})\ .
	\label{eq:sdg_ads_corr3}
	\ee
	The flat-space residue gives the expected square, $\lim_{k_{123}\to0}k_{123}^2\mathcal{M}_3=X_{1,2}^2$. At four points the analogous $s$-channel representation is
	\begin{align}
		\mathcal{M}_4^{(s)}(1^+,2^+,3^+,4^-)
		&=\frac{1}{4}X_{1,2}X_{3,4}
		\hat{\mathcal D}_{1,2}^{(s)}\hat{\mathcal D}_{3,4}^{(s)}
		\left(\frac{1}{E E_L^{(s)}E_R^{(s)}}\right),
		\label{eq:sdg_ads_corr4s}\\
		\hat{\mathcal D}_{1,2}^{(s)}
		&=X_{1,2}\partial_{k_{12}}-i(k_{1w}-k_{2w}),\nonumber\\
		\hat{\mathcal D}_{3,4}^{(s)}
		&=X_{3,4}\partial_{k_{34}}+i(3k_{3w}+k_{4w})\ .
		\nonumber
	\end{align}
	The derivatives act on the external energy sums while $k_I^{(s)}$ is held fixed. This prescription can be justified by first treating the external energies as independent of the exchanged momentum and imposing the on-shell relations after the derivatives have acted. The term with two energy derivatives controls the leading total-energy singularity,
	\be
	\lim_{E\to0}E^3\mathcal{M}_4^{(s)}
	=-\frac{1}{2}\frac{X_{1,2}^2X_{3,4}^2}{(k_I^{(s)})^2-k_{12}^2}\ .
	\label{eq:sdg_ads_corr4s_flat}
	\ee
	It therefore squares the SDYM numerator in this channel. The full gravity correlator requires all three exchange channels. As for SDYM, the leading flat-space residue vanishes, as required by the vanishing four-point tree amplitude of self-dual gravity, while subleading poles and finite terms remain in the AdS correlator.
	
	Equation~\eqref{eq:sdg_ads_corr4s} is therefore a channel-by-channel differential dressing, but it is not yet a complete four-point double copy for the summed correlator. In particular, the two operators depend on the routing of the internal line and on the off-shell continuation used to separate the external and internal energies; the operators in different channels are not related by a single universal relabeling. What is established is that individual scalar exchange diagrams generate the SDYM and self-dual gravity diagrams by kinematic and differential dressings, and that the leading pole of each gravity channel contains the squared SDYM numerator. A universal kernel or color-kinematics prescription assembling the full four-point correlator has not been identified \cite{Chowdhury:2024dcy}. Moreover, the three-point SDYM scalar correlator has been lifted to the spinning YM correlator only up to a boundary contact term, while the analogous four-point uplift and the lift of the self-dual gravity scalar correlators remain open. These qualifications are important when interpreting the scalar relations above as a double copy of physical boundary observables.
	
	\subsection{Chiral higher-spin extensions}
	
	The final direction in this self-dual section is the chiral higher-spin double copy  \cite{Ponomarev:2024jyg}. Chiral higher-spin theories are not parity-invariant completions of higher-spin gravity; rather, they are chiral lightcone theories whose interactions are purely cubic \cite{Ponomarev:2022vjb,Ponomarev:2017nrr}. This makes them natural relatives of SDYM and self-dual gravity. Their cubic vertices factorize as
	\be
	V_3^{\rm chiral}\sim n_{\rm SDYM}\,c_{\rm HS}\ ,
	\label{eq:chs_factorization}
	\ee
	where $n_{\rm SDYM}$ is the kinematic part of the SDYM vertex and $c_{\rm HS}$ is a structure constant of a higher-spin algebra. The factorization follows from Lorentz invariance in the chiral lightcone setup, not from a post hoc rearrangement of amplitudes.
	
	The double-copy spectrum is obtained by pairing a field from each single-copy theory. Thus, fields of helicities $h_1$ and $h_2$ combine into a field $\Phi^{h_1,h_2}$ of total helicity $h=h_1+h_2$. Different pairs $(h_1,h_2)$ can have the same total helicity, with $h_1-h_2$ distinguishing the resulting states; consequently, the double-copy theory generally contains infinitely many fields of a given helicity. At cubic order, the vertex factorizes into contributions associated separately with the two copies, and the light-cone consistency conditions determine the allowed interactions. For chiral higher-spin theories with adjoint internal symmetry, this has the familiar color-to-kinematics interpretation: the internal color structure constant of one copy is replaced by the corresponding kinematic or higher-spin structure constant of the second. More general products involving chiral higher-spin theory, SDYM, and self-dual gravity extend the same structure \cite{Ponomarev:2024jyg}.
	
	This higher-spin extension reinforces the main lesson of the self-dual sector. The double copy is most transparent when the theory admits a cubic formulation in which the vertices are products of Lie-algebraic structure constants. In addition, a classical higher-spin extension of the self-dual double copy was recently constructed in \cite{Misuna:2026has}. For any self-dual spacetime admitting a Kerr--Schild form, the gravitational solution can be used to generate an infinite tower of solutions of chiral higher-spin theory. At the level of the corresponding higher-spin curvatures, the multicopy structure depends on the algebraic type of the background. A full interacting AdS extension of these double copies is still lacking. A starting point towards this double copy could use existing AdS/CFT correlator results for three- and four-point functions in Higher-Spin-SDYM, a higher-spin self-dual Yang--Mills contraction and truncation of chiral higher-spin theory \cite{Skvortsov:2026ofl}.
	
	\section{Outlook}
	Double copy techniques have proved both useful and conceptually illuminating for scattering processes around trivial backgrounds. As reviewed above, scattering on non-trivial backgrounds introduces a number of new technical and conceptual complications, but it also provides an arena in which the double copy may offer new ways of organizing gravitational observables and identifying structures that are otherwise difficult to uncover. Considerable progress has been made in recent years towards extending color--kinematics duality and the double copy to curved spacetimes. Nevertheless, the subject remains highly active, and its status is not yet as well established as in flat space.
	
	A central question is how the double copy should be formulated in the presence of a non-trivial background. One possibility is to start from gauge-theory fluctuations propagating on a non-trivial gauge background, while another is to consider gauge fields propagating directly on a curved spacetime. Much of the literature reviewed here has focused on the latter, although the former more closely parallels the usual interpretation of the double copy as relating gauge-theory and gravitational degrees of freedom. It remains to be understood whether consistent double copy constructions can be formulated in both settings, whether they are equivalent in appropriate cases, or whether one of them provides the more fundamental framework.
	
	On a related note, as briefly discussed in this review, in flat space the amplitude and classical formulations of the double copy have been connected in a broad range of examples, suggesting that they capture different manifestations of the same underlying structure. It remains to be understood whether an analogous relation persists in curved spacetime and, in the absence of a conventional S-matrix, which curved-space observables play the corresponding role, such as boundary correlators, wavefunction coefficients, or a more general notion of scattering data.
	
	Higher-point and loop-level observables will provide important tests of the structures identified so far. To perform these tests on proposed double copies, it is important to make progress on the implications of causality and unitarity on curved space scattering observables. Equally important will be the development of exact and non-perturbative constructions beyond the particularly structured examples currently understood. Together, these directions should help determine which ingredients of the flat-space double copy survive generically in curved spacetimes, which require modification, and which rely essentially on the special properties of flat-space scattering.

	\section{Acknowledgments}
	I am grateful to Tim Adamo, Kymani Armstrong-Williams, John Joseph Carrasco, Henrik Johansson, Théo Keseman, Donal O'Connell, Facundo Rost, Charlotte Sleight, and Chris D. White for insightful comments on the manuscript and useful feedback and discussions on aspects of the review. I also thank my collaborators on double copy and related topics for many useful discussions and insights throughout our work together, as well as the participants of \textit{Journey through Modern Explorations in QFT and Beyond (JME-QFT25)} for stimulating discussions and feedback on parts of the material presented in this review. I am also grateful to the students of the MSc Special Topics courses I taught at Imperial College London, whose questions and discussions helped shape the introductory material on the flat-space and classical double copy.
	
	I acknowledge the use of AI-assisted tools for checking conventions, consistency of notation, and typographical errors throughout the manuscript.
	
	MCG is supported by the Imperial College Research Fellowship.
	
	\appendix
	\numberwithin{equation}{section}
	\renewcommand{\theequation}{\Alph{section}.\arabic{equation}}
	
	\section{Conformal field theory background}
	\label{app:cft_background}
	
	Since many of the observables reviewed in this work are boundary correlators, it is useful to briefly summarize the basic conformal-field-theory facts that underlie the notation used in the main text. Standard references include \cite{DiFrancesco:1997nk,Rychkov:2016iqz,Simmons-Duffin:2016gjk}.
	
	A conformal field theory (CFT) is a quantum field theory invariant under translations, rotations, dilatations, and special conformal transformations. In $d$ dimensions, the corresponding algebra is $\mathfrak{so}(d+1,1)$ in Euclidean signature, or equivalently $\mathfrak{so}(d,2)$ after analytic continuation to Lorentzian signature. Local operators organize into irreducible representations of this algebra. A primary operator $\mathcal O_{\Delta,s}$ is characterized by its conformal dimension $\Delta$ and spin $s$, while its descendants are obtained by acting with translations. Under a scale transformation $x^\mu\to \lambda x^\mu$, a scalar primary transforms as
	\begin{equation}
		\mathcal O(\lambda x)=\lambda^{-\Delta}\mathcal O(x), \label{eq:scaling}
	\end{equation}
	and similarly for spinning operators, with the appropriate tensor structure. The operator product expansion (OPE) is another basic consequence. When two operators approach each other, their product can be expanded as
	\begin{equation}
		\mathcal O_i(x)\mathcal O_j(0)
		=
		\sum_k C_{ij}{}^{k}\,
		|x|^{\Delta_k-\Delta_i-\Delta_j}
		\left(\mathcal O_k(0)+\text{descendants}\right).
	\end{equation}
	Thus the CFT data consist of the spectrum of primary operators and the OPE coefficients. In the AdS/CFT context, exchanged bulk fields are dual to primaries appearing in this expansion, and their dimensions and spins determine the location of poles or exchanged conformal blocks, depending on the representation being used.
	
	In momentum space, the same scaling information of Eq.~\eqref{eq:scaling} appears in a different way. If $\widetilde{\mathcal O}(k)=\int d^dx\, e^{-ik\cdot x}\mathcal O(x)$, then scale invariance implies
	\begin{equation}
		\langle \widetilde{\mathcal O}_1(r k_1)\cdots \widetilde{\mathcal O}_n(r k_n)\rangle'
		=
		r^{\,\Delta_t-(n-1)d}
		\langle \widetilde{\mathcal O}_1(k_1)\cdots \widetilde{\mathcal O}_n(k_n)\rangle',
		\qquad
		\Delta_t\equiv \sum_{i=1}^n \Delta_i ,
	\end{equation}
	after factoring out the overall momentum-conserving delta function,
	\begin{equation}
		\langle \widetilde{\mathcal O}_1(k_1)\cdots \widetilde{\mathcal O}_n(k_n)\rangle
		=(2\pi)^d\delta^{(d)}\!\left(\sum_{i=1}^n k_i\right)\,
		\langle \widetilde{\mathcal O}_1(k_1)\cdots \widetilde{\mathcal O}_n(k_n)\rangle'.
	\end{equation}
	The prime will be used throughout the review to denote correlators with the momentum-conserving delta function removed. For example, the scalar two-point function scales as
	\begin{equation}
		\langle \widetilde{\mathcal O}(k)\widetilde{\mathcal O}(-k)\rangle'
		\propto k^{\,2\Delta-d},
	\end{equation}
	up to normalization and possible logarithms when $2\Delta-d$ is an even non-negative integer \cite{Bzowski:2013sza}.
	
	The consequences of conformal symmetry are encoded in Ward identities. Translation invariance implies momentum conservation, while dilatation invariance gives the scaling equation~Eq.~\eqref{eq:scaling}. Special conformal symmetry imposes additional differential constraints on momentum-space correlators. For scalar operators, and away from coincident-point contact terms, the reduced correlator satisfies
	\begin{equation}
		\sum_{i=1}^n
		\left[
		2(\Delta_i-d)\frac{\partial}{\partial k_i^\mu}
		-2k_i^\nu \frac{\partial}{\partial k_i^\nu}\frac{\partial}{\partial k_i^\mu}
		+k_{i\mu}\frac{\partial}{\partial k_i^\nu}\frac{\partial}{\partial k_{i\nu}}
		\right]
		\langle \widetilde{\mathcal O}_1(k_1)\cdots \widetilde{\mathcal O}_n(k_n)\rangle'
		=0 ,
	\end{equation}
	with analogous expressions for spinning operators \cite{Bzowski:2013sza}. More precisely, for full correlators these Ward identities hold as distributional equations and can receive local contributions supported at coincident points. In momentum space these appear as polynomial terms in the external momenta, reflecting the Fourier transform of contact terms in position space \cite{Bzowski:2018fql}. Thus the homogeneous equation above is the Ward identity for the non-local part of the reduced correlator, while local contact terms must be treated separately.
	
	Finally, conserved currents and stress tensors play a central role in this review. A spin-$s$ symmetric traceless primary in $d>2$ obeys the unitarity bound $\Delta\geq s+d-2$ for $s\geq 1$, while a scalar obeys $\Delta\geq (d-2)/2$ \cite{Simmons-Duffin:2016gjk}. Saturating the bound for $s\geq 1$ gives a conserved operator. Away from coincident points, this means
	\begin{equation}
		\partial\cdot J_{(s)}=0,
		\qquad
		\Delta=s+d-2.
	\end{equation}
	For example, for a conserved spin-one current $J_\mu^a(x)$ one has, for separated insertions,
	\begin{equation}
		\partial_x^\mu \big\langle J_\mu^a(x)\,\mathcal O_1(x_1)\cdots \mathcal O_n(x_n)\big\rangle =0,
		\qquad x\neq x_i .
	\end{equation}
	At coincident points this equation is modified by contact terms, and the corresponding Ward identity reads
	\begin{equation}
		\begin{aligned}
			\partial_x^\mu \big\langle J_\mu^a(x)\,\mathcal O_1(x_1)\cdots \mathcal O_n(x_n)\big\rangle
			= \\
			-\sum_{i=1}^n \delta^{(d)}(x-x_i)\,
			\big\langle \mathcal O_1(x_1)\cdots (\delta^a \mathcal O_i)(x_i)\cdots \mathcal O_n(x_n)\big\rangle \ , \label{eq:Ward_id}
		\end{aligned}
	\end{equation}
	where $\delta^a\mathcal O_i$ is the infinitesimal action of the symmetry on $\mathcal O_i$ \cite{Osborn:1993cr,Simmons-Duffin:2016gjk}. In particular, a conserved current has $(\Delta,s)=(d-1,1)$ and the stress tensor has $(\Delta,s)=(d,2)$. These are precisely the operators dual to bulk gauge fields and gravitons, and this is why they appear repeatedly in the correlator-based versions of the double copy reviewed in the main text. There are solutions to the Ward identity where the RHS vanishes identically, these are referred to as homogeneous solutions. 
	
	In momentum space, these contact terms appear as non-vanishing right-hand sides in the Ward identities. For example, for a $U(1)$ conserved current and stress tensor one finds \cite{Baumann:2020dch}
	\begin{equation}
		k_{1\,i}\,\langle J^i_{\vec k_1}\,\mathcal O_{\vec k_2}\cdots \mathcal O_{\vec k_n}\rangle
		=
		-\sum_{a=2}^n e_a\,
		\langle \mathcal O_{\vec k_2}\cdots \mathcal O_{\vec k_a+\vec k_1}\cdots \mathcal O_{\vec k_n}\rangle \ ,
	\end{equation}
	where $e_a$ are the charges associated with the operator $\mathcal O_a$ and
	\begin{equation}
		z_{1\,i}k_{1\,j}\,\langle T^{ij}_{\vec k_1}\,\mathcal O_{\vec k_2}\cdots \mathcal O_{\vec k_n}\rangle
		=
		-\sum_{a=2}^n \kappa_a(\vec z_1\!\cdot\!\vec k_a)\,
		\langle \mathcal O_{\vec k_2}\cdots \mathcal O_{\vec k_a+\vec k_1}\cdots \mathcal O_{\vec k_n}\rangle  \ ,
	\end{equation}
	where $\kappa_a$ are the couplings between the stress tensor and the operator $\mathcal O_a$.

	\section{Spinorial formalism of GR and the Petrov classification}
	\label{app:spinorial_gr_petrov}
	In this Appendix, we give a brief summary of the conventions used for the spinorial formalism in General Relativity used in the main text \cite{Penrose:1985bww, Stephani:2003tm}. In four dimensions, it is often convenient to rewrite tensors in terms of
	two-component spinors by using the two-to-one homomorphism
	$SL(2,\mathbb C)\rightarrow SO(1,3)$. Since throughout the review we work with mostly plus signature, we take
	\begin{equation}
		\eta_{\mathrm{ab}} = {\rm diag}(-,+,+,+),
		\qquad
		\sigma_{\mathrm{a}}=\{\mathrm{I},\sigma_i\},
	\end{equation}
	and use a vielbein $e^{\mathrm{a}}_\mu$ to convert spacetime indices into local Lorentz
	indices. A vector then corresponds to a bi-spinor
	\begin{equation}
		v_{\alpha\dot\alpha}=v^\mu e^{\mathrm{a}}_\mu (\sigma_{\mathrm{a}})_{\alpha\dot\alpha},
	\end{equation}
	with spinor indices raised and lowered using
	$\epsilon_{\alpha\beta}$ and $\epsilon_{\dot\alpha\dot\beta}$.
	The covariant derivative is written as
	\begin{equation}
		\nabla_{\alpha\dot\alpha}=e^\mu{}_{\alpha\dot\alpha}\nabla_\mu .
	\end{equation}
	This is the curved-space analogue of the spinorial notation reviewed in
	Sec.~\ref{sec:spinorhel}.
	
	For later use it is helpful to recall the decomposition of antisymmetric
	tensors into self-dual and anti-self-dual parts. The Maxwell field
	strength takes the form
	\begin{equation}
		F_{\alpha\dot\alpha\,\beta\dot\beta}
		=
		f_{\alpha\beta}\,\epsilon_{\dot\alpha\dot\beta}
		+\bar f_{\dot\alpha\dot\beta}\,\epsilon_{\alpha\beta},
	\end{equation}
	where $f_{\alpha\beta}=f_{(\alpha\beta)}$ and
	$\bar f_{\dot\alpha\dot\beta}=\bar f_{(\dot\alpha\dot\beta)}$ are symmetric.
	For real field strengths they are complex conjugates of each other. In the
	same way, the Weyl tensor decomposes as
	\begin{equation}
		C_{\alpha\dot\alpha\,\beta\dot\beta\,\gamma\dot\gamma\,\delta\dot\delta}
		=
		\Psi_{\alpha\beta\gamma\delta}\,
		\epsilon_{\dot\alpha\dot\beta}\epsilon_{\dot\gamma\dot\delta}
		+
		\bar\Psi_{\dot\alpha\dot\beta\dot\gamma\dot\delta}\,
		\epsilon_{\alpha\beta}\epsilon_{\gamma\delta},
	\end{equation}
	with
	$\Psi_{\alpha\beta\gamma\delta}=\Psi_{(\alpha\beta\gamma\delta)}$ and
	$\bar\Psi_{\dot\alpha\dot\beta\dot\gamma\dot\delta}
	=\bar\Psi_{(\dot\alpha\dot\beta\dot\gamma\dot\delta)}$.
	Thus the local gravitational degrees of freedom are encoded in the totally
	symmetric Weyl spinor $\Psi_{\alpha\beta\gamma\delta}$ and its complex
	conjugate. In vacuum, the Bianchi identity implies
	\begin{equation}
		\nabla^{\alpha\dot\alpha}\Psi_{\alpha\beta\gamma\delta}=0 .
	\end{equation}
	This is the direct analogue of the source-free Maxwell equation
	$\nabla^{\alpha\dot\alpha}f_{\alpha\beta}=0$.
	
	The spinorial form of the Weyl tensor is particularly useful because it makes
	the algebraic classification of four-dimensional geometries transparent.
	Indeed, the Petrov classification is simply the classification of the totally
	symmetric rank-four spinor $\Psi_{\alpha\beta\gamma\delta}$ under
	$SL(2,\mathbb C)$ transformations. Equivalently, one studies the quartic
	equation
	\begin{equation}
		\Psi_{\alpha\beta\gamma\delta}\,\xi^\alpha\xi^\beta\xi^\gamma\xi^\delta=0 ,
	\end{equation}
	whose roots determine the principal spinors. Since
	$\Psi_{\alpha\beta\gamma\delta}$ is totally symmetric, it can always be
	factorized as
	\begin{equation}
		\Psi_{\alpha\beta\gamma\delta}
		\propto
		\kappa_{(\alpha}\lambda_\beta\mu_\gamma\nu_{\delta)} ,
	\end{equation}
	and the Petrov type is determined by the multiplicities of the principal
	spinors:
	\begin{align}
		{\rm Type~I}:&\qquad
		\Psi_{\alpha\beta\gamma\delta}
		\sim
		\kappa_{(\alpha}\lambda_\beta\mu_\gamma\nu_{\delta)} \ ,\\
		{\rm Type~II}:&\qquad
		\Psi_{\alpha\beta\gamma\delta}
		\sim
		\kappa_{(\alpha}\kappa_\beta\lambda_\gamma\mu_{\delta)}\ ,\\
		{\rm Type~D}:&\qquad
		\Psi_{\alpha\beta\gamma\delta}
		\sim
		\kappa_{(\alpha}\kappa_\beta\lambda_\gamma\lambda_{\delta)} \ ,\\
		{\rm Type~III}:&\qquad
		\Psi_{\alpha\beta\gamma\delta}
		\sim
		\kappa_{(\alpha}\kappa_\beta\kappa_\gamma\lambda_{\delta)} \ ,\\
		{\rm Type~N}:&\qquad
		\Psi_{\alpha\beta\gamma\delta}
		\sim
		\kappa_\alpha\kappa_\beta\kappa_\gamma\kappa_\delta \ ,\\
		{\rm Type~O}:&\qquad
		\Psi_{\alpha\beta\gamma\delta}=0 \ .
	\end{align}
	Petrov type~I is the algebraically general case, in which the Weyl spinor has four distinct principal spinors. A spacetime is \emph{algebraically special} precisely when the Weyl spinor has at least one repeated principal spinor, equivalently when the Weyl tensor has a repeated principal null direction. Thus types II, D, III, N and O are all algebraically special, while type~D is distinguished by the existence of two repeated principal null directions; this is the case for many exact vacuum solutions relevant for the classical double copy, including Schwarzschild and Kerr.
	
	In three dimensions, the spinorial description simplifies further because the Lorentz algebra is $\mathfrak{so}(1,2)\cong \mathfrak{sl}(2,\mathbb R)$ in Lorentzian signature, so vectors can be written as symmetric bi-spinors \cite{Milson:2012ry,castillo20033,doi:10.1063/1.1592611}. Symmetric traceless rank-$s$ tensors are therefore encoded by completely symmetric spinors with $2s$ indices. In particular, the analogue of the four-dimensional Weyl spinor is replaced in three dimensions by the Cotton spinor $C_{abcd}=C_{(abcd)}$, since the Weyl tensor vanishes identically and the Riemann tensor is determined algebraically by the Ricci tensor. Thus, in three-dimensional gravity the conformal geometry is encoded in $C_{abcd}$, and conformal flatness is equivalent to $C_{abcd}=0$. 
	
	\bibliographystyle{elsarticle-num}
	\bibliography{references}

\begin{thebibliography}{100}
\expandafter\ifx\csname url\endcsname\relax
  \def\url#1{\texttt{#1}}\fi
\expandafter\ifx\csname urlprefix\endcsname\relax\def\urlprefix{URL }\fi
\expandafter\ifx\csname href\endcsname\relax
  \def\href#1#2{#2} \def\path#1{#1}\fi

\bibitem{Bern:2019prr}
Z.~Bern, J.~J. Carrasco, M.~Chiodaroli, H.~Johansson, R.~Roiban, {The duality
  between color and kinematics and its applications}, J. Phys. A 57~(33) (2024)
  333002.
\newblock \href {http://arxiv.org/abs/1909.01358} {\path{arXiv:1909.01358}},
  \href {https://doi.org/10.1088/1751-8121/ad5fd0}
  {\path{doi:10.1088/1751-8121/ad5fd0}}.

\bibitem{Bern:2022wqg}
Z.~Bern, J.~J. Carrasco, M.~Chiodaroli, H.~Johansson, R.~Roiban, {The SAGEX
  review on scattering amplitudes Chapter 2: An invitation to color-kinematics
  duality and the double copy}, J. Phys. A 55~(44) (2022) 443003.
\newblock \href {http://arxiv.org/abs/2203.13013} {\path{arXiv:2203.13013}},
  \href {https://doi.org/10.1088/1751-8121/ac93cf}
  {\path{doi:10.1088/1751-8121/ac93cf}}.

\bibitem{White:2024pve}
C.~D. White, {The Classical Double Copy}, World Scientific, 2024.
\newblock \href {https://doi.org/10.1142/q0457} {\path{doi:10.1142/q0457}}.

\bibitem{Bern:2012uf}
Z.~Bern, J.~J.~M. Carrasco, L.~J. Dixon, H.~Johansson, R.~Roiban, {Simplifying
  Multiloop Integrands and Ultraviolet Divergences of Gauge Theory and Gravity
  Amplitudes}, Phys. Rev. D 85 (2012) 105014.
\newblock \href {http://arxiv.org/abs/1201.5366} {\path{arXiv:1201.5366}},
  \href {https://doi.org/10.1103/PhysRevD.85.105014}
  {\path{doi:10.1103/PhysRevD.85.105014}}.

\bibitem{Bern:2013uka}
Z.~Bern, S.~Davies, T.~Dennen, A.~V. Smirnov, V.~A. Smirnov, {Ultraviolet
  Properties of N=4 Supergravity at Four Loops}, Phys. Rev. Lett. 111~(23)
  (2013) 231302.
\newblock \href {http://arxiv.org/abs/1309.2498} {\path{arXiv:1309.2498}},
  \href {https://doi.org/10.1103/PhysRevLett.111.231302}
  {\path{doi:10.1103/PhysRevLett.111.231302}}.

\bibitem{Bern:2012cd}
Z.~Bern, S.~Davies, T.~Dennen, Y.-t. Huang, {Absence of Three-Loop Four-Point
  Divergences in N=4 Supergravity}, Phys. Rev. Lett. 108 (2012) 201301.
\newblock \href {http://arxiv.org/abs/1202.3423} {\path{arXiv:1202.3423}},
  \href {https://doi.org/10.1103/PhysRevLett.108.201301}
  {\path{doi:10.1103/PhysRevLett.108.201301}}.

\bibitem{Bern:2018jmv}
Z.~Bern, J.~J. Carrasco, W.-M. Chen, A.~Edison, H.~Johansson,
  J.~Parra-Martinez, R.~Roiban, M.~Zeng, {Ultraviolet Properties of $\mathcal N
  = 8$ Supergravity at Five Loops}, Phys. Rev. D 98~(8) (2018) 086021.
\newblock \href {http://arxiv.org/abs/1804.09311} {\path{arXiv:1804.09311}},
  \href {https://doi.org/10.1103/PhysRevD.98.086021}
  {\path{doi:10.1103/PhysRevD.98.086021}}.

\bibitem{Bern:2017ucb}
Z.~Bern, J.~J.~M. Carrasco, W.-M. Chen, H.~Johansson, R.~Roiban, M.~Zeng,
  {Five-loop four-point integrand of $N=8$ supergravity as a generalized double
  copy}, Phys. Rev. D 96~(12) (2017) 126012.
\newblock \href {http://arxiv.org/abs/1708.06807} {\path{arXiv:1708.06807}},
  \href {https://doi.org/10.1103/PhysRevD.96.126012}
  {\path{doi:10.1103/PhysRevD.96.126012}}.

\bibitem{Carrasco:2021bmu}
J.~J.~M. Carrasco, I.~A. Vazquez-Holm, {Extracting Einstein from the loop-level
  double-copy}, JHEP 11 (2021) 088.
\newblock \href {http://arxiv.org/abs/2108.06798} {\path{arXiv:2108.06798}},
  \href {https://doi.org/10.1007/JHEP11(2021)088}
  {\path{doi:10.1007/JHEP11(2021)088}}.

\bibitem{Bern:2019crd}
Z.~Bern, C.~Cheung, R.~Roiban, C.-H. Shen, M.~P. Solon, M.~Zeng, {Black Hole
  Binary Dynamics from the Double Copy and Effective Theory}, JHEP 10 (2019)
  206.
\newblock \href {http://arxiv.org/abs/1908.01493} {\path{arXiv:1908.01493}},
  \href {https://doi.org/10.1007/JHEP10(2019)206}
  {\path{doi:10.1007/JHEP10(2019)206}}.

\bibitem{Buonanno:2022pgc}
A.~Buonanno, M.~Khalil, D.~O'Connell, R.~Roiban, M.~P. Solon, M.~Zeng,
  {Snowmass White Paper: Gravitational Waves and Scattering Amplitudes}, in:
  {Snowmass 2021}, 2022.
\newblock \href {http://arxiv.org/abs/2204.05194} {\path{arXiv:2204.05194}}.

\bibitem{Kosower:2022yvp}
D.~A. Kosower, R.~Monteiro, D.~O'Connell, {The SAGEX review on scattering
  amplitudes Chapter 14: Classical gravity from scattering amplitudes}, J.
  Phys. A 55~(44) (2022) 443015.
\newblock \href {http://arxiv.org/abs/2203.13025} {\path{arXiv:2203.13025}},
  \href {https://doi.org/10.1088/1751-8121/ac8846}
  {\path{doi:10.1088/1751-8121/ac8846}}.

\bibitem{Chen:2023dcx}
A.~S.-K. Chen, H.~Elvang, A.~Herderschee, {Bootstrapping the String
  Kawai-Lewellen-Tye Kernel}, Phys. Rev. Lett. 131~(3) (2023) 031602.
\newblock \href {http://arxiv.org/abs/2302.04895} {\path{arXiv:2302.04895}},
  \href {https://doi.org/10.1103/PhysRevLett.131.031602}
  {\path{doi:10.1103/PhysRevLett.131.031602}}.

\bibitem{Maldacena:2011nz}
J.~M. Maldacena, G.~L. Pimentel, {On graviton non-Gaussianities during
  inflation}, JHEP 09 (2011) 045.
\newblock \href {http://arxiv.org/abs/1104.2846} {\path{arXiv:1104.2846}},
  \href {https://doi.org/10.1007/JHEP09(2011)045}
  {\path{doi:10.1007/JHEP09(2011)045}}.

\bibitem{Weinberg:2005vy}
S.~Weinberg, {Quantum contributions to cosmological correlations}, Phys. Rev. D
  72 (2005) 043514.
\newblock \href {http://arxiv.org/abs/hep-th/0506236}
  {\path{arXiv:hep-th/0506236}}, \href
  {https://doi.org/10.1103/PhysRevD.72.043514}
  {\path{doi:10.1103/PhysRevD.72.043514}}.

\bibitem{Baumann:2020dch}
D.~Baumann, C.~Duaso~Pueyo, A.~Joyce, H.~Lee, G.~L. Pimentel, {The Cosmological
  Bootstrap: Spinning Correlators from Symmetries and Factorization}, SciPost
  Phys. 11 (2021) 071.
\newblock \href {http://arxiv.org/abs/2005.04234} {\path{arXiv:2005.04234}},
  \href {https://doi.org/10.21468/SciPostPhys.11.3.071}
  {\path{doi:10.21468/SciPostPhys.11.3.071}}.

\bibitem{Maldacena:1997re}
J.~M. Maldacena, {The Large $N$ limit of superconformal field theories and
  supergravity}, Adv. Theor. Math. Phys. 2 (1998) 231--252.
\newblock \href {http://arxiv.org/abs/hep-th/9711200}
  {\path{arXiv:hep-th/9711200}}, \href
  {https://doi.org/10.4310/ATMP.1998.v2.n2.a1}
  {\path{doi:10.4310/ATMP.1998.v2.n2.a1}}.

\bibitem{Witten:1998qj}
E.~Witten, {Anti de Sitter space and holography}, Adv. Theor. Math. Phys. 2
  (1998) 253--291.
\newblock \href {http://arxiv.org/abs/hep-th/9802150}
  {\path{arXiv:hep-th/9802150}}, \href
  {https://doi.org/10.4310/ATMP.1998.v2.n2.a2}
  {\path{doi:10.4310/ATMP.1998.v2.n2.a2}}.

\bibitem{Raju:2012zr}
S.~Raju, {New Recursion Relations and a Flat Space Limit for AdS/CFT
  Correlators}, Phys. Rev. D 85 (2012) 126009.
\newblock \href {http://arxiv.org/abs/1201.6449} {\path{arXiv:1201.6449}},
  \href {https://doi.org/10.1103/PhysRevD.85.126009}
  {\path{doi:10.1103/PhysRevD.85.126009}}.

\bibitem{Cachazo:2013iea}
F.~Cachazo, S.~He, E.~Y. Yuan, {Scattering of Massless Particles: Scalars,
  Gluons and Gravitons}, JHEP 07 (2014) 033.
\newblock \href {http://arxiv.org/abs/1309.0885} {\path{arXiv:1309.0885}},
  \href {https://doi.org/10.1007/JHEP07(2014)033}
  {\path{doi:10.1007/JHEP07(2014)033}}.

\bibitem{Eberhardt:2020ewh}
L.~Eberhardt, S.~Komatsu, S.~Mizera, {Scattering equations in AdS: scalar
  correlators in arbitrary dimensions}, JHEP 11 (2020) 158.
\newblock \href {http://arxiv.org/abs/2007.06574} {\path{arXiv:2007.06574}},
  \href {https://doi.org/10.1007/JHEP11(2020)158}
  {\path{doi:10.1007/JHEP11(2020)158}}.

\bibitem{Roehrig:2020kck}
K.~Roehrig, D.~Skinner, {Ambitwistor strings and the scattering equations on
  AdS$_{3}${\texttimes}S$^{3}$}, JHEP 02 (2022) 073.
\newblock \href {http://arxiv.org/abs/2007.07234} {\path{arXiv:2007.07234}},
  \href {https://doi.org/10.1007/JHEP02(2022)073}
  {\path{doi:10.1007/JHEP02(2022)073}}.

\bibitem{Gomez:2021qfd}
H.~Gomez, R.~L. Jusinskas, A.~Lipstein, {Cosmological Scattering Equations},
  Phys. Rev. Lett. 127~(25) (2021) 251604.
\newblock \href {http://arxiv.org/abs/2106.11903} {\path{arXiv:2106.11903}},
  \href {https://doi.org/10.1103/PhysRevLett.127.251604}
  {\path{doi:10.1103/PhysRevLett.127.251604}}.

\bibitem{Gomez:2021ujt}
H.~Gomez, R.~Lipinski~Jusinskas, A.~Lipstein, {Cosmological scattering
  equations at tree-level and one-loop}, JHEP 07 (2022) 004.
\newblock \href {http://arxiv.org/abs/2112.12695} {\path{arXiv:2112.12695}},
  \href {https://doi.org/10.1007/JHEP07(2022)004}
  {\path{doi:10.1007/JHEP07(2022)004}}.

\bibitem{Dixon:2013uaa}
L.~J. Dixon, {A brief introduction to modern amplitude methods}, in:
  {Theoretical Advanced Study Institute in Elementary Particle Physics}:
  {Particle Physics: The Higgs Boson and Beyond}, 2014, pp. 31--67.
\newblock \href {http://arxiv.org/abs/1310.5353} {\path{arXiv:1310.5353}},
  \href {https://doi.org/10.5170/CERN-2014-008.31}
  {\path{doi:10.5170/CERN-2014-008.31}}.

\bibitem{Elvang:2013cua}
H.~Elvang, Y.-t. Huang, {Scattering Amplitudes} (8 2013).
\newblock \href {http://arxiv.org/abs/1308.1697} {\path{arXiv:1308.1697}}.

\bibitem{Brandhuber:2022qbk}
A.~Brandhuber, J.~Plefka, G.~Travaglini, {The SAGEX Review on Scattering
  Amplitudes Chapter 1: Modern Fundamentals of Amplitudes}, J. Phys. A 55~(44)
  (2022) 443002.
\newblock \href {http://arxiv.org/abs/2203.13012} {\path{arXiv:2203.13012}},
  \href {https://doi.org/10.1088/1751-8121/ac8254}
  {\path{doi:10.1088/1751-8121/ac8254}}.

\bibitem{Baumann:2022jpr}
D.~Baumann, D.~Green, A.~Joyce, E.~Pajer, G.~L. Pimentel, C.~Sleight,
  M.~Taronna, {Snowmass White Paper: The Cosmological Bootstrap}, SciPost Phys.
  Comm. Rep. 2024 (2024) 1.
\newblock \href {http://arxiv.org/abs/2203.08121} {\path{arXiv:2203.08121}},
  \href {https://doi.org/10.21468/SciPostPhysCommRep.1}
  {\path{doi:10.21468/SciPostPhysCommRep.1}}.

\bibitem{Jazayeri:2021fvk}
S.~Jazayeri, E.~Pajer, D.~Stefanyszyn, {From locality and unitarity to
  cosmological correlators}, JHEP 10 (2021) 065.
\newblock \href {http://arxiv.org/abs/2103.08649} {\path{arXiv:2103.08649}},
  \href {https://doi.org/10.1007/JHEP10(2021)065}
  {\path{doi:10.1007/JHEP10(2021)065}}.

\bibitem{Lee:2024sks}
M.~H.~G. Lee, E.~Pajer, M.~Giroux, H.~S. Hannesdottir, S.~Mizera,
  C.~Pasiecznik, {Records from the S-Matrix Marathon: A Timeless History of
  Time}, 2024.
\newblock \href {http://arxiv.org/abs/2410.00227} {\path{arXiv:2410.00227}}.

\bibitem{Penedones:2016voo}
J.~Penedones, {TASI lectures on AdS/CFT.}, in: {Theoretical Advanced Study
  Institute in Elementary Particle Physics}: {New Frontiers in Fields and
  Strings}, 2017, pp. 75--136.
\newblock \href {http://arxiv.org/abs/1608.04948} {\path{arXiv:1608.04948}},
  \href {https://doi.org/10.1142/9789813149441_0002}
  {\path{doi:10.1142/9789813149441_0002}}.

\bibitem{Fitzpatrick:2011hu}
A.~L. Fitzpatrick, J.~Kaplan, {Analyticity and the Holographic S-Matrix}, JHEP
  10 (2012) 127.
\newblock \href {http://arxiv.org/abs/1111.6972} {\path{arXiv:1111.6972}},
  \href {https://doi.org/10.1007/JHEP10(2012)127}
  {\path{doi:10.1007/JHEP10(2012)127}}.

\bibitem{Carrasco:2015iwa}
J.~J.~M. Carrasco, {TASI 2014: lectures on gauge and gravity amplitude
  relations.}, in: {Theoretical Advanced Study Institute in Elementary Particle
  Physics}: {Journeys Through the Precision Frontier: Amplitudes for
  Colliders}, WSP, 2015, pp. 477--557.
\newblock \href {http://arxiv.org/abs/1506.00974} {\path{arXiv:1506.00974}},
  \href {https://doi.org/10.1142/9789814678766_0011}
  {\path{doi:10.1142/9789814678766_0011}}.

\bibitem{Conde:2014bf}
E.~Conde, {Physics from the S-matrix: Scattering Amplitudes without
  Lagrangians}, PoS Modave 2013 (2014) 005.
\newblock \href {https://doi.org/10.22323/1.201.0005}
  {\path{doi:10.22323/1.201.0005}}.

\bibitem{tHooft:1973alw}
G.~'t~Hooft, {A Planar Diagram Theory for Strong Interactions}, Nucl. Phys. B
  72 (1974) 461.
\newblock \href {https://doi.org/10.1016/0550-3213(74)90154-0}
  {\path{doi:10.1016/0550-3213(74)90154-0}}.

\bibitem{Mangano:1990by}
M.~L. Mangano, S.~J. Parke, {Multiparton amplitudes in gauge theories}, Phys.
  Rept. 200 (1991) 301--367.
\newblock \href {http://arxiv.org/abs/hep-th/0509223}
  {\path{arXiv:hep-th/0509223}}, \href
  {https://doi.org/10.1016/0370-1573(91)90091-Y}
  {\path{doi:10.1016/0370-1573(91)90091-Y}}.

\bibitem{DelDuca:1999rs}
V.~Del~Duca, L.~J. Dixon, F.~Maltoni, {New color decompositions for gauge
  amplitudes at tree and loop level}, Nucl. Phys. B 571 (2000) 51--70.
\newblock \href {http://arxiv.org/abs/hep-ph/9910563}
  {\path{arXiv:hep-ph/9910563}}, \href
  {https://doi.org/10.1016/S0550-3213(99)00809-3}
  {\path{doi:10.1016/S0550-3213(99)00809-3}}.

\bibitem{Kleiss:1988ne}
R.~Kleiss, H.~Kuijf, {Multi - Gluon Cross-sections and Five Jet Production at
  Hadron Colliders}, Nucl. Phys. B 312 (1989) 616--644.
\newblock \href {https://doi.org/10.1016/0550-3213(89)90574-9}
  {\path{doi:10.1016/0550-3213(89)90574-9}}.

\bibitem{Bern:2008qj}
Z.~Bern, J.~J.~M. Carrasco, H.~Johansson, {New Relations for Gauge-Theory
  Amplitudes}, Phys. Rev. D 78 (2008) 085011.
\newblock \href {http://arxiv.org/abs/0805.3993} {\path{arXiv:0805.3993}},
  \href {https://doi.org/10.1103/PhysRevD.78.085011}
  {\path{doi:10.1103/PhysRevD.78.085011}}.

\bibitem{DHoker2018}
E.~D'Hoker, Lecture notes on string perturbation theory,
  \url{https://www.pa.ucla.edu/faculty-websites/dhoker-lecture-notes/meetings/2018-ucdavis.pdf},
  lecture notes from UC Davis meeting, 2018 (2018).

\bibitem{Schlotterer2021}
O.~Schlotterer, String amplitudes,
  \url{https://www.ictp-saifr.org/wp-content/uploads/2021/06/Oliver-Schlotterer.pdf},
  review talk presented at Strings 2021, ICTP-SAIFR, June 30, 2021 (2021).

\bibitem{Berkovits2017}
N.~Berkovits, H.~Gomez, \href{https://doi.org/10.1007/978-3-319-65427-0_6}{An
  Introduction to Pure Spinor Superstring Theory}, Springer International
  Publishing, Cham, 2017, pp. 221--246.
\newblock \href {https://doi.org/10.1007/978-3-319-65427-0_6}
  {\path{doi:10.1007/978-3-319-65427-0_6}}.
\newline\urlprefix\url{https://doi.org/10.1007/978-3-319-65427-0_6}

\bibitem{Staessens:2010vi}
W.~Staessens, B.~Vercnocke, {Lectures on Scattering Amplitudes in String
  Theory}, in: {5th Modave Summer School in Mathematical Physics}, 2010.
\newblock \href {http://arxiv.org/abs/1011.0456} {\path{arXiv:1011.0456}}.

\bibitem{Bjerrum-Bohr:2009ulz}
N.~E.~J. Bjerrum-Bohr, P.~H. Damgaard, P.~Vanhove, {Minimal Basis for Gauge
  Theory Amplitudes}, Phys. Rev. Lett. 103 (2009) 161602.
\newblock \href {http://arxiv.org/abs/0907.1425} {\path{arXiv:0907.1425}},
  \href {https://doi.org/10.1103/PhysRevLett.103.161602}
  {\path{doi:10.1103/PhysRevLett.103.161602}}.

\bibitem{Stieberger:2009hq}
S.~Stieberger, {Open \& Closed vs. Pure Open String Disk Amplitudes} (7 2009).
\newblock \href {http://arxiv.org/abs/0907.2211} {\path{arXiv:0907.2211}}.

\bibitem{Veneziano:1968yb}
G.~Veneziano, {Construction of a crossing - symmetric, Regge behaved amplitude
  for linearly rising trajectories}, Nuovo Cim. A 57 (1968) 190--197.
\newblock \href {https://doi.org/10.1007/BF02824451}
  {\path{doi:10.1007/BF02824451}}.

\bibitem{Kawai:1985xq}
H.~Kawai, D.~C. Lewellen, S.~H.~H. Tye, {A Relation Between Tree Amplitudes of
  Closed and Open Strings}, Nucl. Phys. B 269 (1986) 1--23.
\newblock \href {https://doi.org/10.1016/0550-3213(86)90362-7}
  {\path{doi:10.1016/0550-3213(86)90362-7}}.

\bibitem{Virasoro:1969me}
M.~A. Virasoro, {Alternative constructions of crossing-symmetric amplitudes
  with regge behavior}, Phys. Rev. 177 (1969) 2309--2311.
\newblock \href {https://doi.org/10.1103/PhysRev.177.2309}
  {\path{doi:10.1103/PhysRev.177.2309}}.

\bibitem{Shapiro:1970gy}
J.~A. Shapiro, {Electrostatic analog for the virasoro model}, Phys. Lett. B 33
  (1970) 361--362.
\newblock \href {https://doi.org/10.1016/0370-2693(70)90255-8}
  {\path{doi:10.1016/0370-2693(70)90255-8}}.

\bibitem{Stieberger:2014hba}
S.~Stieberger, T.~R. Taylor, {Closed String Amplitudes as Single-Valued Open
  String Amplitudes}, Nucl. Phys. B 881 (2014) 269--287.
\newblock \href {http://arxiv.org/abs/1401.1218} {\path{arXiv:1401.1218}},
  \href {https://doi.org/10.1016/j.nuclphysb.2014.02.005}
  {\path{doi:10.1016/j.nuclphysb.2014.02.005}}.

\bibitem{Schlotterer:2018zce}
O.~Schlotterer, O.~Schnetz, {Closed strings as single-valued open strings: A
  genus-zero derivation}, J. Phys. A 52~(4) (2019) 045401.
\newblock \href {http://arxiv.org/abs/1808.00713} {\path{arXiv:1808.00713}},
  \href {https://doi.org/10.1088/1751-8121/aaea14}
  {\path{doi:10.1088/1751-8121/aaea14}}.

\bibitem{Mizera:2016jhj}
S.~Mizera, {Inverse of the String Theory KLT Kernel}, JHEP 06 (2017) 084.
\newblock \href {http://arxiv.org/abs/1610.04230} {\path{arXiv:1610.04230}},
  \href {https://doi.org/10.1007/JHEP06(2017)084}
  {\path{doi:10.1007/JHEP06(2017)084}}.

\bibitem{Arkani-Hamed:2017mur}
N.~Arkani-Hamed, Y.~Bai, S.~He, G.~Yan, {Scattering Forms and the Positive
  Geometry of Kinematics, Color and the Worldsheet}, JHEP 05 (2018) 096.
\newblock \href {http://arxiv.org/abs/1711.09102} {\path{arXiv:1711.09102}},
  \href {https://doi.org/10.1007/JHEP05(2018)096}
  {\path{doi:10.1007/JHEP05(2018)096}}.

\bibitem{White:2016jzc}
C.~D. White, {Exact solutions for the biadjoint scalar field}, Phys. Lett. B
  763 (2016) 365--369.
\newblock \href {http://arxiv.org/abs/1606.04724} {\path{arXiv:1606.04724}},
  \href {https://doi.org/10.1016/j.physletb.2016.10.052}
  {\path{doi:10.1016/j.physletb.2016.10.052}}.

\bibitem{DeSmet:2017rve}
P.-J. De~Smet, C.~D. White, {Extended solutions for the biadjoint scalar
  field}, Phys. Lett. B 775 (2017) 163--167.
\newblock \href {http://arxiv.org/abs/1708.01103} {\path{arXiv:1708.01103}},
  \href {https://doi.org/10.1016/j.physletb.2017.11.007}
  {\path{doi:10.1016/j.physletb.2017.11.007}}.

\bibitem{Bahjat-Abbas:2018vgo}
N.~Bahjat-Abbas, R.~Stark-Much{\~a}o, C.~D. White, {Biadjoint wires}, Phys.
  Lett. B 788 (2019) 274--279.
\newblock \href {http://arxiv.org/abs/1810.08118} {\path{arXiv:1810.08118}},
  \href {https://doi.org/10.1016/j.physletb.2018.11.026}
  {\path{doi:10.1016/j.physletb.2018.11.026}}.

\bibitem{Armstrong-Williams:2025spu}
K.~Armstrong-Williams, C.~D. White, {Time-dependent solutions of biadjoint
  scalar field theories}, Phys. Lett. B 866 (2025) 139517.
\newblock \href {http://arxiv.org/abs/2502.01294} {\path{arXiv:2502.01294}},
  \href {https://doi.org/10.1016/j.physletb.2025.139517}
  {\path{doi:10.1016/j.physletb.2025.139517}}.

\bibitem{Armstrong-Williams:2026dmk}
K.~Armstrong-Williams, C.~D. White, {Non-topological solitons in biadjoint
  scalar field theory} (6 2026).
\newblock \href {http://arxiv.org/abs/2606.27054} {\path{arXiv:2606.27054}}.

\bibitem{Johnson:2020pny}
L.~A. Johnson, C.~R.~T. Jones, S.~Paranjape, {Constraints on a Massive
  Double-Copy and Applications to Massive Gravity}, JHEP 02 (2021) 148.
\newblock \href {http://arxiv.org/abs/2004.12948} {\path{arXiv:2004.12948}},
  \href {https://doi.org/10.1007/JHEP02(2021)148}
  {\path{doi:10.1007/JHEP02(2021)148}}.

\bibitem{Chi:2021mio}
H.-H. Chi, H.~Elvang, A.~Herderschee, C.~R.~T. Jones, S.~Paranjape,
  {Generalizations of the double-copy: the KLT bootstrap}, JHEP 03 (2022) 077.
\newblock \href {http://arxiv.org/abs/2106.12600} {\path{arXiv:2106.12600}},
  \href {https://doi.org/10.1007/JHEP03(2022)077}
  {\path{doi:10.1007/JHEP03(2022)077}}.

\bibitem{Azevedo:2018dgo}
T.~Azevedo, M.~Chiodaroli, H.~Johansson, O.~Schlotterer, {Heterotic and bosonic
  string amplitudes via field theory}, JHEP 10 (2018) 012.
\newblock \href {http://arxiv.org/abs/1803.05452} {\path{arXiv:1803.05452}},
  \href {https://doi.org/10.1007/JHEP10(2018)012}
  {\path{doi:10.1007/JHEP10(2018)012}}.

\bibitem{Chiodaroli:2015rdg}
M.~Chiodaroli, M.~Gunaydin, H.~Johansson, R.~Roiban, {Spontaneously Broken
  Yang-Mills-Einstein Supergravities as Double Copies}, JHEP 06 (2017) 064.
\newblock \href {http://arxiv.org/abs/1511.01740} {\path{arXiv:1511.01740}},
  \href {https://doi.org/10.1007/JHEP06(2017)064}
  {\path{doi:10.1007/JHEP06(2017)064}}.

\bibitem{Momeni:2020hmc}
A.~Momeni, J.~Rumbutis, A.~J. Tolley, {Kaluza-Klein from colour-kinematics
  duality for massive fields}, JHEP 08 (2021) 081.
\newblock \href {http://arxiv.org/abs/2012.09711} {\path{arXiv:2012.09711}},
  \href {https://doi.org/10.1007/JHEP08(2021)081}
  {\path{doi:10.1007/JHEP08(2021)081}}.

\bibitem{Momeni:2020vvr}
A.~Momeni, J.~Rumbutis, A.~J. Tolley, {Massive Gravity from Double Copy}, JHEP
  12 (2020) 030.
\newblock \href {http://arxiv.org/abs/2004.07853} {\path{arXiv:2004.07853}},
  \href {https://doi.org/10.1007/JHEP12(2020)030}
  {\path{doi:10.1007/JHEP12(2020)030}}.

\bibitem{Bern:2017yxu}
Z.~Bern, J.~J. Carrasco, W.-M. Chen, H.~Johansson, R.~Roiban, {Gravity
  Amplitudes as Generalized Double Copies of Gauge-Theory Amplitudes}, Phys.
  Rev. Lett. 118~(18) (2017) 181602.
\newblock \href {http://arxiv.org/abs/1701.02519} {\path{arXiv:1701.02519}},
  \href {https://doi.org/10.1103/PhysRevLett.118.181602}
  {\path{doi:10.1103/PhysRevLett.118.181602}}.

\bibitem{Gonzalez:2021bes}
M.~C. Gonz\'alez, A.~Momeni, J.~Rumbutis, {Massive double copy in three
  spacetime dimensions}, JHEP 08 (2021) 116.
\newblock \href {http://arxiv.org/abs/2107.00611} {\path{arXiv:2107.00611}},
  \href {https://doi.org/10.1007/JHEP08(2021)116}
  {\path{doi:10.1007/JHEP08(2021)116}}.

\bibitem{Cachazo:2014xea}
F.~Cachazo, S.~He, E.~Y. Yuan, {Scattering Equations and Matrices: From
  Einstein To Yang-Mills, DBI and NLSM}, JHEP 07 (2015) 149.
\newblock \href {http://arxiv.org/abs/1412.3479} {\path{arXiv:1412.3479}},
  \href {https://doi.org/10.1007/JHEP07(2015)149}
  {\path{doi:10.1007/JHEP07(2015)149}}.

\bibitem{Cheung:2016prv}
C.~Cheung, C.-H. Shen, {Symmetry for Flavor-Kinematics Duality from an Action},
  Phys. Rev. Lett. 118~(12) (2017) 121601.
\newblock \href {http://arxiv.org/abs/1612.00868} {\path{arXiv:1612.00868}},
  \href {https://doi.org/10.1103/PhysRevLett.118.121601}
  {\path{doi:10.1103/PhysRevLett.118.121601}}.

\bibitem{Cheung:2017yef}
C.~Cheung, G.~N. Remmen, C.-H. Shen, C.~Wen, {Pions as Gluons in Higher
  Dimensions}, JHEP 04 (2018) 129.
\newblock \href {http://arxiv.org/abs/1709.04932} {\path{arXiv:1709.04932}},
  \href {https://doi.org/10.1007/JHEP04(2018)129}
  {\path{doi:10.1007/JHEP04(2018)129}}.

\bibitem{CarrilloGonzalez:2018ejf}
M.~Carrillo~Gonz\'alez, R.~Penco, M.~Trodden, {Radiation of scalar modes and
  the classical double copy}, JHEP 11 (2018) 065.
\newblock \href {http://arxiv.org/abs/1809.04611} {\path{arXiv:1809.04611}},
  \href {https://doi.org/10.1007/JHEP11(2018)065}
  {\path{doi:10.1007/JHEP11(2018)065}}.

\bibitem{CarrilloGonzalez:2019fzc}
M.~Carrillo~Gonz\'alez, R.~Penco, M.~Trodden, {Shift symmetries, soft limits,
  and the double copy beyond leading order}, Phys. Rev. D 102~(10) (2020)
  105011.
\newblock \href {http://arxiv.org/abs/1908.07531} {\path{arXiv:1908.07531}},
  \href {https://doi.org/10.1103/PhysRevD.102.105011}
  {\path{doi:10.1103/PhysRevD.102.105011}}.

\bibitem{Hinterbichler:2015pqa}
K.~Hinterbichler, A.~Joyce, {Hidden symmetry of the Galileon}, Phys. Rev. D
  92~(2) (2015) 023503.
\newblock \href {http://arxiv.org/abs/1501.07600} {\path{arXiv:1501.07600}},
  \href {https://doi.org/10.1103/PhysRevD.92.023503}
  {\path{doi:10.1103/PhysRevD.92.023503}}.

\bibitem{Lehners:2023yrj}
J.-L. Lehners, {Review of the no-boundary wave function}, Phys. Rept. 1022
  (2023) 1--82.
\newblock \href {http://arxiv.org/abs/2303.08802} {\path{arXiv:2303.08802}},
  \href {https://doi.org/10.1016/j.physrep.2023.06.002}
  {\path{doi:10.1016/j.physrep.2023.06.002}}.

\bibitem{Halliwell:1989myn}
J.~J. Halliwell, {INTRODUCTORY LECTURES ON QUANTUM COSMOLOGY}, in: {7th
  Jerusalem Winter School for Theoretical Physics: Quantum Cosmology and Baby
  Universes}, 1989.
\newblock \href {http://arxiv.org/abs/0909.2566} {\path{arXiv:0909.2566}}.

\bibitem{Kiefer:2008sw}
C.~Kiefer, B.~Sandhoefer, {Quantum cosmology}, Z. Naturforsch. A 77~(6) (2022)
  543--559.
\newblock \href {http://arxiv.org/abs/0804.0672} {\path{arXiv:0804.0672}},
  \href {https://doi.org/10.1515/zna-2021-0384}
  {\path{doi:10.1515/zna-2021-0384}}.

\bibitem{Bittermann:2022nfh}
N.~Bittermann, A.~Joyce, {Soft limits of the wavefunction in exceptional scalar
  theories}, JHEP 03 (2023) 092.
\newblock \href {http://arxiv.org/abs/2203.05576} {\path{arXiv:2203.05576}},
  \href {https://doi.org/10.1007/JHEP03(2023)092}
  {\path{doi:10.1007/JHEP03(2023)092}}.

\bibitem{Armstrong:2020woi}
C.~Armstrong, A.~E. Lipstein, J.~Mei, {Color/kinematics duality in AdS$_{4}$},
  JHEP 02 (2021) 194.
\newblock \href {http://arxiv.org/abs/2012.02059} {\path{arXiv:2012.02059}},
  \href {https://doi.org/10.1007/JHEP02(2021)194}
  {\path{doi:10.1007/JHEP02(2021)194}}.

\bibitem{Arkani-Hamed:2017jhn}
N.~Arkani-Hamed, T.-C. Huang, Y.-t. Huang, {Scattering amplitudes for all
  masses and spins}, JHEP 11 (2021) 070.
\newblock \href {http://arxiv.org/abs/1709.04891} {\path{arXiv:1709.04891}},
  \href {https://doi.org/10.1007/JHEP11(2021)070}
  {\path{doi:10.1007/JHEP11(2021)070}}.

\bibitem{Dixon:1996wi}
L.~J. Dixon, {Calculating scattering amplitudes efficiently}, in: {Theoretical
  Advanced Study Institute in Elementary Particle Physics (TASI 95): QCD and
  Beyond}, 1996, pp. 539--584.
\newblock \href {http://arxiv.org/abs/hep-ph/9601359}
  {\path{arXiv:hep-ph/9601359}}.

\bibitem{Conde:2016izb}
E.~Conde, E.~Joung, K.~Mkrtchyan, {Spinor-Helicity Three-Point Amplitudes from
  Local Cubic Interactions}, JHEP 08 (2016) 040.
\newblock \href {http://arxiv.org/abs/1605.07402} {\path{arXiv:1605.07402}},
  \href {https://doi.org/10.1007/JHEP08(2016)040}
  {\path{doi:10.1007/JHEP08(2016)040}}.

\bibitem{Cheung:2009dc}
C.~Cheung, D.~O'Connell, {Amplitudes and Spinor-Helicity in Six Dimensions},
  JHEP 07 (2009) 075.
\newblock \href {http://arxiv.org/abs/0902.0981} {\path{arXiv:0902.0981}},
  \href {https://doi.org/10.1088/1126-6708/2009/07/075}
  {\path{doi:10.1088/1126-6708/2009/07/075}}.

\bibitem{Boels:2012ie}
R.~H. Boels, D.~O'Connell, {Simple superamplitudes in higher dimensions}, JHEP
  06 (2012) 163.
\newblock \href {http://arxiv.org/abs/1201.2653} {\path{arXiv:1201.2653}},
  \href {https://doi.org/10.1007/JHEP06(2012)163}
  {\path{doi:10.1007/JHEP06(2012)163}}.

\bibitem{Chiodaroli:2022ssi}
M.~Chiodaroli, M.~Gunaydin, H.~Johansson, R.~Roiban, {Spinor-helicity formalism
  for massive and massless amplitudes in five dimensions}, JHEP 02 (2023) 040.
\newblock \href {http://arxiv.org/abs/2202.08257} {\path{arXiv:2202.08257}},
  \href {https://doi.org/10.1007/JHEP02(2023)040}
  {\path{doi:10.1007/JHEP02(2023)040}}.

\bibitem{Pokraka:2024fao}
A.~Pokraka, S.~Rajan, L.~Ren, A.~Volovich, W.~W. Zhao, {Five-dimensional spinor
  helicity for all masses and spins}, JHEP 07 (2025) 056.
\newblock \href {http://arxiv.org/abs/2405.09533} {\path{arXiv:2405.09533}},
  \href {https://doi.org/10.1007/JHEP07(2025)056}
  {\path{doi:10.1007/JHEP07(2025)056}}.

\bibitem{Caron-Huot:2010nes}
S.~Caron-Huot, D.~O'Connell, {Spinor Helicity and Dual Conformal Symmetry in
  Ten Dimensions}, JHEP 08 (2011) 014.
\newblock \href {http://arxiv.org/abs/1010.5487} {\path{arXiv:1010.5487}},
  \href {https://doi.org/10.1007/JHEP08(2011)014}
  {\path{doi:10.1007/JHEP08(2011)014}}.

\bibitem{Parke:1986gb}
S.~J. Parke, T.~R. Taylor, {An Amplitude for $n$ Gluon Scattering}, Phys. Rev.
  Lett. 56 (1986) 2459.
\newblock \href {https://doi.org/10.1103/PhysRevLett.56.2459}
  {\path{doi:10.1103/PhysRevLett.56.2459}}.

\bibitem{Britto:2005fq}
R.~Britto, F.~Cachazo, B.~Feng, E.~Witten, {Direct proof of tree-level
  recursion relation in Yang-Mills theory}, Phys. Rev. Lett. 94 (2005) 181602.
\newblock \href {http://arxiv.org/abs/hep-th/0501052}
  {\path{arXiv:hep-th/0501052}}, \href
  {https://doi.org/10.1103/PhysRevLett.94.181602}
  {\path{doi:10.1103/PhysRevLett.94.181602}}.

\bibitem{Basile:2024ydc}
T.~Basile, E.~Joung, K.~Mkrtchyan, M.~Mojaza, {Spinor-helicity representations
  of particles of any mass in dS4 and AdS4 spacetimes}, Phys. Rev. D 109~(12)
  (2024) 125003.
\newblock \href {http://arxiv.org/abs/2401.02007} {\path{arXiv:2401.02007}},
  \href {https://doi.org/10.1103/PhysRevD.109.125003}
  {\path{doi:10.1103/PhysRevD.109.125003}}.

\bibitem{Nagaraj:2018nxq}
B.~Nagaraj, D.~Ponomarev, {Spinor-Helicity Formalism for Massless Fields in
  AdS$_4$}, Phys. Rev. Lett. 122~(10) (2019) 101602.
\newblock \href {http://arxiv.org/abs/1811.08438} {\path{arXiv:1811.08438}},
  \href {https://doi.org/10.1103/PhysRevLett.122.101602}
  {\path{doi:10.1103/PhysRevLett.122.101602}}.

\bibitem{Nagaraj:2019zmk}
B.~Nagaraj, D.~Ponomarev, {Spinor-helicity formalism for massless fields in
  AdS$_{4}$. Part II. Potentials}, JHEP 06 (2020) 068.
\newblock \href {http://arxiv.org/abs/1912.07494} {\path{arXiv:1912.07494}},
  \href {https://doi.org/10.1007/JHEP06(2020)068}
  {\path{doi:10.1007/JHEP06(2020)068}}.

\bibitem{Nagaraj:2020sji}
B.~Nagaraj, D.~Ponomarev, {Spinor-helicity formalism for massless fields in
  AdS$_{4}$ III: contact four-point amplitudes}, JHEP 08~(08) (2020) 012.
\newblock \href {http://arxiv.org/abs/2004.07989} {\path{arXiv:2004.07989}},
  \href {https://doi.org/10.1007/JHEP08(2020)012}
  {\path{doi:10.1007/JHEP08(2020)012}}.

\bibitem{Skvortsov:2022wzo}
E.~Skvortsov, Y.~Yin, {On (spinor)-helicity and bosonization in
  AdS$_{4}$/CFT$_{3}$}, JHEP 03 (2023) 204.
\newblock \href {http://arxiv.org/abs/2207.06976} {\path{arXiv:2207.06976}},
  \href {https://doi.org/10.1007/JHEP03(2023)204}
  {\path{doi:10.1007/JHEP03(2023)204}}.

\bibitem{David:2019mos}
A.~David, N.~Fischer, Y.~Neiman, {Spinor-helicity variables for cosmological
  horizons in de Sitter space}, Phys. Rev. D 100~(4) (2019) 045005.
\newblock \href {http://arxiv.org/abs/1906.01058} {\path{arXiv:1906.01058}},
  \href {https://doi.org/10.1103/PhysRevD.100.045005}
  {\path{doi:10.1103/PhysRevD.100.045005}}.

\bibitem{Sleight:2020obc}
C.~Sleight, M.~Taronna, {From AdS to dS exchanges: Spectral representation,
  Mellin amplitudes, and crossing}, Phys. Rev. D 104~(8) (2021) L081902.
\newblock \href {http://arxiv.org/abs/2007.09993} {\path{arXiv:2007.09993}},
  \href {https://doi.org/10.1103/PhysRevD.104.L081902}
  {\path{doi:10.1103/PhysRevD.104.L081902}}.

\bibitem{Chowdhury:2025nnk}
C.~Chowdhury, A.~Lipstein, J.~Marshall, A.~J. Zhang, {On in-in correlators for
  spinning theories and their shadow formulation} (12 2025).
\newblock \href {http://arxiv.org/abs/2512.14694} {\path{arXiv:2512.14694}}.

\bibitem{Farrow:2018yni}
J.~A. Farrow, A.~E. Lipstein, P.~McFadden, {Double copy structure of CFT
  correlators}, JHEP 02 (2019) 130.
\newblock \href {http://arxiv.org/abs/1812.11129} {\path{arXiv:1812.11129}},
  \href {https://doi.org/10.1007/JHEP02(2019)130}
  {\path{doi:10.1007/JHEP02(2019)130}}.

\bibitem{Lipstein:2019mpu}
A.~E. Lipstein, P.~McFadden, {Double copy structure and the flat space limit of
  conformal correlators in even dimensions}, Phys. Rev. D 101~(12) (2020)
  125006.
\newblock \href {http://arxiv.org/abs/1912.10046} {\path{arXiv:1912.10046}},
  \href {https://doi.org/10.1103/PhysRevD.101.125006}
  {\path{doi:10.1103/PhysRevD.101.125006}}.

\bibitem{Bzowski:2017poo}
A.~Bzowski, P.~McFadden, K.~Skenderis, {Renormalised 3-point functions of
  stress tensors and conserved currents in CFT}, JHEP 11 (2018) 153.
\newblock \href {http://arxiv.org/abs/1711.09105} {\path{arXiv:1711.09105}},
  \href {https://doi.org/10.1007/JHEP11(2018)153}
  {\path{doi:10.1007/JHEP11(2018)153}}.

\bibitem{Coriano:2026zag}
C.~Corian{\`o}, S.~Lionetti, {From chiral to conformal anomalies: a double copy
  perspective for CFT correlators} (8 2026).
\newblock \href {http://arxiv.org/abs/2608.05731} {\path{arXiv:2608.05731}}.

\bibitem{Albayrak:2020fyp}
S.~Albayrak, S.~Kharel, D.~Meltzer, {On duality of color and kinematics in
  (A)dS momentum space}, JHEP 03 (2021) 249.
\newblock \href {http://arxiv.org/abs/2012.10460} {\path{arXiv:2012.10460}},
  \href {https://doi.org/10.1007/JHEP03(2021)249}
  {\path{doi:10.1007/JHEP03(2021)249}}.

\bibitem{Jain:2021qcl}
S.~Jain, R.~R. John, A.~Mehta, A.~A. Nizami, A.~Suresh, {Double copy structure
  of parity-violating CFT correlators}, JHEP 07 (2021) 033.
\newblock \href {http://arxiv.org/abs/2104.12803} {\path{arXiv:2104.12803}},
  \href {https://doi.org/10.1007/JHEP07(2021)033}
  {\path{doi:10.1007/JHEP07(2021)033}}.

\bibitem{Alday:2021odx}
L.~F. Alday, C.~Behan, P.~Ferrero, X.~Zhou, {Gluon Scattering in AdS from CFT},
  JHEP 06 (2021) 020.
\newblock \href {http://arxiv.org/abs/2103.15830} {\path{arXiv:2103.15830}},
  \href {https://doi.org/10.1007/JHEP06(2021)020}
  {\path{doi:10.1007/JHEP06(2021)020}}.

\bibitem{Li:2018wkt}
S.~Y. Li, Y.~Wang, S.~Zhou, {KLT-Like Behaviour of Inflationary Graviton
  Correlators}, JCAP 12 (2018) 023.
\newblock \href {http://arxiv.org/abs/1806.06242} {\path{arXiv:1806.06242}},
  \href {https://doi.org/10.1088/1475-7516/2018/12/023}
  {\path{doi:10.1088/1475-7516/2018/12/023}}.

\bibitem{Fazio:2019iit}
A.~R. Fazio, {Cosmological correlators, In\textendash{}In formalism and double
  copy}, Mod. Phys. Lett. A 35~(11) (2020) 2050076.
\newblock \href {http://arxiv.org/abs/1909.07343} {\path{arXiv:1909.07343}},
  \href {https://doi.org/10.1142/S0217732320500765}
  {\path{doi:10.1142/S0217732320500765}}.

\bibitem{Bzowski:2013sza}
A.~Bzowski, P.~McFadden, K.~Skenderis, {Implications of conformal invariance in
  momentum space}, JHEP 03 (2014) 111.
\newblock \href {http://arxiv.org/abs/1304.7760} {\path{arXiv:1304.7760}},
  \href {https://doi.org/10.1007/JHEP03(2014)111}
  {\path{doi:10.1007/JHEP03(2014)111}}.

\bibitem{Bzowski:2018fql}
A.~Bzowski, P.~McFadden, K.~Skenderis, {Renormalised CFT 3-point functions of
  scalars, currents and stress tensors}, JHEP 11 (2018) 159.
\newblock \href {http://arxiv.org/abs/1805.12100} {\path{arXiv:1805.12100}},
  \href {https://doi.org/10.1007/JHEP11(2018)159}
  {\path{doi:10.1007/JHEP11(2018)159}}.

\bibitem{Baumann:2021fxj}
D.~Baumann, W.-M. Chen, C.~Duaso~Pueyo, A.~Joyce, H.~Lee, G.~L. Pimentel,
  {Linking the singularities of cosmological correlators}, JHEP 09 (2022) 010.
\newblock \href {http://arxiv.org/abs/2106.05294} {\path{arXiv:2106.05294}},
  \href {https://doi.org/10.1007/JHEP09(2022)010}
  {\path{doi:10.1007/JHEP09(2022)010}}.

\bibitem{Bonifacio:2022vwa}
J.~Bonifacio, H.~Goodhew, A.~Joyce, E.~Pajer, D.~Stefanyszyn, {The graviton
  four-point function in de Sitter space}, JHEP 06 (2023) 212.
\newblock \href {http://arxiv.org/abs/2212.07370} {\path{arXiv:2212.07370}},
  \href {https://doi.org/10.1007/JHEP06(2023)212}
  {\path{doi:10.1007/JHEP06(2023)212}}.

\bibitem{Raju:2012zs}
S.~Raju, {Four Point Functions of the Stress Tensor and Conserved Currents in
  AdS$_4$/CFT$_3$}, Phys. Rev. D 85 (2012) 126008.
\newblock \href {http://arxiv.org/abs/1201.6452} {\path{arXiv:1201.6452}},
  \href {https://doi.org/10.1103/PhysRevD.85.126008}
  {\path{doi:10.1103/PhysRevD.85.126008}}.

\bibitem{Armstrong:2023phb}
C.~Armstrong, H.~Goodhew, A.~Lipstein, J.~Mei, {Graviton trispectrum from
  gluons}, JHEP 08 (2023) 206.
\newblock \href {http://arxiv.org/abs/2304.07206} {\path{arXiv:2304.07206}},
  \href {https://doi.org/10.1007/JHEP08(2023)206}
  {\path{doi:10.1007/JHEP08(2023)206}}.

\bibitem{Lee:2022fgr}
H.~Lee, X.~Wang, {Cosmological double-copy relations}, Phys. Rev. D 108~(6)
  (2023) L061702.
\newblock \href {http://arxiv.org/abs/2212.11282} {\path{arXiv:2212.11282}},
  \href {https://doi.org/10.1103/PhysRevD.108.L061702}
  {\path{doi:10.1103/PhysRevD.108.L061702}}.

\bibitem{Costa:2011dw}
M.~S. Costa, J.~Penedones, D.~Poland, S.~Rychkov, {Spinning Conformal Blocks},
  JHEP 11 (2011) 154.
\newblock \href {http://arxiv.org/abs/1109.6321} {\path{arXiv:1109.6321}},
  \href {https://doi.org/10.1007/JHEP11(2011)154}
  {\path{doi:10.1007/JHEP11(2011)154}}.

\bibitem{Costa:2018mcg}
M.~S. Costa, T.~Hansen, {AdS Weight Shifting Operators}, JHEP 09 (2018) 040.
\newblock \href {http://arxiv.org/abs/1805.01492} {\path{arXiv:1805.01492}},
  \href {https://doi.org/10.1007/JHEP09(2018)040}
  {\path{doi:10.1007/JHEP09(2018)040}}.

\bibitem{Karateev:2017jgd}
D.~Karateev, P.~Kravchuk, D.~Simmons-Duffin, {Weight Shifting Operators and
  Conformal Blocks}, JHEP 02 (2018) 081.
\newblock \href {http://arxiv.org/abs/1706.07813} {\path{arXiv:1706.07813}},
  \href {https://doi.org/10.1007/JHEP02(2018)081}
  {\path{doi:10.1007/JHEP02(2018)081}}.

\bibitem{Baumann:2019oyu}
D.~Baumann, C.~Duaso~Pueyo, A.~Joyce, H.~Lee, G.~L. Pimentel, {The cosmological
  bootstrap: weight-shifting operators and scalar seeds}, JHEP 12 (2020) 204.
\newblock \href {http://arxiv.org/abs/1910.14051} {\path{arXiv:1910.14051}},
  \href {https://doi.org/10.1007/JHEP12(2020)204}
  {\path{doi:10.1007/JHEP12(2020)204}}.

\bibitem{Chowdhury:2025ohm}
C.~Chowdhury, A.~Lipstein, J.~Marshall, J.~Mei, I.~Sachs, {Cosmological
  dressing rules}, JHEP 03 (2026) 076.
\newblock \href {http://arxiv.org/abs/2503.10598} {\path{arXiv:2503.10598}},
  \href {https://doi.org/10.1007/JHEP03(2026)076}
  {\path{doi:10.1007/JHEP03(2026)076}}.

\bibitem{Schwinger:1960qe}
J.~S. Schwinger, {Brownian motion of a quantum oscillator}, J. Math. Phys. 2
  (1961) 407--432.
\newblock \href {https://doi.org/10.1063/1.1703727}
  {\path{doi:10.1063/1.1703727}}.

\bibitem{Keldysh:1964ud}
L.~V. Keldysh, {Diagram Technique for Nonequilibrium Processes}, Sov. Phys.
  JETP 20 (1965) 1018--1026.
\newblock \href {https://doi.org/10.1142/9789811279461_0007}
  {\path{doi:10.1142/9789811279461_0007}}.

\bibitem{Sleight:2019mgd}
C.~Sleight, {A Mellin Space Approach to Cosmological Correlators}, JHEP 01
  (2020) 090.
\newblock \href {http://arxiv.org/abs/1906.12302} {\path{arXiv:1906.12302}},
  \href {https://doi.org/10.1007/JHEP01(2020)090}
  {\path{doi:10.1007/JHEP01(2020)090}}.

\bibitem{Sleight:2021plv}
C.~Sleight, M.~Taronna, {From dS to AdS and back}, JHEP 12 (2021) 074.
\newblock \href {http://arxiv.org/abs/2109.02725} {\path{arXiv:2109.02725}},
  \href {https://doi.org/10.1007/JHEP12(2021)074}
  {\path{doi:10.1007/JHEP12(2021)074}}.

\bibitem{DiPietro:2021sjt}
L.~Di~Pietro, V.~Gorbenko, S.~Komatsu, {Analyticity and unitarity for
  cosmological correlators}, JHEP 03 (2022) 023.
\newblock \href {http://arxiv.org/abs/2108.01695} {\path{arXiv:2108.01695}},
  \href {https://doi.org/10.1007/JHEP03(2022)023}
  {\path{doi:10.1007/JHEP03(2022)023}}.

\bibitem{Chowdhury:2023arc}
C.~Chowdhury, A.~Lipstein, J.~Mei, I.~Sachs, P.~Vanhove, {The subtle simplicity
  of cosmological correlators}, JHEP 03 (2025) 007.
\newblock \href {http://arxiv.org/abs/2312.13803} {\path{arXiv:2312.13803}},
  \href {https://doi.org/10.1007/JHEP03(2025)007}
  {\path{doi:10.1007/JHEP03(2025)007}}.

\bibitem{Raju:2010by}
S.~Raju, {BCFW for Witten Diagrams}, Phys. Rev. Lett. 106 (2011) 091601.
\newblock \href {http://arxiv.org/abs/1011.0780} {\path{arXiv:1011.0780}},
  \href {https://doi.org/10.1103/PhysRevLett.106.091601}
  {\path{doi:10.1103/PhysRevLett.106.091601}}.

\bibitem{Zhou:2021gnu}
X.~Zhou, {Double Copy Relation in AdS Space}, Phys. Rev. Lett. 127~(14) (2021)
  141601.
\newblock \href {http://arxiv.org/abs/2106.07651} {\path{arXiv:2106.07651}},
  \href {https://doi.org/10.1103/PhysRevLett.127.141601}
  {\path{doi:10.1103/PhysRevLett.127.141601}}.

\bibitem{Drummond:2022dxd}
J.~M. Drummond, R.~Glew, M.~Santagata, {Bern-Carrasco-Johansson relations in
  AdS5\texttimes{}S3 and the double-trace spectrum of super gluons}, Phys. Rev.
  D 107~(8) (2023) L081901.
\newblock \href {http://arxiv.org/abs/2202.09837} {\path{arXiv:2202.09837}},
  \href {https://doi.org/10.1103/PhysRevD.107.L081901}
  {\path{doi:10.1103/PhysRevD.107.L081901}}.

\bibitem{Mack:2009mi}
G.~Mack, {D-independent representation of Conformal Field Theories in D
  dimensions via transformation to auxiliary Dual Resonance Models. Scalar
  amplitudes} (7 2009).
\newblock \href {http://arxiv.org/abs/0907.2407} {\path{arXiv:0907.2407}}.

\bibitem{Penedones:2010ue}
J.~Penedones, {Writing CFT correlation functions as AdS scattering amplitudes},
  JHEP 03 (2011) 025.
\newblock \href {http://arxiv.org/abs/1011.1485} {\path{arXiv:1011.1485}},
  \href {https://doi.org/10.1007/JHEP03(2011)025}
  {\path{doi:10.1007/JHEP03(2011)025}}.

\bibitem{Nirschl:2004pa}
M.~Nirschl, H.~Osborn, {Superconformal Ward identities and their solution},
  Nucl. Phys. B 711 (2005) 409--479.
\newblock \href {http://arxiv.org/abs/hep-th/0407060}
  {\path{arXiv:hep-th/0407060}}, \href
  {https://doi.org/10.1016/j.nuclphysb.2005.01.013}
  {\path{doi:10.1016/j.nuclphysb.2005.01.013}}.

\bibitem{Eden:2000bk}
B.~Eden, A.~C. Petkou, C.~Schubert, E.~Sokatchev, {Partial nonrenormalization
  of the stress tensor four point function in N=4 SYM and AdS / CFT}, Nucl.
  Phys. B 607 (2001) 191--212.
\newblock \href {http://arxiv.org/abs/hep-th/0009106}
  {\path{arXiv:hep-th/0009106}}, \href
  {https://doi.org/10.1016/S0550-3213(01)00151-1}
  {\path{doi:10.1016/S0550-3213(01)00151-1}}.

\bibitem{Alday:2022lkk}
L.~F. Alday, V.~Gon\c{c}alves, X.~Zhou, {Supersymmetric Five-Point Gluon
  Amplitudes in AdS Space}, Phys. Rev. Lett. 128~(16) (2022) 161601.
\newblock \href {http://arxiv.org/abs/2201.04422} {\path{arXiv:2201.04422}},
  \href {https://doi.org/10.1103/PhysRevLett.128.161601}
  {\path{doi:10.1103/PhysRevLett.128.161601}}.

\bibitem{Albayrak:2019yve}
S.~Albayrak, S.~Kharel, {Towards the higher point holographic momentum space
  amplitudes. Part II. Gravitons}, JHEP 12 (2019) 135.
\newblock \href {http://arxiv.org/abs/1908.01835} {\path{arXiv:1908.01835}},
  \href {https://doi.org/10.1007/JHEP12(2019)135}
  {\path{doi:10.1007/JHEP12(2019)135}}.

\bibitem{Aprile:2020luw}
F.~Aprile, P.~Vieira, {Large $p$ explorations. From SUGRA to big STRINGS in
  Mellin space}, JHEP 12 (2020) 206.
\newblock \href {http://arxiv.org/abs/2007.09176} {\path{arXiv:2007.09176}},
  \href {https://doi.org/10.1007/JHEP12(2020)206}
  {\path{doi:10.1007/JHEP12(2020)206}}.

\bibitem{Alday:2025bjp}
L.~F. Alday, M.~Nocchi, A.~S. Sangar{\'e}, {Stringy KLT Relations on AdS}, JHEP
  02 (2026) 124.
\newblock \href {http://arxiv.org/abs/2504.19973} {\path{arXiv:2504.19973}},
  \href {https://doi.org/10.1007/JHEP02(2026)124}
  {\path{doi:10.1007/JHEP02(2026)124}}.

\bibitem{Alday:2023mvu}
L.~F. Alday, T.~Hansen, {The AdS Virasoro-Shapiro amplitude}, JHEP 10 (2023)
  023.
\newblock \href {http://arxiv.org/abs/2306.12786} {\path{arXiv:2306.12786}},
  \href {https://doi.org/10.1007/JHEP10(2023)023}
  {\path{doi:10.1007/JHEP10(2023)023}}.

\bibitem{Alday:2022uxp}
L.~F. Alday, T.~Hansen, J.~A. Silva, {AdS Virasoro-Shapiro from dispersive sum
  rules}, JHEP 10 (2022) 036.
\newblock \href {http://arxiv.org/abs/2204.07542} {\path{arXiv:2204.07542}},
  \href {https://doi.org/10.1007/JHEP10(2022)036}
  {\path{doi:10.1007/JHEP10(2022)036}}.

\bibitem{Alday:2022xwz}
L.~F. Alday, T.~Hansen, J.~A. Silva, {AdS Virasoro-Shapiro from single-valued
  periods}, JHEP 12 (2022) 010.
\newblock \href {http://arxiv.org/abs/2209.06223} {\path{arXiv:2209.06223}},
  \href {https://doi.org/10.1007/JHEP12(2022)010}
  {\path{doi:10.1007/JHEP12(2022)010}}.

\bibitem{Alday:2023jdk}
L.~F. Alday, T.~Hansen, J.~A. Silva, {Emergent Worldsheet for the AdS
  Virasoro-Shapiro Amplitude}, Phys. Rev. Lett. 131~(16) (2023) 161603.
\newblock \href {http://arxiv.org/abs/2305.03593} {\path{arXiv:2305.03593}},
  \href {https://doi.org/10.1103/PhysRevLett.131.161603}
  {\path{doi:10.1103/PhysRevLett.131.161603}}.

\bibitem{Alday:2024yax}
L.~F. Alday, S.~M. Chester, T.~Hansen, D.-l. Zhong, {The AdS Veneziano
  amplitude at small curvature}, JHEP 05 (2024) 322.
\newblock \href {http://arxiv.org/abs/2403.13877} {\path{arXiv:2403.13877}},
  \href {https://doi.org/10.1007/JHEP05(2024)322}
  {\path{doi:10.1007/JHEP05(2024)322}}.

\bibitem{Alday:2024ksp}
L.~F. Alday, T.~Hansen, {Single-valuedness of the AdS Veneziano amplitude},
  JHEP 08 (2024) 108.
\newblock \href {http://arxiv.org/abs/2404.16084} {\path{arXiv:2404.16084}},
  \href {https://doi.org/10.1007/JHEP08(2024)108}
  {\path{doi:10.1007/JHEP08(2024)108}}.

\bibitem{Alday:2025cxr}
L.~F. Alday, R.~S. Pitombo, A.~Str{\"o}mholm~Sangar{\'e}, {Monodromy relations
  for string amplitudes on AdS space}, Phys. Rev. D 113~(2) (2026) L021903.
\newblock \href {http://arxiv.org/abs/2509.02719} {\path{arXiv:2509.02719}},
  \href {https://doi.org/10.1103/mppb-xh1j} {\path{doi:10.1103/mppb-xh1j}}.

\bibitem{brown2006multiplezetavaluesperiods}
F.~C.~S. Brown, \href{https://arxiv.org/abs/math/0606419}{Multiple zeta values
  and periods of moduli spaces $\mathfrak{M}_{0,n}$} (2006).
\newblock \href {http://arxiv.org/abs/math/0606419}
  {\path{arXiv:math/0606419}}.
\newline\urlprefix\url{https://arxiv.org/abs/math/0606419}

\bibitem{Brown2004SingleValuedHyperlogarithms}
F.~C.~S. Brown, \href{https://www.ihes.fr/~brown/RHpaper5.pdf}{Single-valued
  hyperlogarithms and unipotent differential equations}, unpublished notes
  (2004).
\newline\urlprefix\url{https://www.ihes.fr/~brown/RHpaper5.pdf}

\bibitem{Kakkad:2025klm}
H.~Kakkad, A.~Ochirov, S.~Zhang, {Twisted de Rham theory for string double copy
  in AdS}, JHEP 04 (2026) 112.
\newblock \href {http://arxiv.org/abs/2512.23699} {\path{arXiv:2512.23699}},
  \href {https://doi.org/10.1007/JHEP04(2026)112}
  {\path{doi:10.1007/JHEP04(2026)112}}.

\bibitem{Nocchi:2026yms}
M.~Nocchi, R.~S. Pitombo, A.~Str{\"o}mholm~Sangar{\'e}, Y.-X. Tao,
  {All-multiplicity monodromy and KLT relations for AdS string integrals} (6
  2026).
\newblock \href {http://arxiv.org/abs/2606.09769} {\path{arXiv:2606.09769}}.

\bibitem{Sleight:2019hfp}
C.~Sleight, M.~Taronna, {Bootstrapping Inflationary Correlators in Mellin
  Space}, JHEP 02 (2020) 098.
\newblock \href {http://arxiv.org/abs/1907.01143} {\path{arXiv:1907.01143}},
  \href {https://doi.org/10.1007/JHEP02(2020)098}
  {\path{doi:10.1007/JHEP02(2020)098}}.

\bibitem{Sleight:2021iix}
C.~Sleight, M.~Taronna, {On the consistency of (partially-)massless matter
  couplings in de Sitter space}, JHEP 10 (2021) 156.
\newblock \href {http://arxiv.org/abs/2106.00366} {\path{arXiv:2106.00366}},
  \href {https://doi.org/10.1007/JHEP10(2021)156}
  {\path{doi:10.1007/JHEP10(2021)156}}.

\bibitem{Mei:2023jkb}
J.~Mei, {Amplitude Bootstrap in (Anti) de Sitter Space And The Four-Point
  Graviton from Double Copy} (5 2023).
\newblock \href {http://arxiv.org/abs/2305.13894} {\path{arXiv:2305.13894}}.

\bibitem{Armstrong:2022mfr}
C.~Armstrong, H.~Gomez, R.~Lipinski~Jusinskas, A.~Lipstein, J.~Mei, {New
  recursion relations for tree-level correlators in anti{\textendash}de Sitter
  spacetime}, Phys. Rev. D 106~(12) (2022) L121701.
\newblock \href {http://arxiv.org/abs/2209.02709} {\path{arXiv:2209.02709}},
  \href {https://doi.org/10.1103/PhysRevD.106.L121701}
  {\path{doi:10.1103/PhysRevD.106.L121701}}.

\bibitem{Li:2022tby}
Y.-Z. Li, {Flat-space structure of gluons and gravitons in AdS spacetime},
  Phys. Rev. D 107~(12) (2023) 125018.
\newblock \href {http://arxiv.org/abs/2212.13195} {\path{arXiv:2212.13195}},
  \href {https://doi.org/10.1103/PhysRevD.107.125018}
  {\path{doi:10.1103/PhysRevD.107.125018}}.

\bibitem{Dirac:1936fq}
P.~A.~M. Dirac, {Wave equations in conformal space}, Annals Math. 37 (1936)
  429--442.
\newblock \href {https://doi.org/10.2307/1968455} {\path{doi:10.2307/1968455}}.

\bibitem{Mack:1969rr}
G.~Mack, A.~Salam, {Finite component field representations of the conformal
  group}, Annals Phys. 53 (1969) 174--202.
\newblock \href {https://doi.org/10.1016/0003-4916(69)90278-4}
  {\path{doi:10.1016/0003-4916(69)90278-4}}.

\bibitem{Boulware:1970ty}
D.~G. Boulware, L.~S. Brown, R.~D. Peccei, {Deep-inelastic electroproduction
  and conformal symmetry}, Phys. Rev. D 2 (1970) 293--298.
\newblock \href {https://doi.org/10.1103/PhysRevD.2.293}
  {\path{doi:10.1103/PhysRevD.2.293}}.

\bibitem{Ferrara:1972cq}
S.~Ferrara, A.~F. Grillo, R.~Gatto, {Manifestly conformal-covariant expansion
  on the light cone}, Phys. Rev. D 5 (1972) 3102--3108.
\newblock \href {https://doi.org/10.1103/PhysRevD.5.3102}
  {\path{doi:10.1103/PhysRevD.5.3102}}.

\bibitem{Weinberg:2010fx}
S.~Weinberg, {Six-dimensional Methods for Four-dimensional Conformal Field
  Theories}, Phys. Rev. D 82 (2010) 045031.
\newblock \href {http://arxiv.org/abs/1006.3480} {\path{arXiv:1006.3480}},
  \href {https://doi.org/10.1103/PhysRevD.82.045031}
  {\path{doi:10.1103/PhysRevD.82.045031}}.

\bibitem{Simmons-Duffin:2012juh}
D.~Simmons-Duffin, {Projectors, Shadows, and Conformal Blocks}, JHEP 04 (2014)
  146.
\newblock \href {http://arxiv.org/abs/1204.3894} {\path{arXiv:1204.3894}},
  \href {https://doi.org/10.1007/JHEP04(2014)146}
  {\path{doi:10.1007/JHEP04(2014)146}}.

\bibitem{Costa:2011mg}
M.~S. Costa, J.~Penedones, D.~Poland, S.~Rychkov, {Spinning Conformal
  Correlators}, JHEP 11 (2011) 071.
\newblock \href {http://arxiv.org/abs/1107.3554} {\path{arXiv:1107.3554}},
  \href {https://doi.org/10.1007/JHEP11(2011)071}
  {\path{doi:10.1007/JHEP11(2011)071}}.

\bibitem{Costa:2014kfa}
M.~S. Costa, V.~Gon{\c{c}}alves, J.~Penedones, {Spinning AdS Propagators}, JHEP
  09 (2014) 064.
\newblock \href {http://arxiv.org/abs/1404.5625} {\path{arXiv:1404.5625}},
  \href {https://doi.org/10.1007/JHEP09(2014)064}
  {\path{doi:10.1007/JHEP09(2014)064}}.

\bibitem{Simmons-Duffin:2016gjk}
D.~Simmons-Duffin, {The Conformal Bootstrap}, in: {Theoretical Advanced Study
  Institute in Elementary Particle Physics}: {New Frontiers in Fields and
  Strings}, 2017, pp. 1--74.
\newblock \href {http://arxiv.org/abs/1602.07982} {\path{arXiv:1602.07982}},
  \href {https://doi.org/10.1142/9789813149441_0001}
  {\path{doi:10.1142/9789813149441_0001}}.

\bibitem{Mack:1975je}
G.~Mack, {All unitary ray representations of the conformal group SU(2,2) with
  positive energy}, Commun. Math. Phys. 55 (1977) 1.
\newblock \href {https://doi.org/10.1007/BF01613145}
  {\path{doi:10.1007/BF01613145}}.

\bibitem{Ferrara:1974pt}
S.~Ferrara, R.~Gatto, A.~F. Grillo, {Positivity Restrictions on Anomalous
  Dimensions}, Phys. Rev. D 9 (1974) 3564.
\newblock \href {https://doi.org/10.1103/PhysRevD.9.3564}
  {\path{doi:10.1103/PhysRevD.9.3564}}.

\bibitem{Metsaev:1995re}
R.~R. Metsaev, {Massless mixed symmetry bosonic free fields in d-dimensional
  anti-de Sitter space-time}, Phys. Lett. B 354 (1995) 78--84.
\newblock \href {https://doi.org/10.1016/0370-2693(95)00563-Z}
  {\path{doi:10.1016/0370-2693(95)00563-Z}}.

\bibitem{Minwalla:1997ka}
S.~Minwalla, {Restrictions imposed by superconformal invariance on quantum
  field theories}, Adv. Theor. Math. Phys. 2 (1998) 783--851.
\newblock \href {http://arxiv.org/abs/hep-th/9712074}
  {\path{arXiv:hep-th/9712074}}, \href
  {https://doi.org/10.4310/ATMP.1998.v2.n4.a4}
  {\path{doi:10.4310/ATMP.1998.v2.n4.a4}}.

\bibitem{Diwakar:2021juk}
P.~Diwakar, A.~Herderschee, R.~Roiban, F.~Teng, {BCJ amplitude relations for
  Anti-de Sitter boundary correlators in embedding space}, JHEP 10 (2021) 141.
\newblock \href {http://arxiv.org/abs/2106.10822} {\path{arXiv:2106.10822}},
  \href {https://doi.org/10.1007/JHEP10(2021)141}
  {\path{doi:10.1007/JHEP10(2021)141}}.

\bibitem{Herderschee:2022ntr}
A.~Herderschee, R.~Roiban, F.~Teng, {On the differential representation and
  color-kinematics duality of AdS boundary correlators}, JHEP 05 (2022) 026.
\newblock \href {http://arxiv.org/abs/2201.05067} {\path{arXiv:2201.05067}},
  \href {https://doi.org/10.1007/JHEP05(2022)026}
  {\path{doi:10.1007/JHEP05(2022)026}}.

\bibitem{Cheung:2022pdk}
C.~Cheung, J.~Parra-Martinez, A.~Sivaramakrishnan, {On-shell correlators and
  color-kinematics duality in curved symmetric spacetimes}, JHEP 05 (2022) 027.
\newblock \href {http://arxiv.org/abs/2201.05147} {\path{arXiv:2201.05147}},
  \href {https://doi.org/10.1007/JHEP05(2022)027}
  {\path{doi:10.1007/JHEP05(2022)027}}.

\bibitem{Cheung:2021zvb}
C.~Cheung, J.~Mangan, {Covariant color-kinematics duality}, JHEP 11 (2021) 069.
\newblock \href {http://arxiv.org/abs/2108.02276} {\path{arXiv:2108.02276}},
  \href {https://doi.org/10.1007/JHEP11(2021)069}
  {\path{doi:10.1007/JHEP11(2021)069}}.

\bibitem{Sivaramakrishnan:2021srm}
A.~Sivaramakrishnan, {Towards color-kinematics duality in generic spacetimes},
  JHEP 04 (2022) 036.
\newblock \href {http://arxiv.org/abs/2110.15356} {\path{arXiv:2110.15356}},
  \href {https://doi.org/10.1007/JHEP04(2022)036}
  {\path{doi:10.1007/JHEP04(2022)036}}.

\bibitem{Prabhu:2020avf}
S.~G. Prabhu, {The classical double copy in curved spacetimes: perturbative
  Yang-Mills from the bi-adjoint scalar}, JHEP 05 (2024) 117.
\newblock \href {http://arxiv.org/abs/2011.06588} {\path{arXiv:2011.06588}},
  \href {https://doi.org/10.1007/JHEP05(2024)117}
  {\path{doi:10.1007/JHEP05(2024)117}}.

\bibitem{Bonifacio:2018zex}
J.~Bonifacio, K.~Hinterbichler, A.~Joyce, R.~A. Rosen, {Shift Symmetries in
  (Anti) de Sitter Space}, JHEP 02 (2019) 178.
\newblock \href {http://arxiv.org/abs/1812.08167} {\path{arXiv:1812.08167}},
  \href {https://doi.org/10.1007/JHEP02(2019)178}
  {\path{doi:10.1007/JHEP02(2019)178}}.

\bibitem{Baumann:2024ttn}
D.~Baumann, G.~Mathys, G.~L. Pimentel, F.~Rost, {A new twist on spinning (A)dS
  correlators}, JHEP 01 (2025) 202.
\newblock \href {http://arxiv.org/abs/2408.02727} {\path{arXiv:2408.02727}},
  \href {https://doi.org/10.1007/JHEP01(2025)202}
  {\path{doi:10.1007/JHEP01(2025)202}}.

\bibitem{CarrilloGonzalez:2025qjk}
M.~Carrillo~Gonz{\'a}lez, T.~Keseman, {Spinning Boundary Correlators from
  (A)dS$_4$ Twistors} (9 2025).
\newblock \href {http://arxiv.org/abs/2510.00096} {\path{arXiv:2510.00096}}.

\bibitem{Bala:2025gmz}
A.~Bala, S.~Jain, D.~K. S., D.~Mazumdar, V.~Singh, {3D conformal field theory
  in twistor space}, JHEP 12 (2025) 120.
\newblock \href {http://arxiv.org/abs/2502.18562} {\path{arXiv:2502.18562}},
  \href {https://doi.org/10.1007/JHEP12(2025)120}
  {\path{doi:10.1007/JHEP12(2025)120}}.

\bibitem{Bala:2025jbh}
A.~Bala, S.~Jain, D.~K. S., D.~Mazumdar, V.~Singh, B.~Thakkar, {A Supertwistor
  Formalism for $\mathcal{N}=1,2,3,4\;\text{SCFT}_3$} (3 2025).
\newblock \href {http://arxiv.org/abs/2503.19970} {\path{arXiv:2503.19970}}.

\bibitem{Bala:2025qxr}
A.~Bala, D.~K. S, {An Ode to the Penrose and Witten transforms in twistor space
  for 3D CFT}, JHEP 11 (2025) 056.
\newblock \href {http://arxiv.org/abs/2505.14082} {\path{arXiv:2505.14082}},
  \href {https://doi.org/10.1007/JHEP11(2025)056}
  {\path{doi:10.1007/JHEP11(2025)056}}.

\bibitem{Ansari:2025fvi}
A.~Ansari, S.~Jain, D.~K. S, {AdS$_4$ Boundary Wightman functions in Twistor
  Space: Factorization, Conformal blocks and a Double Copy} (12 2025).
\newblock \href {http://arxiv.org/abs/2512.04172} {\path{arXiv:2512.04172}}.

\bibitem{CarrilloGonzalez:2026eum}
M.~Carrillo~Gonz{\'a}lez, T.~Keseman, {4d CFT Correlators from Ambitwistors} (6
  2026).
\newblock \href {http://arxiv.org/abs/2606.13770} {\path{arXiv:2606.13770}}.

\bibitem{Adamo:2016rtr}
T.~Adamo, D.~Skinner, J.~Williams, {Twistor methods for AdS$_{5}$}, JHEP 08
  (2016) 167.
\newblock \href {http://arxiv.org/abs/1607.03763} {\path{arXiv:1607.03763}},
  \href {https://doi.org/10.1007/JHEP08(2016)167}
  {\path{doi:10.1007/JHEP08(2016)167}}.

\bibitem{Adamo:2017zkm}
T.~Adamo, R.~Monteiro, M.~F. Paulos, {Space-time CFTs from the Riemann sphere},
  JHEP 08 (2017) 067.
\newblock \href {http://arxiv.org/abs/1703.04589} {\path{arXiv:1703.04589}},
  \href {https://doi.org/10.1007/JHEP08(2017)067}
  {\path{doi:10.1007/JHEP08(2017)067}}.

\bibitem{Binder:2020raz}
D.~J. Binder, D.~Z. Freedman, S.~S. Pufu, {A bispinor formalism for spinning
  Witten diagrams}, JHEP 02 (2022) 040.
\newblock \href {http://arxiv.org/abs/2003.07448} {\path{arXiv:2003.07448}},
  \href {https://doi.org/10.1007/JHEP02(2022)040}
  {\path{doi:10.1007/JHEP02(2022)040}}.

\bibitem{Adamo:2017qyl}
T.~Adamo, {Lectures on twistor theory}, PoS Modave2017 (2018) 003.
\newblock \href {http://arxiv.org/abs/1712.02196} {\path{arXiv:1712.02196}},
  \href {https://doi.org/10.22323/1.323.0003} {\path{doi:10.22323/1.323.0003}}.

\bibitem{Witten:2003nn}
E.~Witten, {Perturbative gauge theory as a string theory in twistor space},
  Commun. Math. Phys. 252 (2004) 189--258.
\newblock \href {http://arxiv.org/abs/hep-th/0312171}
  {\path{arXiv:hep-th/0312171}}, \href
  {https://doi.org/10.1007/s00220-004-1187-3}
  {\path{doi:10.1007/s00220-004-1187-3}}.

\bibitem{Roiban:2004yf}
R.~Roiban, M.~Spradlin, A.~Volovich, {On the tree level S matrix of Yang-Mills
  theory}, Phys. Rev. D 70 (2004) 026009.
\newblock \href {http://arxiv.org/abs/hep-th/0403190}
  {\path{arXiv:hep-th/0403190}}, \href
  {https://doi.org/10.1103/PhysRevD.70.026009}
  {\path{doi:10.1103/PhysRevD.70.026009}}.

\bibitem{Mason:2009sa}
L.~J. Mason, D.~Skinner, {Scattering Amplitudes and BCFW Recursion in Twistor
  Space}, JHEP 01 (2010) 064.
\newblock \href {http://arxiv.org/abs/0903.2083} {\path{arXiv:0903.2083}},
  \href {https://doi.org/10.1007/JHEP01(2010)064}
  {\path{doi:10.1007/JHEP01(2010)064}}.

\bibitem{Adamo:2011pv}
T.~Adamo, M.~Bullimore, L.~Mason, D.~Skinner, {Scattering Amplitudes and Wilson
  Loops in Twistor Space}, J. Phys. A 44 (2011) 454008.
\newblock \href {http://arxiv.org/abs/1104.2890} {\path{arXiv:1104.2890}},
  \href {https://doi.org/10.1088/1751-8113/44/45/454008}
  {\path{doi:10.1088/1751-8113/44/45/454008}}.

\bibitem{Cachazo:2012kg}
F.~Cachazo, D.~Skinner, {Gravity from Rational Curves in Twistor Space}, Phys.
  Rev. Lett. 110~(16) (2013) 161301.
\newblock \href {http://arxiv.org/abs/1207.0741} {\path{arXiv:1207.0741}},
  \href {https://doi.org/10.1103/PhysRevLett.110.161301}
  {\path{doi:10.1103/PhysRevLett.110.161301}}.

\bibitem{Arkani-Hamed:2009hub}
N.~Arkani-Hamed, F.~Cachazo, C.~Cheung, J.~Kaplan, {The S-Matrix in Twistor
  Space}, JHEP 03 (2010) 110.
\newblock \href {http://arxiv.org/abs/0903.2110} {\path{arXiv:0903.2110}},
  \href {https://doi.org/10.1007/JHEP03(2010)110}
  {\path{doi:10.1007/JHEP03(2010)110}}.

\bibitem{Penrose:1986ca}
R.~Penrose, W.~Rindler, {SPINORS AND SPACE-TIME. VOL. 2: SPINOR AND TWISTOR
  METHODS IN SPACE-TIME GEOMETRY}, Cambridge Monographs on Mathematical
  Physics, Cambridge University Press, 1988.
\newblock \href {https://doi.org/10.1017/CBO9780511524486}
  {\path{doi:10.1017/CBO9780511524486}}.

\bibitem{Bala:2026hag}
A.~Bala, S.~Jain, D.~Mazumdar, A.~A. Rao, {$\mathbb{R}$eal Ambitwistors {\&}
  Massive Twistors for CFT$_4$} (8 2026).
\newblock \href {http://arxiv.org/abs/2608.18210} {\path{arXiv:2608.18210}}.

\bibitem{Adamo:2024hme}
T.~Adamo, S.~Klisch, {The KLT Kernel in Twistor Space}, Commun. Math. Phys.
  406~(4) (2025) 79.
\newblock \href {http://arxiv.org/abs/2406.04539} {\path{arXiv:2406.04539}},
  \href {https://doi.org/10.1007/s00220-025-05254-0}
  {\path{doi:10.1007/s00220-025-05254-0}}.

\bibitem{Adamo:2015ina}
T.~Adamo, {Gravity with a cosmological constant from rational curves}, JHEP 11
  (2015) 098.
\newblock \href {http://arxiv.org/abs/1508.02554} {\path{arXiv:1508.02554}},
  \href {https://doi.org/10.1007/JHEP11(2015)098}
  {\path{doi:10.1007/JHEP11(2015)098}}.

\bibitem{Griffiths:1994prl}
P.~Griffiths, J.~Harris, {Principles of Algebraic Geometry}, 1994.
\newblock \href {https://doi.org/10.1002/9781118032527}
  {\path{doi:10.1002/9781118032527}}.

\bibitem{harris1992algebraic}
J.~Harris, Algebraic Geometry: A First Course, Vol. 133 of Graduate Texts in
  Mathematics, Springer, New York, 1992.
\newblock \href {https://doi.org/10.1007/978-1-4757-2189-8}
  {\path{doi:10.1007/978-1-4757-2189-8}}.

\bibitem{Arundine:2026fbr}
M.~Arundine, D.~Baumann, M.~H.~G. Lee, G.~L. Pimentel, F.~Rost, {The
  Cosmological Grassmannian} (2 2026).
\newblock \href {http://arxiv.org/abs/2602.07117} {\path{arXiv:2602.07117}}.

\bibitem{ElMaazouz2024}
Y.~{El Maazouz}, Y.~{Mandelshtam}, {The positive orthogonal Grassmannian},
  arXiv e-prints (2024) arXiv:2412.14091\href {http://arxiv.org/abs/2412.14091}
  {\path{arXiv:2412.14091}}, \href {https://doi.org/10.48550/arXiv.2412.14091}
  {\path{doi:10.48550/arXiv.2412.14091}}.

\bibitem{Goodhew:2021oqg}
H.~Goodhew, S.~Jazayeri, M.~H.~G. Lee, E.~Pajer, {Cutting cosmological
  correlators}, JCAP 08 (2021) 003.
\newblock \href {http://arxiv.org/abs/2104.06587} {\path{arXiv:2104.06587}},
  \href {https://doi.org/10.1088/1475-7516/2021/08/003}
  {\path{doi:10.1088/1475-7516/2021/08/003}}.

\bibitem{De:2026shn}
S.~De, H.~Lee, {The Vasiliev Grassmannian} (3 2026).
\newblock \href {http://arxiv.org/abs/2603.24656} {\path{arXiv:2603.24656}}.

\bibitem{Bala:2026lvw}
A.~Bala, S.~Jain, V.~Singh, {A Cosmological BCFW Bridge and Its Canonical
  Geometry} (6 2026).
\newblock \href {http://arxiv.org/abs/2606.25032} {\path{arXiv:2606.25032}}.

\bibitem{Bala:2026bdx}
A.~Bala, S.~Jain, D.~K. S., A.~A. Rao, {Super-Grassmannians for $\mathcal{N}=2$
  to $4$ SCFT$_3$: From AdS$_4$ Correlators to $\mathcal{N}=4$ SYM scattering
  Amplitudes} (4 2026).
\newblock \href {http://arxiv.org/abs/2604.07503} {\path{arXiv:2604.07503}}.

\bibitem{Bala:2026hdm}
A.~Bala, S.~Jain, D.~K. S., A.~A. Rao, {The $\mathcal{N}=1$ Super-Grassmannian
  for CFT$_3$ and a Foray on AdS and Cosmological Correlators} (4 2026).
\newblock \href {http://arxiv.org/abs/2604.07446} {\path{arXiv:2604.07446}}.

\bibitem{Huang:2026tsh}
Y.-t. Huang, C.-K. Kuo, Y.~Liu, J.~Mei, {Beyond Discontinuities: Cosmological
  WFCs from the Supersymmetric Orthogonal Grassmannian} (4 2026).
\newblock \href {http://arxiv.org/abs/2604.08512} {\path{arXiv:2604.08512}}.

\bibitem{Arundine:2026myr}
M.~Arundine, G.~L. Pimentel, {Cosmological Collider in the Grassmannian} (5
  2026).
\newblock \href {http://arxiv.org/abs/2605.21581} {\path{arXiv:2605.21581}}.

\bibitem{Bala:2026trw}
A.~Bala, S.~Jain, D.~K. S, {The Conformal Grassmannian: A Symplectic
  Bi-Grassmannian for $CFT_ 4$ Correlators} (5 2026).
\newblock \href {http://arxiv.org/abs/2605.06811} {\path{arXiv:2605.06811}}.

\bibitem{S:2026qwn}
D.~K. S, {Four-point functions, Twistors and Supersymmetry in the Symplectic
  Bi-Grassmannian for CFT$_4$ and AdS$_5$} (8 2026).
\newblock \href {http://arxiv.org/abs/2608.14451} {\path{arXiv:2608.14451}}.

\bibitem{Adamo:2017nia}
T.~Adamo, E.~Casali, L.~Mason, S.~Nekovar, {Scattering on plane waves and the
  double copy}, Class. Quant. Grav. 35~(1) (2018) 015004.
\newblock \href {http://arxiv.org/abs/1706.08925} {\path{arXiv:1706.08925}},
  \href {https://doi.org/10.1088/1361-6382/aa9961}
  {\path{doi:10.1088/1361-6382/aa9961}}.

\bibitem{Adamo:2017sze}
T.~Adamo, E.~Casali, L.~Mason, S.~Nekovar, {Amplitudes on plane waves from
  ambitwistor strings}, JHEP 11 (2017) 160.
\newblock \href {http://arxiv.org/abs/1708.09249} {\path{arXiv:1708.09249}},
  \href {https://doi.org/10.1007/JHEP11(2017)160}
  {\path{doi:10.1007/JHEP11(2017)160}}.

\bibitem{Adamo:2018mpq}
T.~Adamo, E.~Casali, L.~Mason, S.~Nekovar, {Plane wave backgrounds and
  colour-kinematics duality}, JHEP 02 (2019) 198.
\newblock \href {http://arxiv.org/abs/1810.05115} {\path{arXiv:1810.05115}},
  \href {https://doi.org/10.1007/JHEP02(2019)198}
  {\path{doi:10.1007/JHEP02(2019)198}}.

\bibitem{Adamo:2020qru}
T.~Adamo, A.~Ilderton, {Classical and quantum double copy of back-reaction},
  JHEP 09 (2020) 200.
\newblock \href {http://arxiv.org/abs/2005.05807} {\path{arXiv:2005.05807}},
  \href {https://doi.org/10.1007/JHEP09(2020)200}
  {\path{doi:10.1007/JHEP09(2020)200}}.

\bibitem{Ilderton:2024oly}
A.~Ilderton, W.~Lindved, {Toward double copy on arbitrary backgrounds}, JHEP 11
  (2024) 100.
\newblock \href {http://arxiv.org/abs/2405.10016} {\path{arXiv:2405.10016}},
  \href {https://doi.org/10.1007/JHEP11(2024)100}
  {\path{doi:10.1007/JHEP11(2024)100}}.

\bibitem{Ilderton:2025gug}
A.~Ilderton, W.~Lindved, {Coherent states, background fields, and double copy},
  JHEP 09 (2025) 156.
\newblock \href {http://arxiv.org/abs/2505.16852} {\path{arXiv:2505.16852}},
  \href {https://doi.org/10.1007/JHEP09(2025)156}
  {\path{doi:10.1007/JHEP09(2025)156}}.

\bibitem{Aoude:2025jvt}
R.~Aoude, D.~O'Connell, M.~Sergola, C.~D. White, {Hawking radiation meets the
  double copy}, JHEP 05 (2026) 166.
\newblock \href {http://arxiv.org/abs/2510.25866} {\path{arXiv:2510.25866}},
  \href {https://doi.org/10.1007/JHEP05(2026)166}
  {\path{doi:10.1007/JHEP05(2026)166}}.

\bibitem{Kosower:2018adc}
D.~A. Kosower, B.~Maybee, D.~O'Connell, {Amplitudes, Observables, and Classical
  Scattering}, JHEP 02 (2019) 137.
\newblock \href {http://arxiv.org/abs/1811.10950} {\path{arXiv:1811.10950}},
  \href {https://doi.org/10.1007/JHEP02(2019)137}
  {\path{doi:10.1007/JHEP02(2019)137}}.

\bibitem{Adamo:2022rmp}
T.~Adamo, A.~Cristofoli, A.~Ilderton, {Classical physics from amplitudes on
  curved backgrounds}, JHEP 08 (2022) 281.
\newblock \href {http://arxiv.org/abs/2203.13785} {\path{arXiv:2203.13785}},
  \href {https://doi.org/10.1007/JHEP08(2022)281}
  {\path{doi:10.1007/JHEP08(2022)281}}.

\bibitem{aoude2024amplitudeshawkingradiation}
R.~Aoude, D.~O'Connell, M.~Sergola,
  \href{https://arxiv.org/abs/2412.05267}{Amplitudes for hawking radiation}
  (2024).
\newblock \href {http://arxiv.org/abs/2412.05267} {\path{arXiv:2412.05267}}.
\newline\urlprefix\url{https://arxiv.org/abs/2412.05267}

\bibitem{Hawking:1975vcx}
S.~W. Hawking, {Particle Creation by Black Holes}, Commun. Math. Phys. 43
  (1975) 199--220, [Erratum: Commun.Math.Phys. 46, 206 (1976)].
\newblock \href {https://doi.org/10.1007/BF02345020}
  {\path{doi:10.1007/BF02345020}}.

\bibitem{Ilderton:2025aql}
A.~Ilderton, W.~Lindved, K.~Rajeev, {Hawking Radiation from the Double Copy},
  Phys. Rev. Lett. 136~(8) (2026) 081603.
\newblock \href {http://arxiv.org/abs/2510.25852} {\path{arXiv:2510.25852}},
  \href {https://doi.org/10.1103/vnm5-jwwv} {\path{doi:10.1103/vnm5-jwwv}}.

\bibitem{Carrasco:2025bgu}
J.~J.~M. Carrasco, Y.~Chen, {Double Copy Root of Hawking Thermality}, Phys.
  Rev. Lett. 136~(8) (2026) 081604.
\newblock \href {http://arxiv.org/abs/2511.01832} {\path{arXiv:2511.01832}},
  \href {https://doi.org/10.1103/6gfs-kxbt} {\path{doi:10.1103/6gfs-kxbt}}.

\bibitem{Anninos:2020ccj}
D.~Anninos, B.~M{\"u}hlmann, {Notes on matrix models (matrix musings)}, J.
  Stat. Mech. 2008 (2020) 083109.
\newblock \href {http://arxiv.org/abs/2004.01171} {\path{arXiv:2004.01171}},
  \href {https://doi.org/10.1088/1742-5468/aba499}
  {\path{doi:10.1088/1742-5468/aba499}}.

\bibitem{Wigner1993}
E.~P. Wigner,
  \href{https://doi.org/10.1007/978-3-662-02781-3_35}{Characteristic Vectors of
  Bordered Matrices with Infinite Dimensions I}, Springer Berlin Heidelberg,
  Berlin, Heidelberg, 1993, pp. 524--540.
\newblock \href {https://doi.org/10.1007/978-3-662-02781-3_35}
  {\path{doi:10.1007/978-3-662-02781-3_35}}.
\newline\urlprefix\url{https://doi.org/10.1007/978-3-662-02781-3_35}

\bibitem{Carrasco:2026ijt}
J.~J.~M. Carrasco, Y.~Chen, N.~H. Pavao, A.~Seifi, {Nonperturbative double
  copy: worldline instantons, color thermality, and backreaction}, JHEP 05
  (2026) 177.
\newblock \href {http://arxiv.org/abs/2601.17884} {\path{arXiv:2601.17884}},
  \href {https://doi.org/10.1007/JHEP05(2026)177}
  {\path{doi:10.1007/JHEP05(2026)177}}.

\bibitem{Parikh:1999mf}
M.~K. Parikh, F.~Wilczek, {Hawking radiation as tunneling}, Phys. Rev. Lett. 85
  (2000) 5042--5045.
\newblock \href {http://arxiv.org/abs/hep-th/9907001}
  {\path{arXiv:hep-th/9907001}}, \href
  {https://doi.org/10.1103/PhysRevLett.85.5042}
  {\path{doi:10.1103/PhysRevLett.85.5042}}.

\bibitem{Didenko:2008va}
V.~E. Didenko, A.~S. Matveev, M.~A. Vasiliev, {Unfolded Description of AdS(4)
  Kerr Black Hole}, Phys. Lett. B 665 (2008) 284--293.
\newblock \href {http://arxiv.org/abs/0801.2213} {\path{arXiv:0801.2213}},
  \href {https://doi.org/10.1016/j.physletb.2008.05.067}
  {\path{doi:10.1016/j.physletb.2008.05.067}}.

\bibitem{Didenko:2009td}
V.~E. Didenko, M.~A. Vasiliev, {Static BPS black hole in 4d higher-spin gauge
  theory}, Phys. Lett. B 682 (2009) 305--315, [Erratum: Phys.Lett.B 722, 389
  (2013)].
\newblock \href {http://arxiv.org/abs/0906.3898} {\path{arXiv:0906.3898}},
  \href {https://doi.org/10.1016/j.physletb.2009.11.023}
  {\path{doi:10.1016/j.physletb.2009.11.023}}.

\bibitem{Monteiro:2014cda}
R.~Monteiro, D.~O'Connell, C.~D. White, Black holes and the double copy, JHEP
  12 (2014) 056.
\newblock \href {http://arxiv.org/abs/1410.0239} {\path{arXiv:1410.0239}},
  \href {https://doi.org/10.1007/JHEP12(2014)056}
  {\path{doi:10.1007/JHEP12(2014)056}}.

\bibitem{Luna:2015paa}
A.~Luna, R.~Monteiro, D.~O'Connell, C.~D. White, The classical double copy for
  taub--nut spacetime, Phys.\ Lett.\ B 750 (2015) 272--277.
\newblock \href {http://arxiv.org/abs/1507.01869} {\path{arXiv:1507.01869}},
  \href {https://doi.org/10.1016/j.physletb.2015.09.021}
  {\path{doi:10.1016/j.physletb.2015.09.021}}.

\bibitem{Luna:2016due}
A.~Luna, R.~Monteiro, I.~Nicholson, D.~O'Connell, C.~D. White, The double copy:
  Bremsstrahlung and accelerating black holes, JHEP 06 (2016) 023.
\newblock \href {http://arxiv.org/abs/1603.05737} {\path{arXiv:1603.05737}},
  \href {https://doi.org/10.1007/JHEP06(2016)023}
  {\path{doi:10.1007/JHEP06(2016)023}}.

\bibitem{Luna:2016hge}
A.~Luna, R.~Monteiro, I.~Nicholson, A.~Ochirov, D.~O'Connell, N.~Westerberg,
  C.~D. White, Perturbative spacetimes from yang--mills theory, JHEP 04 (2017)
  069.
\newblock \href {http://arxiv.org/abs/1611.07508} {\path{arXiv:1611.07508}},
  \href {https://doi.org/10.1007/JHEP04(2017)069}
  {\path{doi:10.1007/JHEP04(2017)069}}.

\bibitem{Ilderton:2018ldc}
A.~Ilderton, Screw-symmetric gravitational waves: A double copy of the vortex,
  Phys.\ Lett.\ B 782 (2018) 22--27.
\newblock \href {http://arxiv.org/abs/1804.07290} {\path{arXiv:1804.07290}},
  \href {https://doi.org/10.1016/j.physletb.2018.04.069}
  {\path{doi:10.1016/j.physletb.2018.04.069}}.

\bibitem{Gurses:2018ckx}
M.~G{\"u}rses, B.~Tekin, Classical double copy: Kerr--schild--kundt metrics
  from yang--mills theory, Phys.\ Rev.\ D 98~(12) (2018) 126017.
\newblock \href {http://arxiv.org/abs/1810.03411} {\path{arXiv:1810.03411}},
  \href {https://doi.org/10.1103/PhysRevD.98.126017}
  {\path{doi:10.1103/PhysRevD.98.126017}}.

\bibitem{Bah:2019sda}
I.~Bah, R.~Dempsey, P.~Weck, Kerr--schild double copy and complex worldlines,
  JHEP 02 (2020) 180.
\newblock \href {http://arxiv.org/abs/1910.04197} {\path{arXiv:1910.04197}},
  \href {https://doi.org/10.1007/JHEP02(2020)180}
  {\path{doi:10.1007/JHEP02(2020)180}}.

\bibitem{CarrilloGonzalez:2019pxo}
M.~Carrillo~Gonz{\'a}lez, B.~Melcher, K.~Ratliff, S.~Watson, C.~D. White, The
  classical double copy in three spacetime dimensions, JHEP 07 (2019) 167.
\newblock \href {http://arxiv.org/abs/1904.11001} {\path{arXiv:1904.11001}},
  \href {https://doi.org/10.1007/JHEP07(2019)167}
  {\path{doi:10.1007/JHEP07(2019)167}}.

\bibitem{BahjatAbbas:2020mrf}
N.~Bahjat-Abbas, R.~Stark-Much{\~a}o, C.~D. White, Monopoles, shockwaves and
  the classical double copy, JHEP 04 (2020) 102.
\newblock \href {http://arxiv.org/abs/2001.09918} {\path{arXiv:2001.09918}},
  \href {https://doi.org/10.1007/JHEP04(2020)102}
  {\path{doi:10.1007/JHEP04(2020)102}}.

\bibitem{Gumus:2020has}
M.~K. G{\"u}m{\"u}{\c s}, G.~Alka{\c c}, More on the classical double copy in
  three spacetime dimensions, Phys.\ Rev.\ D 102 (2020) 024074.
\newblock \href {http://arxiv.org/abs/2006.00552} {\path{arXiv:2006.00552}},
  \href {https://doi.org/10.1103/PhysRevD.102.024074}
  {\path{doi:10.1103/PhysRevD.102.024074}}.

\bibitem{Alkac:2021kut}
G.~Alka{\c c}, M.~K. G{\"u}m{\"u}{\c s}, M.~A. Olpak, Kerr--schild double copy
  of the coulomb solution in three dimensions, Phys.\ Rev.\ D 104 (2021)
  044034.
\newblock \href {http://arxiv.org/abs/2105.11550} {\path{arXiv:2105.11550}},
  \href {https://doi.org/10.1103/PhysRevD.104.044034}
  {\path{doi:10.1103/PhysRevD.104.044034}}.

\bibitem{Alkac:2022wkv}
G.~Alka{\c c}, M.~K. G{\"u}m{\"u}{\c s}, M.~A. Olpak, Generalized black holes
  in 3d kerr--schild double copy, Phys.\ Rev.\ D 106 (2022) 026013.
\newblock \href {http://arxiv.org/abs/2205.08503} {\path{arXiv:2205.08503}},
  \href {https://doi.org/10.1103/PhysRevD.106.026013}
  {\path{doi:10.1103/PhysRevD.106.026013}}.

\bibitem{Dempsey:2023gza}
R.~Dempsey, P.~Weck, Compactifying the kerr--schild double copy, JHEP 05 (2023)
  198.
\newblock \href {http://arxiv.org/abs/2211.14327} {\path{arXiv:2211.14327}},
  \href {https://doi.org/10.1007/JHEP05(2023)198}
  {\path{doi:10.1007/JHEP05(2023)198}}.

\bibitem{Easson:2023ldu}
D.~A. Easson, G.~T. Herczeg, T.~Manton, M.~Pezzelle, Isometries and the double
  copy, JHEP 12 (2023) 074.
\newblock \href {http://arxiv.org/abs/2306.13687} {\path{arXiv:2306.13687}},
  \href {https://doi.org/10.1007/JHEP12(2023)074}
  {\path{doi:10.1007/JHEP12(2023)074}}.

\bibitem{Ortaggio:2024abc}
M.~Ortaggio, V.~Pravda, A.~Pravdov{\'a}, Kerr--schild double copy for kundt
  spacetimes of any dimension, JHEP 02 (2024) 069.
\newblock \href {http://arxiv.org/abs/2312.00706} {\path{arXiv:2312.00706}},
  \href {https://doi.org/10.1007/JHEP02(2024)069}
  {\path{doi:10.1007/JHEP02(2024)069}}.

\bibitem{Ridgway:2015fdl}
A.~K. Ridgway, M.~B. Wise, {Static Spherically Symmetric Kerr-Schild Metrics
  and Implications for the Classical Double Copy}, Phys. Rev. D 94~(4) (2016)
  044023.
\newblock \href {http://arxiv.org/abs/1512.02243} {\path{arXiv:1512.02243}},
  \href {https://doi.org/10.1103/PhysRevD.94.044023}
  {\path{doi:10.1103/PhysRevD.94.044023}}.

\bibitem{Ceresole:2023wxg}
A.~Ceresole, T.~Damour, A.~Nagar, P.~Rettegno, {Double copy,
  Kerr\textendash{}Schild gauges and the effective-one-body formalism}, Class.
  Quant. Grav. 42~(2) (2025) 025004.
\newblock \href {http://arxiv.org/abs/2312.01478} {\path{arXiv:2312.01478}},
  \href {https://doi.org/10.1088/1361-6382/ad987f}
  {\path{doi:10.1088/1361-6382/ad987f}}.

\bibitem{Easson:2020esh}
D.~A. Easson, C.~Keeler, T.~Manton, {Classical double copy of nonsingular black
  holes}, Phys. Rev. D 102~(8) (2020) 086015.
\newblock \href {http://arxiv.org/abs/2007.16186} {\path{arXiv:2007.16186}},
  \href {https://doi.org/10.1103/PhysRevD.102.086015}
  {\path{doi:10.1103/PhysRevD.102.086015}}.

\bibitem{Chawla:2023bsu}
S.~Chawla, C.~Keeler, {Black hole horizons from the double copy}, Class. Quant.
  Grav. 40~(22) (2023) 225004.
\newblock \href {http://arxiv.org/abs/2306.02417} {\path{arXiv:2306.02417}},
  \href {https://doi.org/10.1088/1361-6382/acfe57}
  {\path{doi:10.1088/1361-6382/acfe57}}.

\bibitem{He:2023iew}
J.-L. He, J.-H. Huang, {Cosmological horizons from classical double copy},
  Phys. Lett. B 851 (2024) 138579.
\newblock \href {http://arxiv.org/abs/2312.00972} {\path{arXiv:2312.00972}},
  \href {https://doi.org/10.1016/j.physletb.2024.138579}
  {\path{doi:10.1016/j.physletb.2024.138579}}.

\bibitem{Chacon:2020fmr}
E.~Chac\'on, H.~Garc\'\i{}a-Compe\'an, A.~Luna, R.~Monteiro, C.~D. White, {New
  heavenly double copies}, JHEP 03 (2021) 247.
\newblock \href {http://arxiv.org/abs/2008.09603} {\path{arXiv:2008.09603}},
  \href {https://doi.org/10.1007/JHEP03(2021)247}
  {\path{doi:10.1007/JHEP03(2021)247}}.

\bibitem{Chacon:2025hh}
E.~Chac\'on, H.~Garc\'{\i}a-Compe\'an, G.~Robles, Hyper-hermitian weyl double
  copy, JHEP 05 (2025) 202.
\newblock \href {http://arxiv.org/abs/2410.21610} {\path{arXiv:2410.21610}},
  \href {https://doi.org/10.1007/JHEP05(2025)202}
  {\path{doi:10.1007/JHEP05(2025)202}}.

\bibitem{Alkac:2021bav}
G.~Alkac, M.~K. Gumus, M.~Tek, {The Kerr-Schild Double Copy in Lifshitz
  Spacetime}, JHEP 05 (2021) 214.
\newblock \href {http://arxiv.org/abs/2103.06986} {\path{arXiv:2103.06986}},
  \href {https://doi.org/10.1007/JHEP05(2021)214}
  {\path{doi:10.1007/JHEP05(2021)214}}.

\bibitem{Andrzejewski:2019hub}
K.~Andrzejewski, S.~Prencel, {From polarized gravitational waves to
  analytically solvable electromagnetic beams}, Phys. Rev. D 100~(4) (2019)
  045006.
\newblock \href {http://arxiv.org/abs/1901.05255} {\path{arXiv:1901.05255}},
  \href {https://doi.org/10.1103/PhysRevD.100.045006}
  {\path{doi:10.1103/PhysRevD.100.045006}}.

\bibitem{Bahjat-Abbas:2017htu}
N.~Bahjat-Abbas, A.~Luna, C.~D. White, {The Kerr-Schild double copy in curved
  spacetime}, JHEP 12 (2017) 004.
\newblock \href {http://arxiv.org/abs/1710.01953} {\path{arXiv:1710.01953}},
  \href {https://doi.org/10.1007/JHEP12(2017)004}
  {\path{doi:10.1007/JHEP12(2017)004}}.

\bibitem{Carrillo-Gonzalez:2017iyj}
M.~Carrillo-Gonz\'alez, R.~Penco, M.~Trodden, {The classical double copy in
  maximally symmetric spacetimes}, JHEP 04 (2018) 028.
\newblock \href {http://arxiv.org/abs/1711.01296} {\path{arXiv:1711.01296}},
  \href {https://doi.org/10.1007/JHEP04(2018)028}
  {\path{doi:10.1007/JHEP04(2018)028}}.

\bibitem{Stephani:2003tm}
H.~Stephani, D.~Kramer, M.~A.~H. MacCallum, C.~Hoenselaers, E.~Herlt, {Exact
  solutions of Einstein's field equations}, Cambridge Monographs on
  Mathematical Physics, Cambridge Univ. Press, Cambridge, 2003.
\newblock \href {https://doi.org/10.1017/CBO9780511535185}
  {\path{doi:10.1017/CBO9780511535185}}.

\bibitem{Garcia-Compean:2024uie}
H.~Garc\'\i{}a-Compe\'an, C.~Ramos, {Classical Kerr-Schild double copy in
  bigravity for maximally symmetric spacetimes}, JHEP 07 (2024) 074.
\newblock \href {http://arxiv.org/abs/2403.19608} {\path{arXiv:2403.19608}},
  \href {https://doi.org/10.1007/JHEP07(2024)074}
  {\path{doi:10.1007/JHEP07(2024)074}}.

\bibitem{Garcia-Compean:2024zze}
H.~Garc\'\i{}a-Compe\'an, C.~I. Ramos, {Pleba\'nski-Demia\'nski solutions in
  bigravity and Kerr-Schild double copy relations using an effective metric}
  (12 2024).
\newblock \href {http://arxiv.org/abs/2412.17191} {\path{arXiv:2412.17191}}.

\bibitem{Gonzalez:2021ztm}
M.~C. Gonz\'alez, A.~Momeni, J.~Rumbutis, {Massive double copy in the
  high-energy limit}, JHEP 04 (2022) 094.
\newblock \href {http://arxiv.org/abs/2112.08401} {\path{arXiv:2112.08401}},
  \href {https://doi.org/10.1007/JHEP04(2022)094}
  {\path{doi:10.1007/JHEP04(2022)094}}.

\bibitem{Fronsdal:1978rb}
C.~Fronsdal, {Massless Fields with Integer Spin}, Phys. Rev. D 18 (1978) 3624.
\newblock \href {https://doi.org/10.1103/PhysRevD.18.3624}
  {\path{doi:10.1103/PhysRevD.18.3624}}.

\bibitem{Didenko:2022qxq}
V.~E. Didenko, N.~K. Dosmanbetov, {Classical Double Copy and Higher-Spin
  Fields}, Phys. Rev. Lett. 130~(7) (2023) 071603.
\newblock \href {http://arxiv.org/abs/2210.04704} {\path{arXiv:2210.04704}},
  \href {https://doi.org/10.1103/PhysRevLett.130.071603}
  {\path{doi:10.1103/PhysRevLett.130.071603}}.

\bibitem{Beetar:2024ptv}
C.~Beetar, M.~Carrillo~Gonz\'alez, S.~Jaitly, T.~Keseman, {Double Copy in AdS3
  from Minitwistor Space} (10 2024).
\newblock \href {http://arxiv.org/abs/2410.23342} {\path{arXiv:2410.23342}}.

\bibitem{Holton:2026eha}
B.~P. Holton, {Residual Symmetries and Their Algebras in the Kerr-Schild Double
  Copy} (4 2026).
\newblock \href {http://arxiv.org/abs/2604.05262} {\path{arXiv:2604.05262}}.

\bibitem{Gibbons:2004uw}
G.~W. Gibbons, H.~Lu, D.~N. Page, C.~N. Pope, {The General Kerr-de Sitter
  metrics in all dimensions}, J. Geom. Phys. 53 (2005) 49--73.
\newblock \href {http://arxiv.org/abs/hep-th/0404008}
  {\path{arXiv:hep-th/0404008}}, \href
  {https://doi.org/10.1016/j.geomphys.2004.05.001}
  {\path{doi:10.1016/j.geomphys.2004.05.001}}.

\bibitem{Israel:1970kp}
W.~Israel, {Source of the kerr metric}, Phys. Rev. D 2 (1970) 641--646.
\newblock \href {https://doi.org/10.1103/PhysRevD.2.641}
  {\path{doi:10.1103/PhysRevD.2.641}}.

\bibitem{Arkani-Hamed:2019ymq}
N.~Arkani-Hamed, Y.-t. Huang, D.~O'Connell, {Kerr black holes as elementary
  particles}, JHEP 01 (2020) 046.
\newblock \href {http://arxiv.org/abs/1906.10100} {\path{arXiv:1906.10100}},
  \href {https://doi.org/10.1007/JHEP01(2020)046}
  {\path{doi:10.1007/JHEP01(2020)046}}.

\bibitem{Ortaggio:2023cdz}
M.~Ortaggio, V.~Pravda, A.~Pravdova, {Kerr-Schild double copy for Kundt
  spacetimes of any dimension}, JHEP 2024~(02) (2024) 069.
\newblock \href {http://arxiv.org/abs/2312.00706} {\path{arXiv:2312.00706}},
  \href {https://doi.org/10.1007/JHEP02(2024)069}
  {\path{doi:10.1007/JHEP02(2024)069}}.

\bibitem{Ortaggio:2023rzp}
M.~Ortaggio, A.~Srinivasan, {Charging Kerr-Schild spacetimes in higher
  dimensions}, Phys. Rev. D 110~(4) (2024) 044035.
\newblock \href {http://arxiv.org/abs/2309.02900} {\path{arXiv:2309.02900}},
  \href {https://doi.org/10.1103/PhysRevD.110.044035}
  {\path{doi:10.1103/PhysRevD.110.044035}}.

\bibitem{Hassaine:2024mfs}
M.~Hassaine, D.~Kubiznak, A.~Srinivasan, {Extremal Kerr-Schild Form} (11 2024).
\newblock \href {http://arxiv.org/abs/2411.17805} {\path{arXiv:2411.17805}}.

\bibitem{Srinivasan:2025hro}
A.~Srinivasan, {Generalized Kerr-Schild spacetimes in arbitrary dimensions} (1
  2025).
\newblock \href {http://arxiv.org/abs/2501.00847} {\path{arXiv:2501.00847}}.

\bibitem{Alencar:2026zdz}
G.~Alencar, C.~R. Muniz, M.~S. Oliveira, {Classical double copy of black
  strings in an Anti-de Sitter background} (1 2026).
\newblock \href {http://arxiv.org/abs/2601.22383} {\path{arXiv:2601.22383}}.

\bibitem{Rodriguez:2026zxm}
J.~A. Rodr{\'\i}guez, {Kerr-Schild Double Copy of the Randall-Sundrum Black
  String} (4 2026).
\newblock \href {http://arxiv.org/abs/2604.05447} {\path{arXiv:2604.05447}}.

\bibitem{Luna:2018dpt}
A.~Luna, R.~Monteiro, I.~Nicholson, D.~O'Connell, {Type D Spacetimes and the
  Weyl Double Copy}, Class. Quant. Grav. 36 (2019) 065003.
\newblock \href {http://arxiv.org/abs/1810.08183} {\path{arXiv:1810.08183}},
  \href {https://doi.org/10.1088/1361-6382/ab03e6}
  {\path{doi:10.1088/1361-6382/ab03e6}}.

\bibitem{Godazgar:2020zbv}
H.~Godazgar, M.~Godazgar, R.~Monteiro, D.~Peinador~Veiga, C.~N. Pope, {Weyl
  Double Copy for Gravitational Waves}, Phys. Rev. Lett. 126~(10) (2021)
  101103.
\newblock \href {http://arxiv.org/abs/2010.02925} {\path{arXiv:2010.02925}},
  \href {https://doi.org/10.1103/PhysRevLett.126.101103}
  {\path{doi:10.1103/PhysRevLett.126.101103}}.

\bibitem{Alawadhi:2019urr}
R.~Alawadhi, D.~S. Berman, B.~Spence, D.~Peinador~Veiga, {S-duality and the
  double copy}, JHEP 03 (2020) 059.
\newblock \href {http://arxiv.org/abs/1911.06797} {\path{arXiv:1911.06797}},
  \href {https://doi.org/10.1007/JHEP03(2020)059}
  {\path{doi:10.1007/JHEP03(2020)059}}.

\bibitem{Alawadhi:2020jrv}
R.~Alawadhi, D.~S. Berman, B.~Spence, {Weyl doubling}, JHEP 09 (2020) 127.
\newblock \href {http://arxiv.org/abs/2007.03264} {\path{arXiv:2007.03264}},
  \href {https://doi.org/10.1007/JHEP09(2020)127}
  {\path{doi:10.1007/JHEP09(2020)127}}.

\bibitem{ArmstrongWilliams:2025src}
K.~Armstrong\textendash~Williams, N.~Moynihan, C.~D. White, Deriving {W}eyl
  {D}ouble {C}opies with sources, JHEP 03 (2025) 121.
\newblock \href {http://arxiv.org/abs/2407.18107} {\path{arXiv:2407.18107}},
  \href {https://doi.org/10.1007/JHEP03(2025)121}
  {\path{doi:10.1007/JHEP03(2025)121}}.

\bibitem{Godazgar:2021iae}
H.~Godazgar, M.~Godazgar, R.~Monteiro, D.~Peinador~Veiga, C.~N. Pope,
  {Asymptotic Weyl double copy}, JHEP 11 (2021) 126.
\newblock \href {http://arxiv.org/abs/2109.07866} {\path{arXiv:2109.07866}},
  \href {https://doi.org/10.1007/JHEP11(2021)126}
  {\path{doi:10.1007/JHEP11(2021)126}}.

\bibitem{Chawla:2024mse}
S.~Chawla, K.~Fransen, C.~Keeler, {The Penrose limit of the Weyl double copy},
  Class. Quant. Grav. 41~(24) (2024) 245015.
\newblock \href {http://arxiv.org/abs/2406.14601} {\path{arXiv:2406.14601}},
  \href {https://doi.org/10.1088/1361-6382/ad8f8c}
  {\path{doi:10.1088/1361-6382/ad8f8c}}.

\bibitem{Liu:2024byr}
Y.-R. Liu, J.-R. Zhang, Y.-L. Zhang, {Slowly rotating charges from Weyl double
  copy for Kerr black hole with Chern\textendash{}Simons correction}, Commun.
  Theor. Phys. 76~(8) (2024) 085405.
\newblock \href {http://arxiv.org/abs/2404.07888} {\path{arXiv:2404.07888}},
  \href {https://doi.org/10.1088/1572-9494/ad4a37}
  {\path{doi:10.1088/1572-9494/ad4a37}}.

\bibitem{Sabharwal:2019ngs}
S.~Sabharwal, J.~W. Dalhuisen, {Anti-Self-Dual Spacetimes, Gravitational
  Instantons and Knotted Zeros of the Weyl Tensor}, JHEP 07 (2019) 004.
\newblock \href {http://arxiv.org/abs/1904.06030} {\path{arXiv:1904.06030}},
  \href {https://doi.org/10.1007/JHEP07(2019)004}
  {\path{doi:10.1007/JHEP07(2019)004}}.

\bibitem{Monteiro:2020plf}
R.~Monteiro, D.~O'Connell, D.~Peinador~Veiga, M.~Sergola, {Classical solutions
  and their double copy in split signature}, JHEP 05 (2021) 268.
\newblock \href {http://arxiv.org/abs/2012.11190} {\path{arXiv:2012.11190}},
  \href {https://doi.org/10.1007/JHEP05(2021)268}
  {\path{doi:10.1007/JHEP05(2021)268}}.

\bibitem{Easson:2021asd}
D.~A. Easson, T.~Manton, A.~Svesko, {Sources in the Weyl Double Copy}, Phys.
  Rev. Lett. 127~(27) (2021) 271101.
\newblock \href {http://arxiv.org/abs/2110.02293} {\path{arXiv:2110.02293}},
  \href {https://doi.org/10.1103/PhysRevLett.127.271101}
  {\path{doi:10.1103/PhysRevLett.127.271101}}.

\bibitem{Keeler:2020rcv}
C.~Keeler, T.~Manton, N.~Monga, {From Navier-Stokes to Maxwell via Einstein},
  JHEP 08 (2020) 147.
\newblock \href {http://arxiv.org/abs/2005.04242} {\path{arXiv:2005.04242}},
  \href {https://doi.org/10.1007/JHEP08(2020)147}
  {\path{doi:10.1007/JHEP08(2020)147}}.

\bibitem{Chawla:2025uwu}
S.~Chawla, K.~Fransen, C.~Keeler, {Expanding on the double copy in null Fermi
  coordinates} (4 2025).
\newblock \href {http://arxiv.org/abs/2504.07224} {\path{arXiv:2504.07224}}.

\bibitem{Keeler:2024bdt}
C.~Keeler, N.~Monga, {On type-II Spacetimes and the Double Copy for Fluids
  Metrics} (4 2024).
\newblock \href {http://arxiv.org/abs/2404.03195} {\path{arXiv:2404.03195}}.

\bibitem{Monteiro:2021ztt}
R.~Monteiro, S.~Nagy, D.~O'Connell, D.~Peinador~Veiga, M.~Sergola, {NS-NS
  spacetimes from amplitudes}, JHEP 06 (2022) 021.
\newblock \href {http://arxiv.org/abs/2112.08336} {\path{arXiv:2112.08336}},
  \href {https://doi.org/10.1007/JHEP06(2022)021}
  {\path{doi:10.1007/JHEP06(2022)021}}.

\bibitem{Han:2022ubu}
S.~Han, {Weyl double copy and massless free-fields in curved spacetimes},
  Class. Quant. Grav. 39~(22) (2022) 225009.
\newblock \href {http://arxiv.org/abs/2204.01907} {\path{arXiv:2204.01907}},
  \href {https://doi.org/10.1088/1361-6382/ac96c2}
  {\path{doi:10.1088/1361-6382/ac96c2}}.

\bibitem{Han:2022wdc}
S.~Han, The weyl double copy in vacuum spacetimes with a cosmological constant,
  JHEP 09 (2022) 238.
\newblock \href {http://arxiv.org/abs/2205.08654} {\path{arXiv:2205.08654}},
  \href {https://doi.org/10.1007/JHEP09(2022)238}
  {\path{doi:10.1007/JHEP09(2022)238}}.

\bibitem{Alkac:2024rwdc}
G.~Alka{\c c}, M.~K. G{\"u}m{\"u}{\c s}, O.~Ka{\c s}{\i}kc{\i}, M.~A. Olpak,
  M.~Tek, Regularized weyl double copy, Phys.\ Rev.\ D 109 (2024) 084047.
\newblock \href {http://arxiv.org/abs/2310.06048} {\path{arXiv:2310.06048}},
  \href {https://doi.org/10.1103/PhysRevD.109.084047}
  {\path{doi:10.1103/PhysRevD.109.084047}}.

\bibitem{Zhao:2025fiveD}
W.~Zhao, P.-J. Mao, J.-B. Wu, Five dimensional weyl double copy, Phys.\ Rev.\ D
  111 (2025) L081902.
\newblock \href {http://arxiv.org/abs/2409.06786} {\path{arXiv:2409.06786}},
  \href {https://doi.org/10.1103/PhysRevD.111.L081902}
  {\path{doi:10.1103/PhysRevD.111.L081902}}.

\bibitem{Zhao:2025typeD}
W.~Zhao, P.-J. Mao, J.-B. Wu, Weyl double copy in type d spacetime in four and
  five dimensions, Phys.\ Rev.\ D 111 (2025) 066005.
\newblock \href {http://arxiv.org/abs/2411.04774} {\path{arXiv:2411.04774}},
  \href {https://doi.org/10.1103/PhysRevD.111.066005}
  {\path{doi:10.1103/PhysRevD.111.066005}}.

\bibitem{Kent:2025godel}
B.~Kent, T.~Manton, S.~Shashi, Background ambiguity and the g\"odel double
  copy, JHEP 03 (2025) 033.
\newblock \href {http://arxiv.org/abs/2411.04207} {\path{arXiv:2411.04207}},
  \href {https://doi.org/10.1007/JHEP03(2025)033}
  {\path{doi:10.1007/JHEP03(2025)033}}.

\bibitem{Chawla:2023aligned}
S.~Chawla, C.~Keeler, Aligned fields double copy to kerr--nut--(a)ds, JHEP 04
  (2023) 005.
\newblock \href {http://arxiv.org/abs/2209.09275} {\path{arXiv:2209.09275}},
  \href {https://doi.org/10.1007/JHEP04(2023)005}
  {\path{doi:10.1007/JHEP04(2023)005}}.

\bibitem{Alkac:2025iyw}
G.~Alkac, M.~K. Gumus, M.~A. Olpak, {Weyl double copy in Lifshitz spacetimes},
  Phys. Rev. D 113~(8) (2026) 084053.
\newblock \href {http://arxiv.org/abs/2511.00632} {\path{arXiv:2511.00632}},
  \href {https://doi.org/10.1103/wdcz-mrhp} {\path{doi:10.1103/wdcz-mrhp}}.

\bibitem{Zhao:2024wtn}
W.~Zhao, P.-J. Mao, J.-B. Wu, {Weyl double copy in type D spacetime in four and
  five dimensions} (11 2024).
\newblock \href {http://arxiv.org/abs/2411.04774} {\path{arXiv:2411.04774}}.

\bibitem{Zhao:2024ljb}
W.~Zhao, P.-J. Mao, J.-B. Wu, {Five dimensional Weyl double copy} (9 2024).
\newblock \href {http://arxiv.org/abs/2409.06786} {\path{arXiv:2409.06786}}.

\bibitem{Emond:2020lwi}
W.~T. Emond, Y.-T. Huang, U.~Kol, N.~Moynihan, D.~O'Connell, {Amplitudes from
  Coulomb to Kerr-Taub-NUT}, JHEP 05 (2022) 055.
\newblock \href {http://arxiv.org/abs/2010.07861} {\path{arXiv:2010.07861}},
  \href {https://doi.org/10.1007/JHEP05(2022)055}
  {\path{doi:10.1007/JHEP05(2022)055}}.

\bibitem{CarrilloGonzalez:2022mxx}
M.~Carrillo~Gonz\'alez, A.~Momeni, J.~Rumbutis, {Cotton double copy for
  gravitational waves}, Phys. Rev. D 106~(2) (2022) 025006.
\newblock \href {http://arxiv.org/abs/2202.10476} {\path{arXiv:2202.10476}},
  \href {https://doi.org/10.1103/PhysRevD.106.025006}
  {\path{doi:10.1103/PhysRevD.106.025006}}.

\bibitem{Emond:2022uaf}
W.~T. Emond, N.~Moynihan, {Scattering amplitudes and the Cotton double copy},
  JHEP 07 (2024) 009.
\newblock \href {http://arxiv.org/abs/2202.10499} {\path{arXiv:2202.10499}},
  \href {https://doi.org/10.1007/JHEP07(2024)009}
  {\path{doi:10.1007/JHEP07(2024)009}}.

\bibitem{CarrilloGonzalez:2022ggn}
M.~Carrillo~Gonz\'alez, W.~T. Emond, N.~Moynihan, J.~Rumbutis, C.~D. White,
  {Mini-twistors and the Cotton double copy}, JHEP 03 (2023) 177.
\newblock \href {http://arxiv.org/abs/2212.04783} {\path{arXiv:2212.04783}},
  \href {https://doi.org/10.1007/JHEP03(2023)177}
  {\path{doi:10.1007/JHEP03(2023)177}}.

\bibitem{Deser:1981wh}
S.~Deser, R.~Jackiw, S.~Templeton, {Topologically Massive Gauge Theories},
  Annals Phys. 140 (1982) 372--411, [Erratum: Annals Phys. 185, 406 (1988)].
\newblock \href {https://doi.org/10.1016/0003-4916(82)90164-6}
  {\path{doi:10.1016/0003-4916(82)90164-6}}.

\bibitem{Milson:2012ry}
R.~Milson, L.~Wylleman, {Three-dimensional spacetimes of maximal order}, Class.
  Quant. Grav. 30 (2013) 095004.
\newblock \href {http://arxiv.org/abs/1210.6920} {\path{arXiv:1210.6920}},
  \href {https://doi.org/10.1088/0264-9381/30/9/095004}
  {\path{doi:10.1088/0264-9381/30/9/095004}}.

\bibitem{castillo20033}
G.~Castillo, \href{https://books.google.co.uk/books?id=kgBAAQAAIAAJ}{3-D
  Spinors, Spin-Weighted Functions and their Applications}, Progress in
  Mathematical Physics, Birkh{\"a}user Boston, 2003.
\newline\urlprefix\url{https://books.google.co.uk/books?id=kgBAAQAAIAAJ}

\bibitem{doi:10.1063/1.1592611}
G.~F. Torres~del Castillo, L.~F. Gómez-Ceballos,
  \href{https://aip.scitation.org/doi/abs/10.1063/1.1592611}{Algebraic
  classification of the curvature of three-dimensional manifolds with
  indefinite metric}, Journal of Mathematical Physics 44~(9) (2003) 4374--4380.
\newblock \href
  {http://arxiv.org/abs/https://aip.scitation.org/doi/pdf/10.1063/1.1592611}
  {\path{arXiv:https://aip.scitation.org/doi/pdf/10.1063/1.1592611}}, \href
  {https://doi.org/10.1063/1.1592611} {\path{doi:10.1063/1.1592611}}.
\newline\urlprefix\url{https://aip.scitation.org/doi/abs/10.1063/1.1592611}

\bibitem{Gibbons:2008vi}
G.~W. Gibbons, C.~N. Pope, E.~Sezgin, {The General Supersymmetric Solution of
  Topologically Massive Supergravity}, Class. Quant. Grav. 25 (2008) 205005.
\newblock \href {http://arxiv.org/abs/0807.2613} {\path{arXiv:0807.2613}},
  \href {https://doi.org/10.1088/0264-9381/25/20/205005}
  {\path{doi:10.1088/0264-9381/25/20/205005}}.

\bibitem{Chow:2009km}
D.~D.~K. Chow, C.~N. Pope, E.~Sezgin, {Classification of solutions in
  topologically massive gravity}, Class. Quant. Grav. 27 (2010) 105001.
\newblock \href {http://arxiv.org/abs/0906.3559} {\path{arXiv:0906.3559}},
  \href {https://doi.org/10.1088/0264-9381/27/10/105001}
  {\path{doi:10.1088/0264-9381/27/10/105001}}.

\bibitem{Anninos:2008fx}
D.~Anninos, W.~Li, M.~Padi, W.~Song, A.~Strominger, {Warped AdS(3) Black
  Holes}, JHEP 03 (2009) 130.
\newblock \href {http://arxiv.org/abs/0807.3040} {\path{arXiv:0807.3040}},
  \href {https://doi.org/10.1088/1126-6708/2009/03/130}
  {\path{doi:10.1088/1126-6708/2009/03/130}}.

\bibitem{White:2020sfn}
C.~D. White, {Twistorial Foundation for the Classical Double Copy}, Phys. Rev.
  Lett. 126~(6) (2021) 061602.
\newblock \href {http://arxiv.org/abs/2012.02479} {\path{arXiv:2012.02479}},
  \href {https://doi.org/10.1103/PhysRevLett.126.061602}
  {\path{doi:10.1103/PhysRevLett.126.061602}}.

\bibitem{Chacon:2021wbr}
E.~Chac\'on, S.~Nagy, C.~D. White, {The Weyl double copy from twistor space},
  JHEP 05 (2021) 2239.
\newblock \href {http://arxiv.org/abs/2103.16441} {\path{arXiv:2103.16441}},
  \href {https://doi.org/10.1007/JHEP05(2021)239}
  {\path{doi:10.1007/JHEP05(2021)239}}.

\bibitem{Chacon:2021hfe}
E.~Chac\'on, A.~Luna, C.~D. White, {Double copy of the multipole expansion},
  Phys. Rev. D 106~(8) (2022) 086020.
\newblock \href {http://arxiv.org/abs/2108.07702} {\path{arXiv:2108.07702}},
  \href {https://doi.org/10.1103/PhysRevD.106.086020}
  {\path{doi:10.1103/PhysRevD.106.086020}}.

\bibitem{Chacon:2021lox}
E.~Chac\'on, S.~Nagy, C.~D. White, {Alternative formulations of the twistor
  double copy}, JHEP 03 (2022) 180.
\newblock \href {http://arxiv.org/abs/2112.06764} {\path{arXiv:2112.06764}},
  \href {https://doi.org/10.1007/JHEP03(2022)180}
  {\path{doi:10.1007/JHEP03(2022)180}}.

\bibitem{Adamo:2021dfg}
T.~Adamo, U.~Kol, {Classical double copy at null infinity}, Class. Quant. Grav.
  39~(10) (2022) 105007.
\newblock \href {http://arxiv.org/abs/2109.07832} {\path{arXiv:2109.07832}},
  \href {https://doi.org/10.1088/1361-6382/ac635e}
  {\path{doi:10.1088/1361-6382/ac635e}}.

\bibitem{Armstrong-Williams:2023ssz}
K.~Armstrong-Williams, C.~D. White, {A spinorial double copy for $ \mathcal{N}
  $ = 0 supergravity}, JHEP 05 (2023) 047.
\newblock \href {http://arxiv.org/abs/2303.04631} {\path{arXiv:2303.04631}},
  \href {https://doi.org/10.1007/JHEP05(2023)047}
  {\path{doi:10.1007/JHEP05(2023)047}}.

\bibitem{Guevara:2021yud}
A.~Guevara, {Reconstructing Classical Spacetimes from the S-Matrix in Twistor
  Space} (12 2021).
\newblock \href {http://arxiv.org/abs/2112.05111} {\path{arXiv:2112.05111}}.

\bibitem{Luna:2022dxo}
A.~Luna, N.~Moynihan, C.~D. White, {Why is the Weyl double copy local in
  position space?}, JHEP 12 (2022) 046.
\newblock \href {http://arxiv.org/abs/2208.08548} {\path{arXiv:2208.08548}},
  \href {https://doi.org/10.1007/JHEP12(2022)046}
  {\path{doi:10.1007/JHEP12(2022)046}}.

\bibitem{tsai}
C.~C. Tsai, The penrose transform for einstein-weyl and related spaces, Phd
  thesis, University of Edinburgh, Edinburgh, UK, available at
  \url{https://era.ed.ac.uk/bitstream/handle/1842/14584/Tsai1996.Pdf?sequence=1}
  (June 1996).

\bibitem{Seet:2024vmh}
S.~Seet, {Twistor Space and Celestial Holography}, Ph.D. thesis, Cambridge U.,
  DAMTP (2024).
\newblock \href {https://doi.org/10.17863/CAM.112793}
  {\path{doi:10.17863/CAM.112793}}.

\bibitem{alma990162587480107026}
P.~E. Jones, R.~Penrose, Minitwistors (1984).

\bibitem{Kessel:2018zqm}
P.~Kessel, J.~Raeymaekers, {Simple unfolded equations for massive higher spins
  in AdS$_{3}$}, JHEP 08 (2018) 076.
\newblock \href {http://arxiv.org/abs/1805.07279} {\path{arXiv:1805.07279}},
  \href {https://doi.org/10.1007/JHEP08(2018)076}
  {\path{doi:10.1007/JHEP08(2018)076}}.

\bibitem{Anastasiou:2018rdx}
A.~Anastasiou, L.~Borsten, M.~J. Duff, S.~Nagy, M.~Zoccali, {Gravity as Gauge
  Theory Squared: A Ghost Story}, Phys. Rev. Lett. 121~(21) (2018) 211601.
\newblock \href {http://arxiv.org/abs/1807.02486} {\path{arXiv:1807.02486}},
  \href {https://doi.org/10.1103/PhysRevLett.121.211601}
  {\path{doi:10.1103/PhysRevLett.121.211601}}.

\bibitem{Anastasiou:2014qba}
A.~Anastasiou, L.~Borsten, M.~J. Duff, L.~J. Hughes, S.~Nagy, {Yang-Mills
  origin of gravitational symmetries}, Phys. Rev. Lett. 113~(23) (2014) 231606.
\newblock \href {http://arxiv.org/abs/1408.4434} {\path{arXiv:1408.4434}},
  \href {https://doi.org/10.1103/PhysRevLett.113.231606}
  {\path{doi:10.1103/PhysRevLett.113.231606}}.

\bibitem{Godazgar:2022gfw}
M.~Godazgar, C.~N. Pope, A.~Saha, H.~Zhang, {BRST symmetry and the
  convolutional double copy}, JHEP 11 (2022) 038.
\newblock \href {http://arxiv.org/abs/2208.06903} {\path{arXiv:2208.06903}},
  \href {https://doi.org/10.1007/JHEP11(2022)038}
  {\path{doi:10.1007/JHEP11(2022)038}}.

\bibitem{Beneke:2021ilf}
M.~Beneke, P.~Hager, A.~F. Sanfilippo, {Double copy for Lagrangians at
  trilinear order}, JHEP 02 (2022) 083.
\newblock \href {http://arxiv.org/abs/2106.09054} {\path{arXiv:2106.09054}},
  \href {https://doi.org/10.1007/JHEP02(2022)083}
  {\path{doi:10.1007/JHEP02(2022)083}}.

\bibitem{Ferrero:2020vww}
P.~Ferrero, D.~Francia, {On the Lagrangian formulation of the double copy to
  cubic order}, JHEP 02 (2021) 213.
\newblock \href {http://arxiv.org/abs/2012.00713} {\path{arXiv:2012.00713}},
  \href {https://doi.org/10.1007/JHEP02(2021)213}
  {\path{doi:10.1007/JHEP02(2021)213}}.

\bibitem{Borsten:2020xbt}
L.~Borsten, S.~Nagy, {The pure BRST Einstein-Hilbert Lagrangian from the
  double-copy to cubic order}, JHEP 07 (2020) 093.
\newblock \href {http://arxiv.org/abs/2004.14945} {\path{arXiv:2004.14945}},
  \href {https://doi.org/10.1007/JHEP07(2020)093}
  {\path{doi:10.1007/JHEP07(2020)093}}.

\bibitem{Luna:2020adi}
A.~Luna, S.~Nagy, C.~White, {The convolutional double copy: a case study with a
  point}, JHEP 09 (2020) 062.
\newblock \href {http://arxiv.org/abs/2004.11254} {\path{arXiv:2004.11254}},
  \href {https://doi.org/10.1007/JHEP09(2020)062}
  {\path{doi:10.1007/JHEP09(2020)062}}.

\bibitem{LopesCardoso:2018xes}
G.~Lopes~Cardoso, G.~Inverso, S.~Nagy, S.~Nampuri, {Comments on the double copy
  construction for gravitational theories}, PoS CORFU2017 (2018) 177.
\newblock \href {http://arxiv.org/abs/1803.07670} {\path{arXiv:1803.07670}},
  \href {https://doi.org/10.22323/1.318.0177} {\path{doi:10.22323/1.318.0177}}.

\bibitem{Liang:2023zxo}
Q.~Liang, S.~Nagy, {Convolutional double copy in (anti) de Sitter space}, JHEP
  04 (2024) 139.
\newblock \href {http://arxiv.org/abs/2311.14319} {\path{arXiv:2311.14319}},
  \href {https://doi.org/10.1007/JHEP04(2024)139}
  {\path{doi:10.1007/JHEP04(2024)139}}.

\bibitem{Borsten:2019prq}
L.~Borsten, I.~Jubb, V.~Makwana, S.~Nagy, {Gauge {\texttimes} gauge on
  spheres}, JHEP 06 (2020) 096.
\newblock \href {http://arxiv.org/abs/1911.12324} {\path{arXiv:1911.12324}},
  \href {https://doi.org/10.1007/JHEP06(2020)096}
  {\path{doi:10.1007/JHEP06(2020)096}}.

\bibitem{Borsten:2021zir}
L.~Borsten, I.~Jubb, V.~Makwana, S.~Nagy, {Gauge {\texttimes} gauge = gravity
  on homogeneous spaces using tensor convolutions}, JHEP 06 (2021) 117.
\newblock \href {http://arxiv.org/abs/2104.01135} {\path{arXiv:2104.01135}},
  \href {https://doi.org/10.1007/JHEP06(2021)117}
  {\path{doi:10.1007/JHEP06(2021)117}}.

\bibitem{Pasarin:2020qoa}
O.~Pasarin, A.~A. Tseytlin, {Generalised Schwarzschild metric from double copy
  of point-like charge solution in Born-Infeld theory}, Phys. Lett. B 807
  (2020) 135594.
\newblock \href {http://arxiv.org/abs/2005.12396} {\path{arXiv:2005.12396}},
  \href {https://doi.org/10.1016/j.physletb.2020.135594}
  {\path{doi:10.1016/j.physletb.2020.135594}}.

\bibitem{Mkrtchyan:2022ulc}
K.~Mkrtchyan, M.~Svazas, {Solutions in Nonlinear Electrodynamics and their
  double copy regular black holes}, JHEP 09 (2022) 012.
\newblock \href {http://arxiv.org/abs/2205.14187} {\path{arXiv:2205.14187}},
  \href {https://doi.org/10.1007/JHEP09(2022)012}
  {\path{doi:10.1007/JHEP09(2022)012}}.

\bibitem{Lescano:2021ooe}
E.~Lescano, J.~A. Rodr{\'\i}guez, {Higher-derivative heterotic Double Field
  Theory and classical double copy}, JHEP 07 (2021) 072.
\newblock \href {http://arxiv.org/abs/2101.03376} {\path{arXiv:2101.03376}},
  \href {https://doi.org/10.1007/JHEP07(2021)072}
  {\path{doi:10.1007/JHEP07(2021)072}}.

\bibitem{Armstrong-Williams:2026wzi}
K.~Armstrong-Williams, {Towards a Non-Perturbative Classical Double Copy} (8
  2026).
\newblock \href {http://arxiv.org/abs/2608.14512} {\path{arXiv:2608.14512}}.

\bibitem{Lee:2018gxc}
K.~Lee, {Kerr-Schild Double Field Theory and Classical Double Copy}, JHEP 10
  (2018) 027.
\newblock \href {http://arxiv.org/abs/1807.08443} {\path{arXiv:1807.08443}},
  \href {https://doi.org/10.1007/JHEP10(2018)027}
  {\path{doi:10.1007/JHEP10(2018)027}}.

\bibitem{Angus:2021zhy}
S.~Angus, K.~Cho, K.~Lee, {The classical double copy for half-maximal
  supergravities and T-duality}, JHEP 10 (2021) 211.
\newblock \href {http://arxiv.org/abs/2105.12857} {\path{arXiv:2105.12857}},
  \href {https://doi.org/10.1007/JHEP10(2021)211}
  {\path{doi:10.1007/JHEP10(2021)211}}.

\bibitem{Cho:2021nim}
K.~Cho, K.~Kim, K.~Lee, {The off-shell recursion for gravity and the classical
  double copy for currents}, JHEP 01 (2022) 186.
\newblock \href {http://arxiv.org/abs/2109.06392} {\path{arXiv:2109.06392}},
  \href {https://doi.org/10.1007/JHEP01(2022)186}
  {\path{doi:10.1007/JHEP01(2022)186}}.

\bibitem{Berman:2020xvs}
D.~S. Berman, K.~Kim, K.~Lee, {The classical double copy for M-theory from a
  Kerr-Schild ansatz for exceptional field theory}, JHEP 04 (2021) 071.
\newblock \href {http://arxiv.org/abs/2010.08255} {\path{arXiv:2010.08255}},
  \href {https://doi.org/10.1007/JHEP04(2021)071}
  {\path{doi:10.1007/JHEP04(2021)071}}.

\bibitem{Kim:2019jwm}
K.~Kim, K.~Lee, R.~Monteiro, I.~Nicholson, D.~Peinador~Veiga, {The Classical
  Double Copy of a Point Charge}, JHEP 02 (2020) 046.
\newblock \href {http://arxiv.org/abs/1912.02177} {\path{arXiv:1912.02177}},
  \href {https://doi.org/10.1007/JHEP02(2020)046}
  {\path{doi:10.1007/JHEP02(2020)046}}.

\bibitem{Cho:2019ype}
W.~Cho, K.~Lee, {Heterotic Kerr-Schild Double Field Theory and Classical Double
  Copy}, JHEP 07 (2019) 030.
\newblock \href {http://arxiv.org/abs/1904.11650} {\path{arXiv:1904.11650}},
  \href {https://doi.org/10.1007/JHEP07(2019)030}
  {\path{doi:10.1007/JHEP07(2019)030}}.

\bibitem{Lescano:2020nve}
E.~Lescano, J.~A. Rodr{\'\i}guez, {$ \mathcal{N} $ = 1 supersymmetric Double
  Field Theory and the generalized Kerr-Schild ansatz}, JHEP 10 (2020) 148.
\newblock \href {http://arxiv.org/abs/2002.07751} {\path{arXiv:2002.07751}},
  \href {https://doi.org/10.1007/JHEP10(2020)148}
  {\path{doi:10.1007/JHEP10(2020)148}}.

\bibitem{Rodriguez:2025hiu}
J.~A. Rodriguez, {Supersymmetric {\ensuremath{\alpha}}'-corrections to the
  generalized Kerr-Schild ansatz}, JHEP 01 (2026) 038.
\newblock \href {http://arxiv.org/abs/2510.15130} {\path{arXiv:2510.15130}},
  \href {https://doi.org/10.1007/JHEP01(2026)038}
  {\path{doi:10.1007/JHEP01(2026)038}}.

\bibitem{Martinez:2004nb}
C.~Martinez, R.~Troncoso, J.~Zanelli, {Exact black hole solution with a
  minimally coupled scalar field}, Phys. Rev. D 70 (2004) 084035.
\newblock \href {http://arxiv.org/abs/hep-th/0406111}
  {\path{arXiv:hep-th/0406111}}, \href
  {https://doi.org/10.1103/PhysRevD.70.084035}
  {\path{doi:10.1103/PhysRevD.70.084035}}.

\bibitem{Kent:2025pvu}
B.~Kent, A.~Zimmerman, {A new framework for classical double copies} (5 2025).
\newblock \href {http://arxiv.org/abs/2505.03887} {\path{arXiv:2505.03887}}.

\bibitem{Huang:2019cja}
Y.-T. Huang, U.~Kol, D.~O'Connell, {Double copy of electric-magnetic duality},
  Phys. Rev. D 102~(4) (2020) 046005.
\newblock \href {http://arxiv.org/abs/1911.06318} {\path{arXiv:1911.06318}},
  \href {https://doi.org/10.1103/PhysRevD.102.046005}
  {\path{doi:10.1103/PhysRevD.102.046005}}.

\bibitem{Banerjee:2019saj}
A.~Banerjee, E.~O. Colg\'ain, J.~A. Rosabal, H.~Yavartanoo, {Ehlers as EM
  duality in the double copy}, Phys. Rev. D 102 (2020) 126017.
\newblock \href {http://arxiv.org/abs/1912.02597} {\path{arXiv:1912.02597}},
  \href {https://doi.org/10.1103/PhysRevD.102.126017}
  {\path{doi:10.1103/PhysRevD.102.126017}}.

\bibitem{Easson:2022zoh}
D.~A. Easson, T.~Manton, A.~Svesko, {Einstein-Maxwell theory and the Weyl
  double copy}, Phys. Rev. D 107~(4) (2023) 044063.
\newblock \href {http://arxiv.org/abs/2210.16339} {\path{arXiv:2210.16339}},
  \href {https://doi.org/10.1103/PhysRevD.107.044063}
  {\path{doi:10.1103/PhysRevD.107.044063}}.

\bibitem{Papapetrou:1966zz}
A.~Papapetrou, {Champs gravitationnels stationnaires a symetrie axiale}, Ann.
  Inst. H. Poincare Phys. Theor. 4 (1966) 83--105.

\bibitem{KYano}
K.~Yano, \href{http://www.jstor.org/stable/1969782}{Some remarks on tensor
  fields and curvature}, Annals of Mathematics 55~(2) (1952) 328--347.
\newline\urlprefix\url{http://www.jstor.org/stable/1969782}

\bibitem{Didenko:2026nip}
V.~E. Didenko, N.~K. Dosmanbetov, {Classifying double copies and multicopies in
  AdS} (5 2026).
\newblock \href {http://arxiv.org/abs/2605.12359} {\path{arXiv:2605.12359}}.

\bibitem{Elor:2020nqe}
G.~Elor, K.~Farnsworth, M.~L. Graesser, G.~Herczeg, {The Newman-Penrose Map and
  the Classical Double Copy}, JHEP 12 (2020) 121.
\newblock \href {http://arxiv.org/abs/2006.08630} {\path{arXiv:2006.08630}},
  \href {https://doi.org/10.1007/JHEP12(2020)121}
  {\path{doi:10.1007/JHEP12(2020)121}}.

\bibitem{Farnsworth:2021wvs}
K.~Farnsworth, M.~L. Graesser, G.~Herczeg, {Twistor space origins of the
  Newman-Penrose map}, SciPost Phys. 13~(4) (2022) 099.
\newblock \href {http://arxiv.org/abs/2104.09525} {\path{arXiv:2104.09525}},
  \href {https://doi.org/10.21468/SciPostPhys.13.4.099}
  {\path{doi:10.21468/SciPostPhys.13.4.099}}.

\bibitem{Farnsworth:2023mff}
K.~Farnsworth, M.~L. Graesser, G.~Herczeg, {Double Kerr-Schild spacetimes and
  the Newman-Penrose map}, JHEP 10 (2023) 010.
\newblock \href {http://arxiv.org/abs/2306.16445} {\path{arXiv:2306.16445}},
  \href {https://doi.org/10.1007/JHEP10(2023)010}
  {\path{doi:10.1007/JHEP10(2023)010}}.

\bibitem{Chalmers:1996rq}
G.~Chalmers, W.~Siegel, {The Selfdual sector of QCD amplitudes}, Phys. Rev. D
  54 (1996) 7628--7633.
\newblock \href {http://arxiv.org/abs/hep-th/9606061}
  {\path{arXiv:hep-th/9606061}}, \href
  {https://doi.org/10.1103/PhysRevD.54.7628}
  {\path{doi:10.1103/PhysRevD.54.7628}}.

\bibitem{Monteiro:2011pc}
R.~Monteiro, D.~O'Connell, {The Kinematic Algebra From the Self-Dual Sector},
  JHEP 07 (2011) 007.
\newblock \href {http://arxiv.org/abs/1105.2565} {\path{arXiv:1105.2565}},
  \href {https://doi.org/10.1007/JHEP07(2011)007}
  {\path{doi:10.1007/JHEP07(2011)007}}.

\bibitem{Monteiro:2022xwq}
R.~Monteiro, {From Moyal deformations to chiral higher-spin theories and to
  celestial algebras}, JHEP 03 (2023) 062.
\newblock \href {http://arxiv.org/abs/2212.11266} {\path{arXiv:2212.11266}},
  \href {https://doi.org/10.1007/JHEP03(2023)062}
  {\path{doi:10.1007/JHEP03(2023)062}}.

\bibitem{Bu:2022iak}
W.~Bu, S.~Heuveline, D.~Skinner, {Moyal deformations, W$_{1+\infty}$ and
  celestial holography}, JHEP 12 (2022) 011.
\newblock \href {http://arxiv.org/abs/2208.13750} {\path{arXiv:2208.13750}},
  \href {https://doi.org/10.1007/JHEP12(2022)011}
  {\path{doi:10.1007/JHEP12(2022)011}}.

\bibitem{Mason:2025pbz}
L.~Mason, A.~Sharma, {Chiral higher-spin theories from twistor space} (5 2025).
\newblock \href {http://arxiv.org/abs/2505.09419} {\path{arXiv:2505.09419}}.

\bibitem{Lipstein:2023pih}
A.~Lipstein, S.~Nagy, {Self-Dual Gravity and Color-Kinematics Duality in AdS4},
  Phys. Rev. Lett. 131~(8) (2023) 081501.
\newblock \href {http://arxiv.org/abs/2304.07141} {\path{arXiv:2304.07141}},
  \href {https://doi.org/10.1103/PhysRevLett.131.081501}
  {\path{doi:10.1103/PhysRevLett.131.081501}}.

\bibitem{CarrilloGonzalez:2024sto}
M.~Carrillo~Gonz\'alez, A.~Lipstein, S.~Nagy, {Self-dual cosmology}, JHEP 10
  (2024) 183.
\newblock \href {http://arxiv.org/abs/2407.12905} {\path{arXiv:2407.12905}},
  \href {https://doi.org/10.1007/JHEP10(2024)183}
  {\path{doi:10.1007/JHEP10(2024)183}}.

\bibitem{Brown:2023zxm}
G.~R. Brown, J.~Gowdy, B.~Spence, {Self-dual fields on self-dual backgrounds
  and the double copy}, Phys. Rev. D 109~(2) (2024) 026009.
\newblock \href {http://arxiv.org/abs/2307.11063} {\path{arXiv:2307.11063}},
  \href {https://doi.org/10.1103/PhysRevD.109.026009}
  {\path{doi:10.1103/PhysRevD.109.026009}}.

\bibitem{Dunajski:2000iq}
M.~Dunajski, L.~J. Mason, {HyperKahler hierarchies and their twistor theory},
  Commun. Math. Phys. 213 (2000) 641--672.
\newblock \href {http://arxiv.org/abs/math/0001008}
  {\path{arXiv:math/0001008}}, \href {https://doi.org/10.1007/PL00005532}
  {\path{doi:10.1007/PL00005532}}.

\bibitem{Adamo:2022mev}
T.~Adamo, L.~Mason, A.~Sharma, {Graviton scattering in self-dual radiative
  space-times}, Class. Quant. Grav. 40~(9) (2023) 095002.
\newblock \href {http://arxiv.org/abs/2203.02238} {\path{arXiv:2203.02238}},
  \href {https://doi.org/10.1088/1361-6382/acc233}
  {\path{doi:10.1088/1361-6382/acc233}}.

\bibitem{Adamo:2020yzi}
T.~Adamo, L.~Mason, A.~Sharma, {Gluon Scattering on Self-Dual Radiative Gauge
  Fields}, Commun. Math. Phys. 399 (2023) 1731--1771.
\newblock \href {http://arxiv.org/abs/2010.14996} {\path{arXiv:2010.14996}},
  \href {https://doi.org/10.1007/s00220-022-04582-9}
  {\path{doi:10.1007/s00220-022-04582-9}}.

\bibitem{Tod:1982}
K.~P. Tod, \href{https://doi.org/10.1063/1.525482}{Self‐dual kerr–schild
  metrics and null maxwell fields}, Journal of Mathematical Physics 23~(6)
  (1982) 1147--1148.
\newblock \href
  {http://arxiv.org/abs/https://pubs.aip.org/aip/jmp/article-pdf/23/6/1147/19271875/1147_1_online.pdf}
  {\path{arXiv:https://pubs.aip.org/aip/jmp/article-pdf/23/6/1147/19271875/1147_1_online.pdf}},
  \href {https://doi.org/10.1063/1.525482} {\path{doi:10.1063/1.525482}}.
\newline\urlprefix\url{https://doi.org/10.1063/1.525482}

\bibitem{Albertini:2025ptks}
E.~Albertini, M.~L. Graesser, G.~Herczeg, The penrose transform and the
  kerr--schild double copy (11 2025).
\newblock \href {http://arxiv.org/abs/2511.14854} {\path{arXiv:2511.14854}}.

\bibitem{Berman:2018hfs}
D.~S. Berman, E.~Chac\'on, A.~Luna, C.~D. White, The self-dual classical double
  copy, and the eguchi--hanson instanton, JHEP 01 (2019) 107.
\newblock \href {http://arxiv.org/abs/1809.04063} {\path{arXiv:1809.04063}},
  \href {https://doi.org/10.1007/JHEP01(2019)107}
  {\path{doi:10.1007/JHEP01(2019)107}}.

\bibitem{ArmstrongWilliams:2022npd}
K.~Armstrong-Williams, C.~D. White, S.~Wikeley, Non-perturbative aspects of the
  self-dual double copy, JHEP 08 (2022) 160.
\newblock \href {http://arxiv.org/abs/2205.02136} {\path{arXiv:2205.02136}},
  \href {https://doi.org/10.1007/JHEP08(2022)160}
  {\path{doi:10.1007/JHEP08(2022)160}}.

\bibitem{Chowdhury:2024dcy}
C.~Chowdhury, G.~Doran, A.~Lipstein, R.~Monteiro, S.~Nagy, K.~Singh,
  {Light-cone actions and correlators of self-dual theories in AdS$_{4}$}, JHEP
  01 (2025) 172.
\newblock \href {http://arxiv.org/abs/2411.04172} {\path{arXiv:2411.04172}},
  \href {https://doi.org/10.1007/JHEP01(2025)172}
  {\path{doi:10.1007/JHEP01(2025)172}}.

\bibitem{Ponomarev:2024jyg}
D.~Ponomarev, {Chiral higher-spin double copy}, JHEP 01 (2025) 143.
\newblock \href {http://arxiv.org/abs/2409.19449} {\path{arXiv:2409.19449}},
  \href {https://doi.org/10.1007/JHEP01(2025)143}
  {\path{doi:10.1007/JHEP01(2025)143}}.

\bibitem{Ponomarev:2022vjb}
D.~Ponomarev, {Basic Introduction to Higher-Spin Theories}, Int. J. Theor.
  Phys. 62~(7) (2023) 146.
\newblock \href {http://arxiv.org/abs/2206.15385} {\path{arXiv:2206.15385}},
  \href {https://doi.org/10.1007/s10773-023-05399-5}
  {\path{doi:10.1007/s10773-023-05399-5}}.

\bibitem{Ponomarev:2017nrr}
D.~Ponomarev, {Chiral Higher Spin Theories and Self-Duality}, JHEP 12 (2017)
  141.
\newblock \href {http://arxiv.org/abs/1710.00270} {\path{arXiv:1710.00270}},
  \href {https://doi.org/10.1007/JHEP12(2017)141}
  {\path{doi:10.1007/JHEP12(2017)141}}.

\bibitem{Misuna:2026has}
N.~Misuna, D.~Ponomarev, A.~Solomin, {Self-dual classical higher-spin
  multicopy}, Phys. Lett. B 878 (2026) 140525.
\newblock \href {http://arxiv.org/abs/2604.13646} {\path{arXiv:2604.13646}},
  \href {https://doi.org/10.1016/j.physletb.2026.140525}
  {\path{doi:10.1016/j.physletb.2026.140525}}.

\bibitem{Skvortsov:2026ofl}
E.~Skvortsov, R.~Van~Dongen, {Self-dual holography: four-point AdS/CFT
  correlators in higher-spin gravity} (5 2026).
\newblock \href {http://arxiv.org/abs/2605.30276} {\path{arXiv:2605.30276}}.

\bibitem{DiFrancesco:1997nk}
P.~Di~Francesco, P.~Mathieu, D.~Senechal, {Conformal Field Theory}, Graduate
  Texts in Contemporary Physics, Springer-Verlag, New York, 1997.
\newblock \href {https://doi.org/10.1007/978-1-4612-2256-9}
  {\path{doi:10.1007/978-1-4612-2256-9}}.

\bibitem{Rychkov:2016iqz}
S.~Rychkov, {EPFL Lectures on Conformal Field Theory in D{\ensuremath{>}}= 3
  Dimensions}, SpringerBriefs in Physics, 2016.
\newblock \href {http://arxiv.org/abs/1601.05000} {\path{arXiv:1601.05000}},
  \href {https://doi.org/10.1007/978-3-319-43626-5}
  {\path{doi:10.1007/978-3-319-43626-5}}.

\bibitem{Osborn:1993cr}
H.~Osborn, A.~C. Petkou, {Implications of conformal invariance in field
  theories for general dimensions}, Annals Phys. 231 (1994) 311--362.
\newblock \href {http://arxiv.org/abs/hep-th/9307010}
  {\path{arXiv:hep-th/9307010}}, \href {https://doi.org/10.1006/aphy.1994.1045}
  {\path{doi:10.1006/aphy.1994.1045}}.

\bibitem{Penrose:1985bww}
R.~Penrose, W.~Rindler, {Spinors and Space-Time}, Cambridge Monographs on
  Mathematical Physics, Cambridge Univ. Press, Cambridge, UK, 2011.
\newblock \href {https://doi.org/10.1017/CBO9780511564048}
  {\path{doi:10.1017/CBO9780511564048}}.

\end{thebibliography}
	
\end{document}